%% file: arXiv.tex
\documentclass[twoside,a4paper,11pt]{sea22}
\usepackage{graphicx}
\usepackage{hyperref}
\usepackage{movie15}
\usepackage{color}
\usepackage{booktabs} 
\usepackage{siunitx}  
\usepackage{xcolor}
\usepackage{natbib}    
\usepackage[export]{adjustbox}
\usepackage{animate}
\usepackage{pdflscape}
\input{mlaeff}

\input{isolatin.sty}

\newcommand{\BPRP}{\mbox{$G_{\rm BP}-G_{\rm RP}$}} 
\newcommand{\GG}{\mbox{$G$}}
\newcommand{\GBP}{\mbox{$G_{\rm BP}$}} 
\newcommand{\GRP}{\mbox{$G_{\rm RP}$}} 
\newcommand{\pic}{\mbox{$\varpi_{\rm c}$}}
\newcommand{\spic}{\mbox{$\sigma_{\varpi_{\rm c}}$}}

\newcommand{\Teff}{\mbox{$T_{\rm eff}$}}

\newcommand{\RV}{\mbox{$R_{5495}$}}

\newcommand{\cre}{\color{red}}

\def\lesssim{\mathrel{\hbox{\rlap{\hbox{\lower4pt\hbox{$\sim$}}}\hbox{$<$}}}}
\def\gtrsim{\mathrel{\hbox{\rlap{\hbox{\lower4pt\hbox{$\sim$}}}\hbox{$>$}}}}
\begin{document}
\pagenumbering{arabic}
\pagestyle{myheadings}
\thispagestyle{empty}
\vspace*{-4.0cm}
\textit{\flushleft\small
Highlights of Spanish Astrophysics XIII\\
Proceedings of the XVII Scientific Meeting of the Spanish Astronomical Society \\
\hspace{5mm}held on 13-17 July 2026, in Tarragona, Spain.}
\vspace*{1.0cm}


\begin{flushleft}
{\bf {\LARGE
%
A new method to cleanly separate hot and cool populations using just Gaia and 2MASS photometry
%
}\\
\vspace*{1cm}
%
J. Maíz Apellániz$^1$ \& M. Pantaleoni González$^2$
%
}\\
\vspace*{0.5cm}
%
$^1$ Centro de Astrobiología, CSIC-INTA, Spain\\
$^2$ Universität Wien, Austria\\
%
\end{flushleft}
%
\markboth{
Separating hot and cool populations
}{ 
%
 Ma{\'\i}z Apell\'aniz  \& Pantaleoni Gonz\'alez
%
}
\thispagestyle{empty}
\vspace*{0.4cm}
\begin{minipage}[l]{0.09\textwidth}
\ 
\end{minipage}
\begin{minipage}[r]{0.9\textwidth}
\vspace{1cm}
\section*{Abstract}{\small
We have developed a photometric method that produces clean hot (intrinsically blue) populations using Gaia and 2MASS 
photometry alone with [a] a very small (false positives $<$~1\%) contamination from cool (intrinsically red) luminous sources, 
[b] independent of extinction (limited only by S/N), and [c] with only small losses (false negatives $<$~10\%) of real OB stars. 
Applying the method to Gaia DR3 data we are able to recover extinguished OB stars down to $E(G_{\rm BP}-G_{\rm RP}) = 5$, producing
a catalog of $\sim 2\cdot 10^5$ candidate OB stars that reaches beyond the Scutum-Centaurus arm. This method can potentially 
extend the sample of known OB stars by an order of magnitude, which would yield a large leap in our knowledge of the populations 
of compact-object progenitors, including gravitational-wave sources. 
%
\normalsize}
\end{minipage}
%
%


\section{What do we want to do?}      

$\,\!$\indent The overall Galactic stellar population is dominated (in number) by cool stars (defined here as those
with \Teff~$\lesssim$~7~kK) due to the combination of the
IMF and the longer lifetimes of low-mass stars. Hotter populations are less numerous but are, in general, more luminous (with the
exception of WDs and other evolved phases of low-mass stars) and more massive. The higher luminosities increase their proportions
with respect to cooler populations in mixed, magnitude-limited samples, by reaching longer distances but for Galactic disk 
populations extinction enters into play by making them fainter and redder. The diminished brightness decreases their proportions
in such samples and the reddening introduces confusion with cool populations, especially with those that are also luminous (RGB,
RC, and AGB stars). The final result is that photometric surveys of the Galaxy yield a mixture of hot and cool populations that
cannot be easily distinguished without additional ``expensive'' means such as spectroscopy. The two goals of this work are: 
[a] to generate a method that uses available \textit{Gaia}+2MASS photometry to separate hot and cool populations and to test its
limits and [b] to apply it to the specific problem of detecting early-type massive stars (mostly OB stars but also WR stars and AF
supergiants).


\section{The full sample}      

$\,\!$\indent We build our sample from the five-astrometric-parameter, $G \le 17.0$~mag, \BPRP\ between $-$1.0 and 8.0~mag
\textit{Gaia}~DR3 sample of \citet{Maizetal23}, applying several cuts, with the final one based on the astrometric calibration of 
\citet{Maiz22}, as detailed in Table~\ref{table1}.
The first frame of the animated Fig.~\ref{fig1} shows three colour-colour and one colour-absolute
magnitude diagrams with the full \num{85404897}-star sample.

\begin{table}
\label{table1}
\caption{Cuts applied to select our full sample.}
\centerline{
\begin{tabular}{lrr}
                                                                 &                 &               \\
Initial sample:                                                  & \num{145677450} & \num{100.0}\% \\
In the Milky Way:                                                & \num{145180963} &  \num{99.7}\% \\
With 2MASS PSC counterpart:                                      & \num{141144229} &  \num{96.9}\% \\
AAA 2MASS PSC phot. with $\sigma_J,\sigma_H,\sigma_K < 0.1$ mag: & \num{113494889} &  \num{77.9}\% \\
\textit{Gaia}~DR3 astrometry with $\pic/\spic \ge 5.0$:          & \num{85404897}  &  \num{58.6}\% \\
\end{tabular}
}
\end{table}


\section{Colour-colour selection criteria}      

$\,\!$\indent Our goal is to develop colour-colour criteria with the available \textit{Gaia}~DR3 \GG\GBP\GRP + 
2MASS $JHK$ photometry to select hot stars ($\Teff \gtrsim 7$~kK) that satisfy three conditions for the 
OB-star subsample:

\begin{itemize}
 \item A very small (false positives $<$ 1\%) contamination from cool sources.
 \item Independent of extinction (though magnitude limited).
 \item Small losses (false negatives $<$ 10\%) of hot stars with $\Teff \gtrsim 10$~kK.
\end{itemize}

\begin{figure}
 \centerline{
  \begin{animateinline}[loop,controls,buttonsize=1em]{0.4}
  \begin{minipage}{\linewidth}
  \includegraphics*[width=0.49\linewidth]{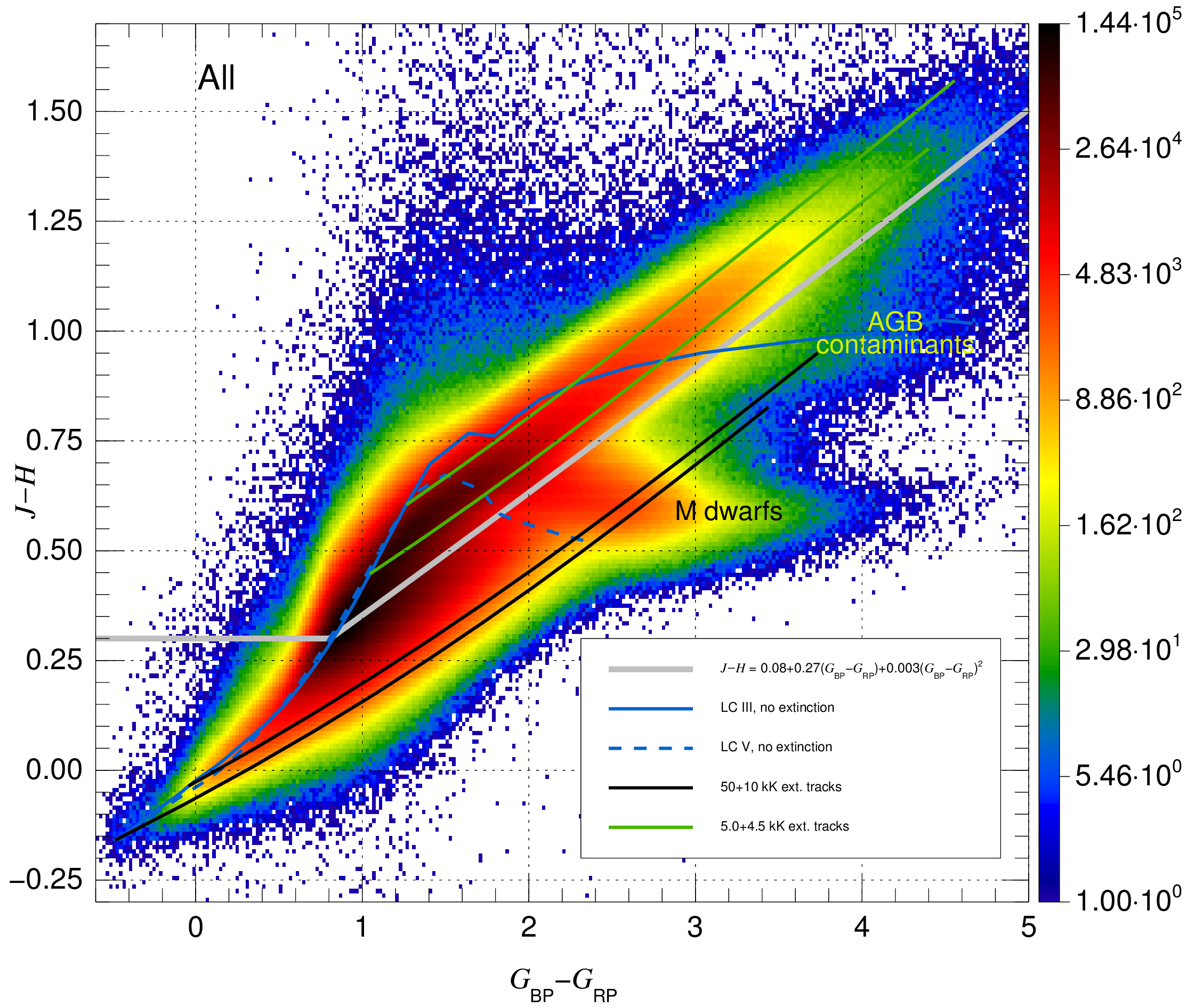} \
  \includegraphics*[width=0.49\linewidth]{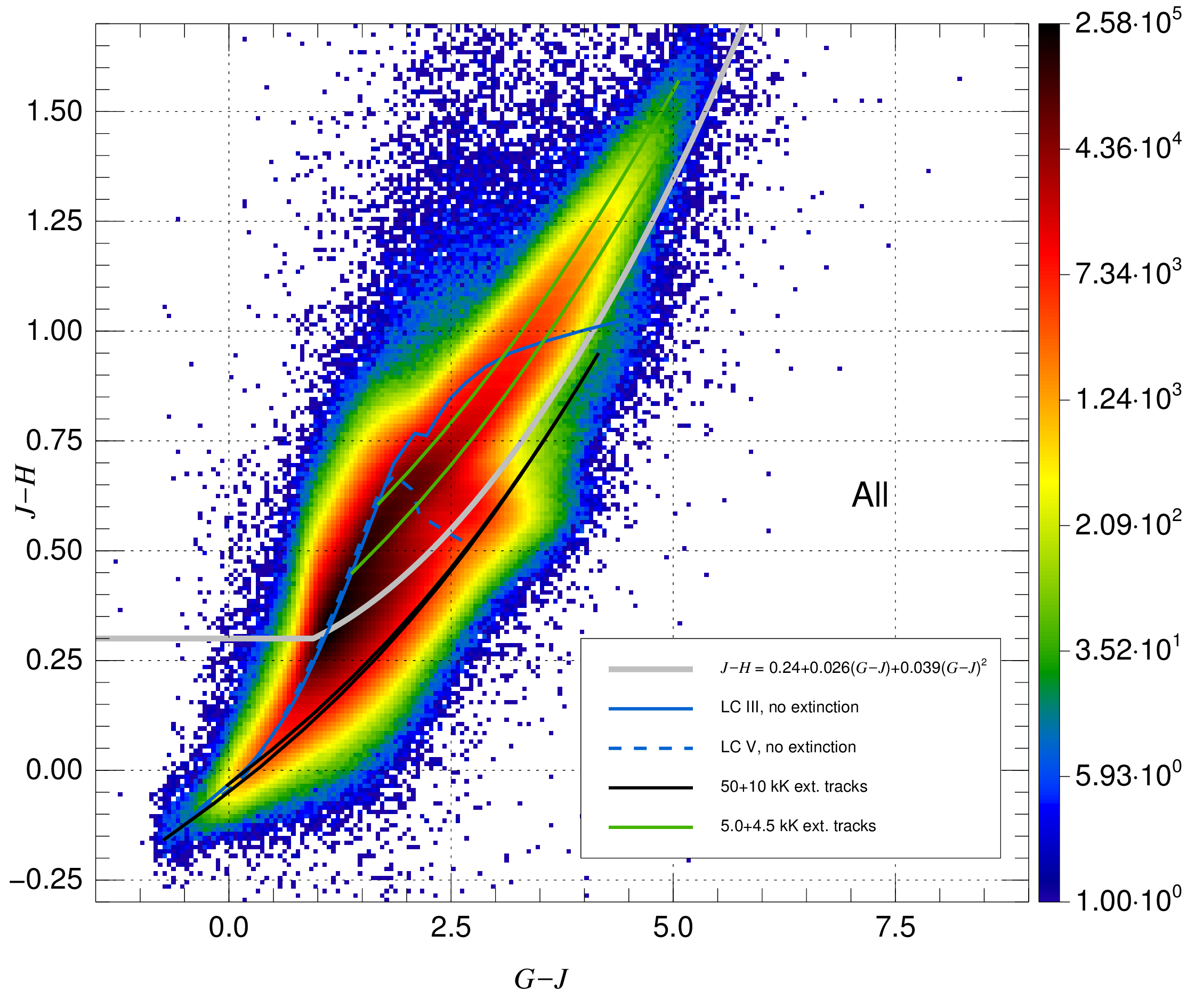}
  \\           
  \includegraphics*[width=0.49\linewidth]{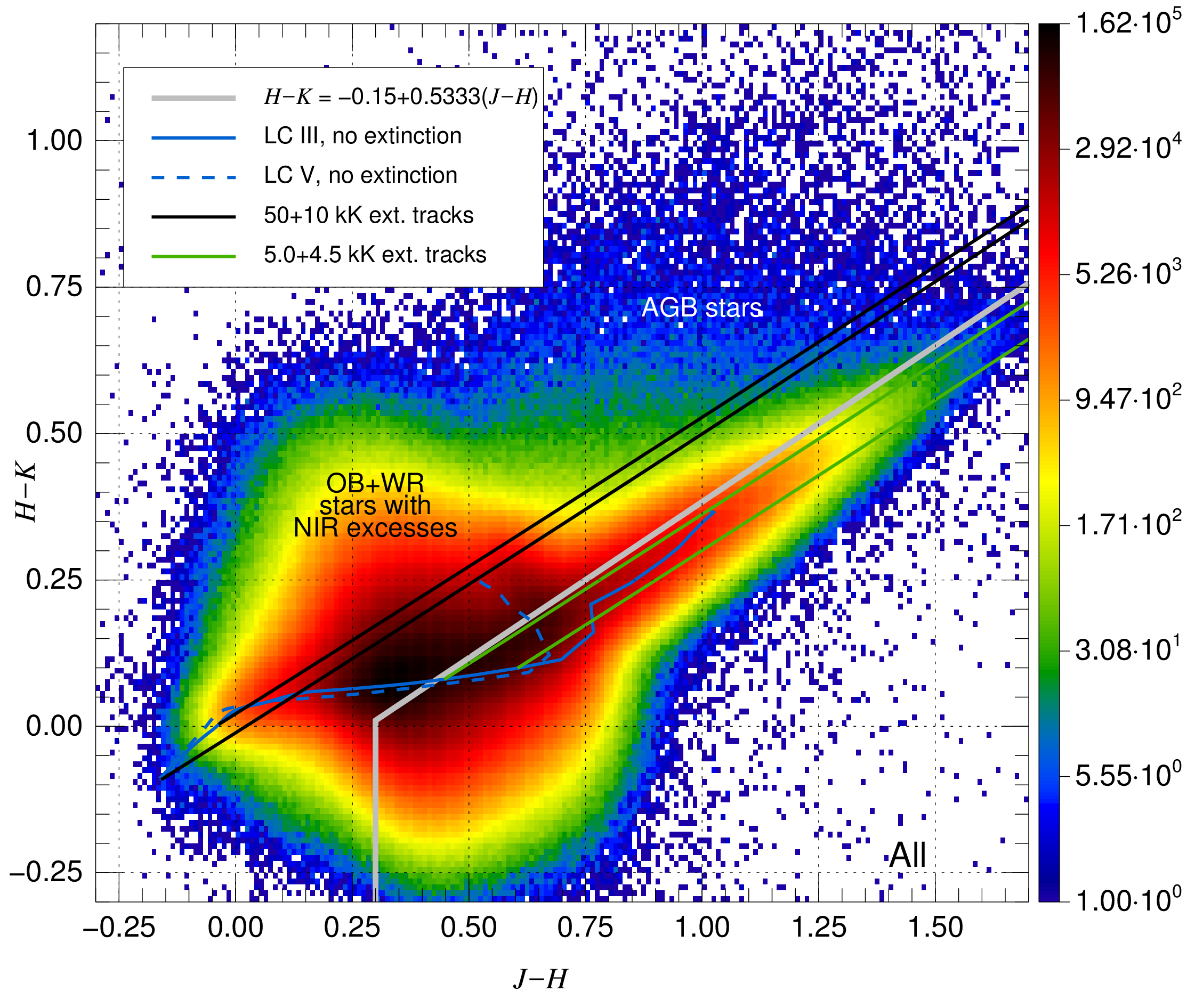} \
  \includegraphics*[width=0.49\linewidth]{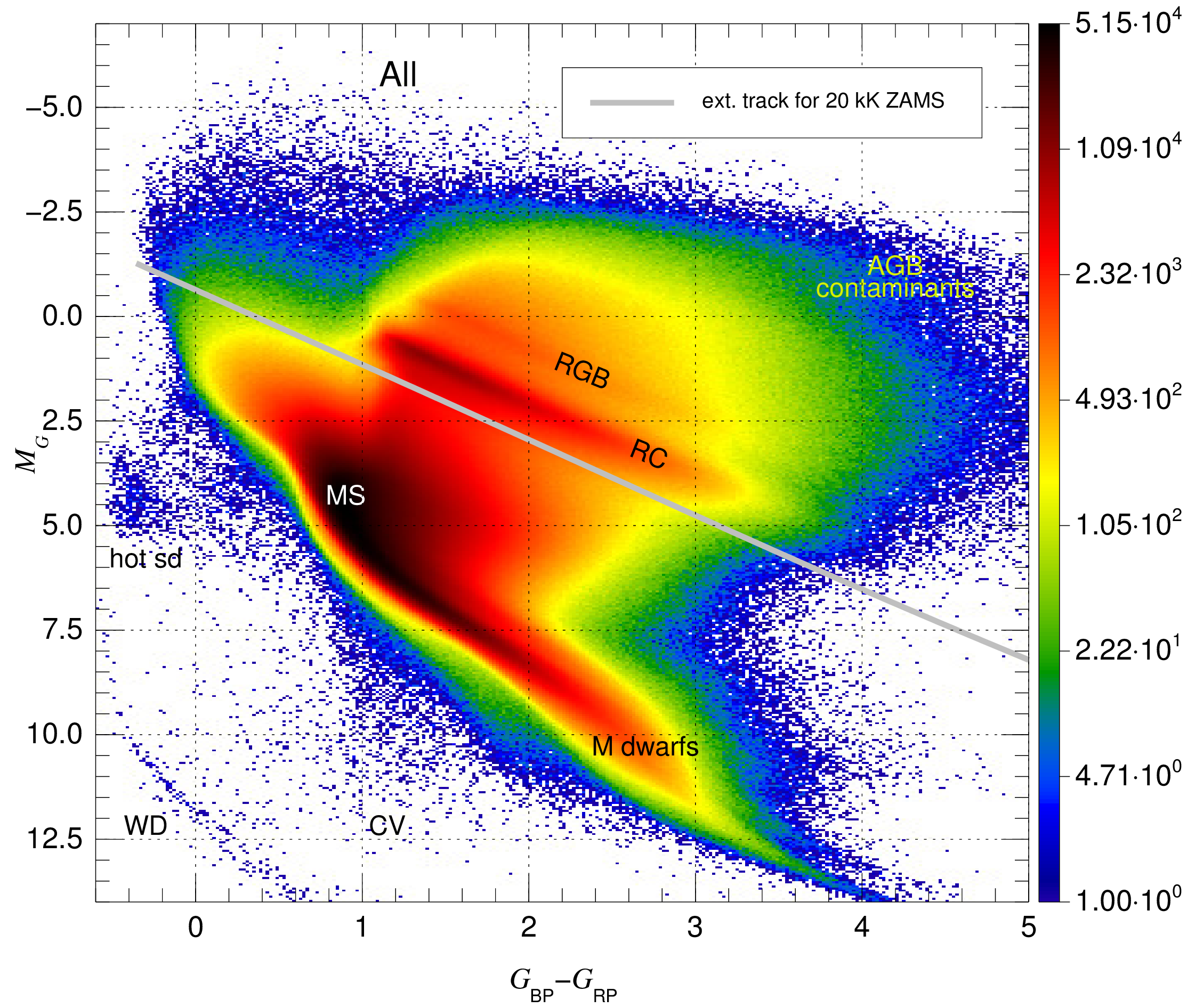}  
  \end{minipage}
  \newframe[5]
  \begin{minipage}{\linewidth}
  \includegraphics*[width=0.49\linewidth]{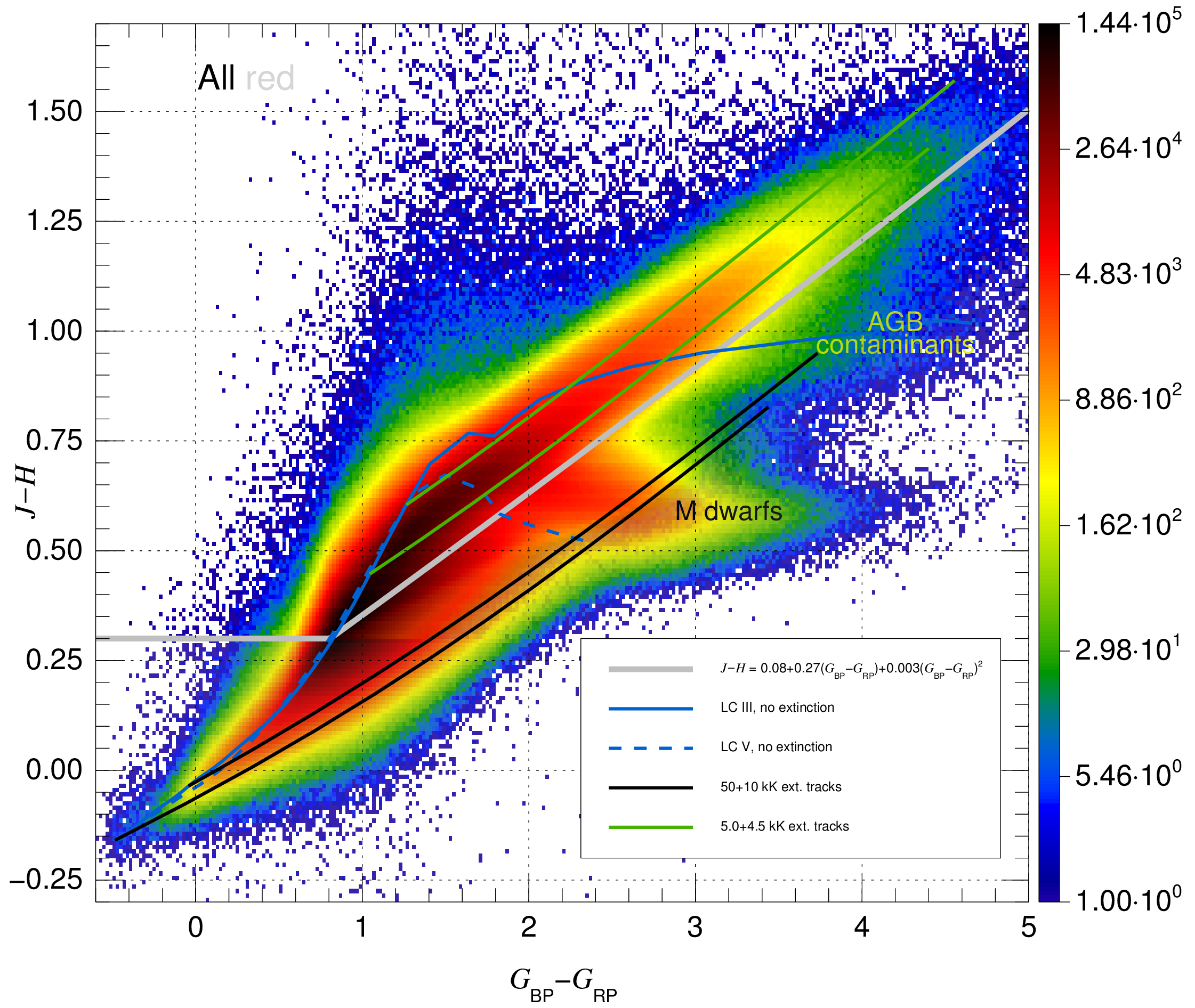} \
  \includegraphics*[width=0.49\linewidth]{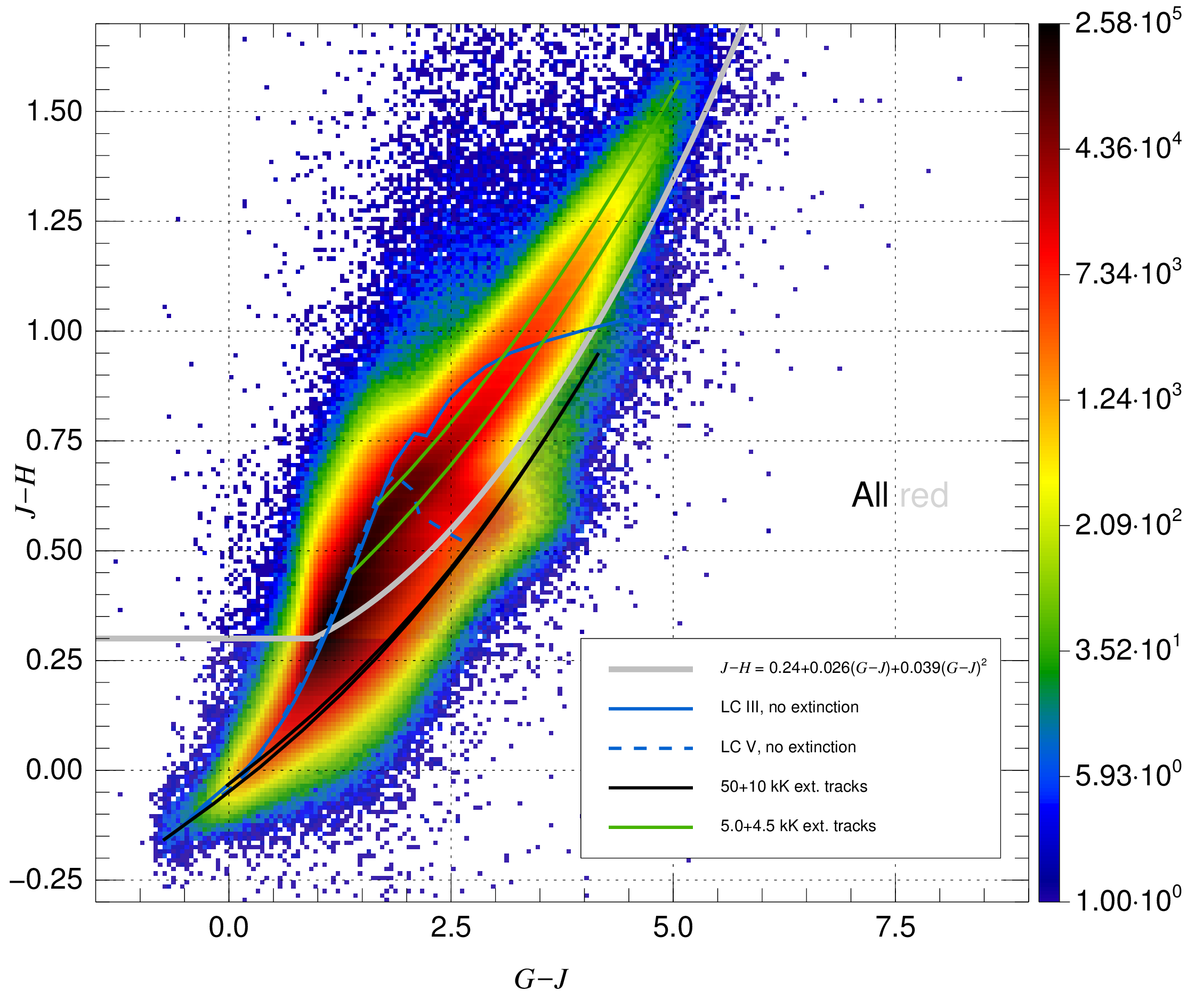}
  \\           
  \includegraphics*[width=0.49\linewidth]{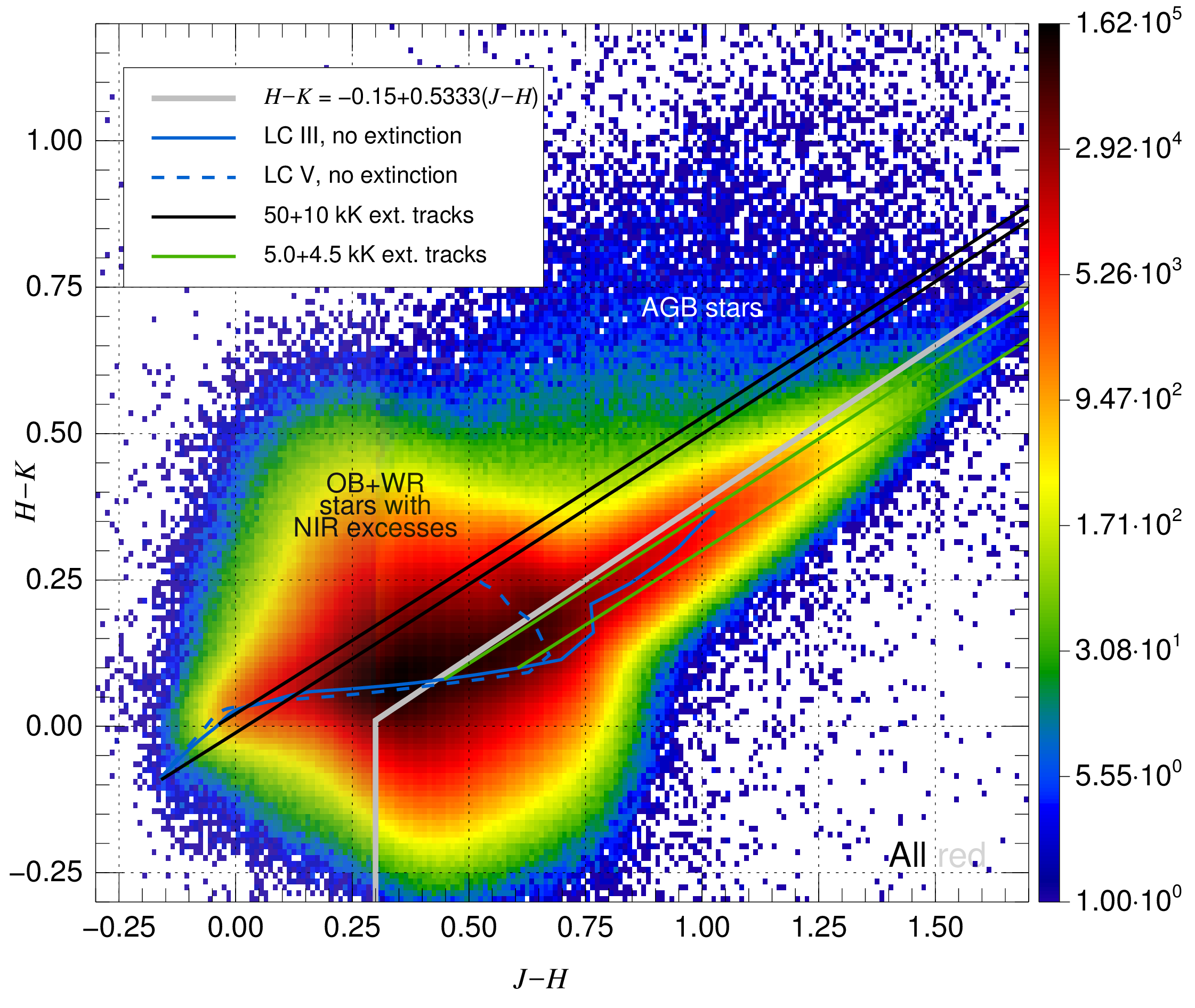} \
  \includegraphics*[width=0.49\linewidth]{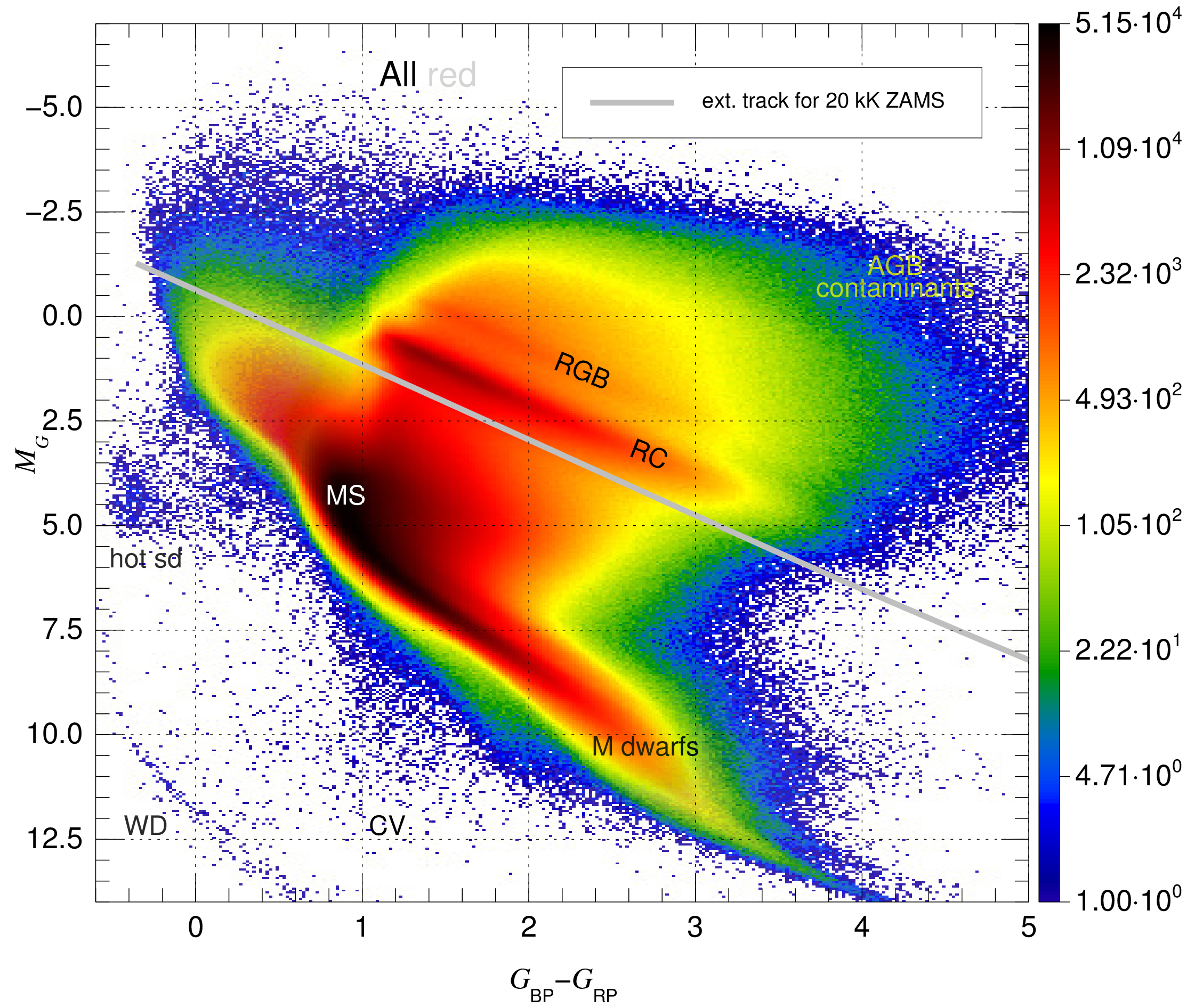}  
  \end{minipage}
  \newframe
  \begin{minipage}{\linewidth}
  \includegraphics*[width=0.49\linewidth]{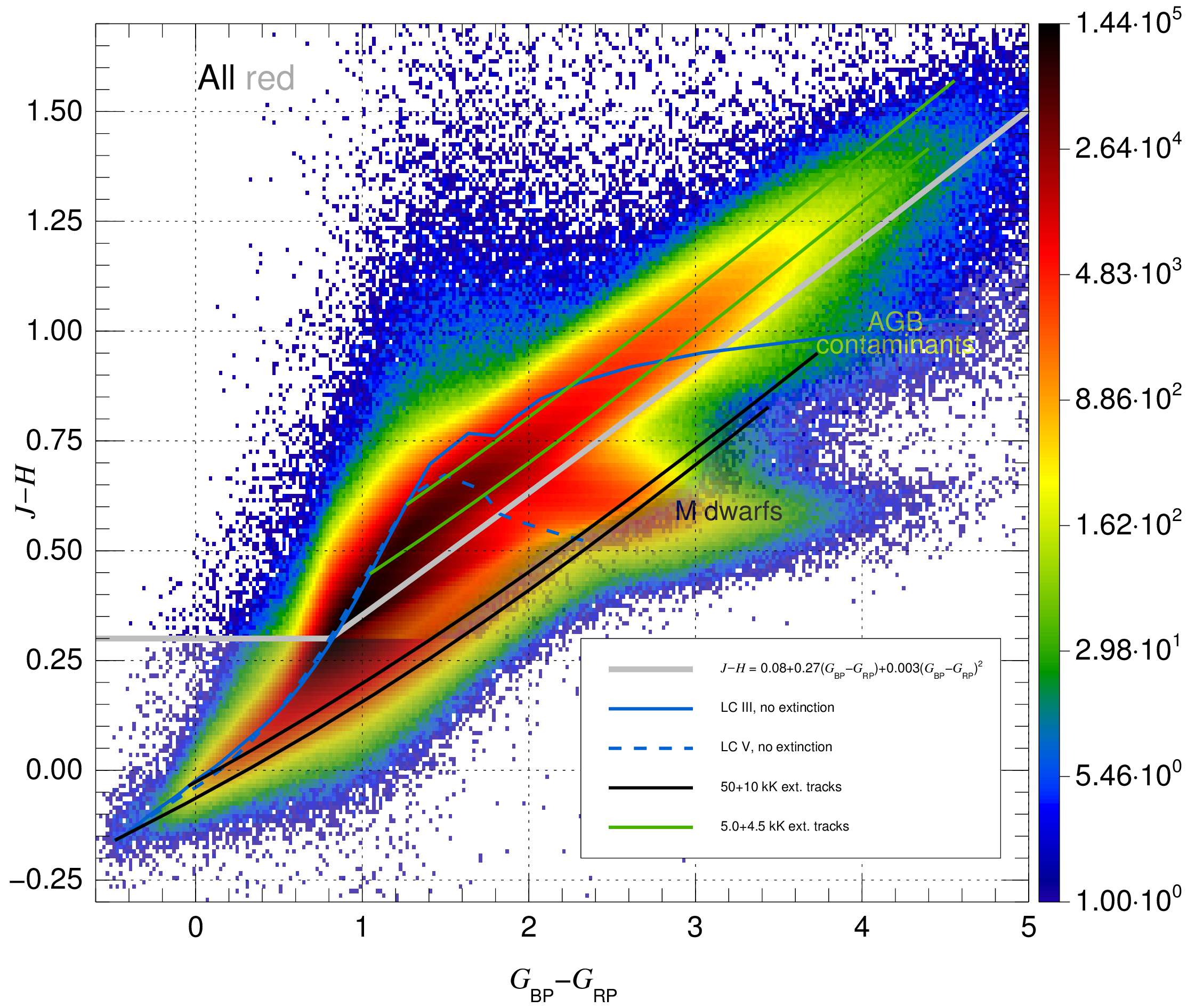} \
  \includegraphics*[width=0.49\linewidth]{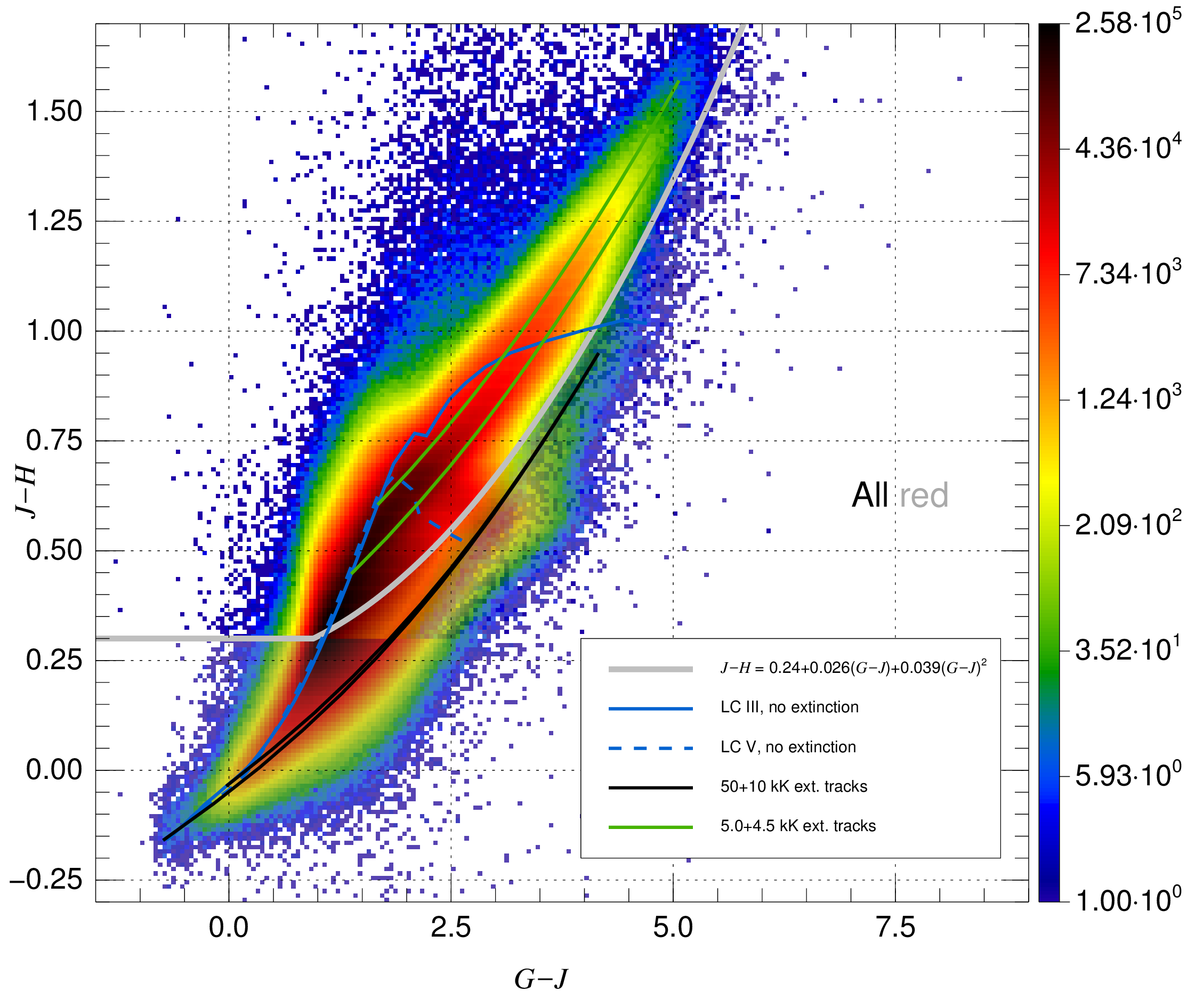}
  \\           
  \includegraphics*[width=0.49\linewidth]{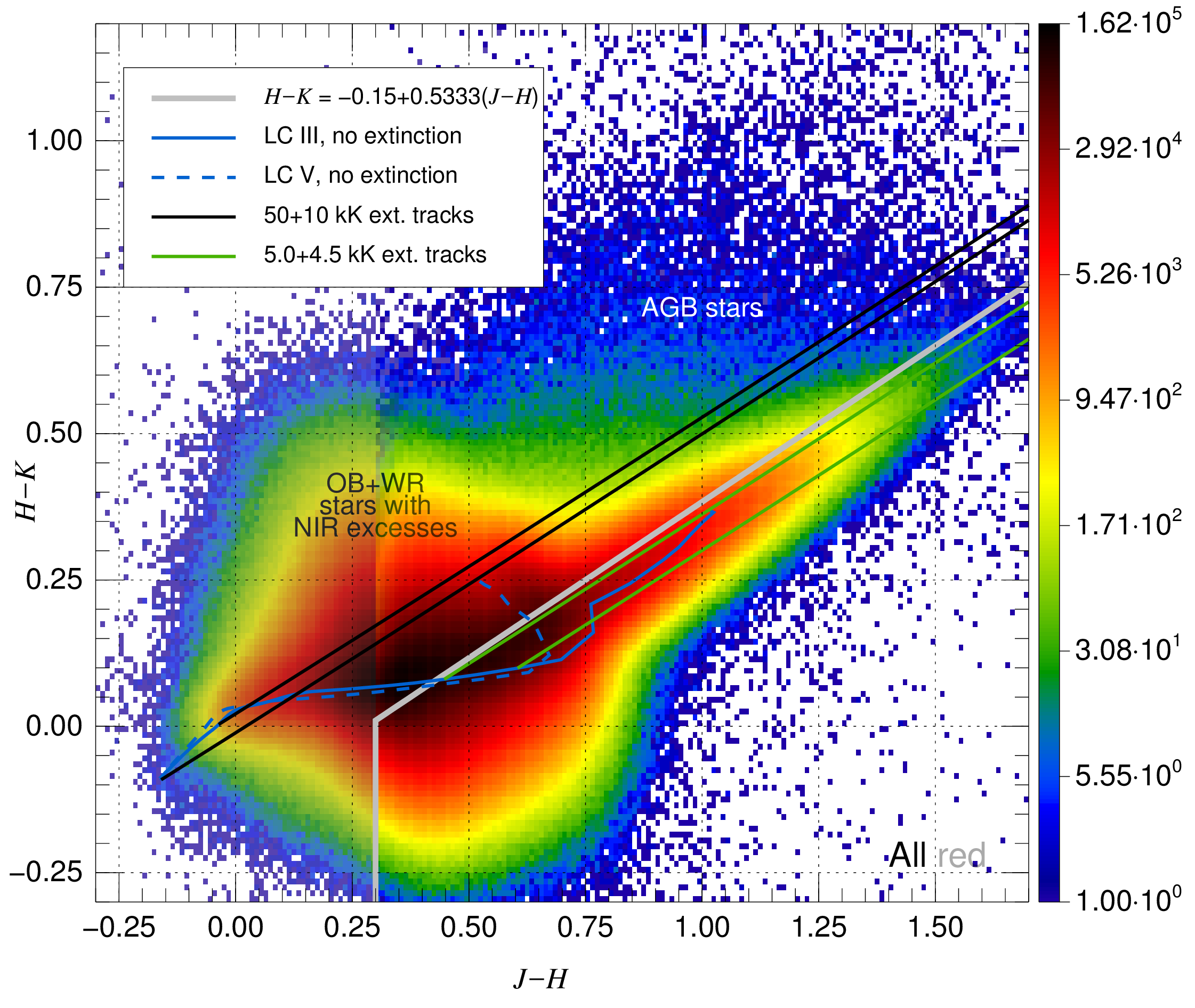} \
  \includegraphics*[width=0.49\linewidth]{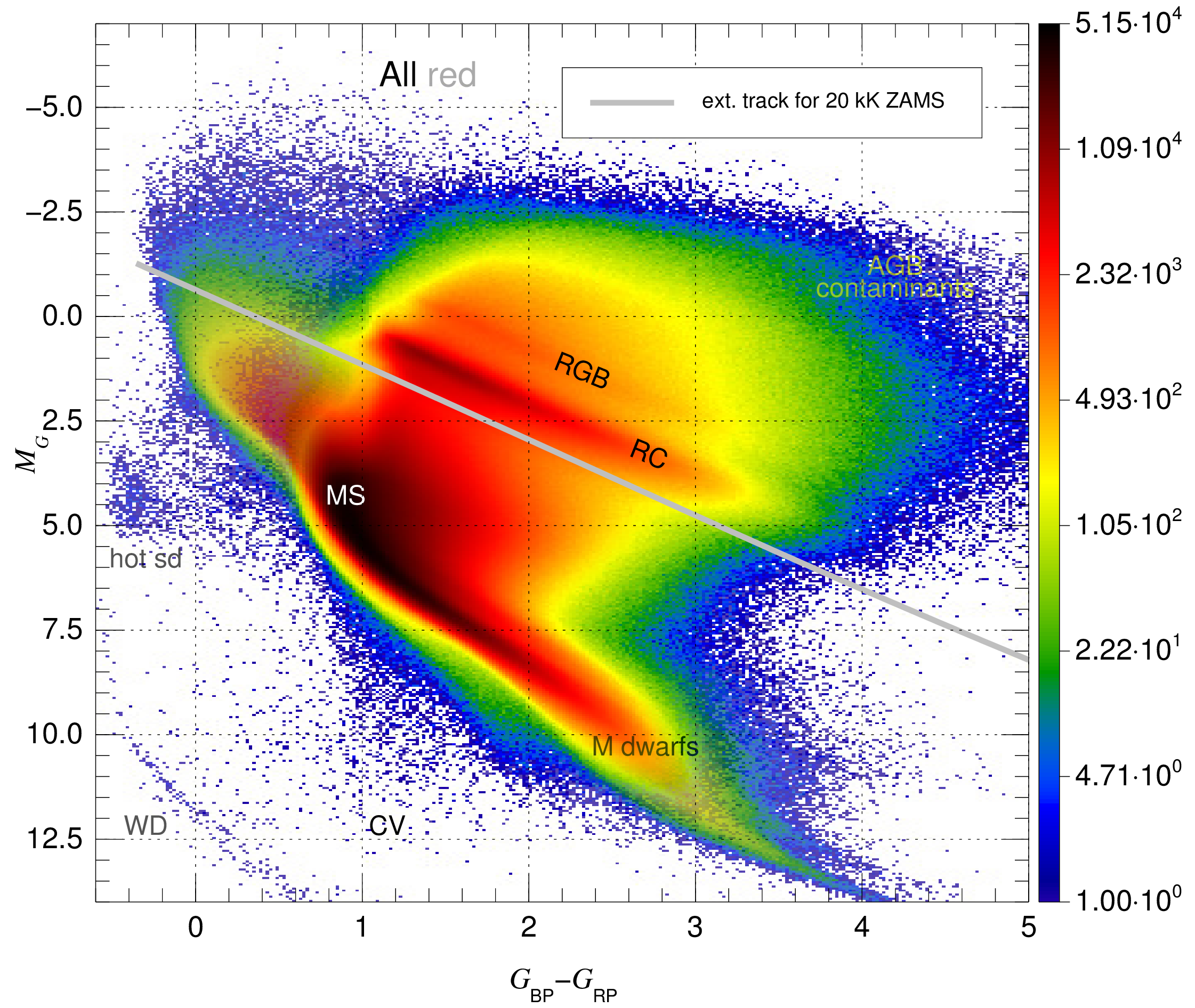}  
  \end{minipage}
  \newframe
  \begin{minipage}{\linewidth}
  \includegraphics*[width=0.49\linewidth]{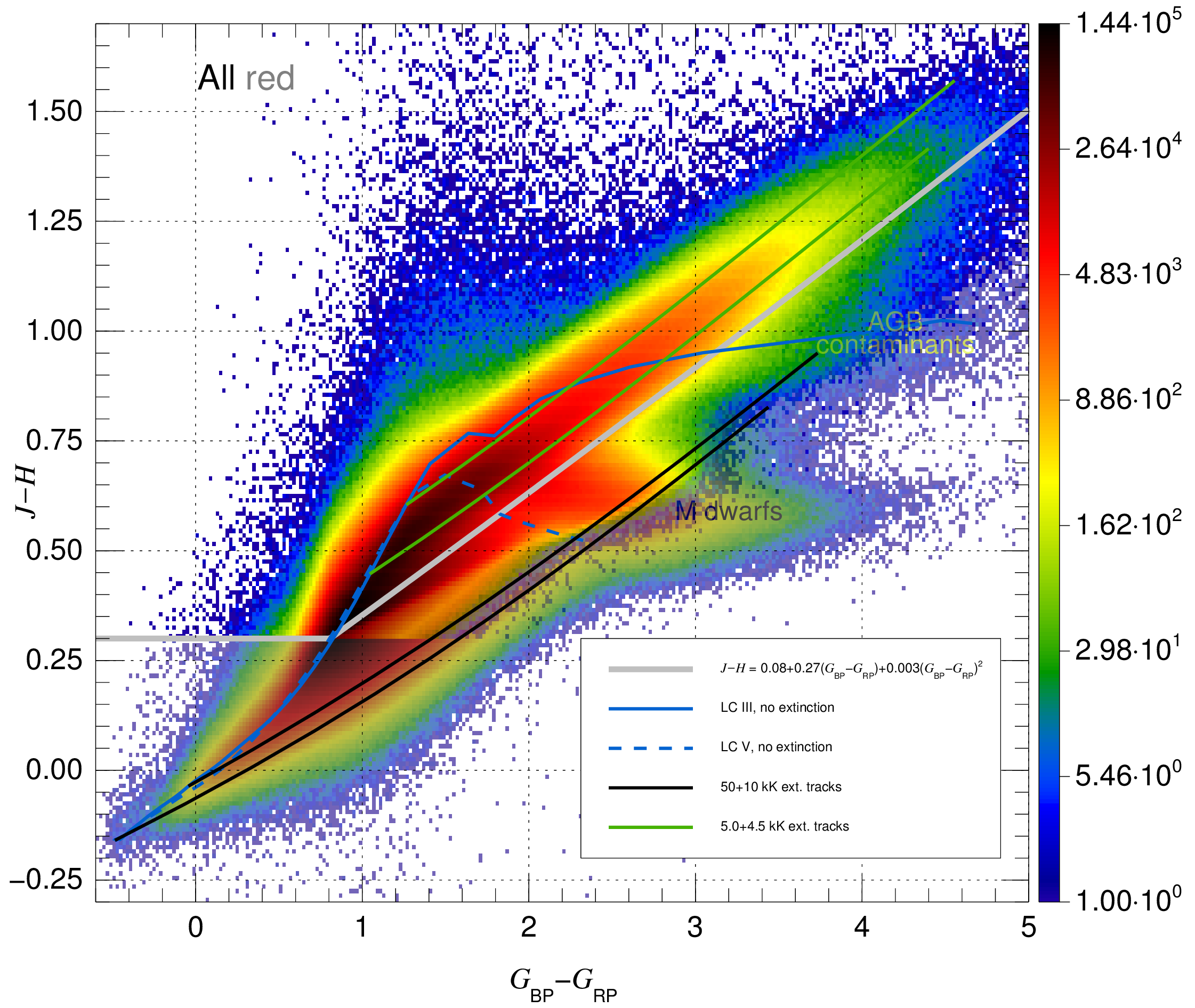} \
  \includegraphics*[width=0.49\linewidth]{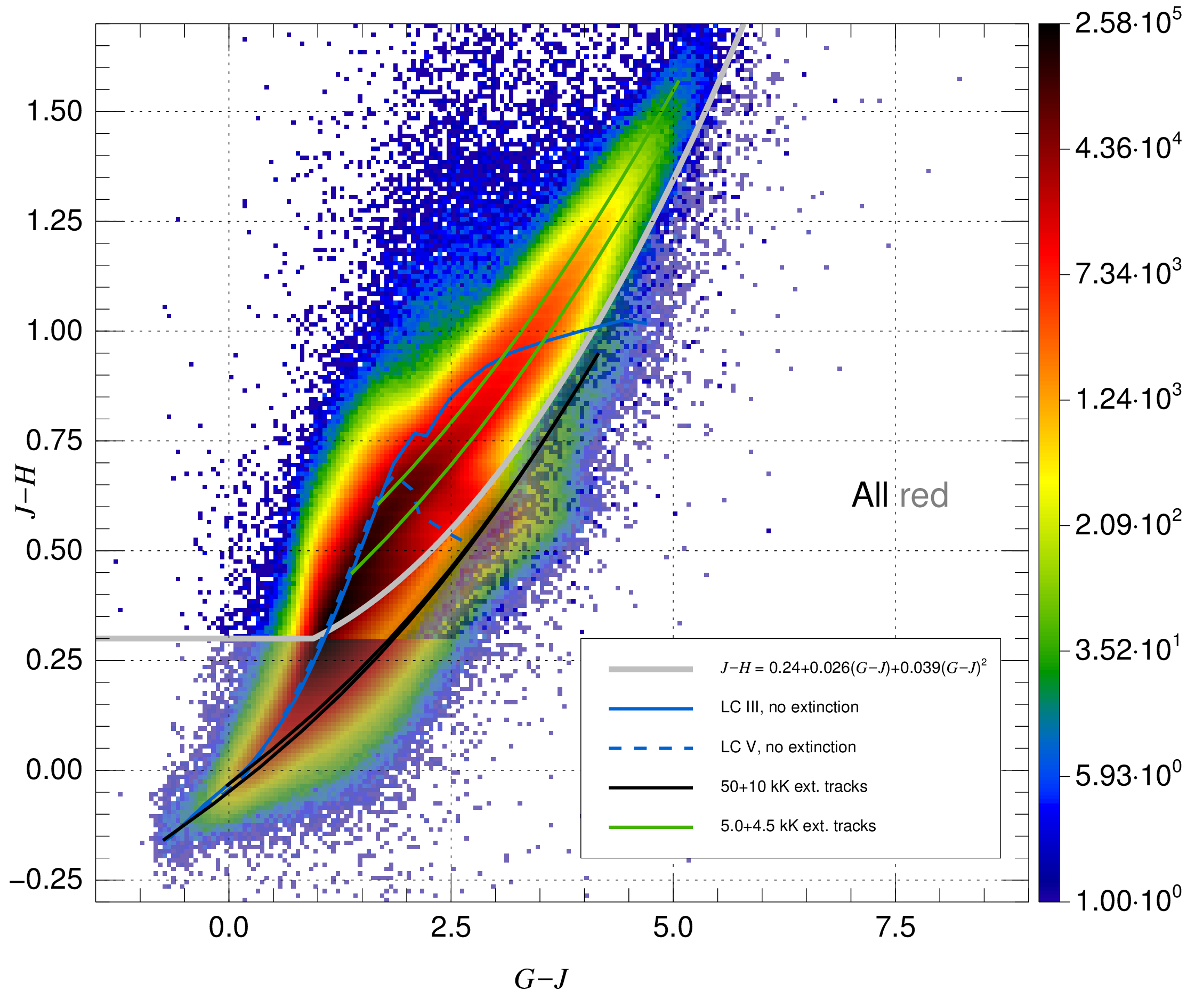}
  \\           
  \includegraphics*[width=0.49\linewidth]{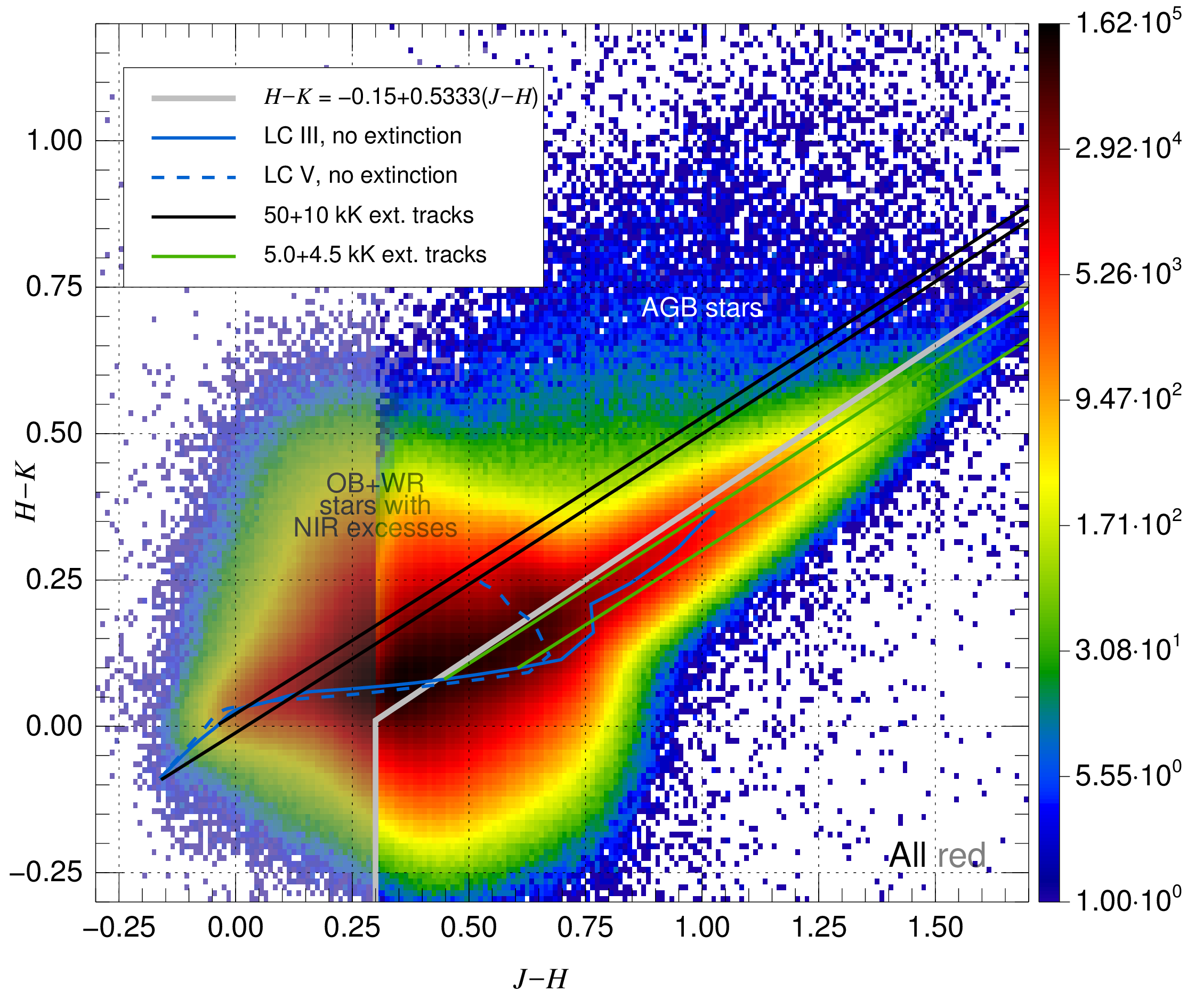} \
  \includegraphics*[width=0.49\linewidth]{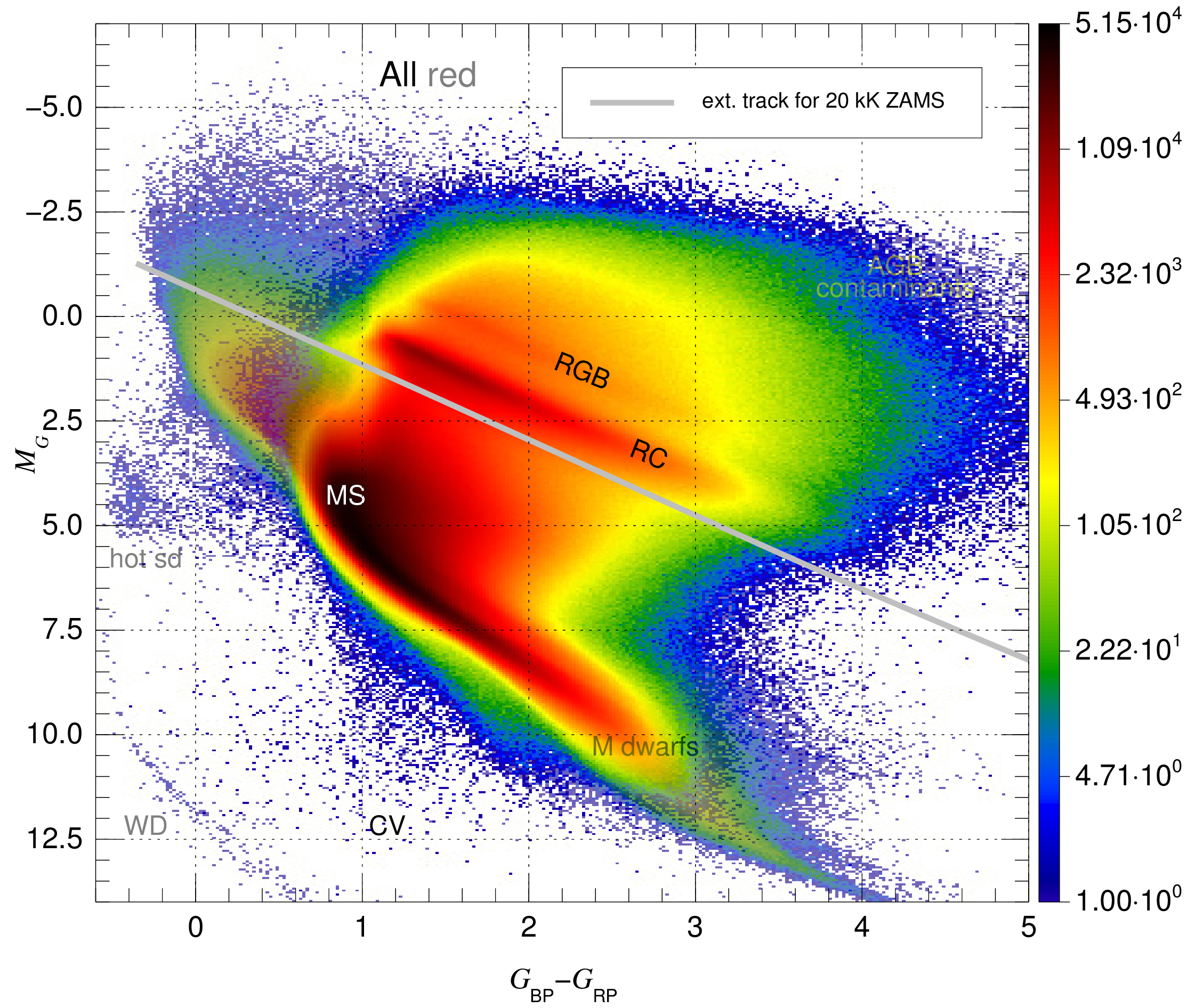}  
  \end{minipage}
  \newframe
  \begin{minipage}{\linewidth}
  \includegraphics*[width=0.49\linewidth]{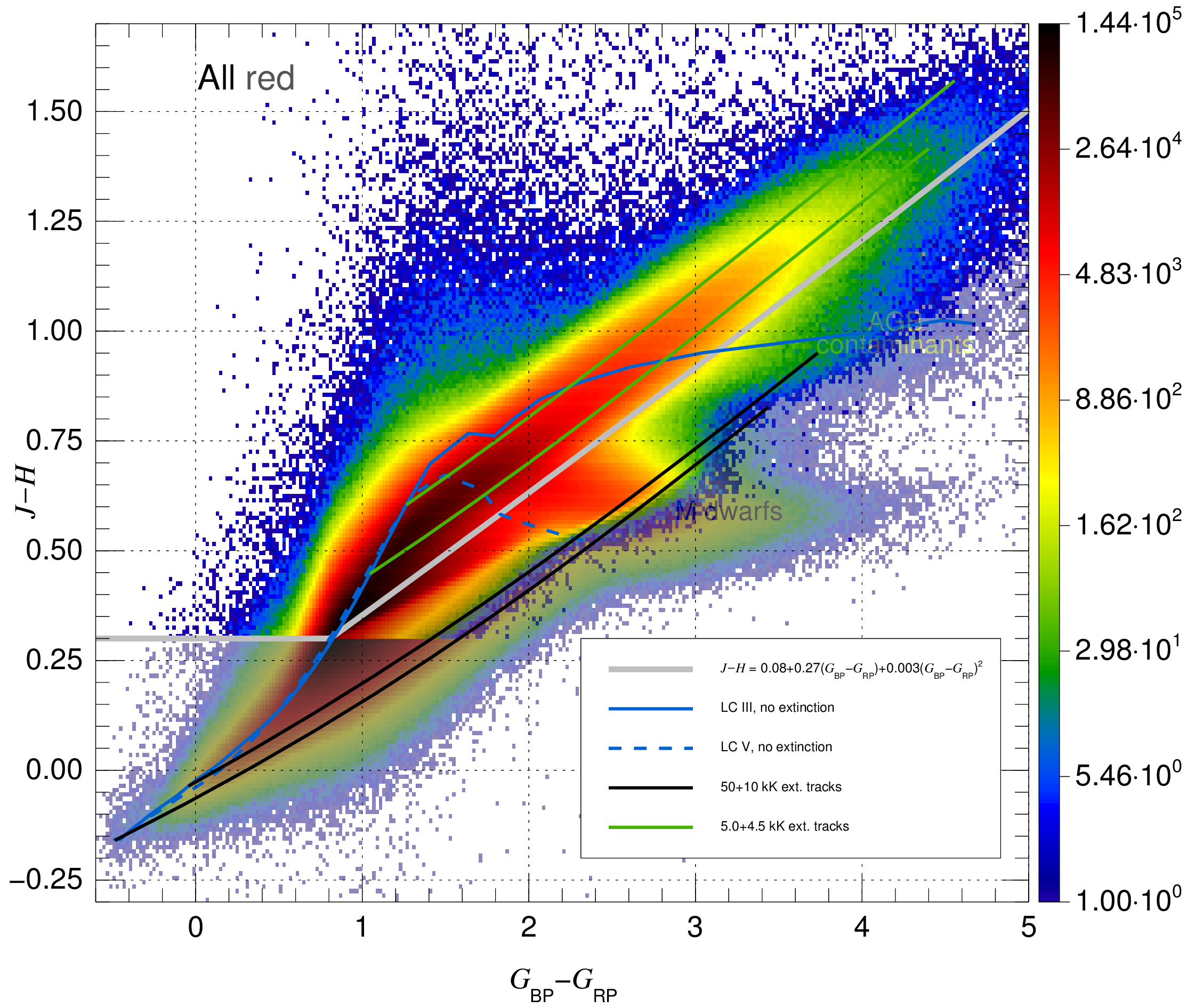} \
  \includegraphics*[width=0.49\linewidth]{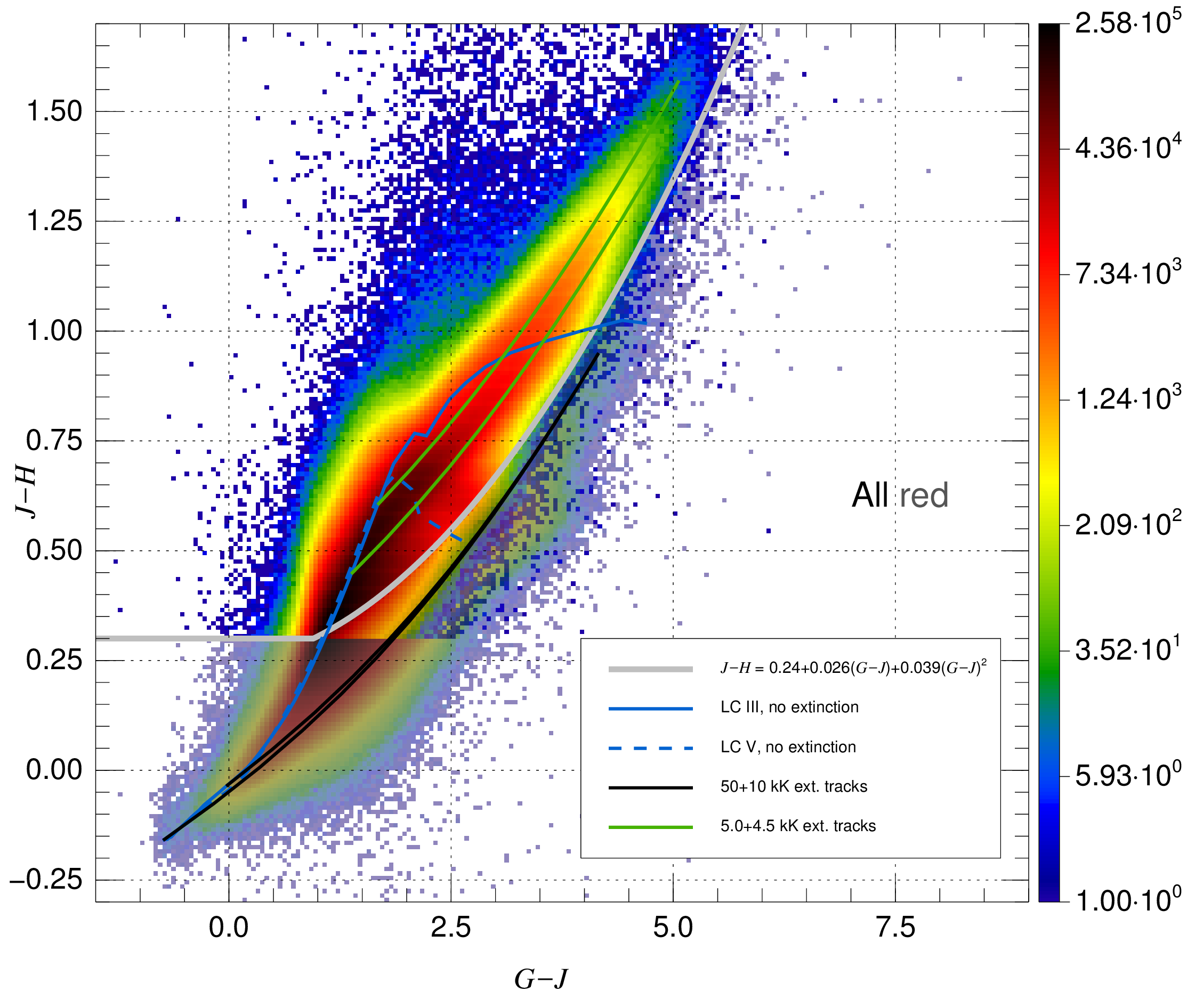}
  \\           
  \includegraphics*[width=0.49\linewidth]{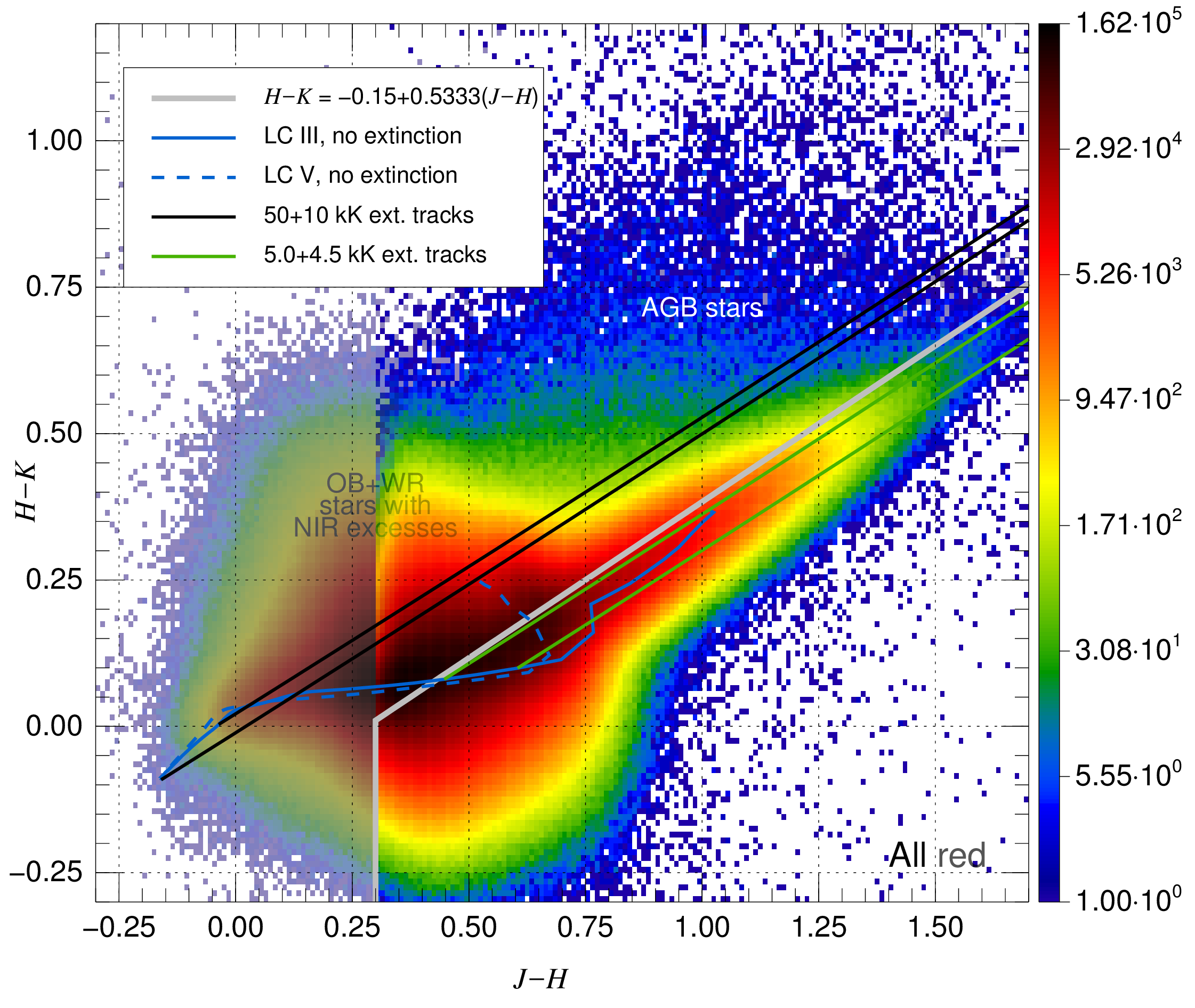} \
  \includegraphics*[width=0.49\linewidth]{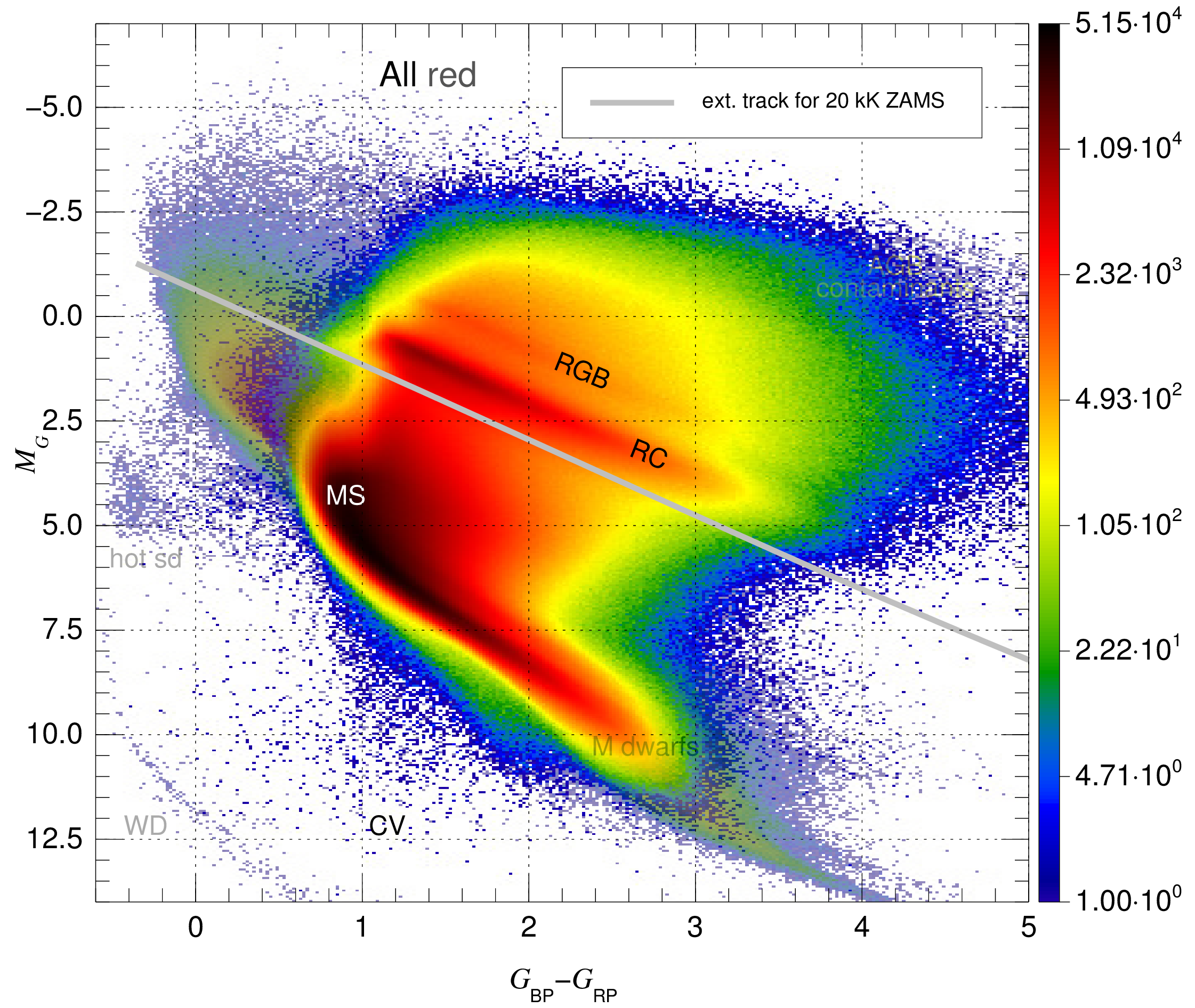}  
  \end{minipage}
  \newframe
  \begin{minipage}{\linewidth}
  \includegraphics*[width=0.49\linewidth]{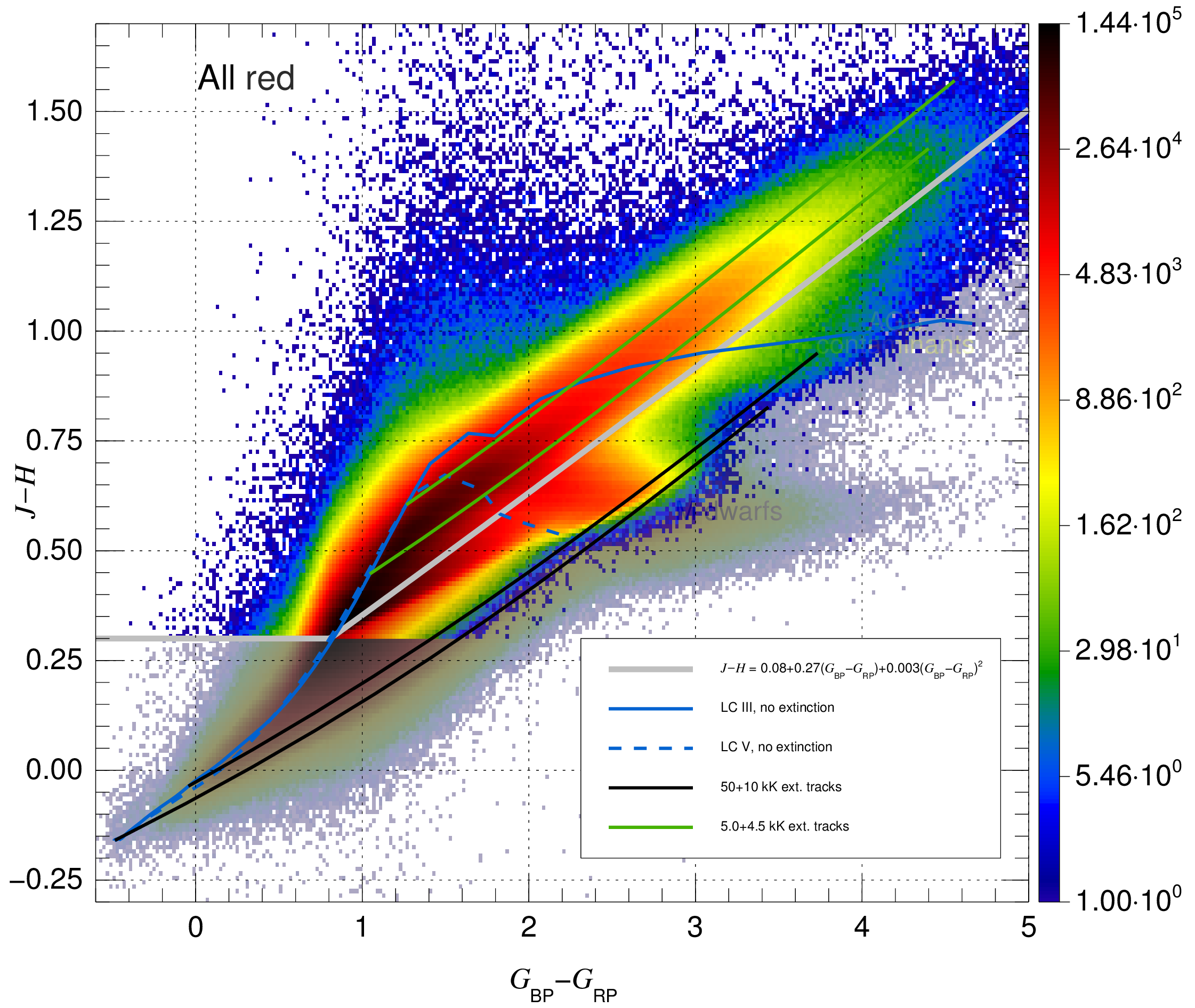} \
  \includegraphics*[width=0.49\linewidth]{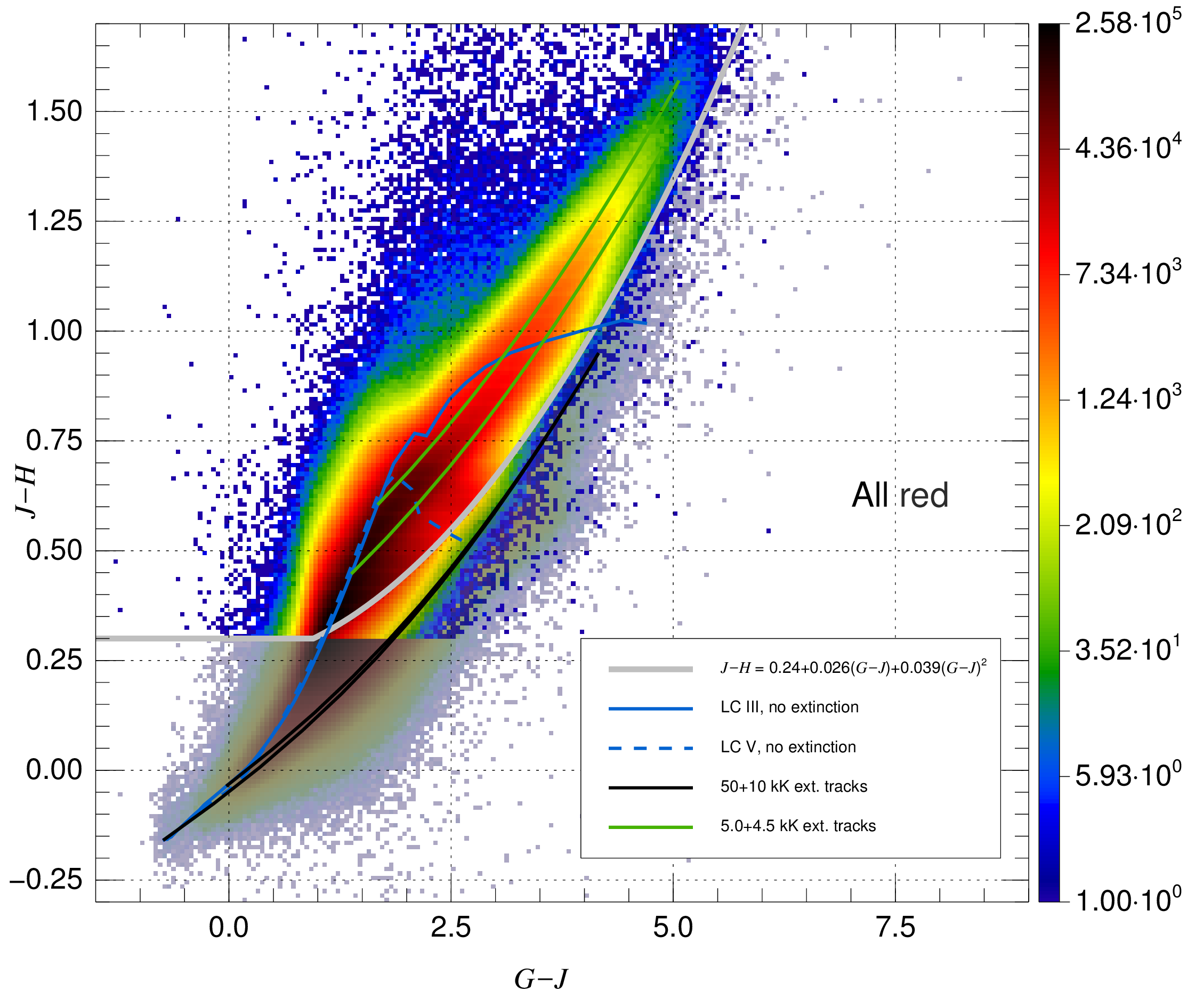}
  \\           
  \includegraphics*[width=0.49\linewidth]{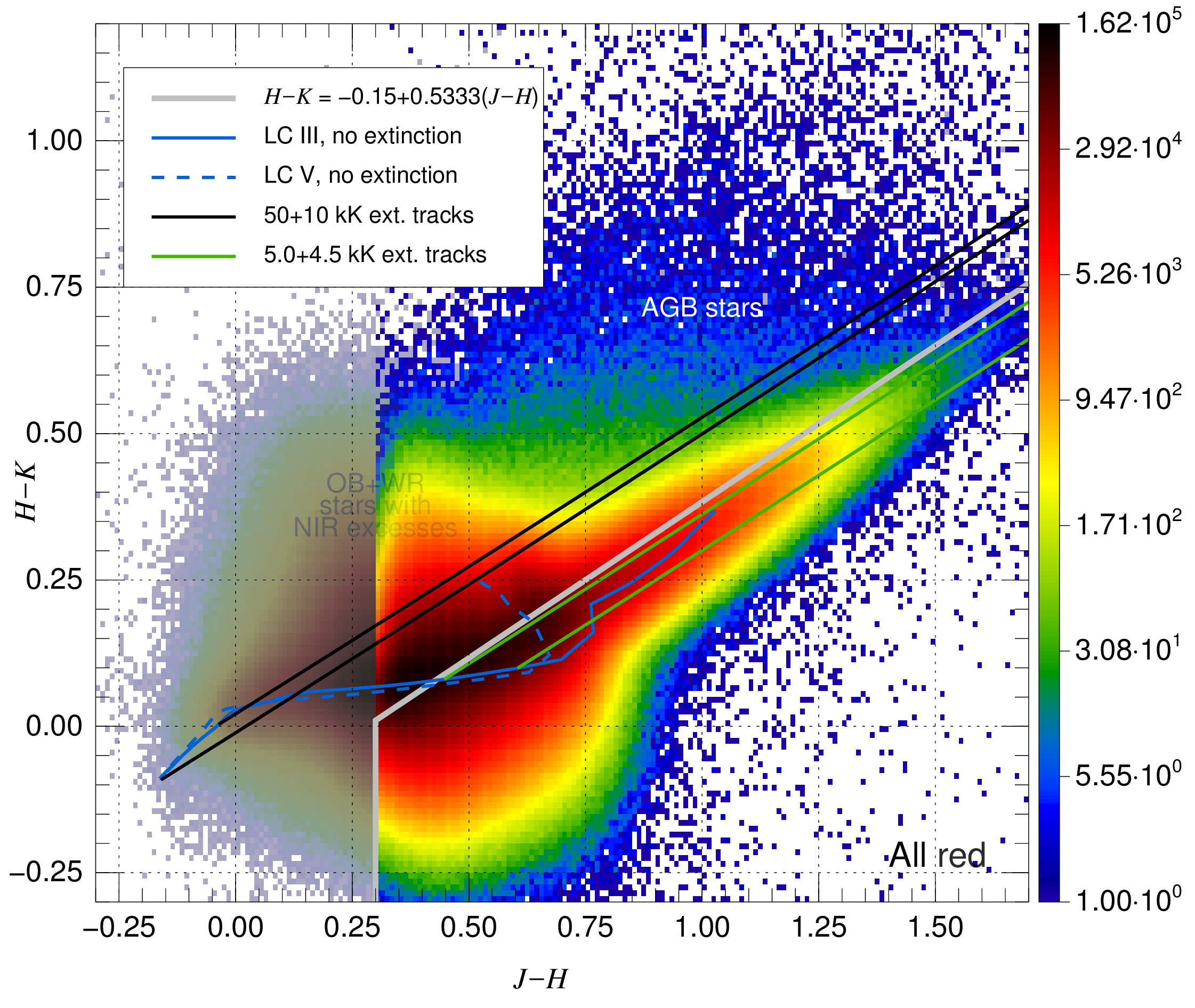} \
  \includegraphics*[width=0.49\linewidth]{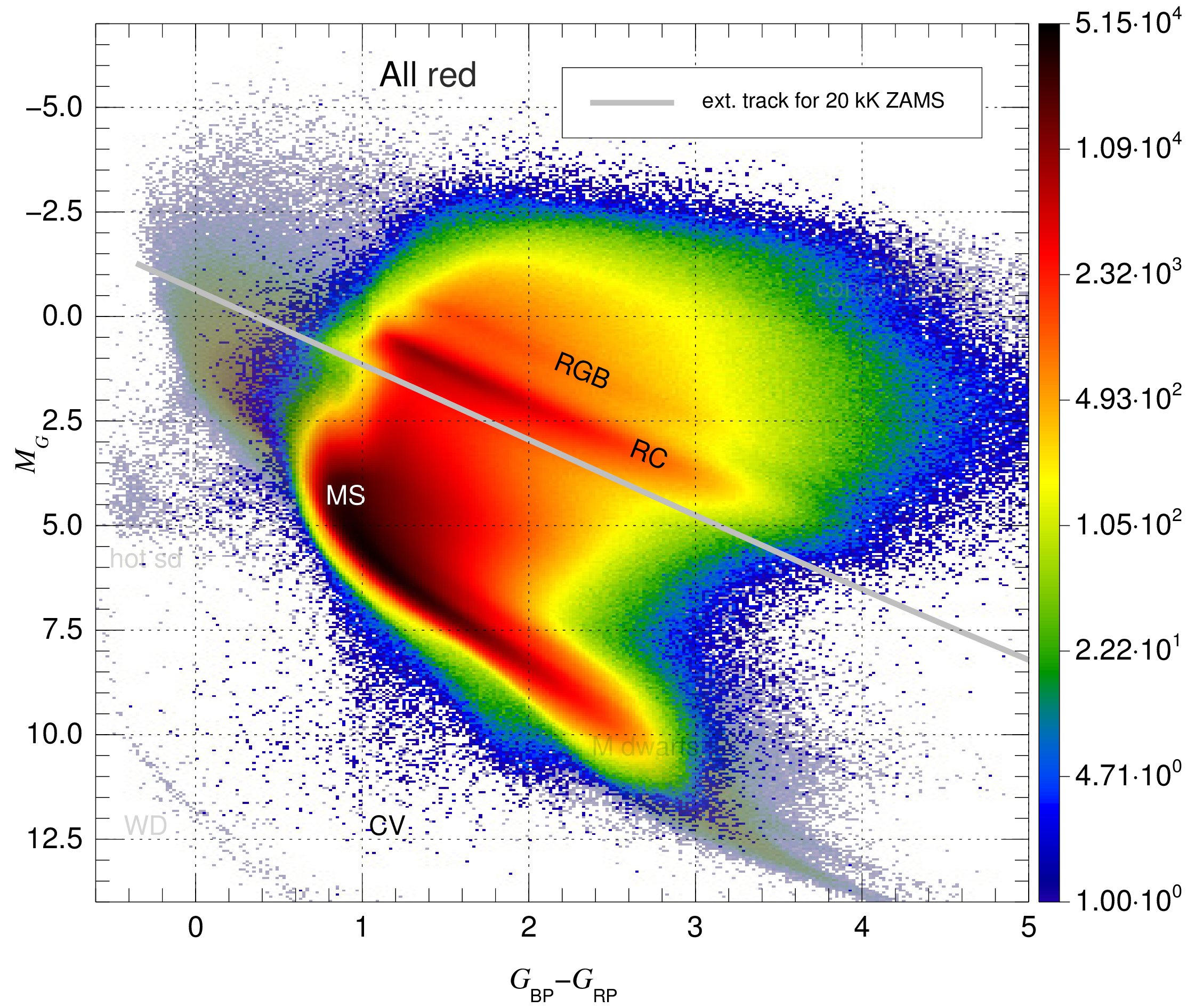}  
  \end{minipage}
  \newframe[0.4]
  \begin{minipage}{\linewidth}
  \includegraphics*[width=0.49\linewidth]{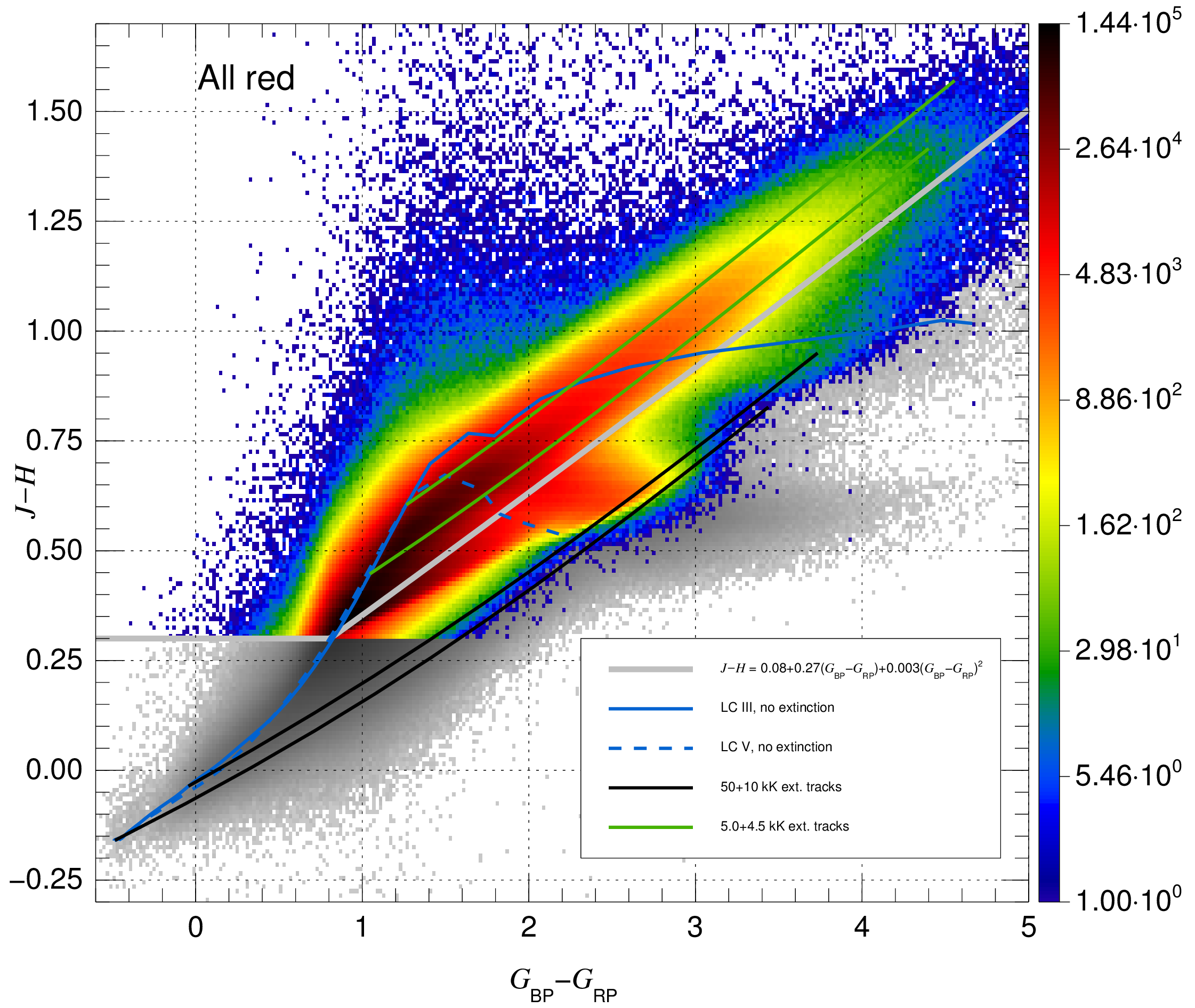} \
  \includegraphics*[width=0.49\linewidth]{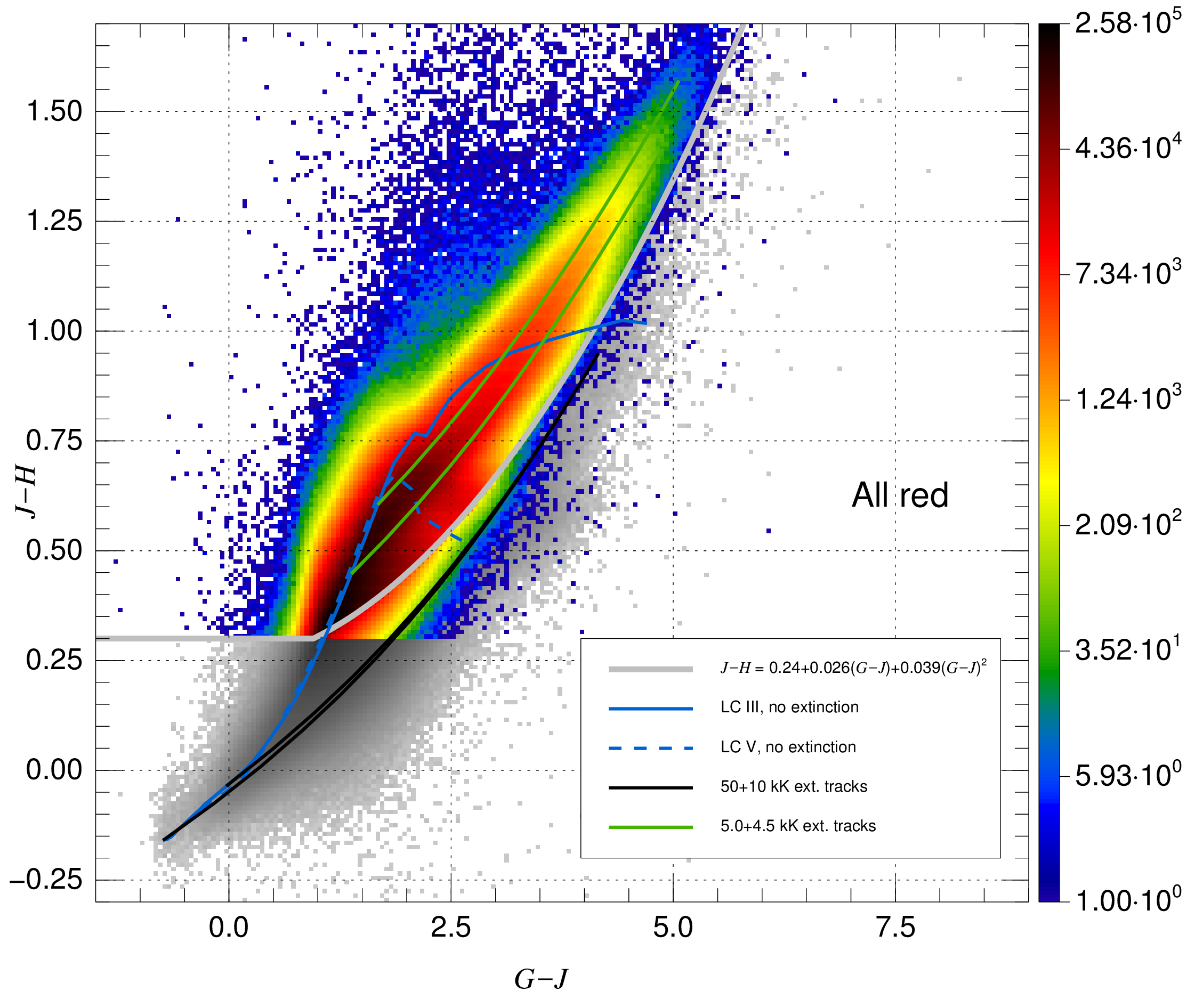}
  \\           
  \includegraphics*[width=0.49\linewidth]{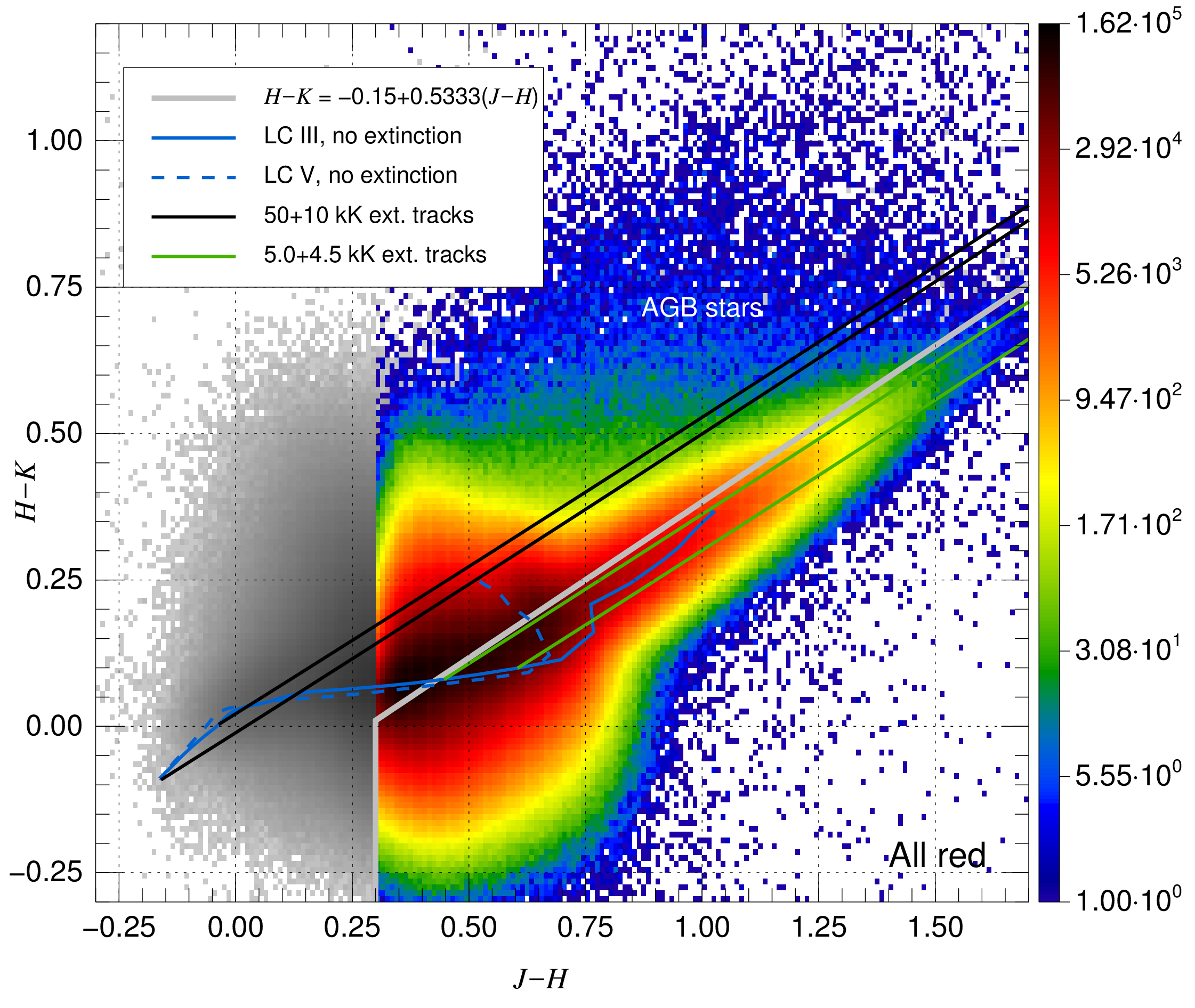} \
  \includegraphics*[width=0.49\linewidth]{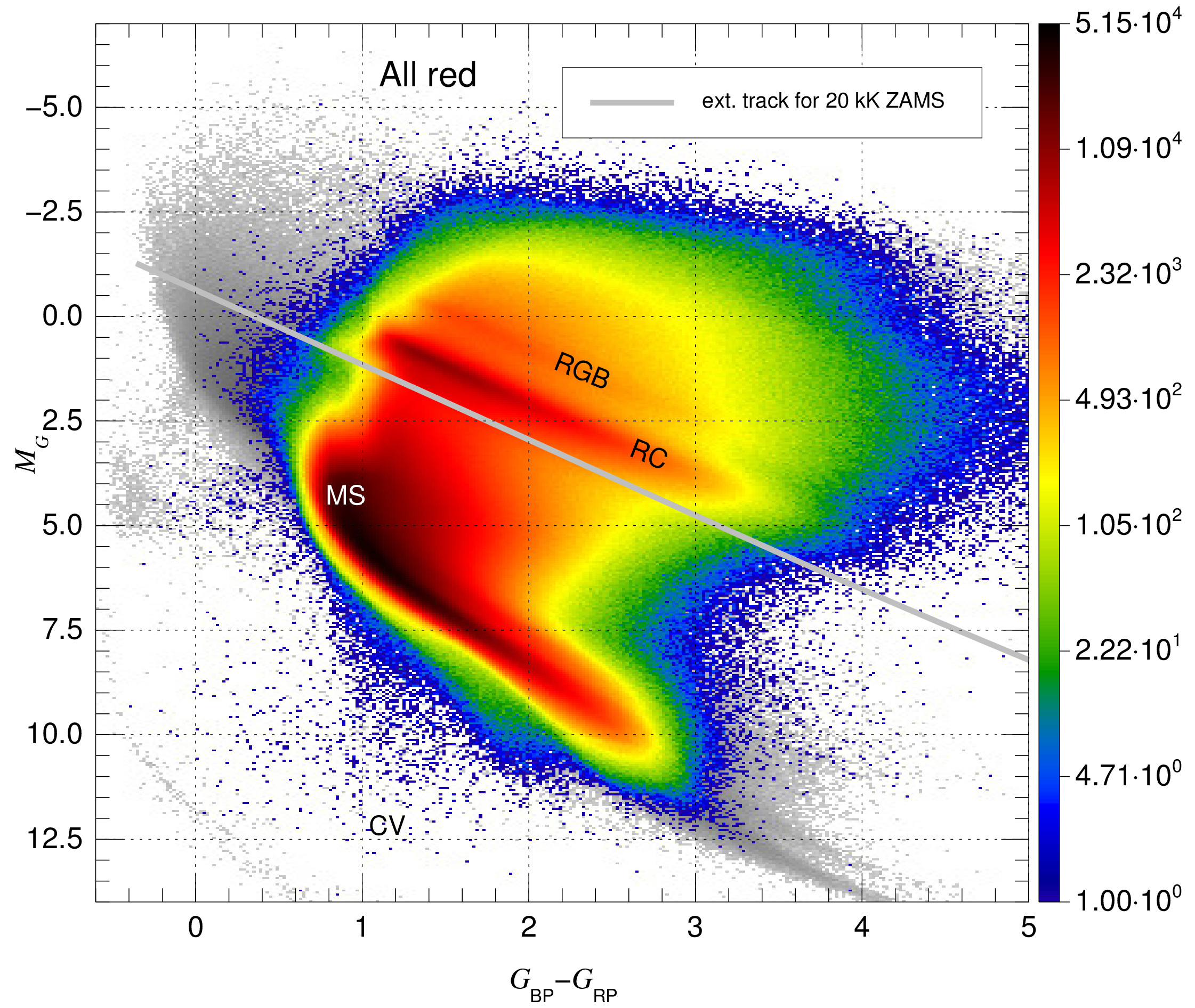}  
  \end{minipage}
  \newframe[5]
  \begin{minipage}{\linewidth}
  \includegraphics*[width=0.49\linewidth]{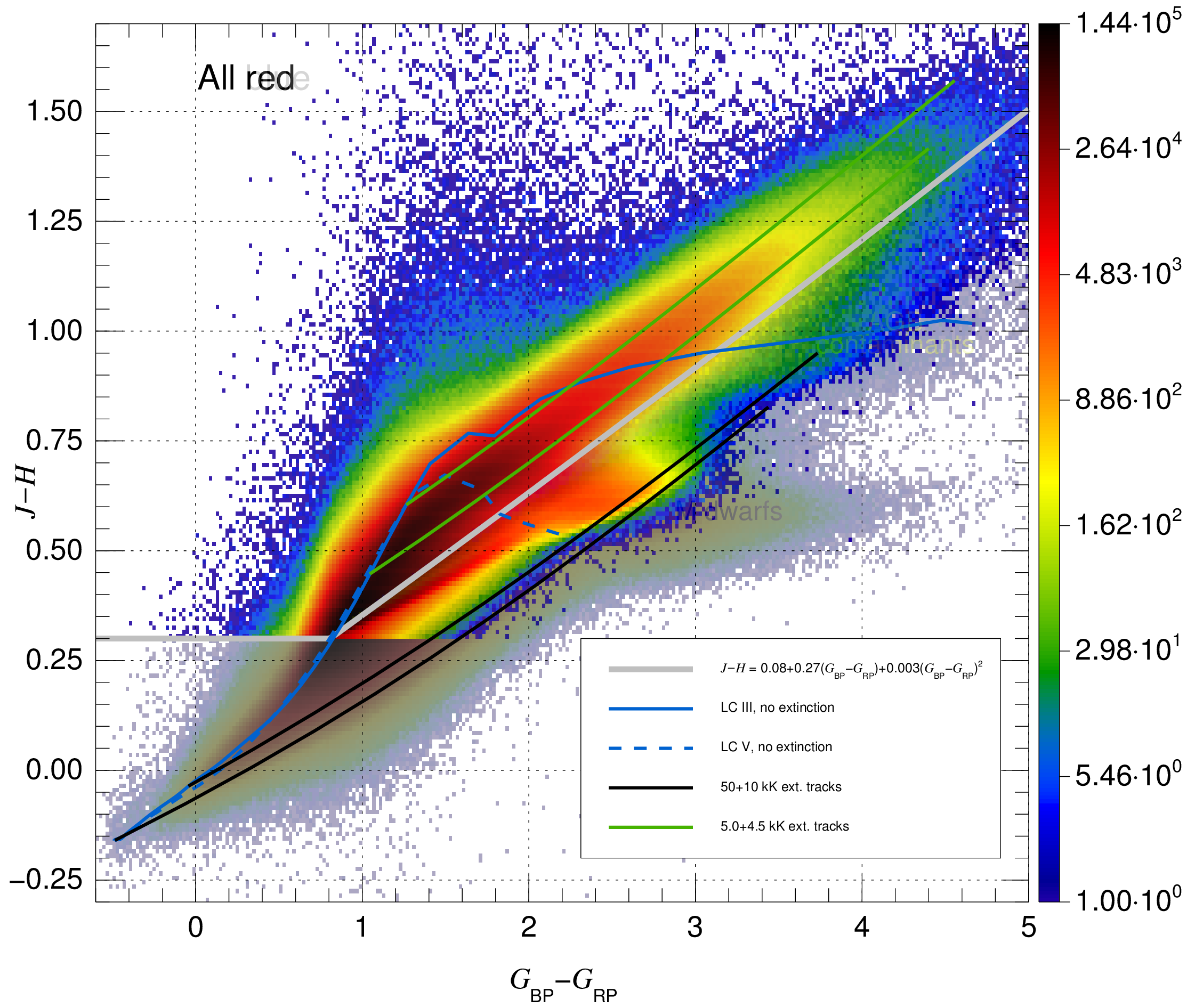} \
  \includegraphics*[width=0.49\linewidth]{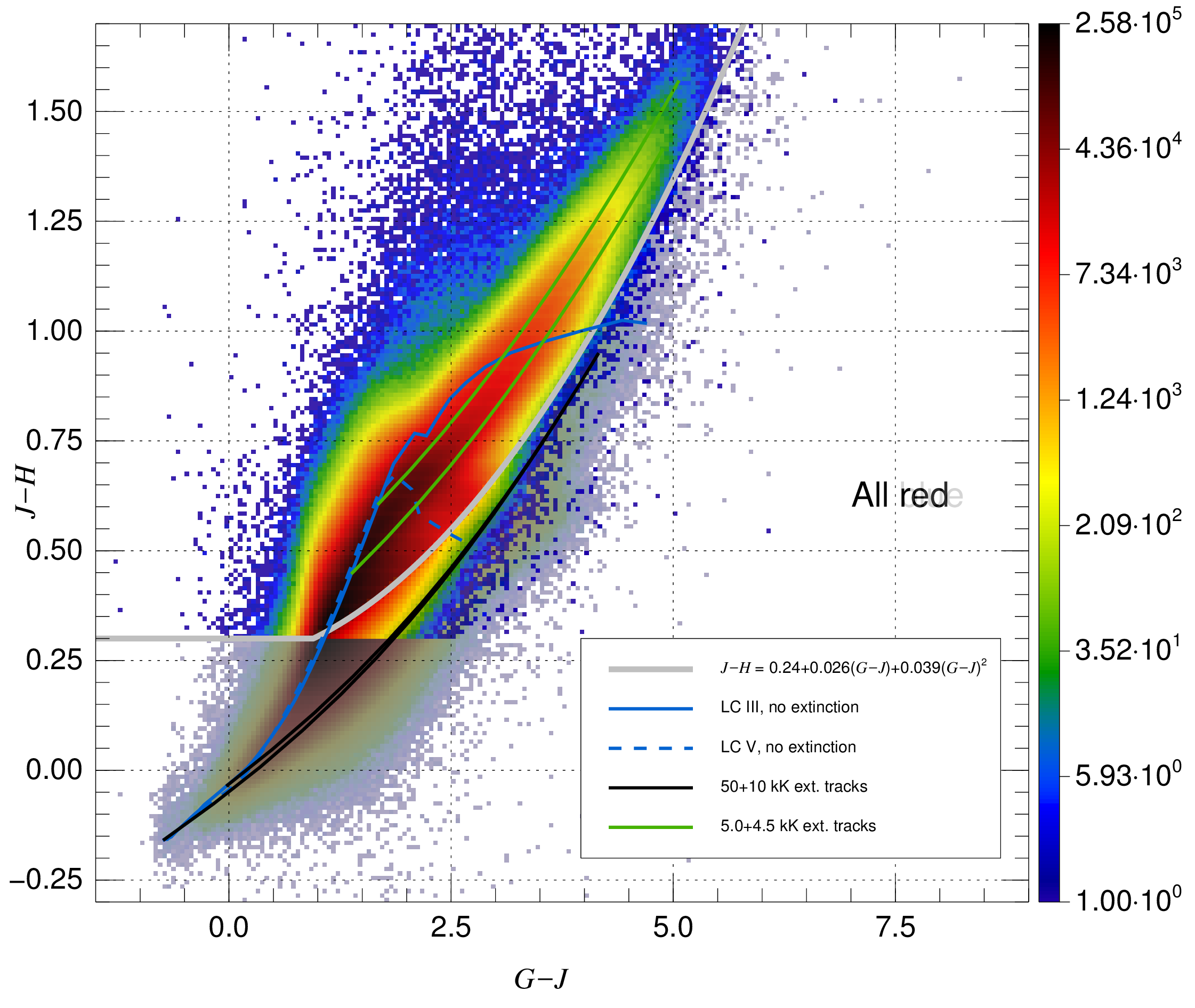}
  \\           
  \includegraphics*[width=0.49\linewidth]{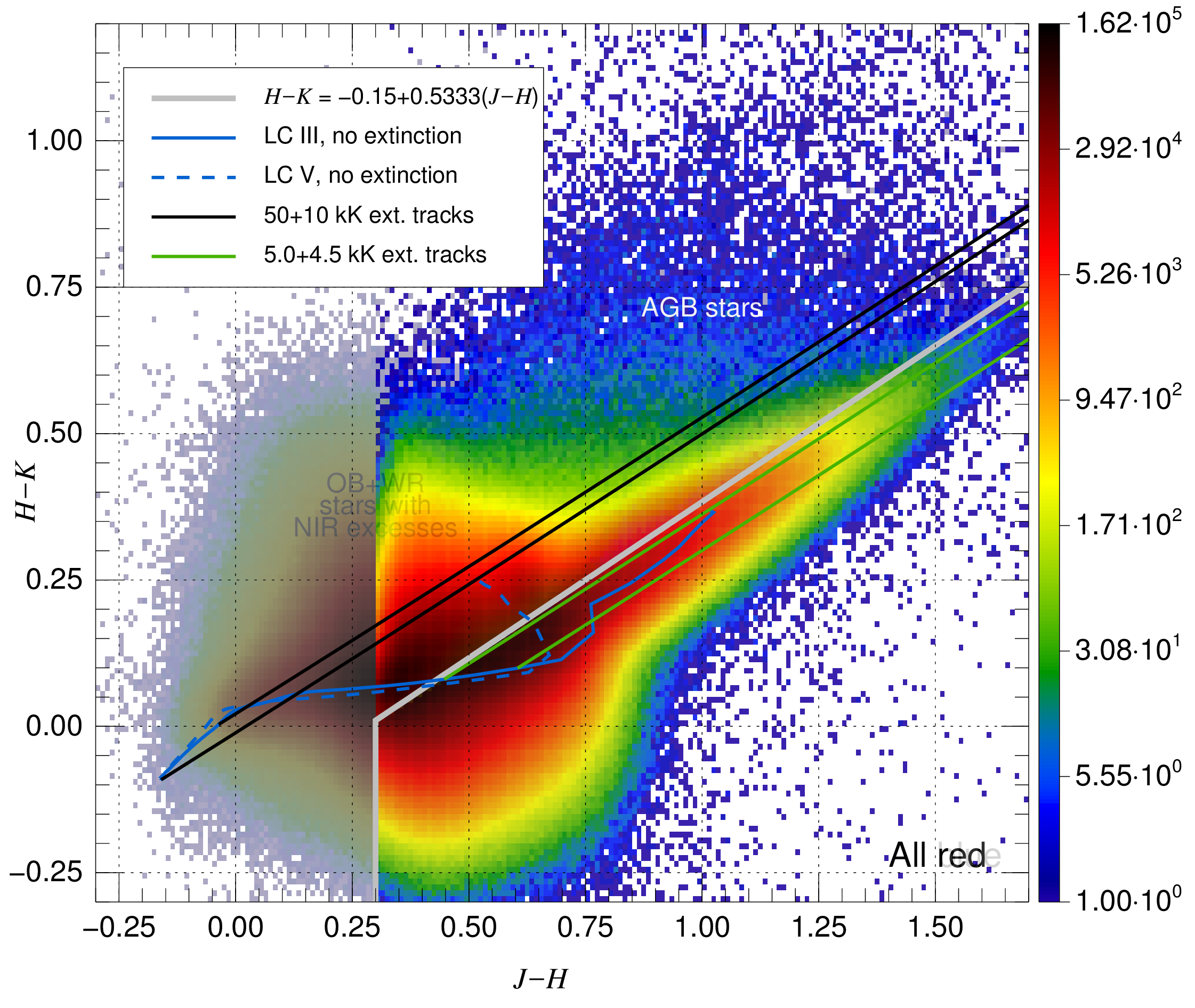} \
  \includegraphics*[width=0.49\linewidth]{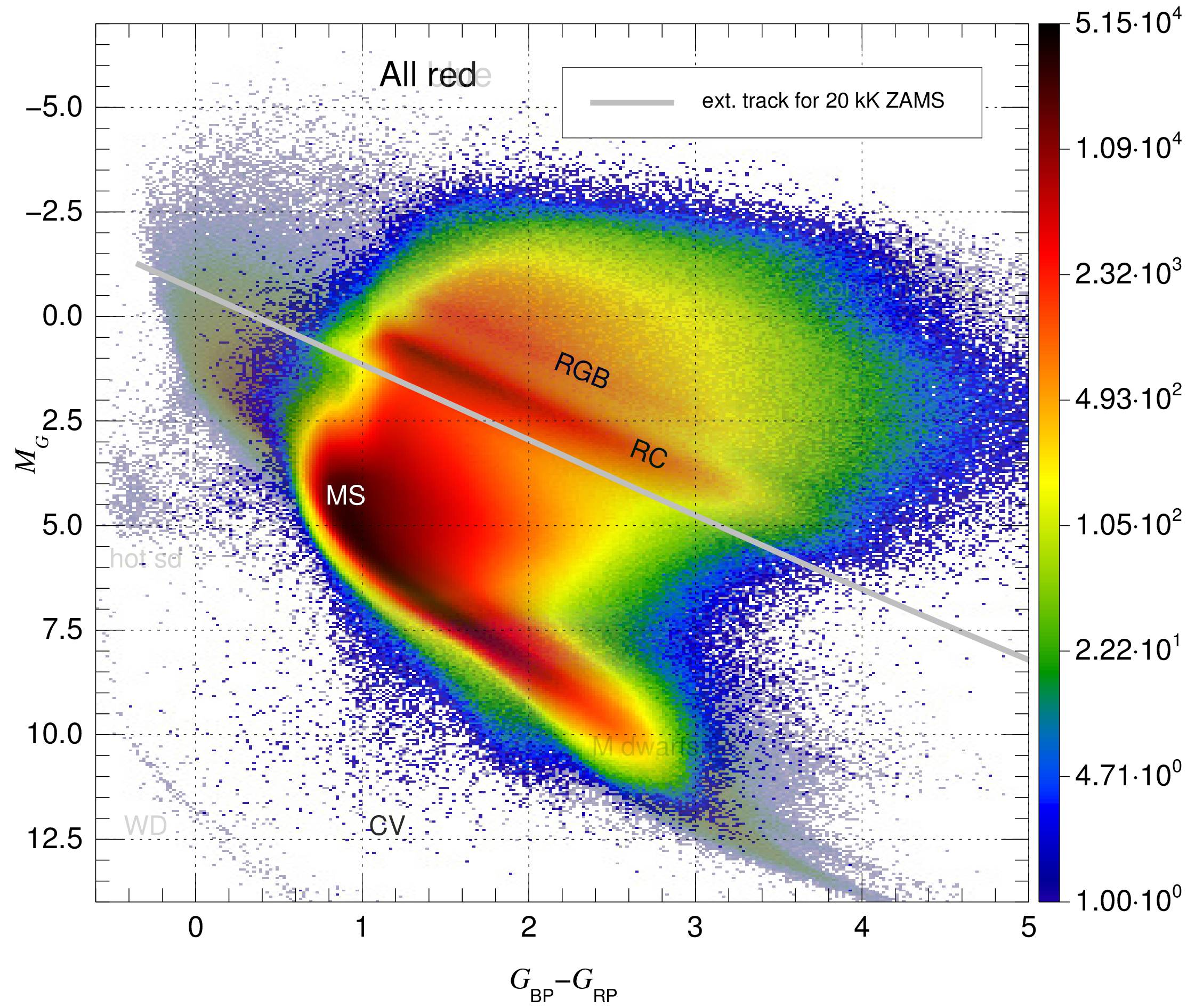}  
  \end{minipage}
  \newframe
  \begin{minipage}{\linewidth}
  \includegraphics*[width=0.49\linewidth]{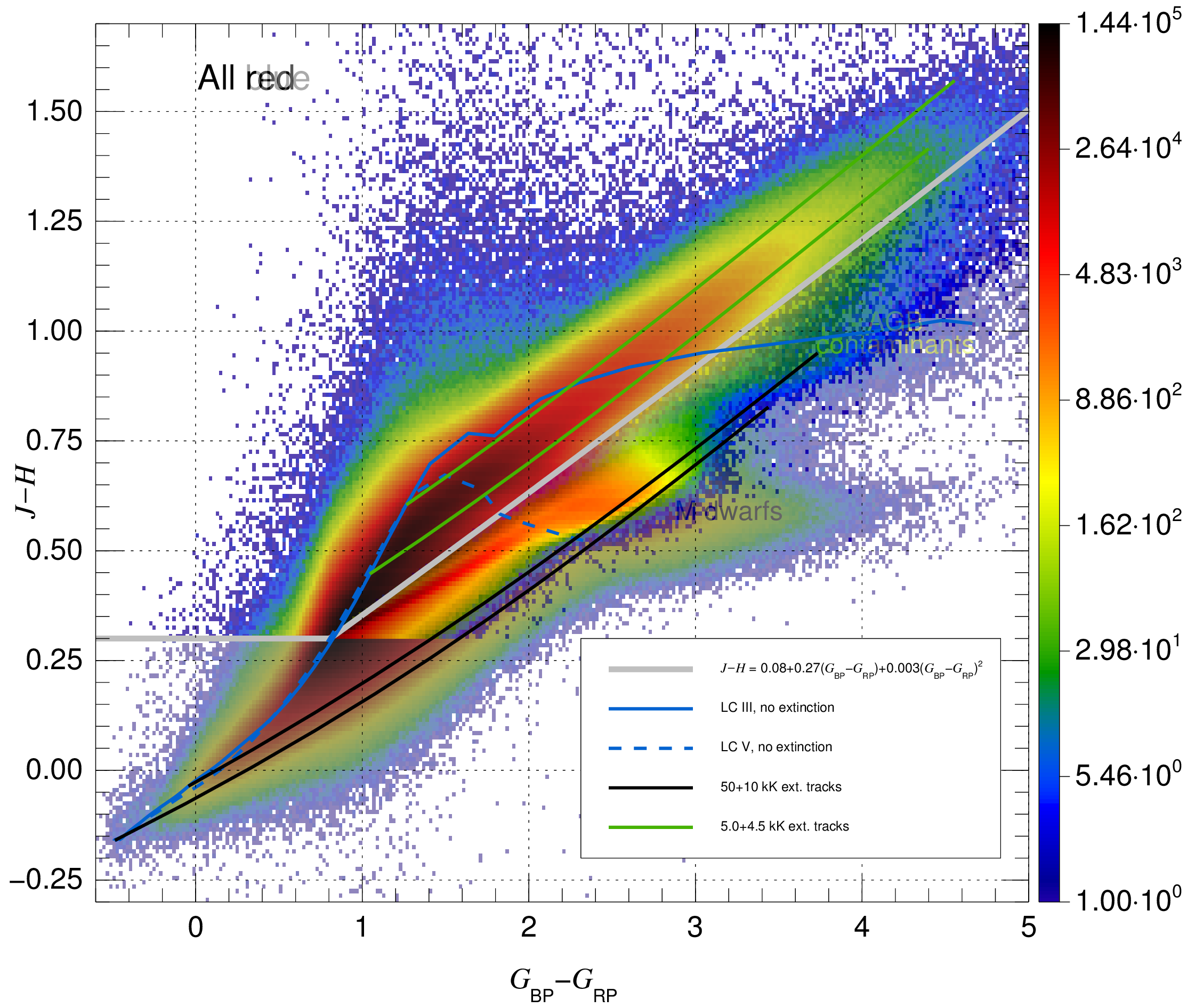} \
  \includegraphics*[width=0.49\linewidth]{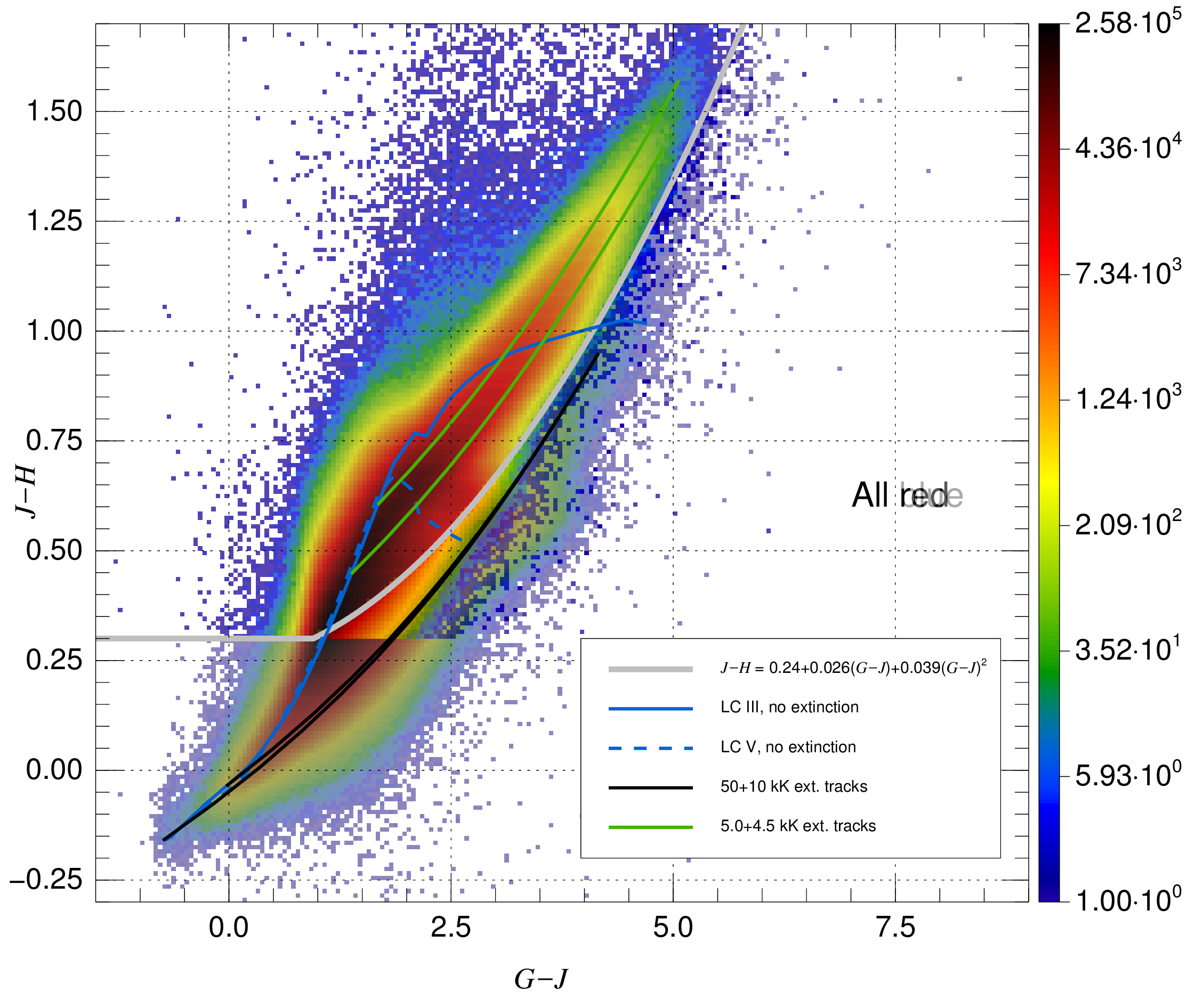}
  \\           
  \includegraphics*[width=0.49\linewidth]{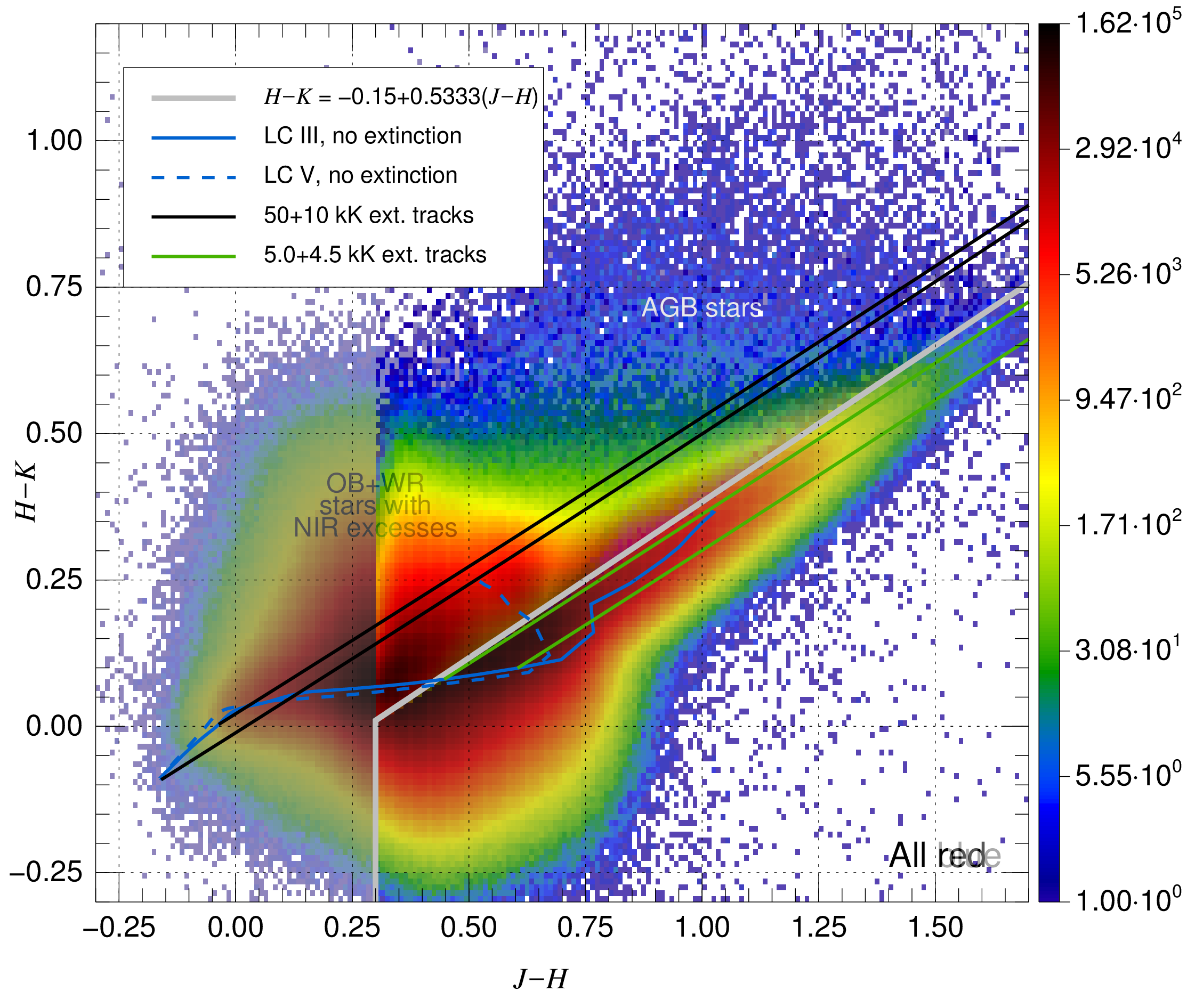} \
  \includegraphics*[width=0.49\linewidth]{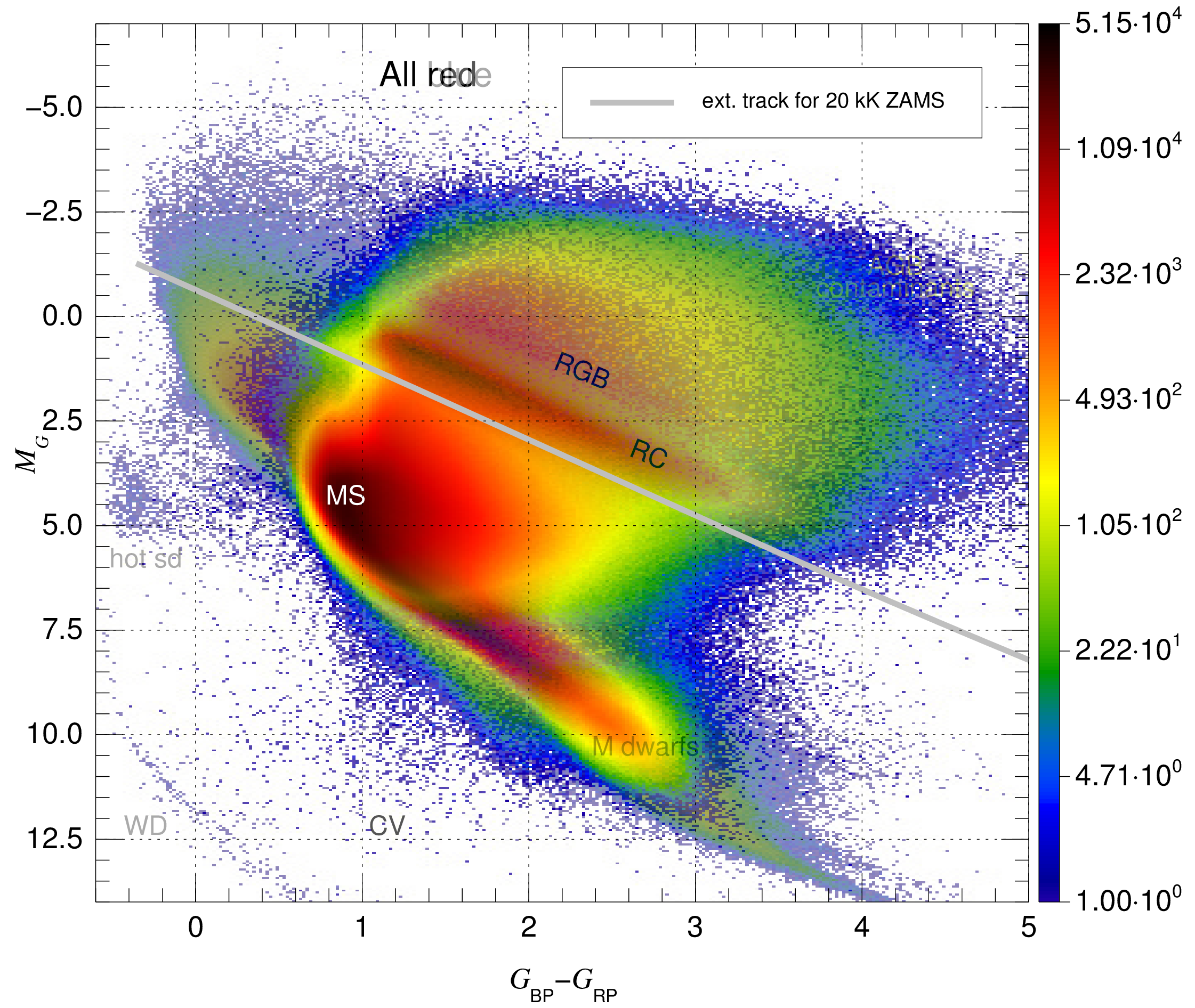}  
  \end{minipage}
  \newframe
  \begin{minipage}{\linewidth}
  \includegraphics*[width=0.49\linewidth]{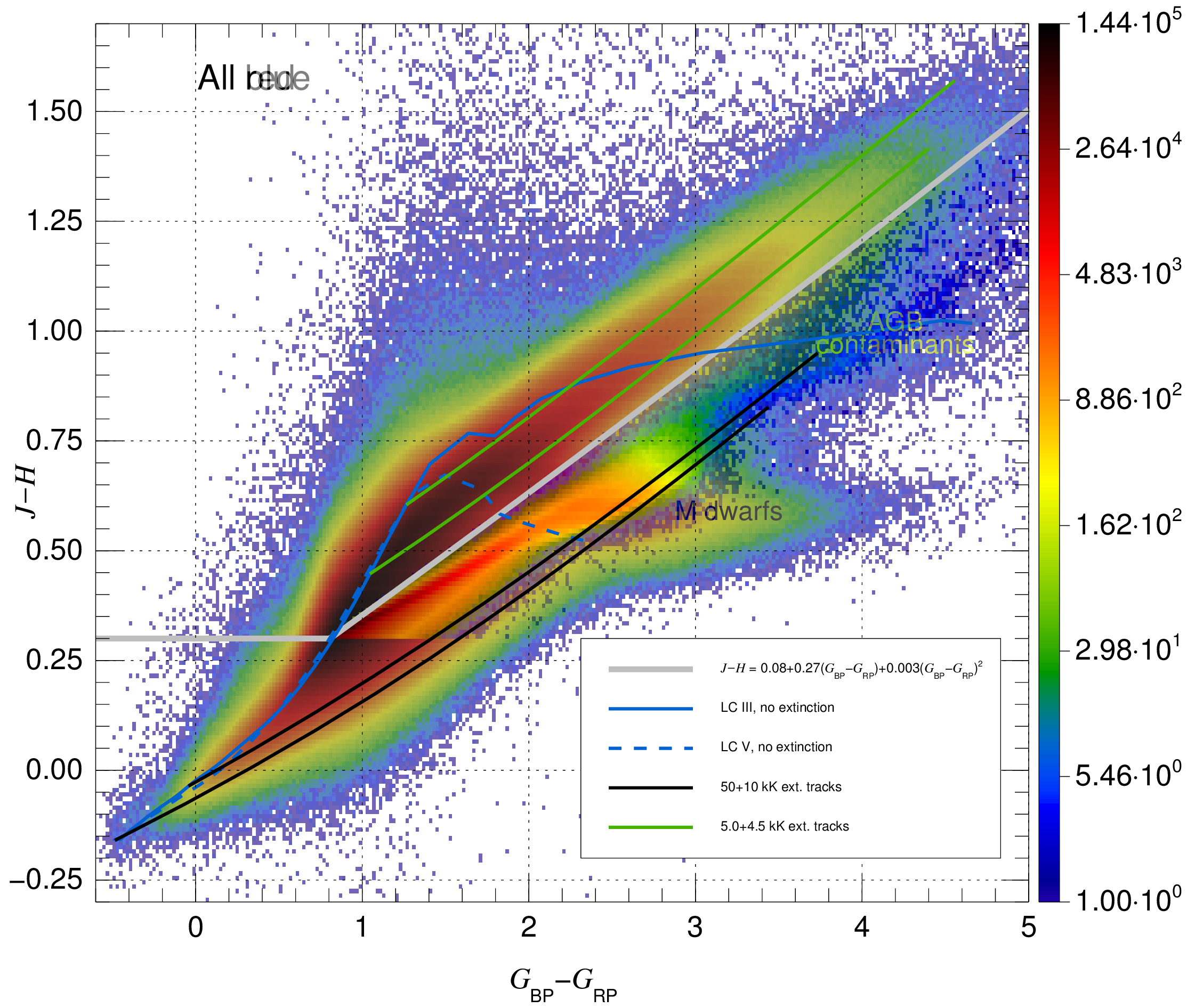} \
  \includegraphics*[width=0.49\linewidth]{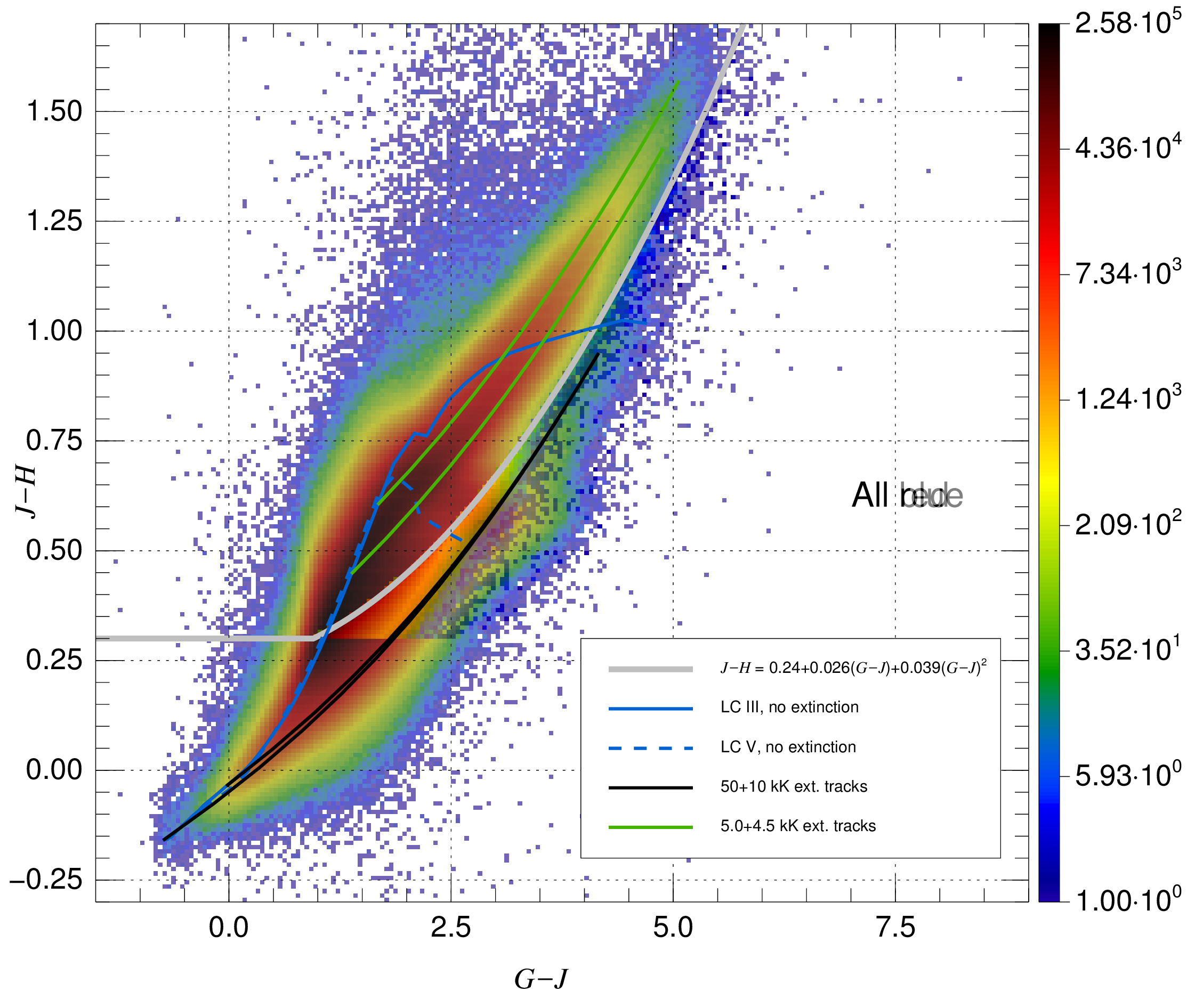}
  \\           
  \includegraphics*[width=0.49\linewidth]{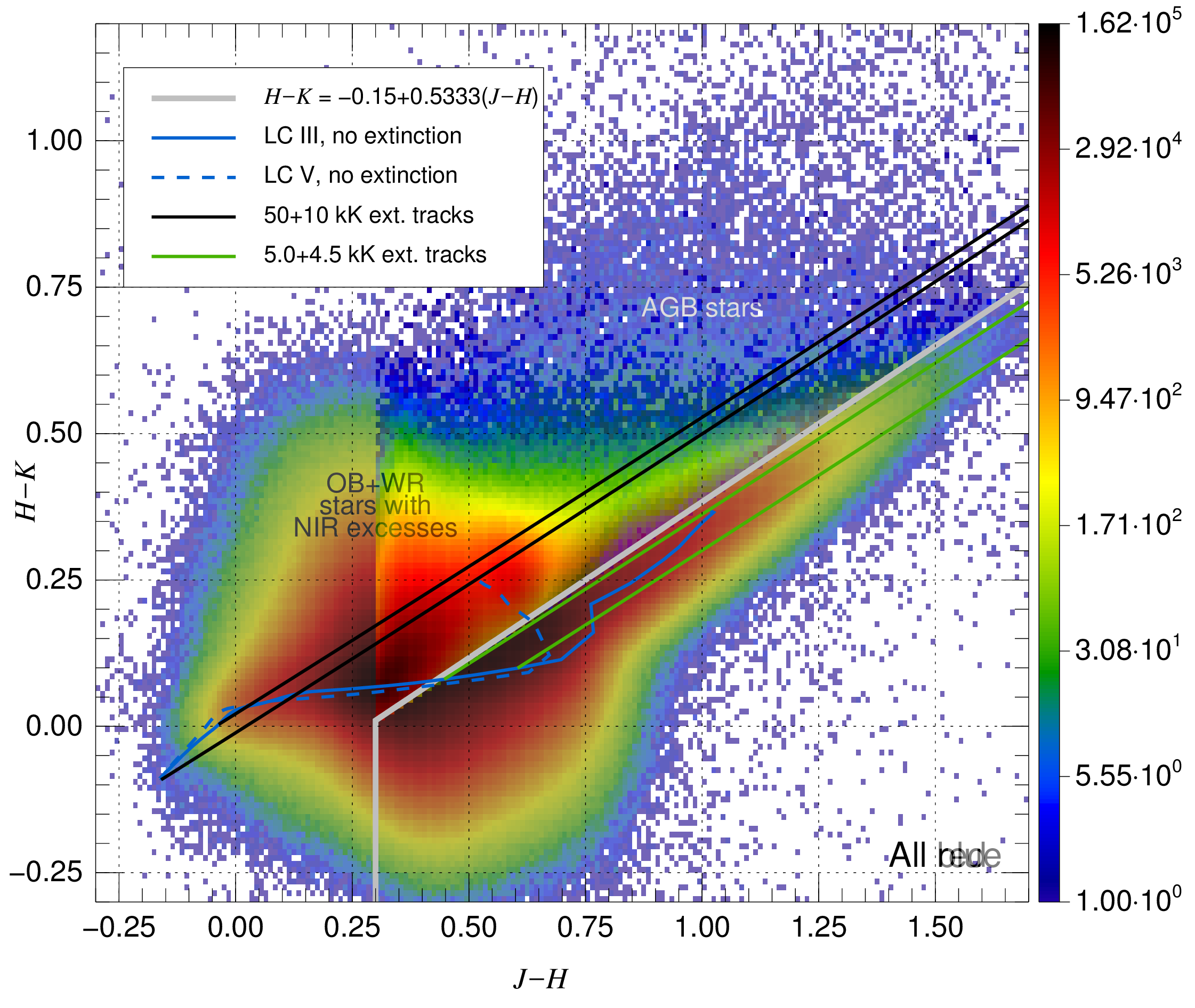} \
  \includegraphics*[width=0.49\linewidth]{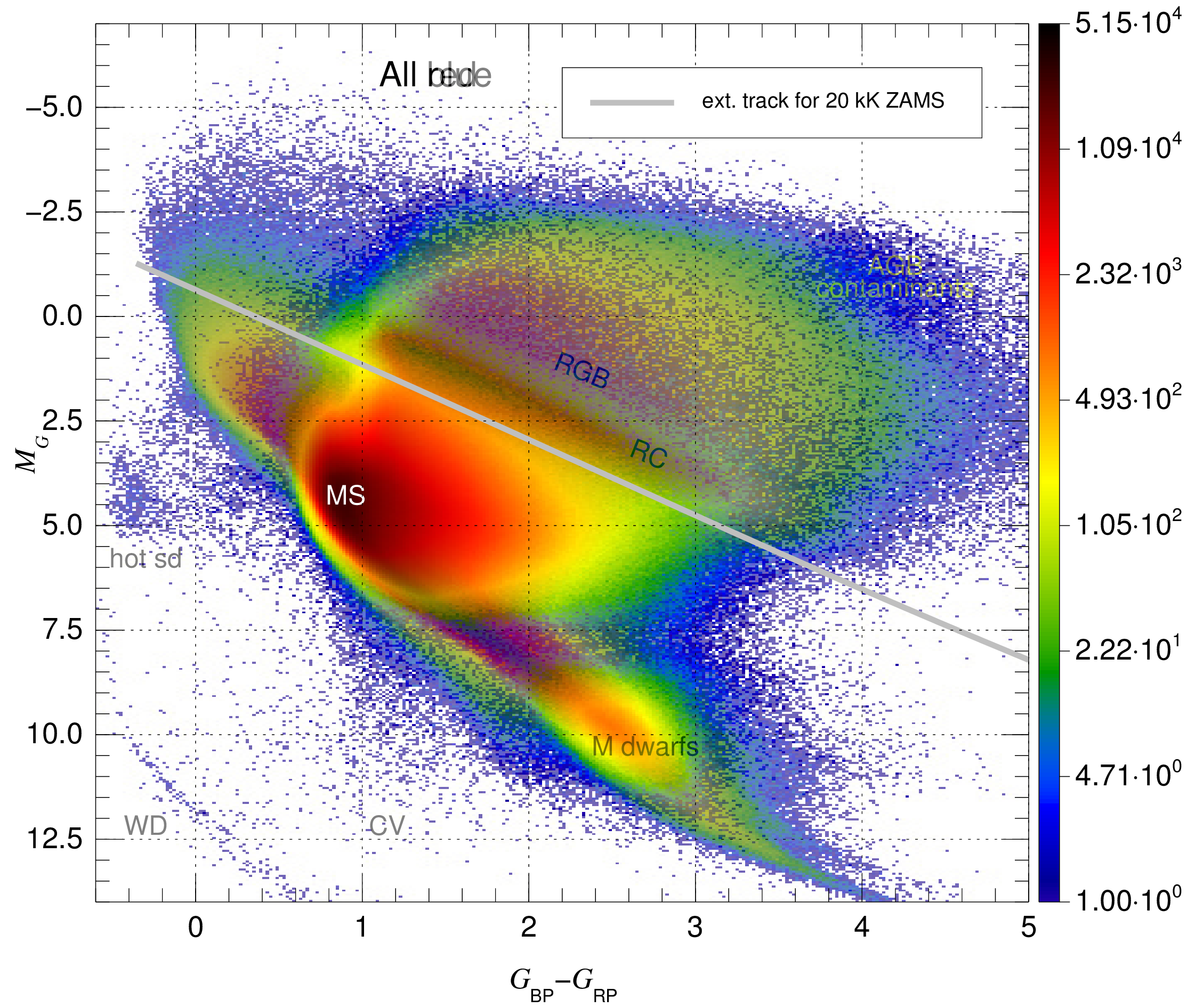}  
  \end{minipage}
  \newframe
  \begin{minipage}{\linewidth}
  \includegraphics*[width=0.49\linewidth]{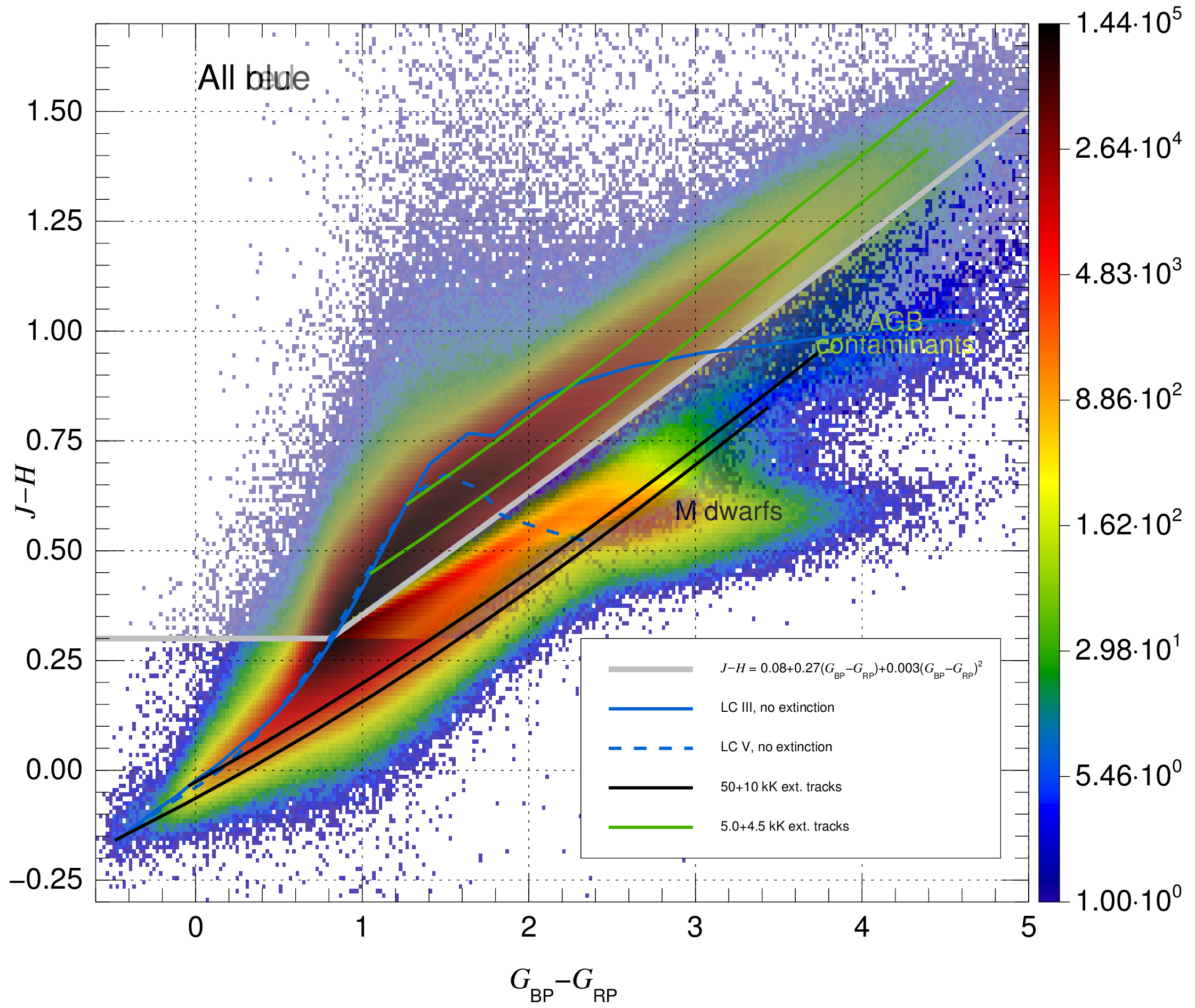} \
  \includegraphics*[width=0.49\linewidth]{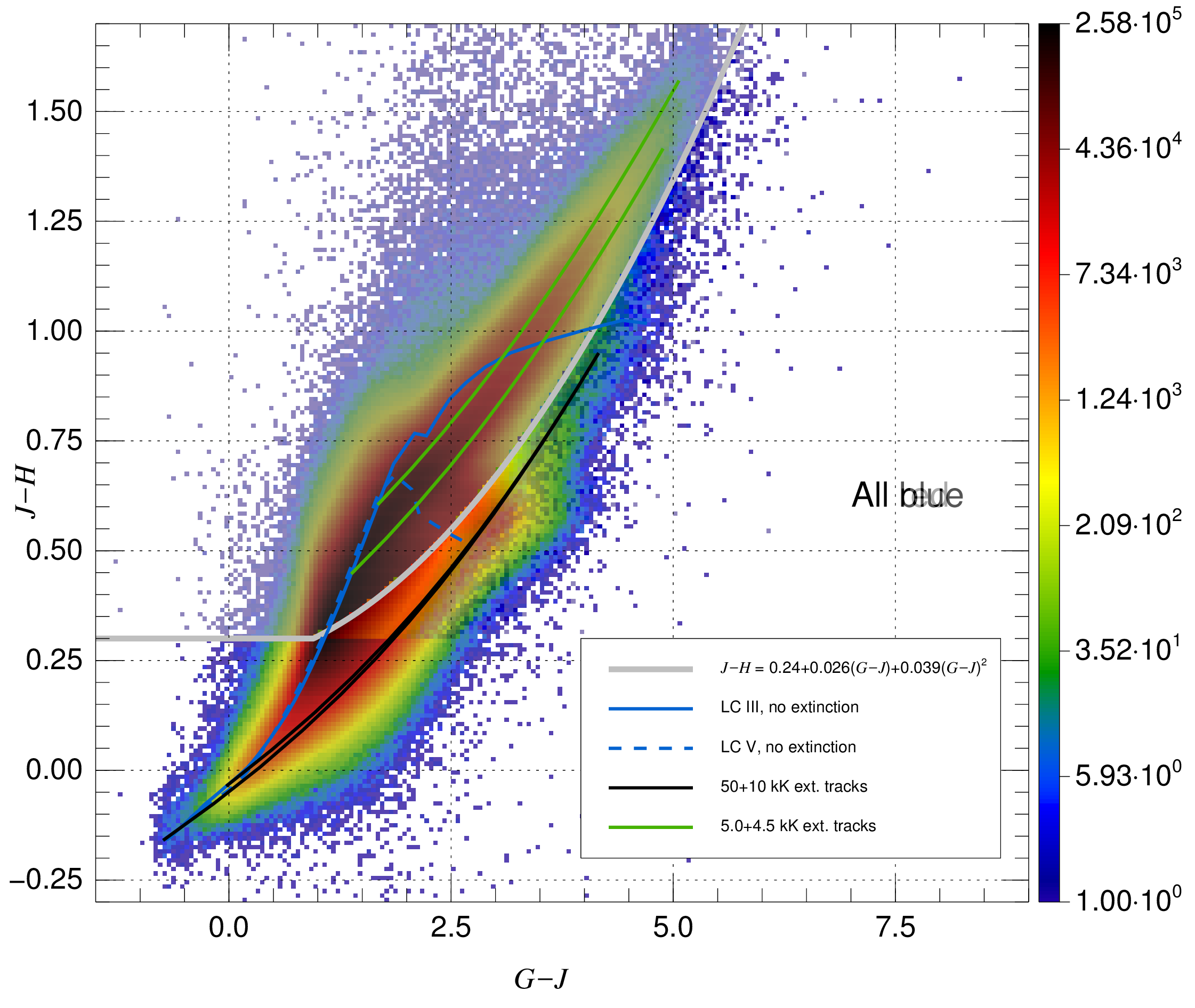}
  \\           
  \includegraphics*[width=0.49\linewidth]{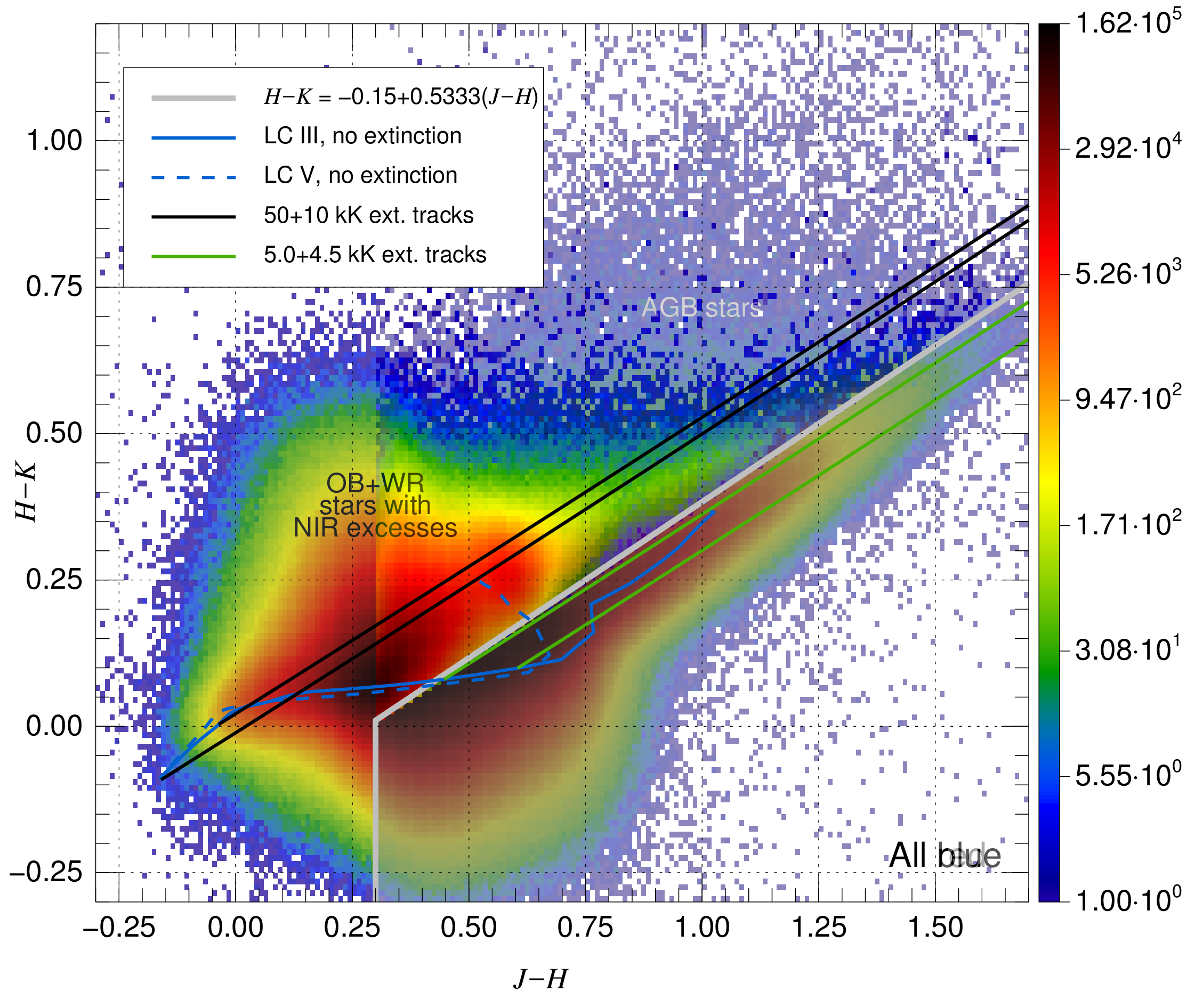} \
  \includegraphics*[width=0.49\linewidth]{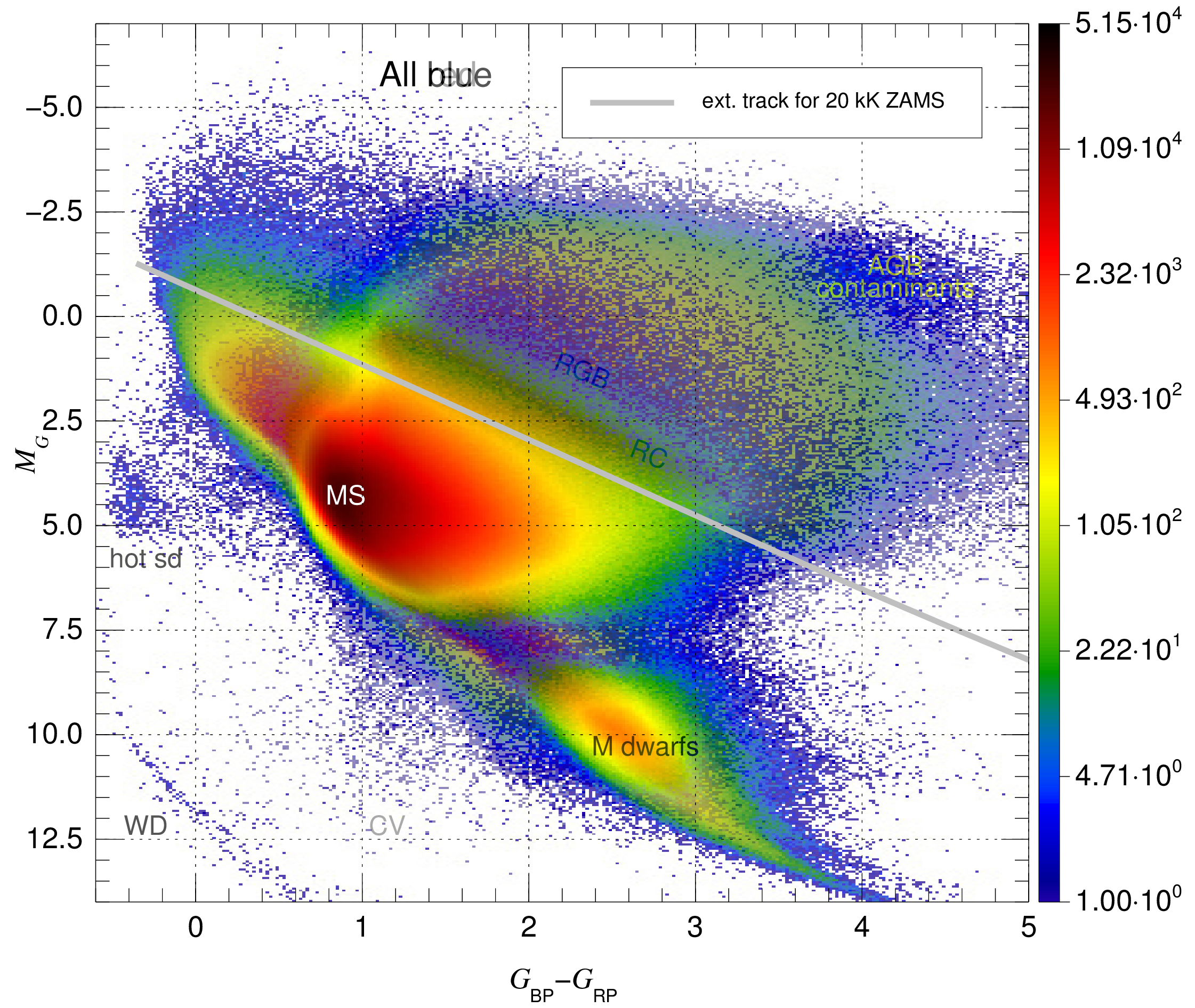}  
  \end{minipage}
  \newframe
  \begin{minipage}{\linewidth}
  \includegraphics*[width=0.49\linewidth]{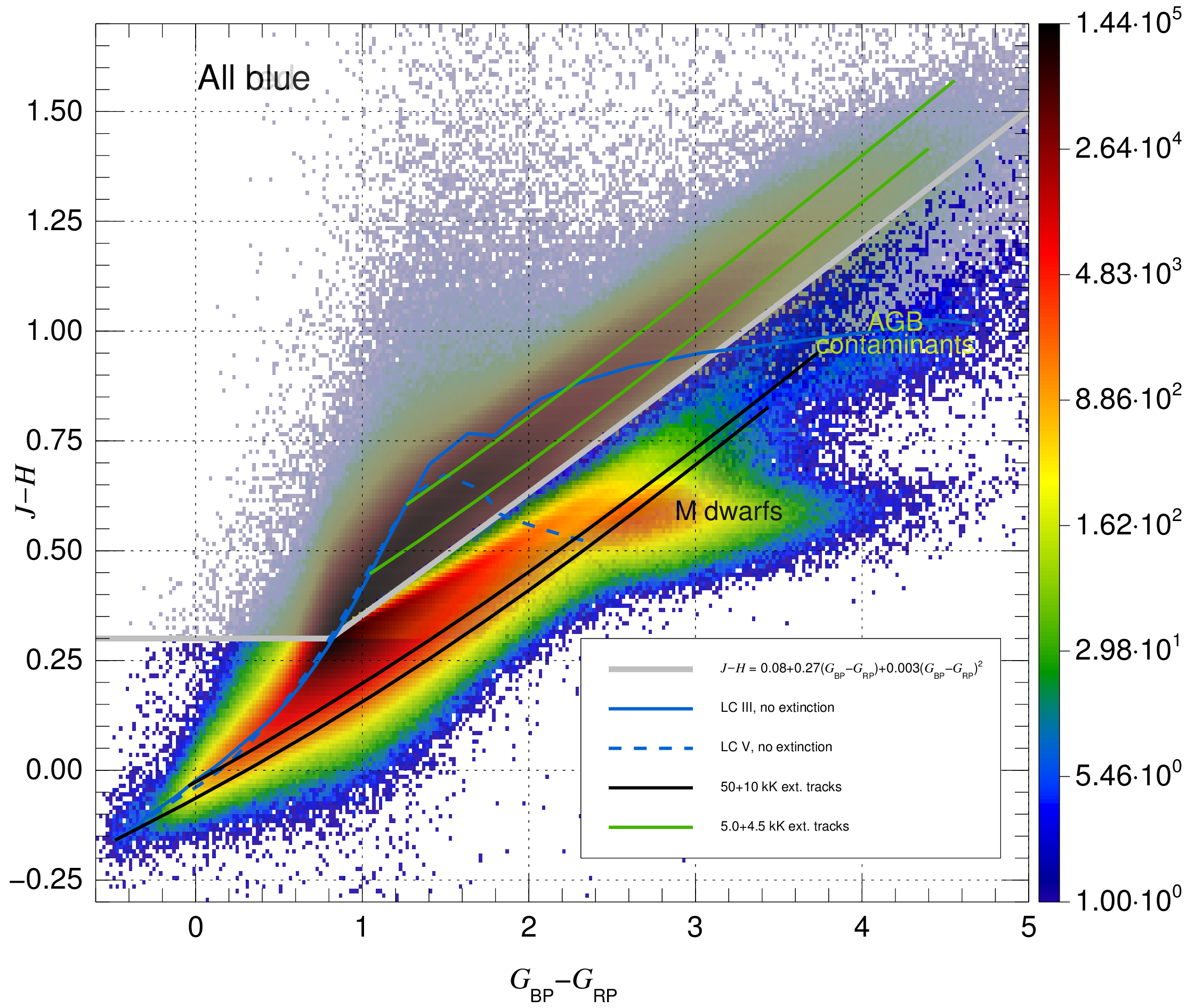} \
  \includegraphics*[width=0.49\linewidth]{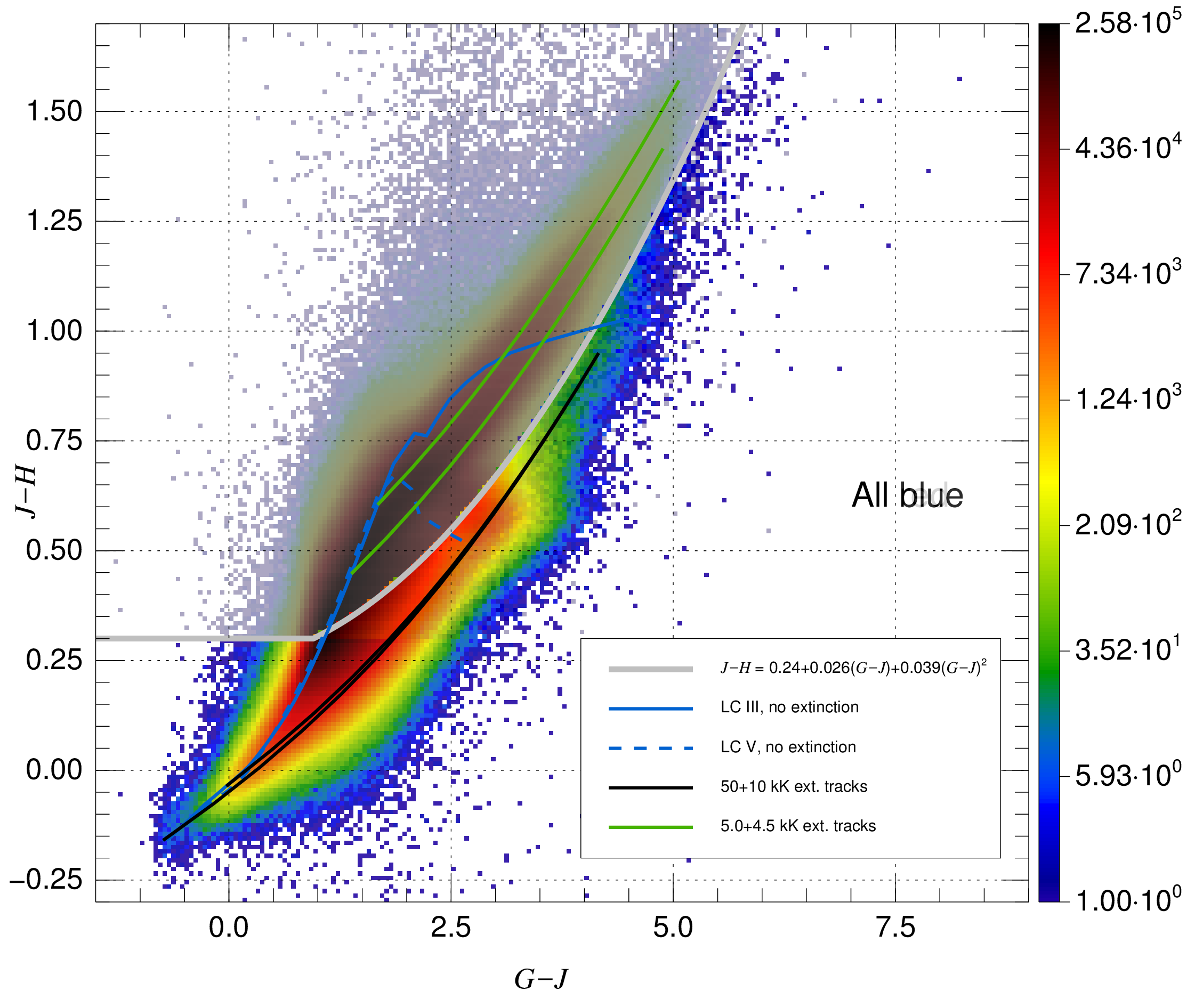}
  \\           
  \includegraphics*[width=0.49\linewidth]{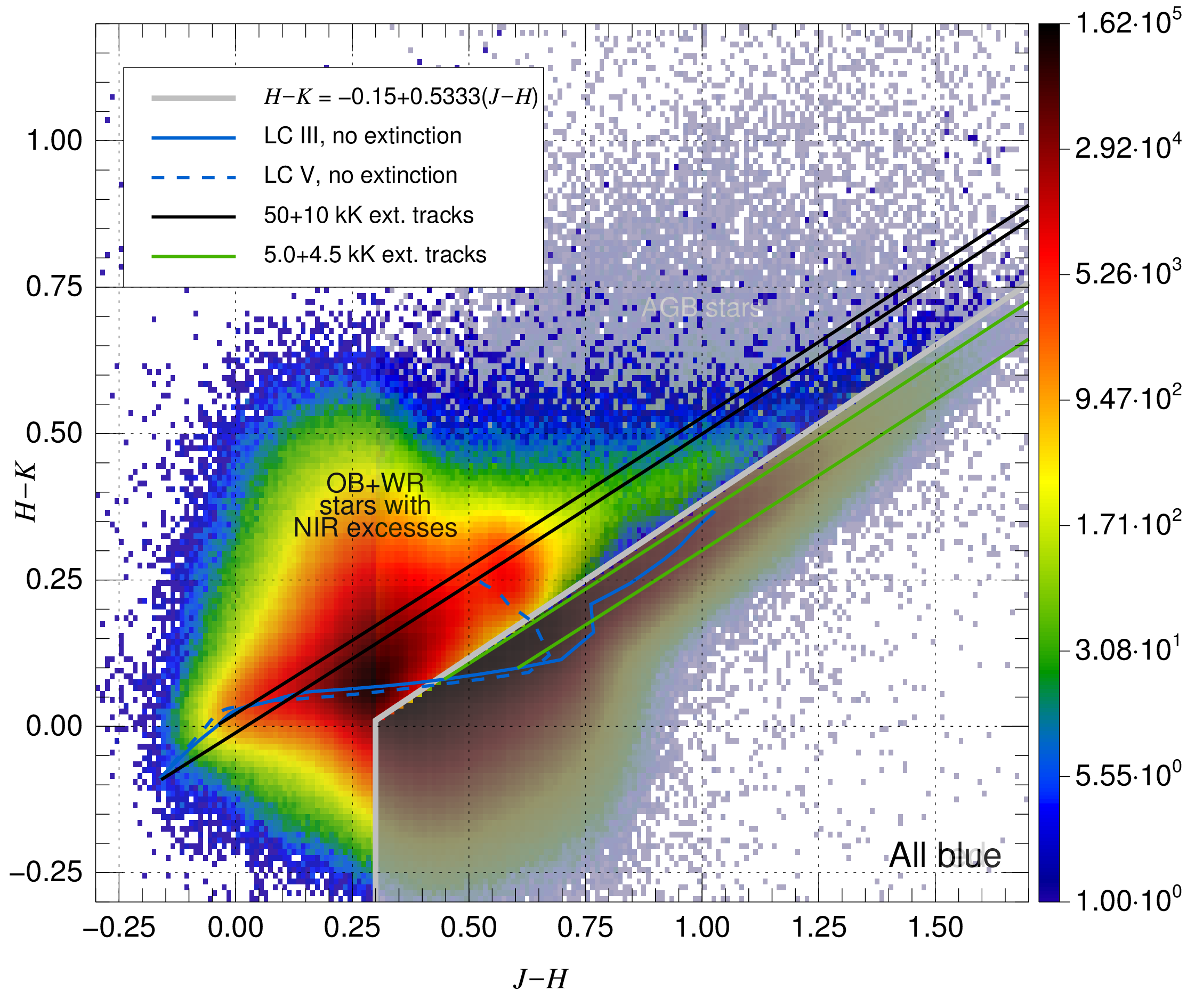} \
  \includegraphics*[width=0.49\linewidth]{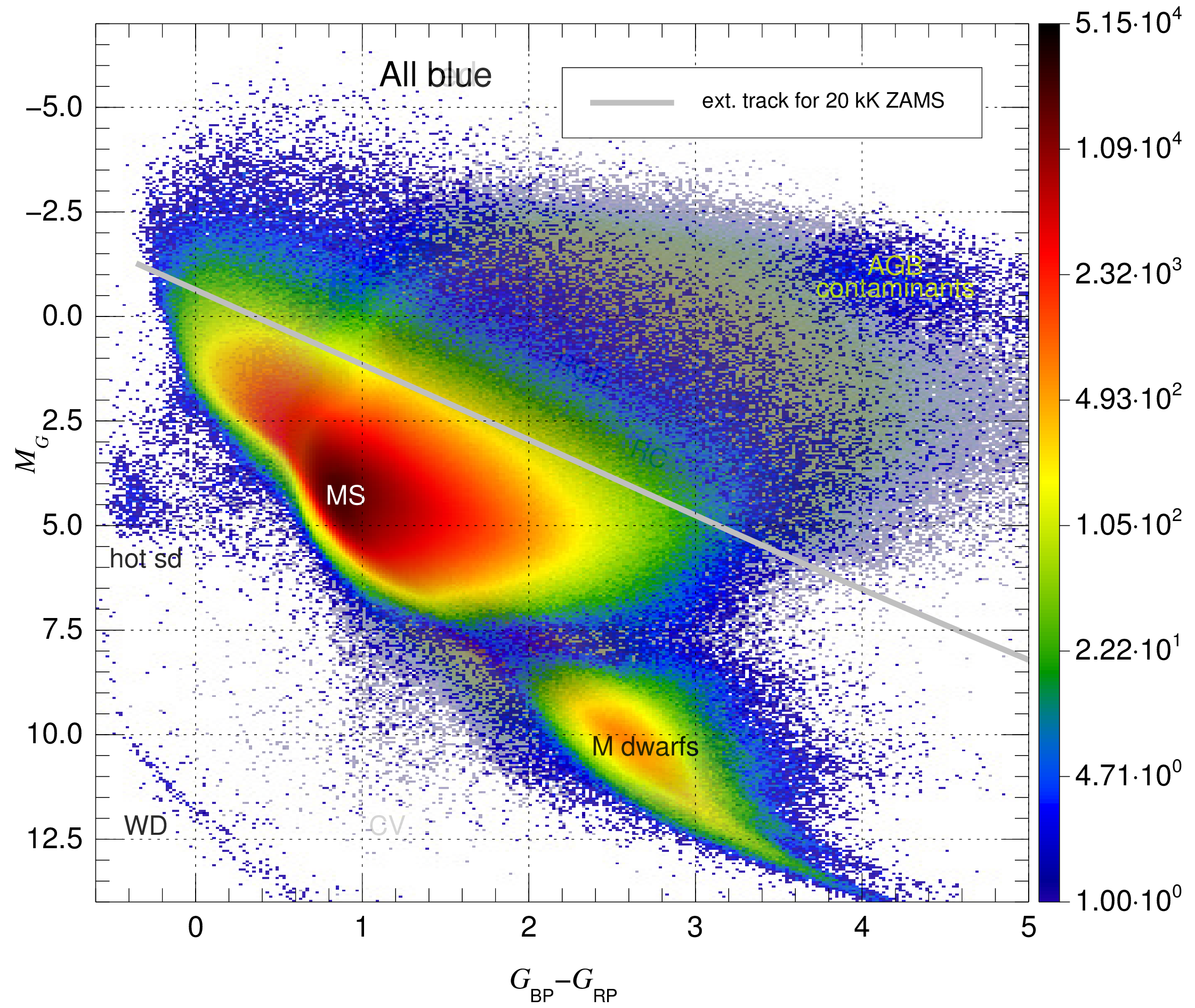}  
  \end{minipage}
  \newframe[0.4]
  \begin{minipage}{\linewidth}
  \includegraphics*[width=0.49\linewidth]{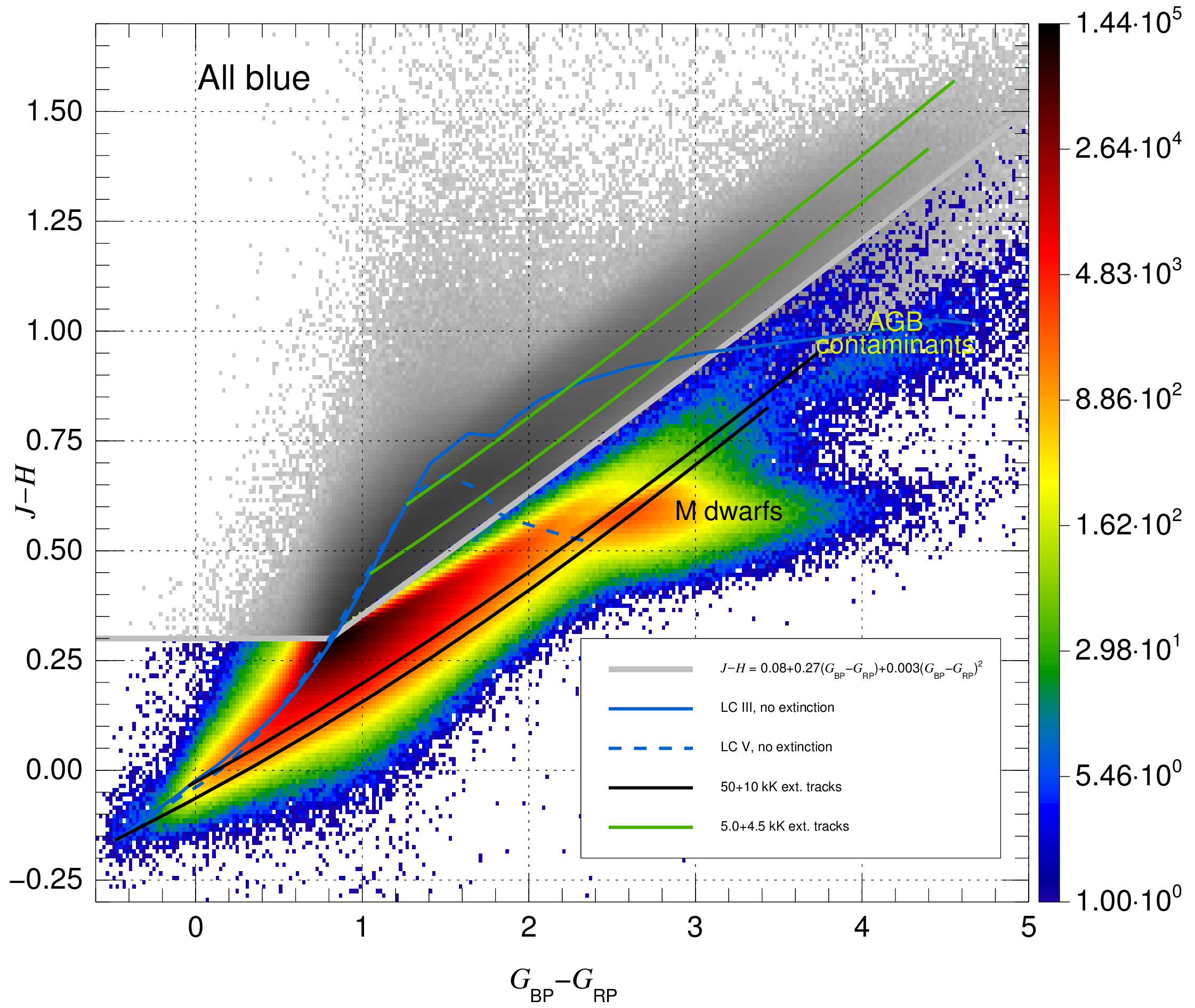} \
  \includegraphics*[width=0.49\linewidth]{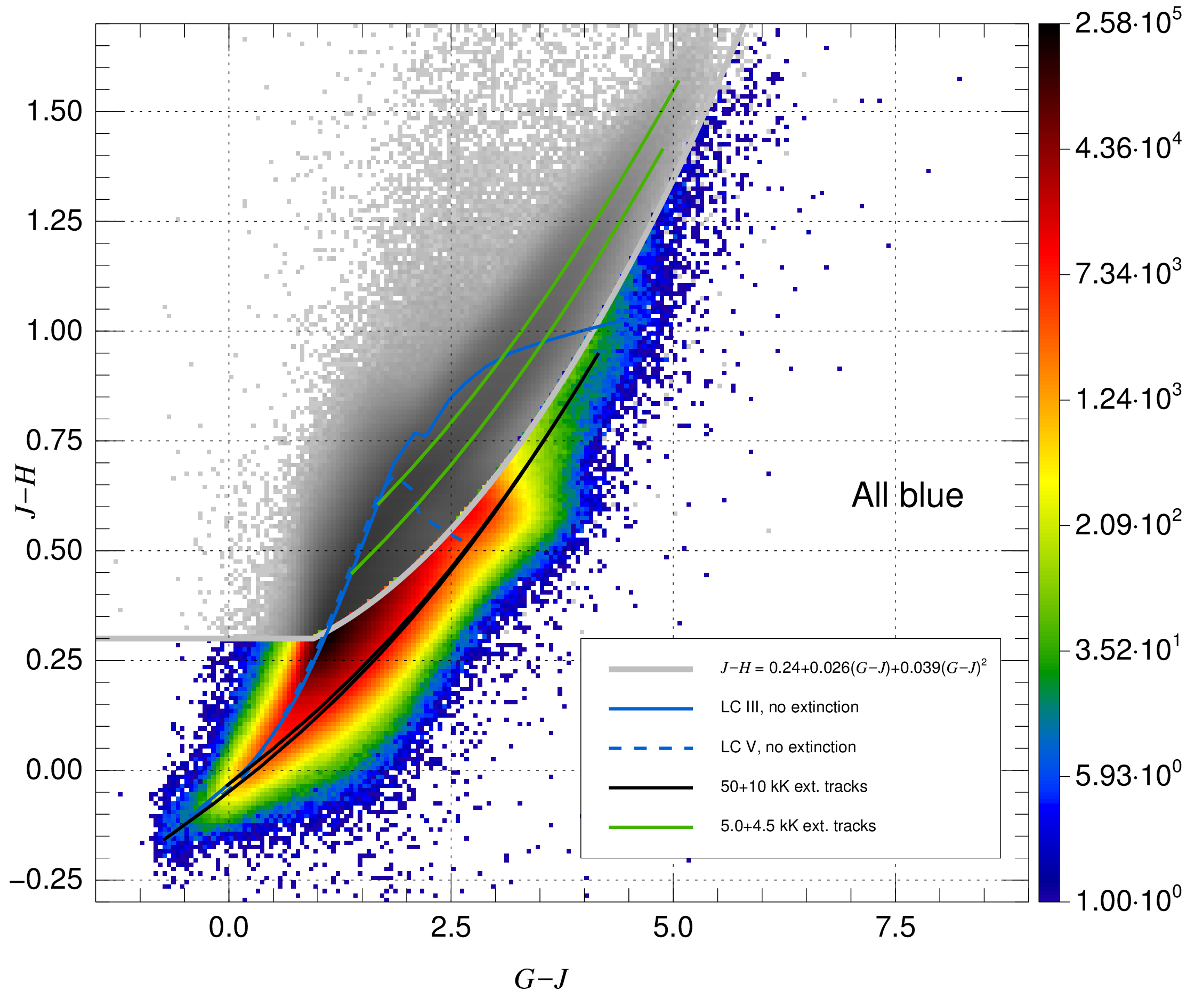}
  \\           
  \includegraphics*[width=0.49\linewidth]{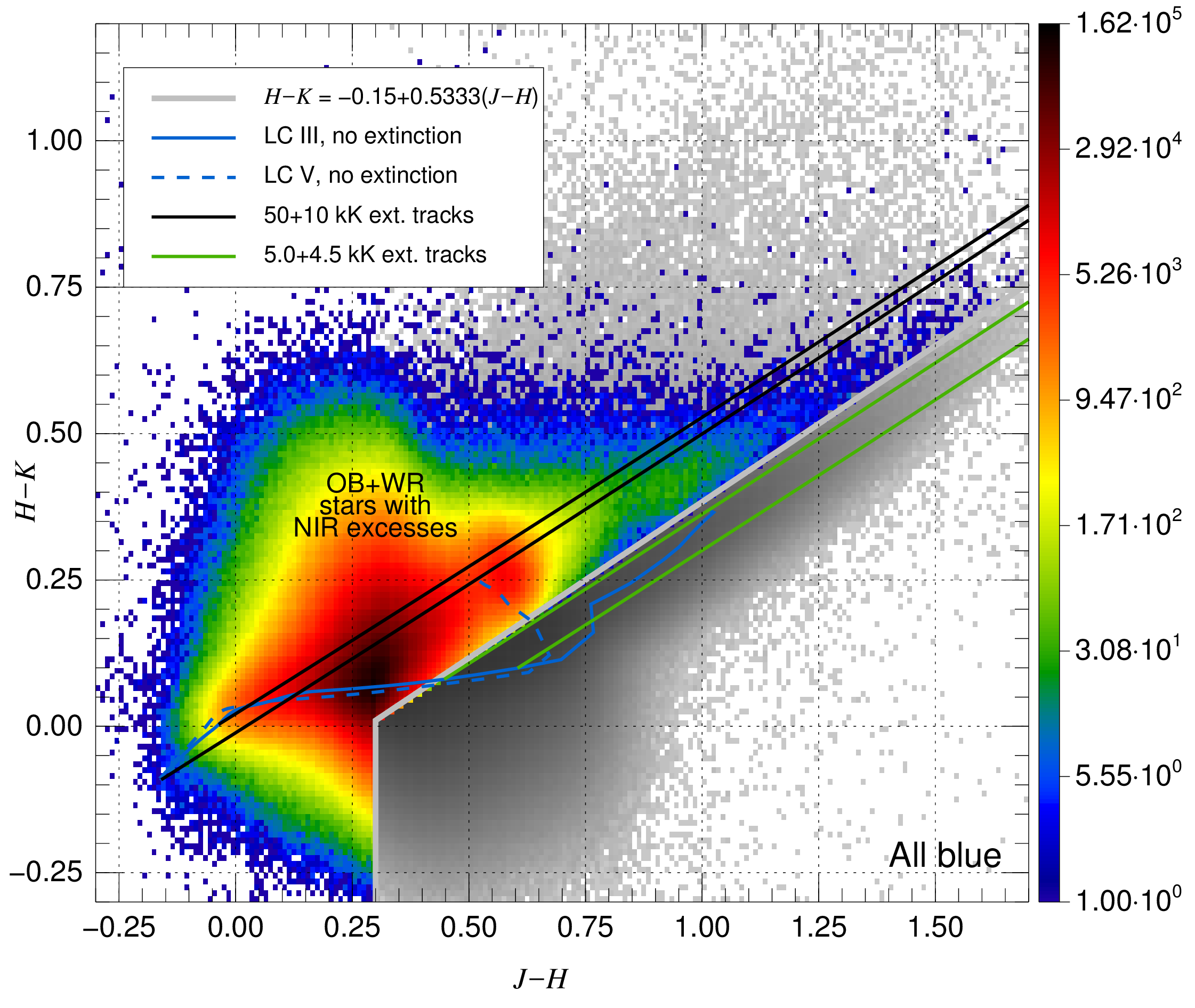} \
  \includegraphics*[width=0.49\linewidth]{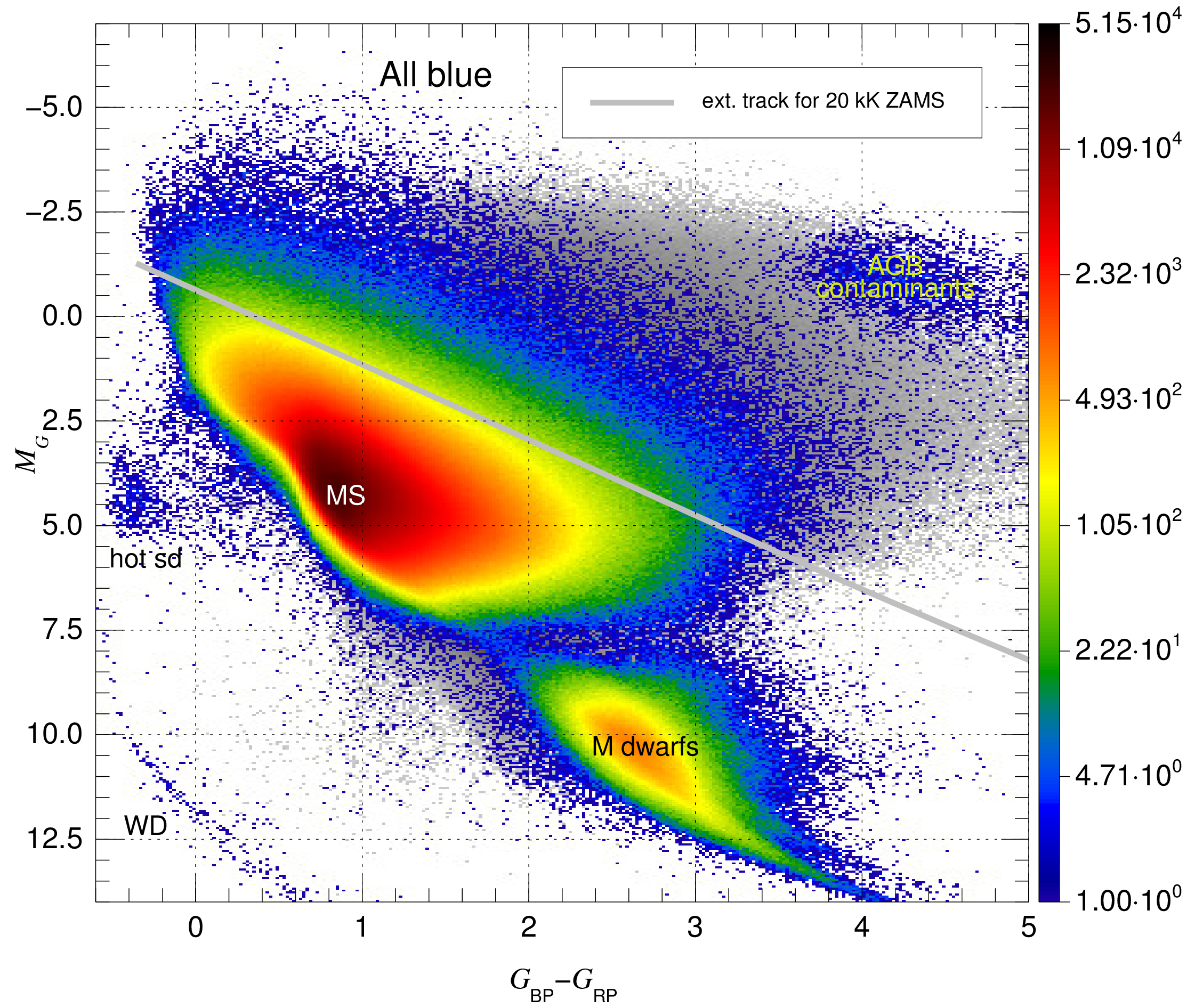}  
  \end{minipage}
  \newframe[5]
  \begin{minipage}{\linewidth}
  \includegraphics*[width=0.49\linewidth]{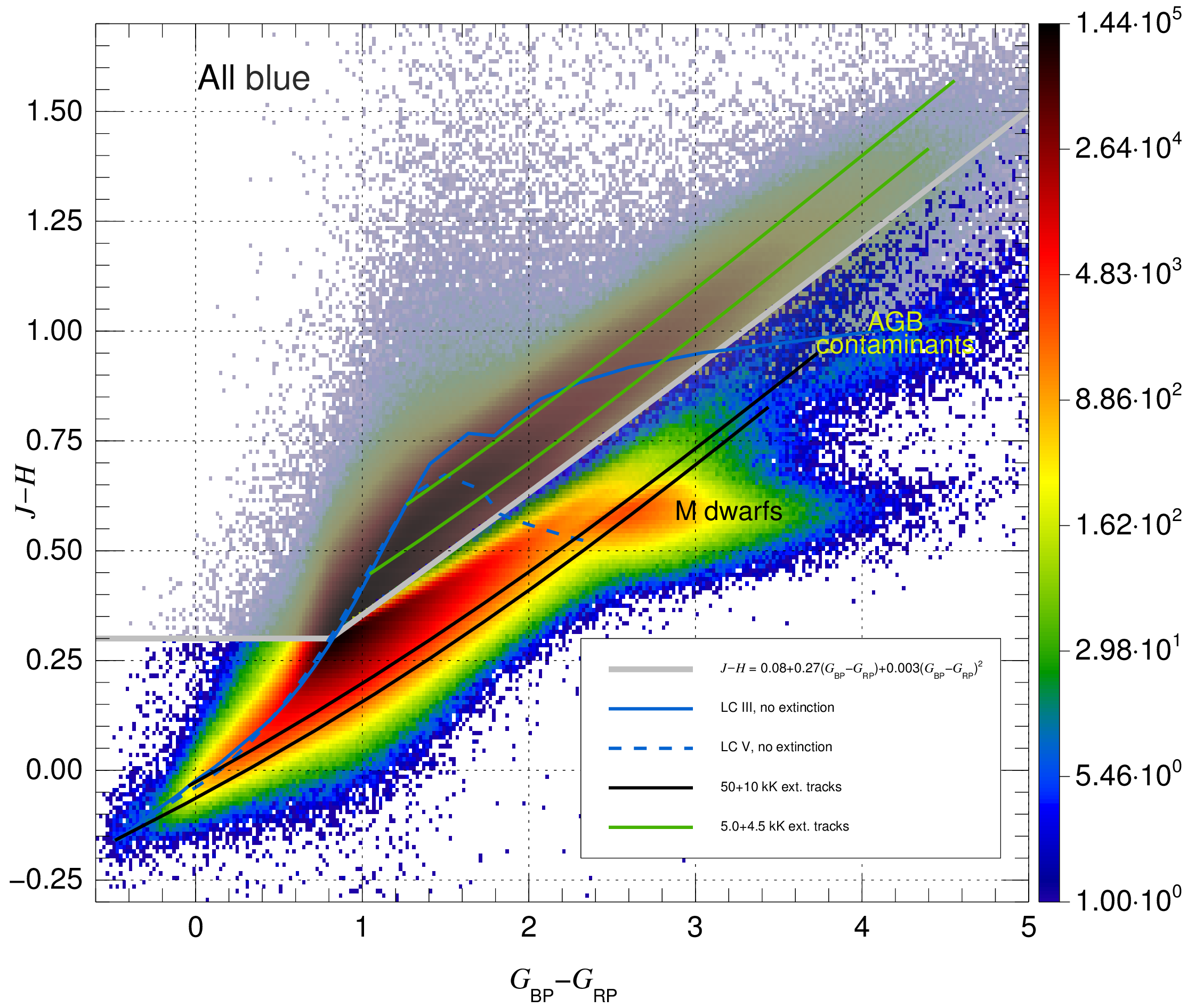} \
  \includegraphics*[width=0.49\linewidth]{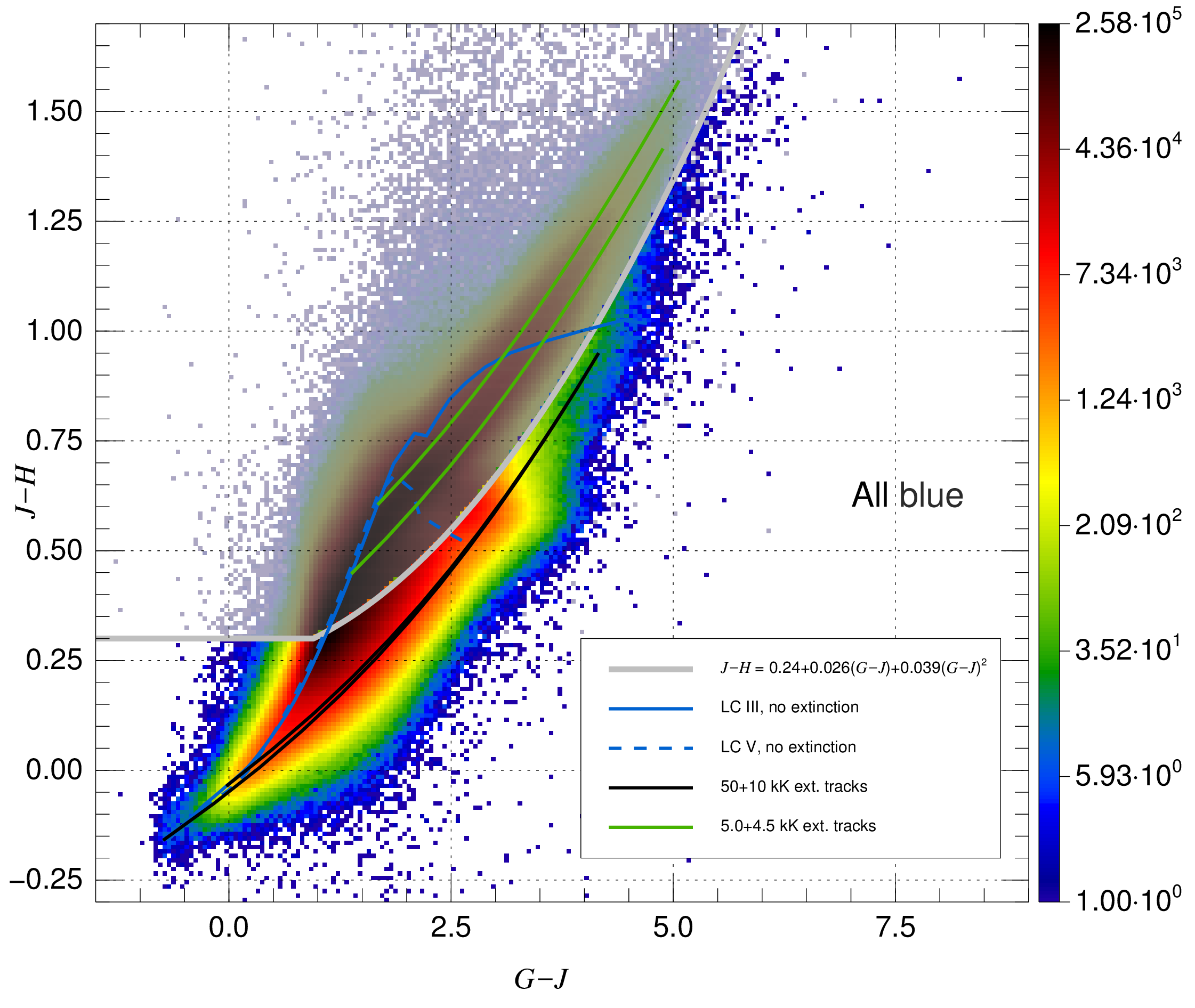}
  \\           
  \includegraphics*[width=0.49\linewidth]{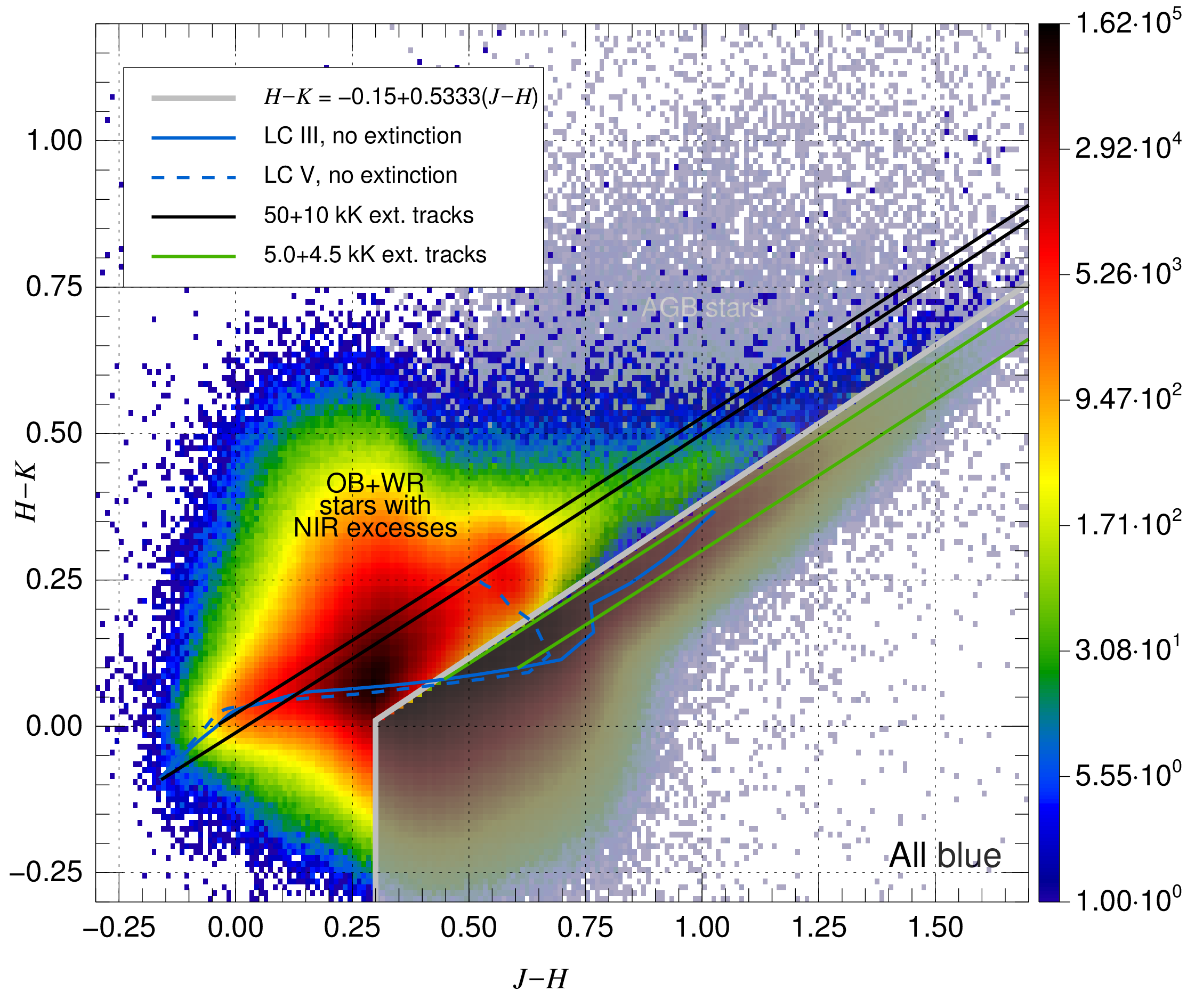} \
  \includegraphics*[width=0.49\linewidth]{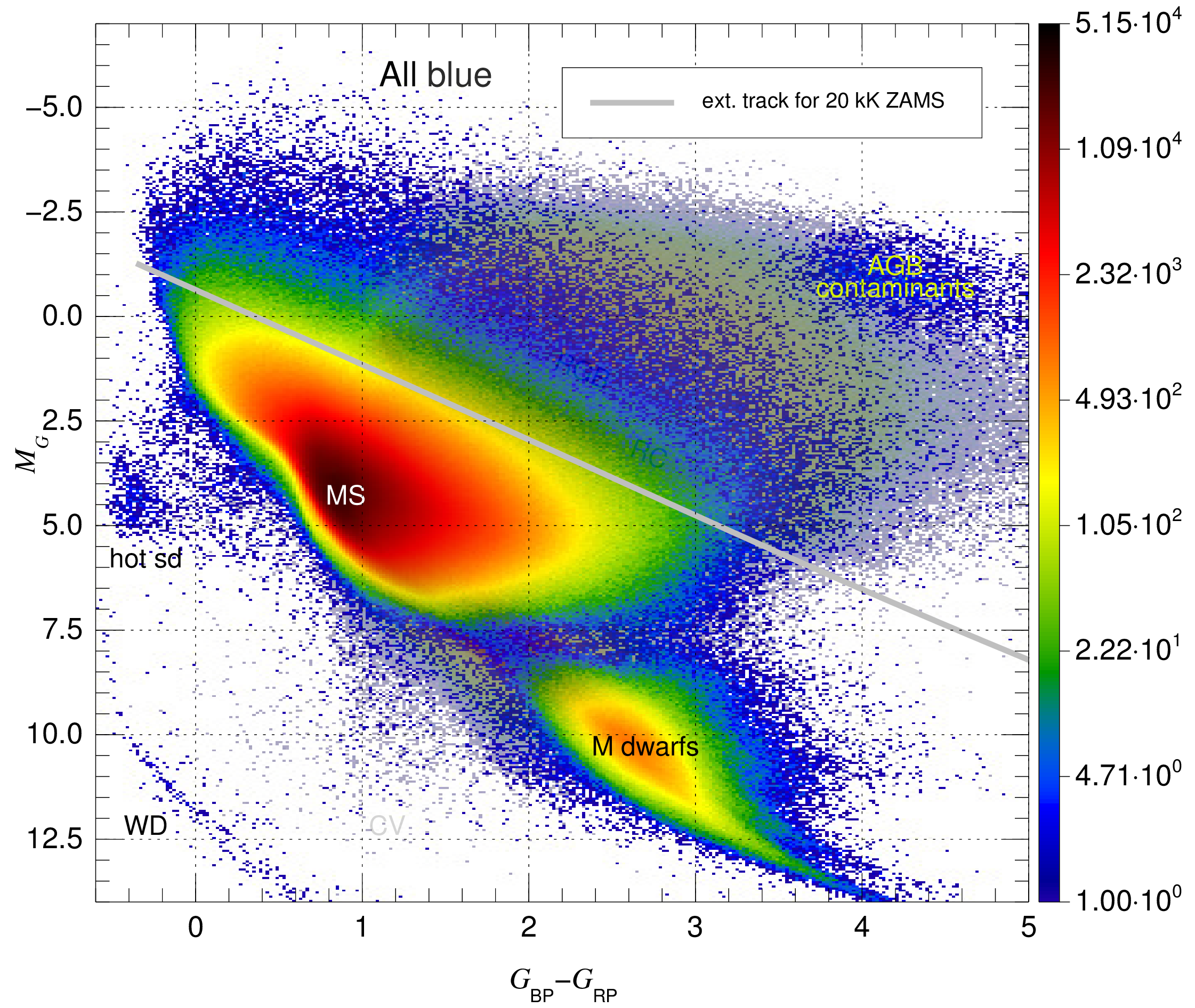}  
  \end{minipage}
  \newframe
  \begin{minipage}{\linewidth}
  \includegraphics*[width=0.49\linewidth]{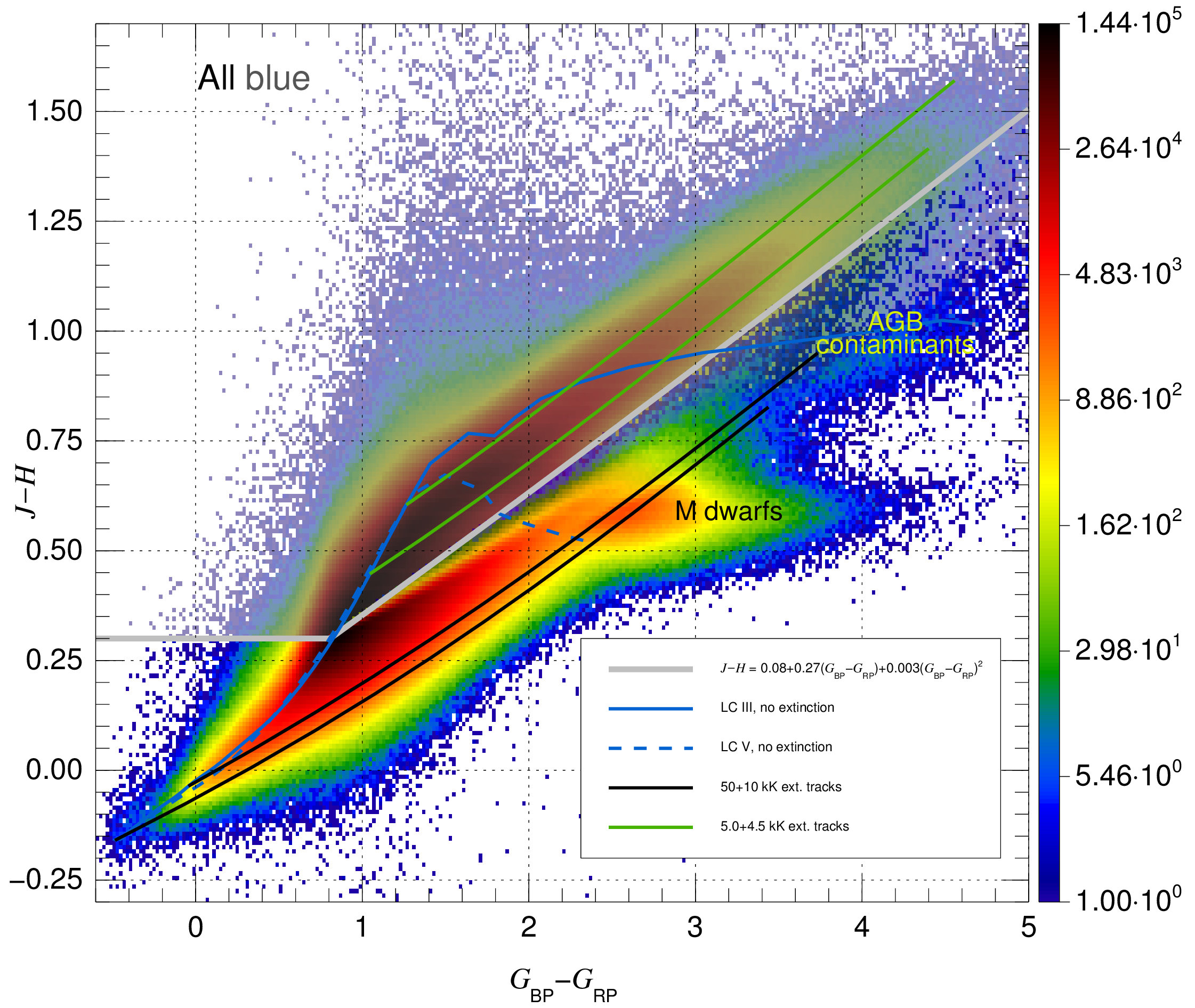} \
  \includegraphics*[width=0.49\linewidth]{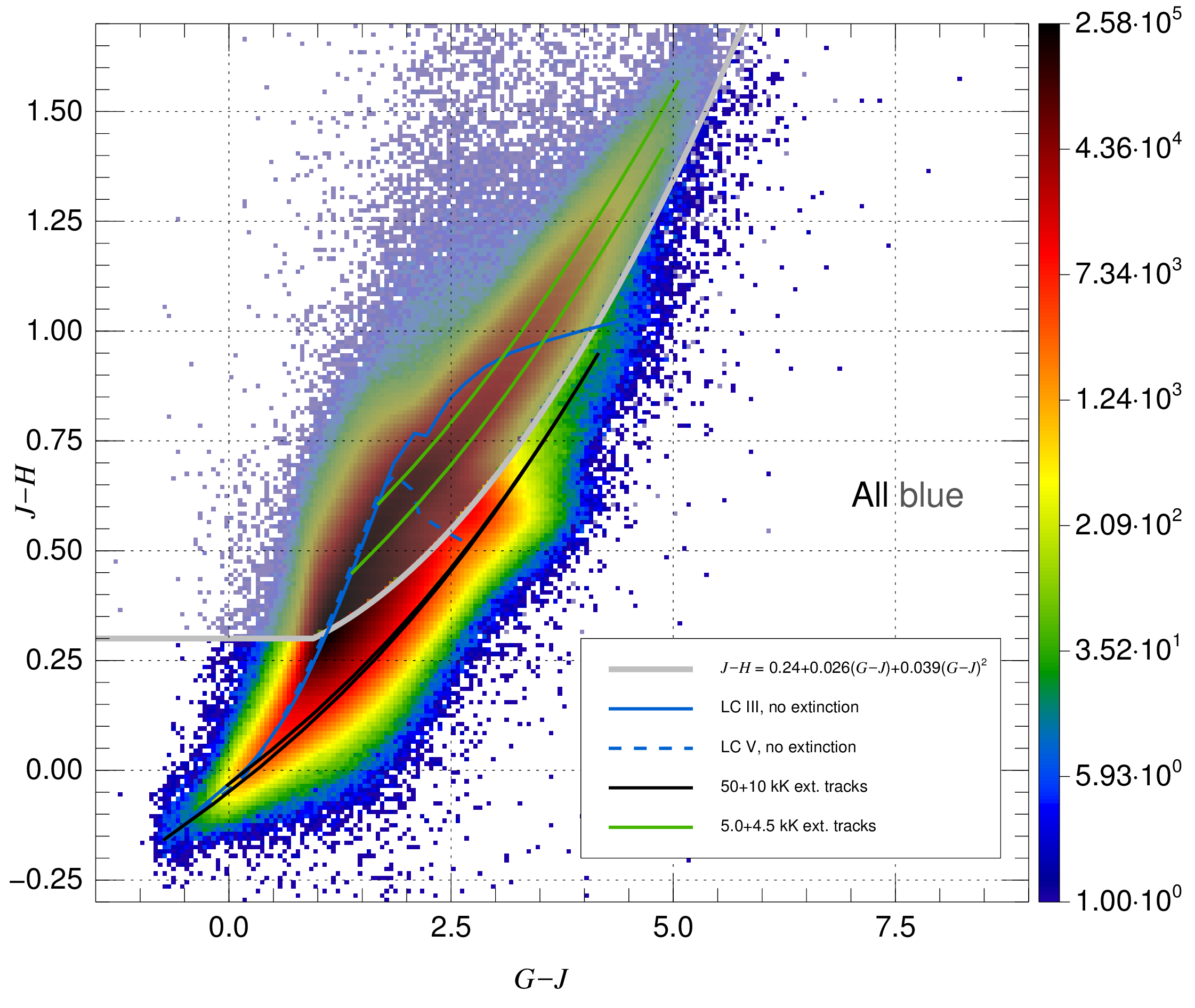}
  \\           
  \includegraphics*[width=0.49\linewidth]{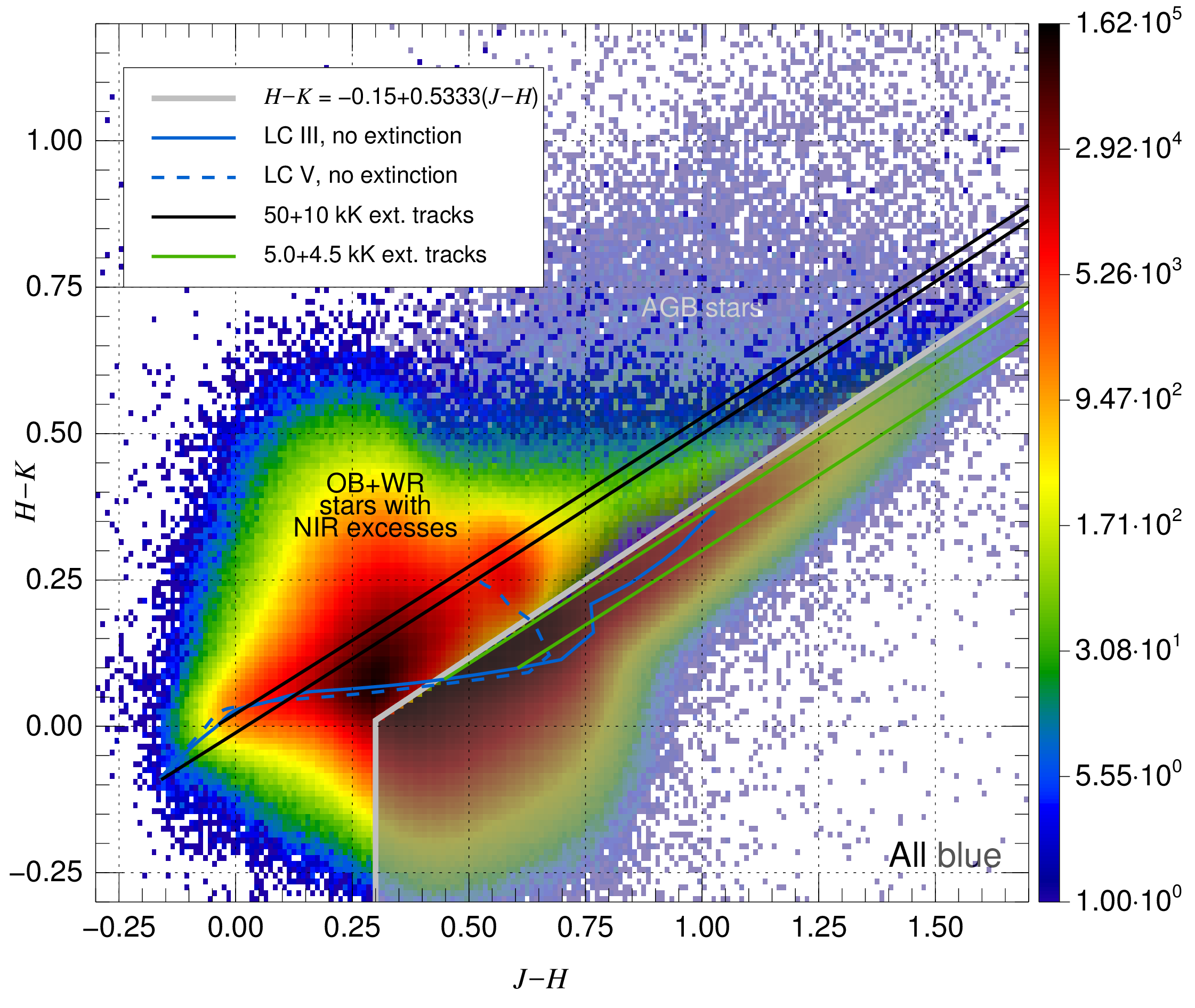} \
  \includegraphics*[width=0.49\linewidth]{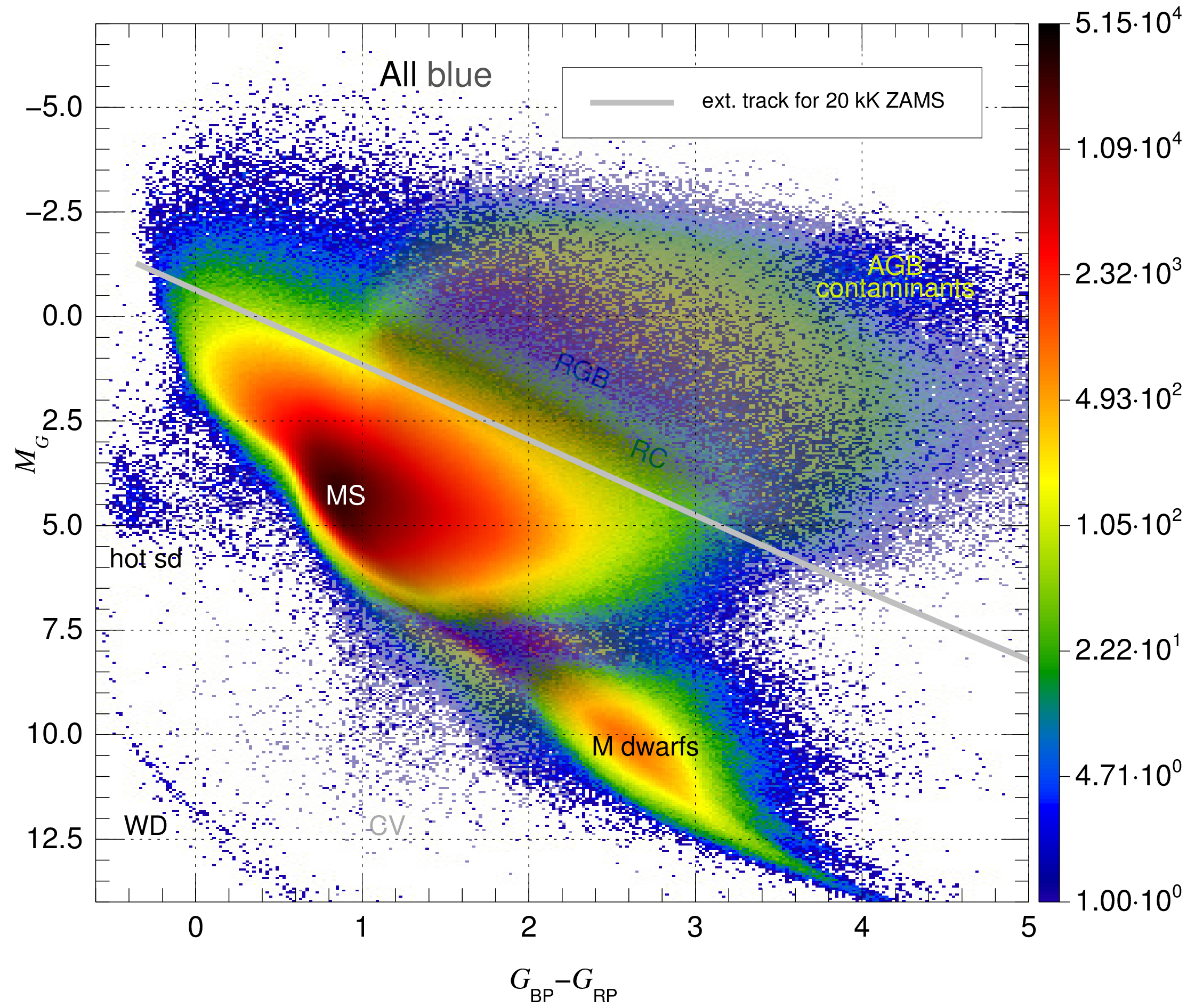}  
  \end{minipage}
  \newframe
  \begin{minipage}{\linewidth}
  \includegraphics*[width=0.49\linewidth]{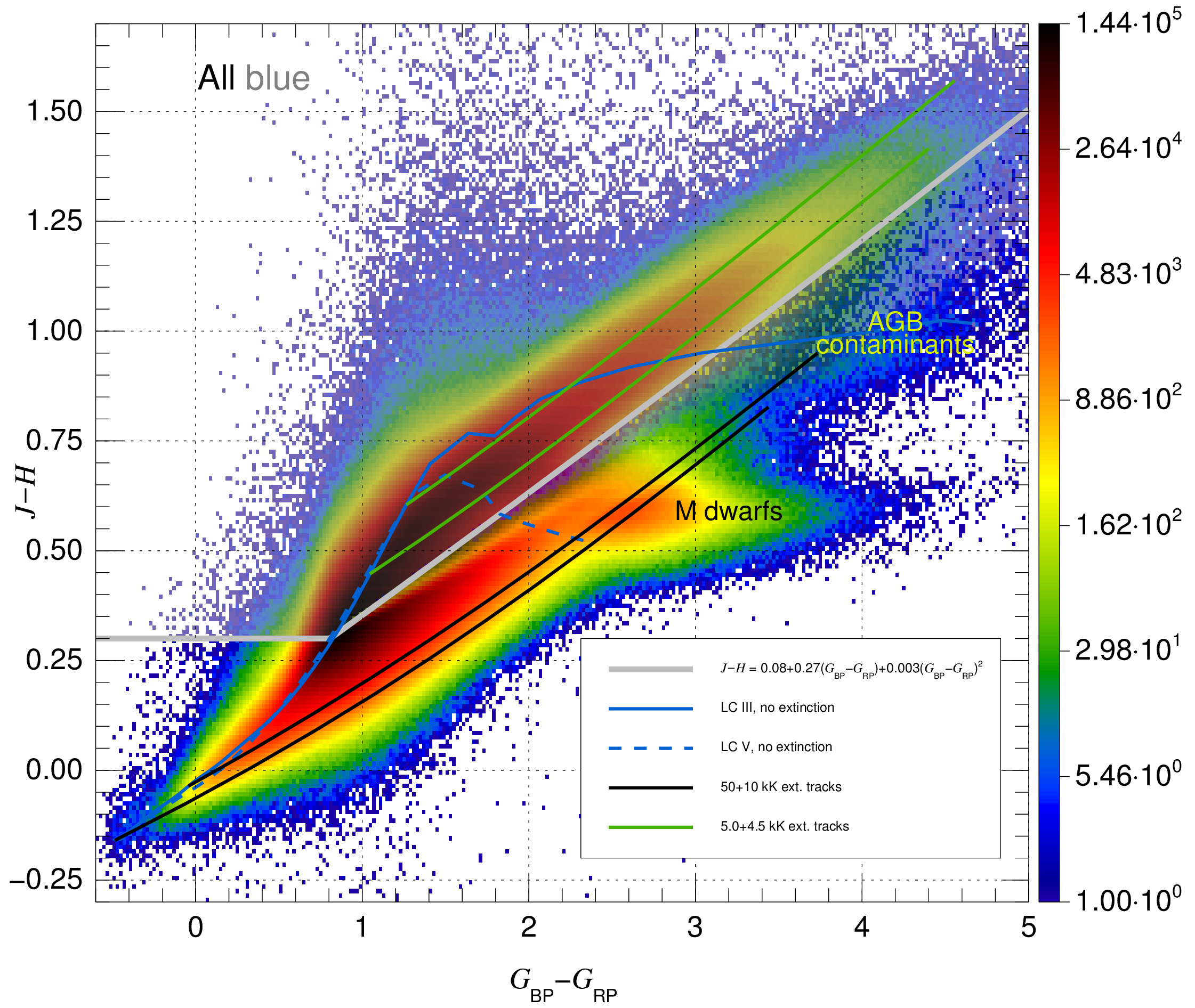} \
  \includegraphics*[width=0.49\linewidth]{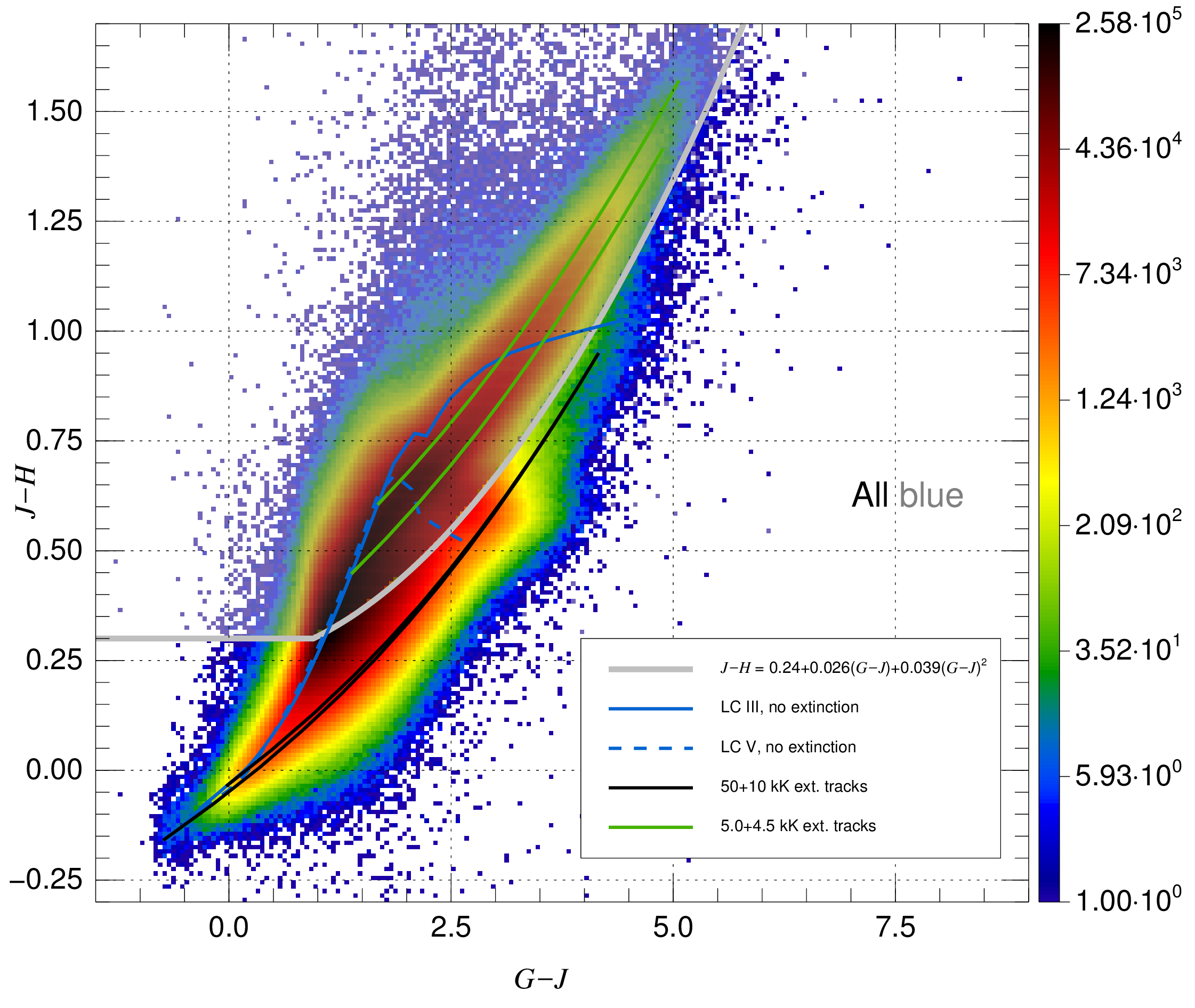}
  \\           
  \includegraphics*[width=0.49\linewidth]{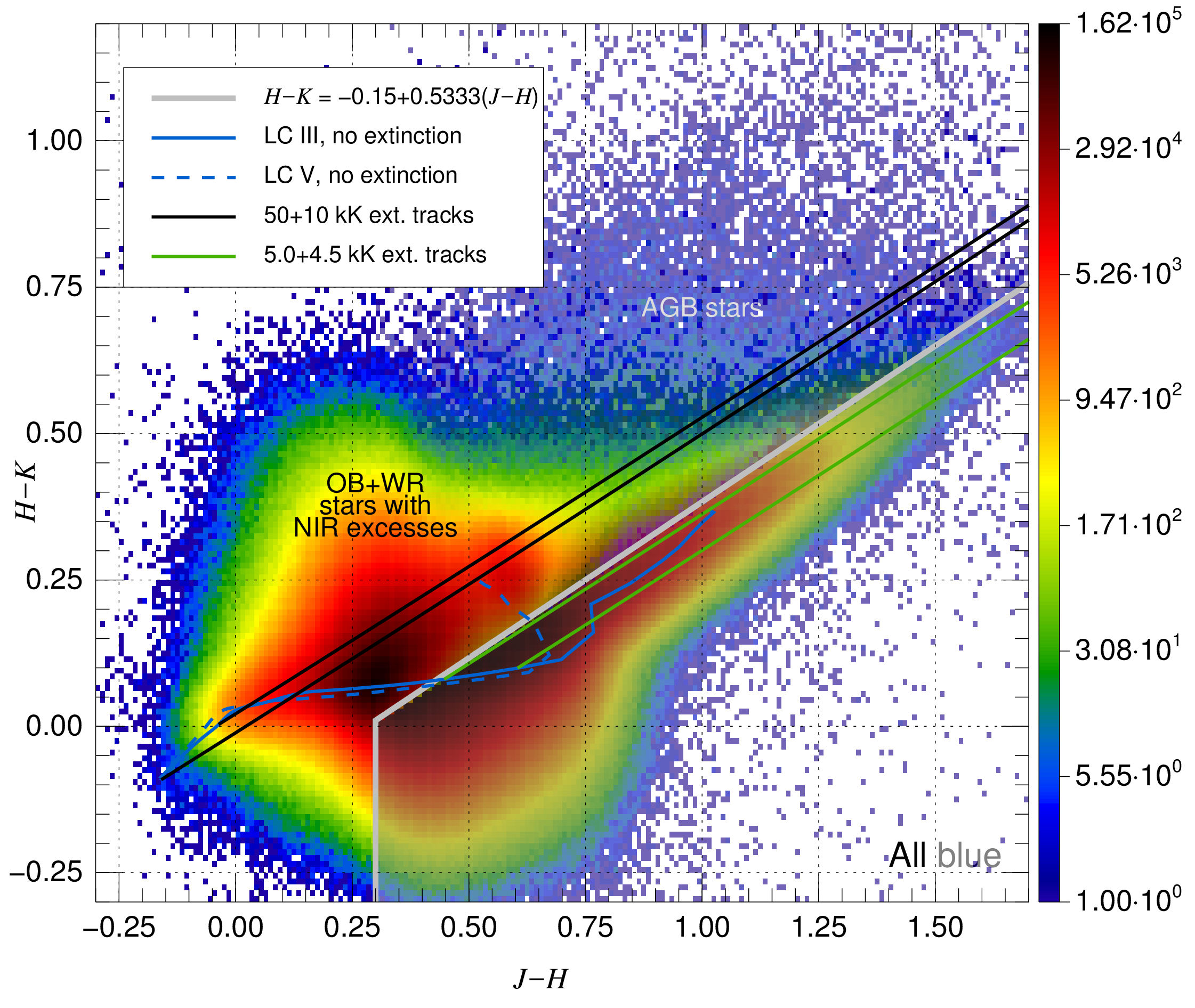} \
  \includegraphics*[width=0.49\linewidth]{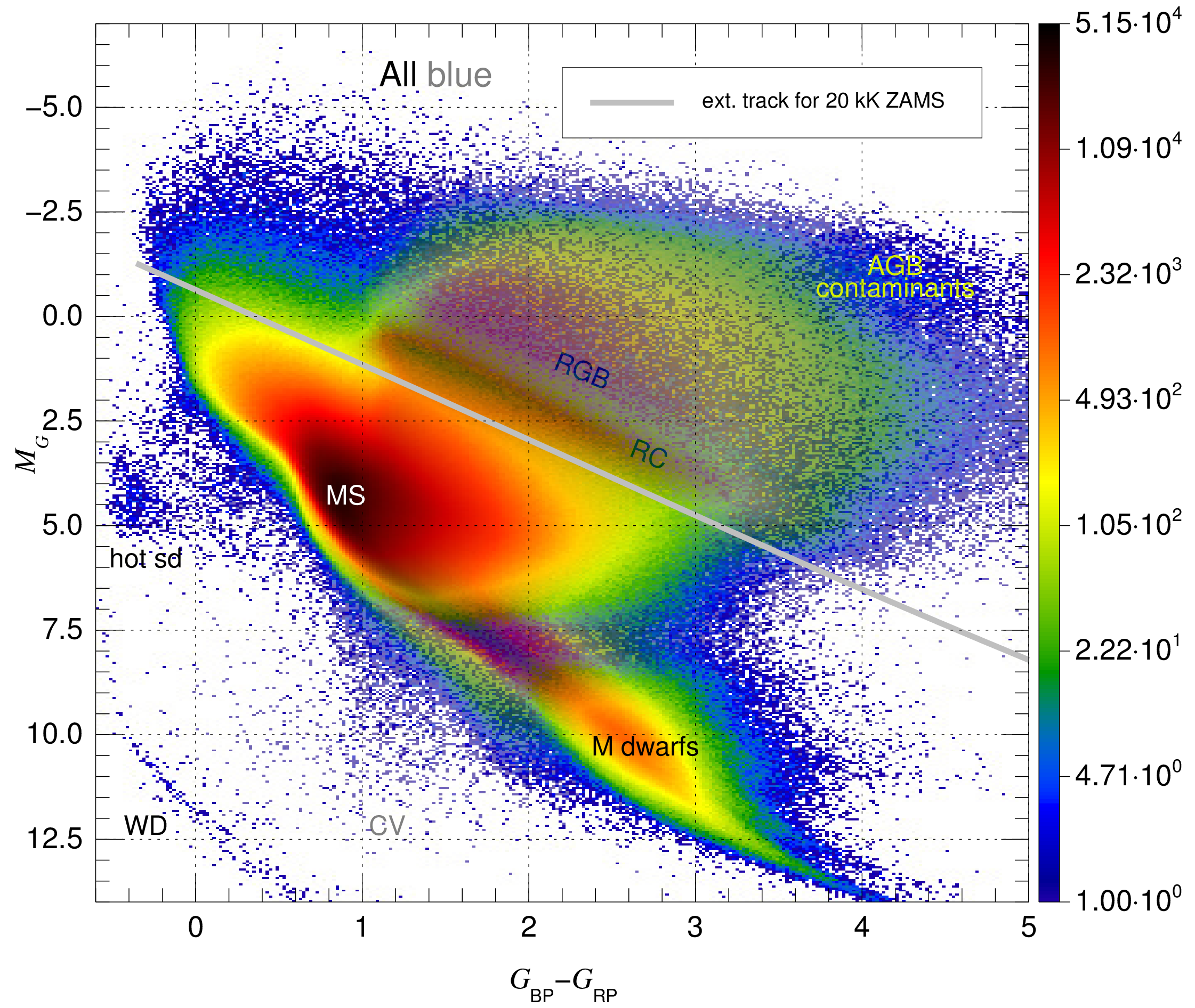}  
  \end{minipage}
  \newframe
  \begin{minipage}{\linewidth}
  \includegraphics*[width=0.49\linewidth]{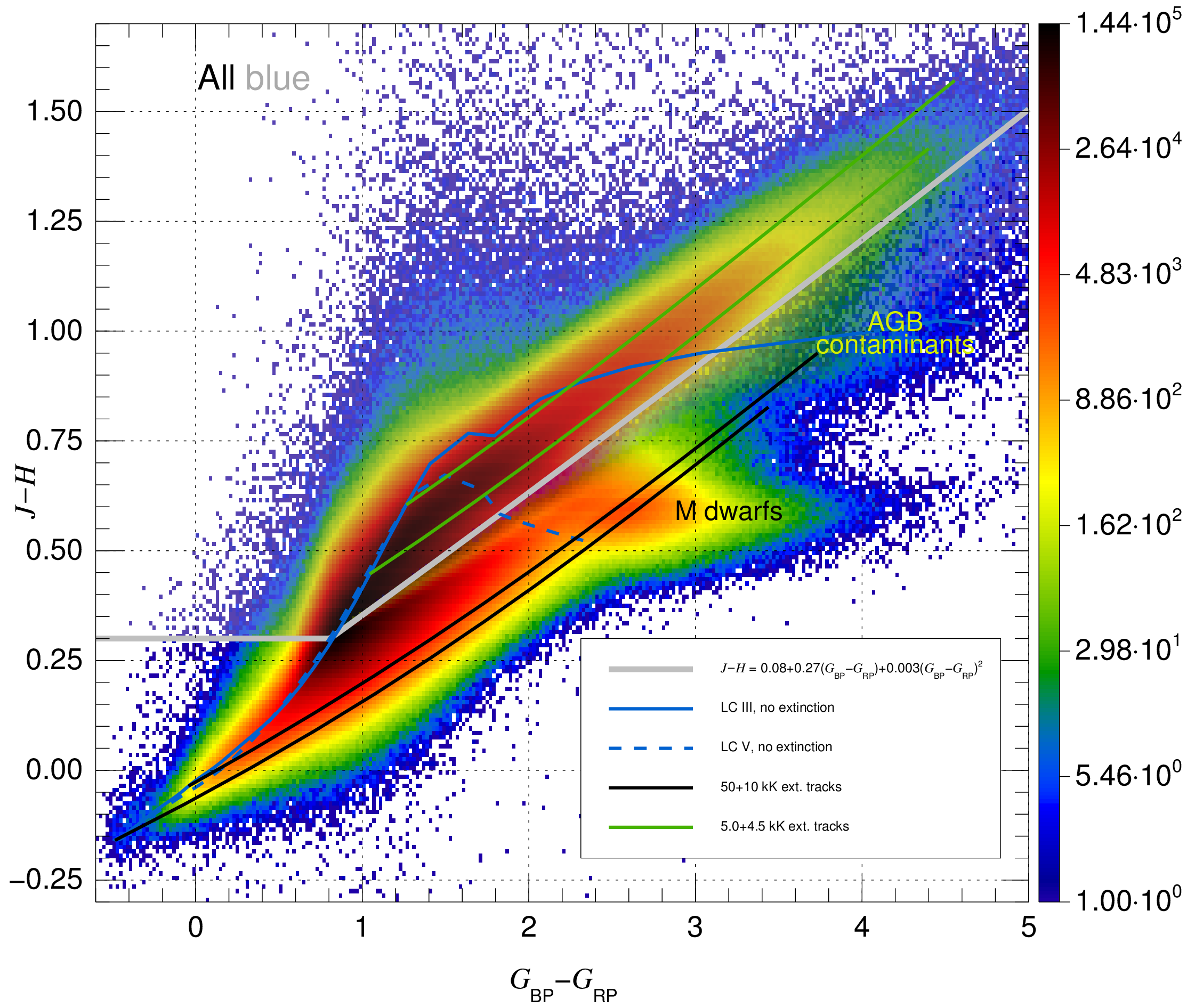} \
  \includegraphics*[width=0.49\linewidth]{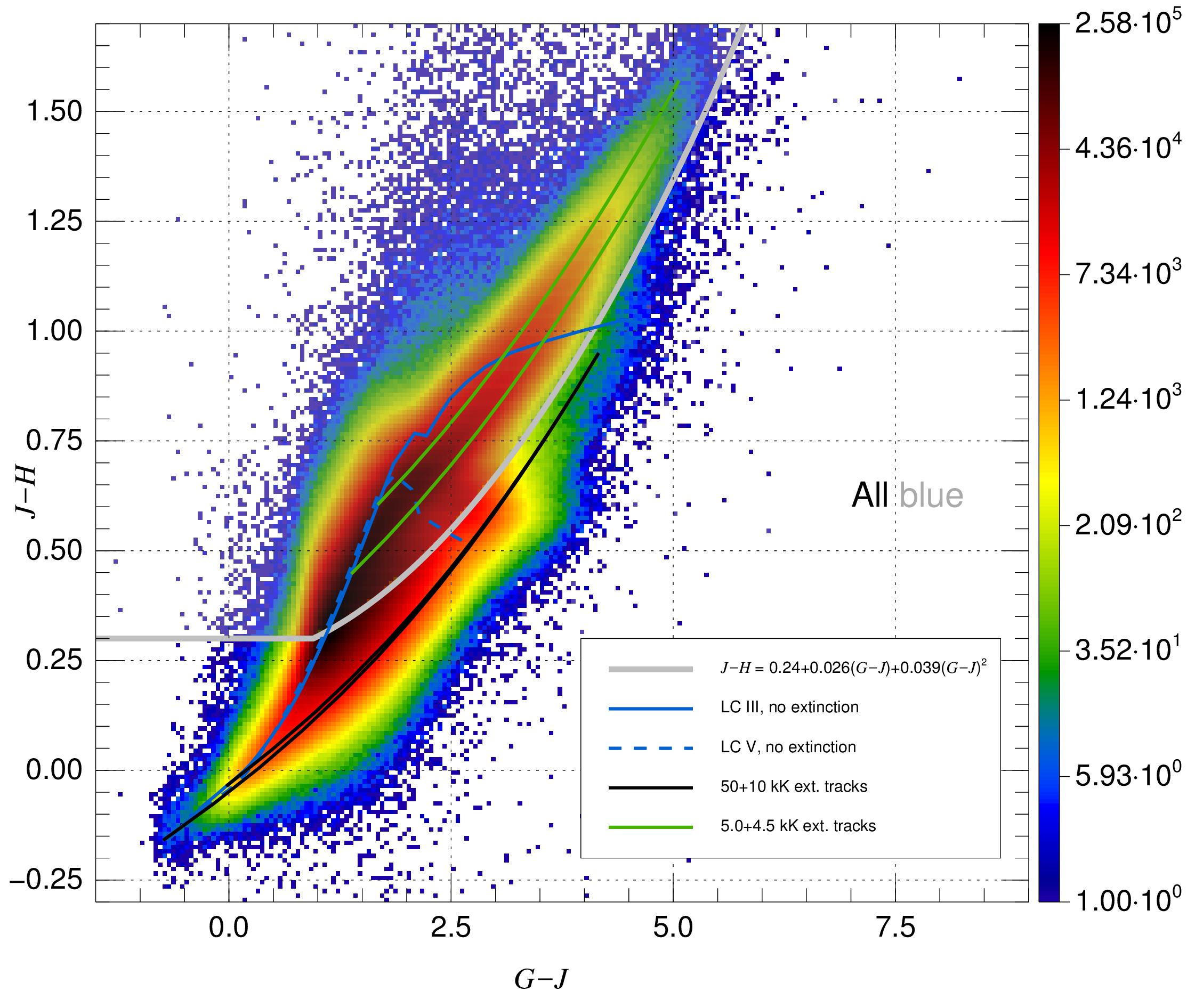}
  \\           
  \includegraphics*[width=0.49\linewidth]{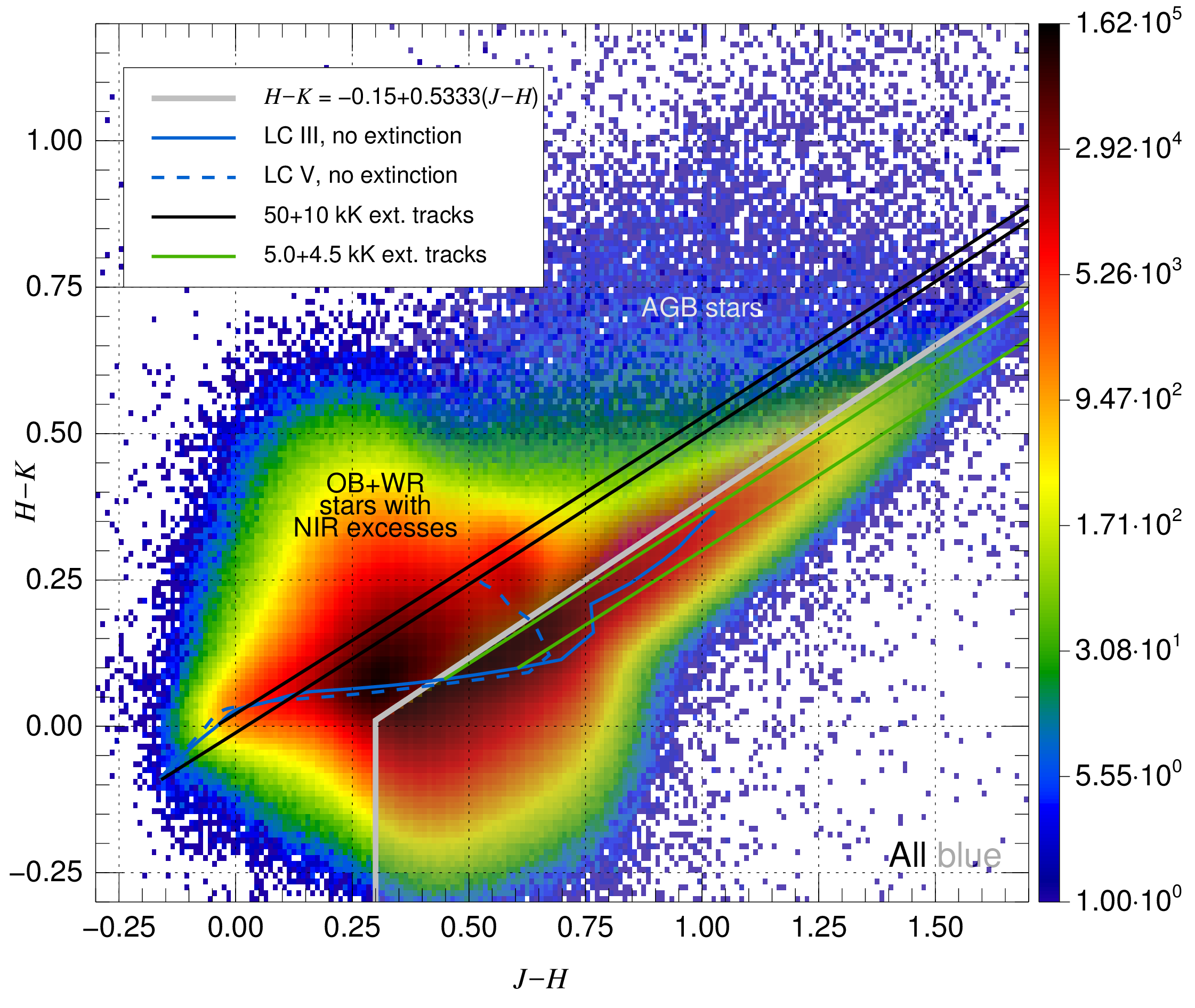} \
  \includegraphics*[width=0.49\linewidth]{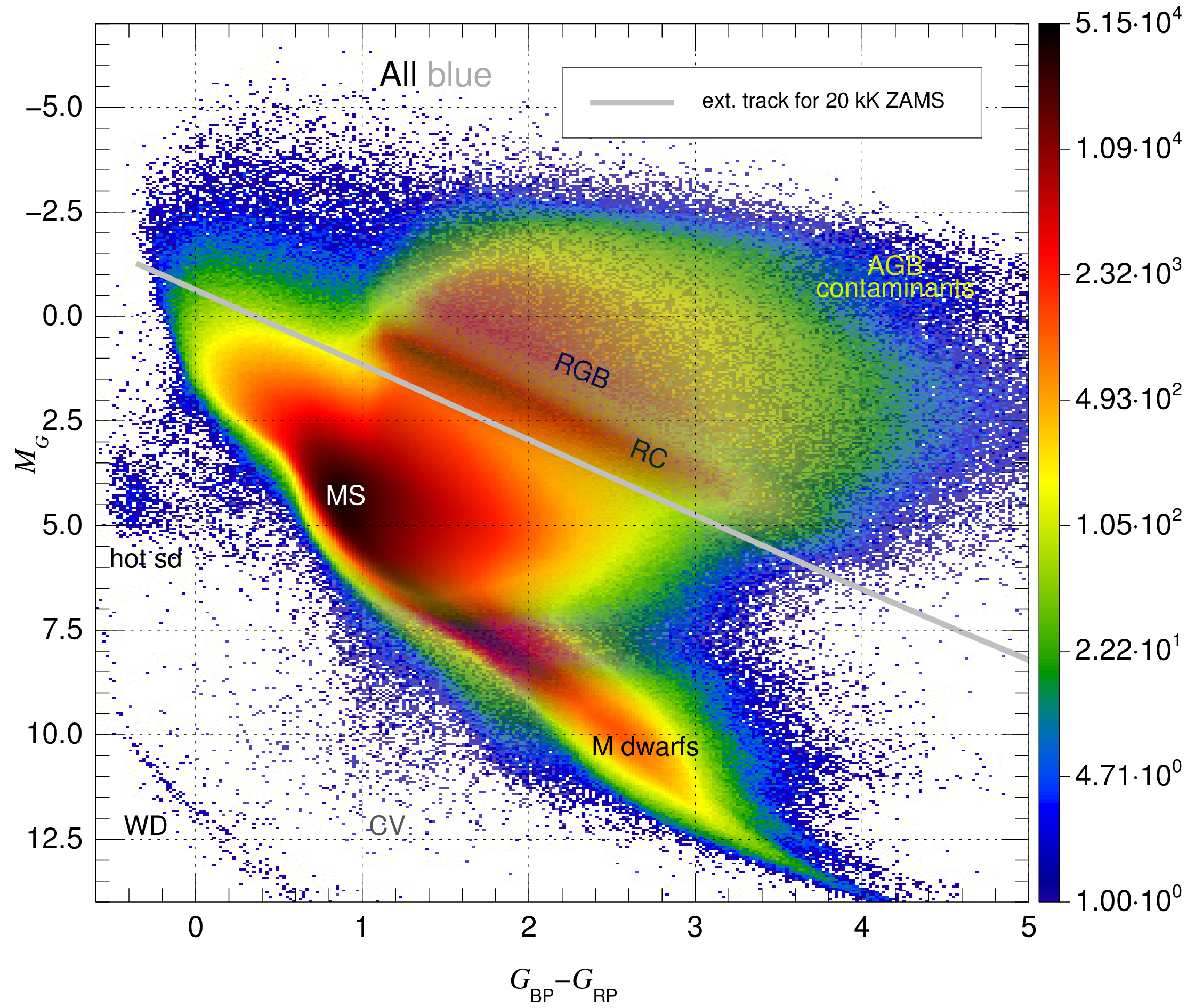}  
  \end{minipage}
  \newframe
  \begin{minipage}{\linewidth}
  \includegraphics*[width=0.49\linewidth]{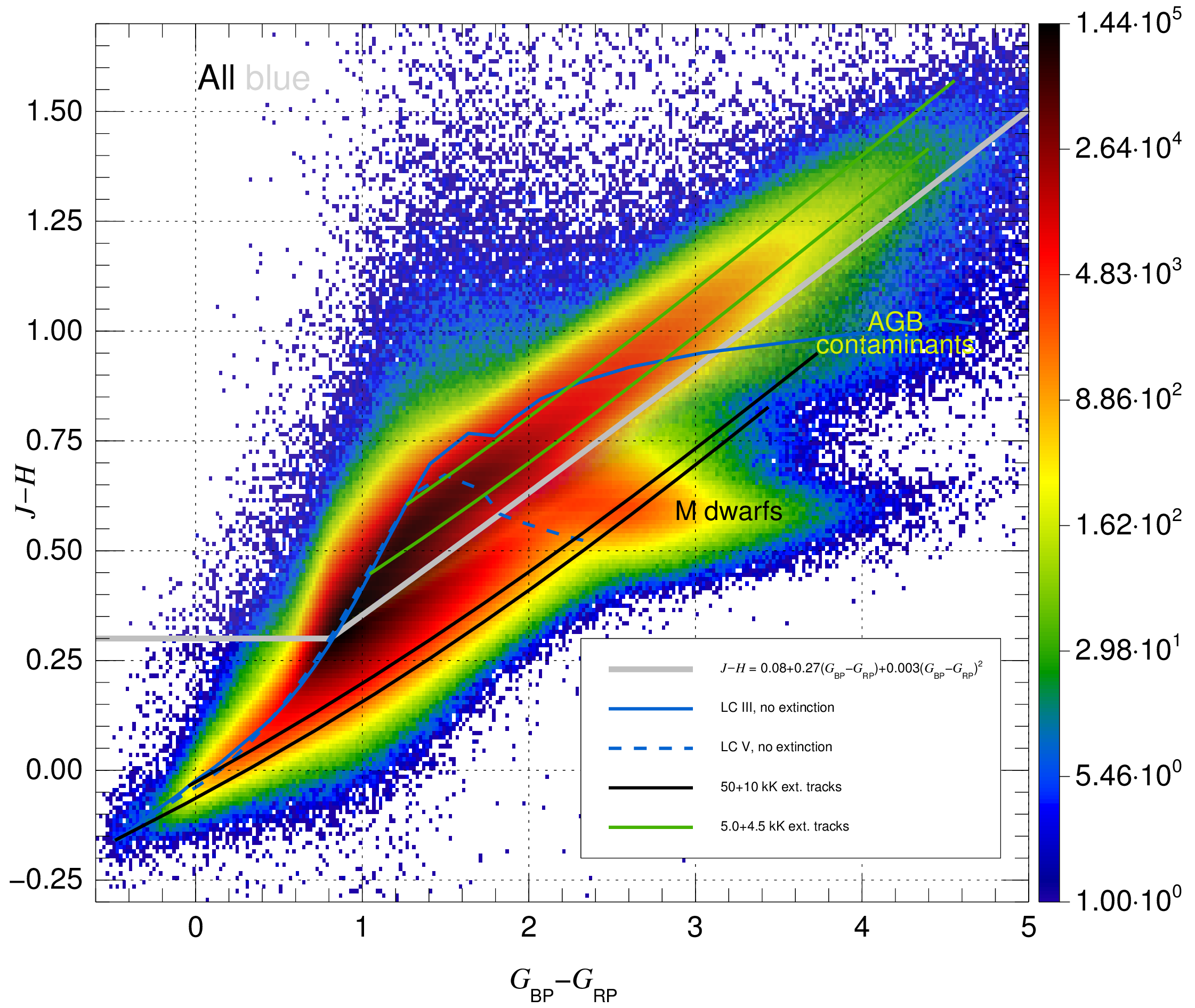} \
  \includegraphics*[width=0.49\linewidth]{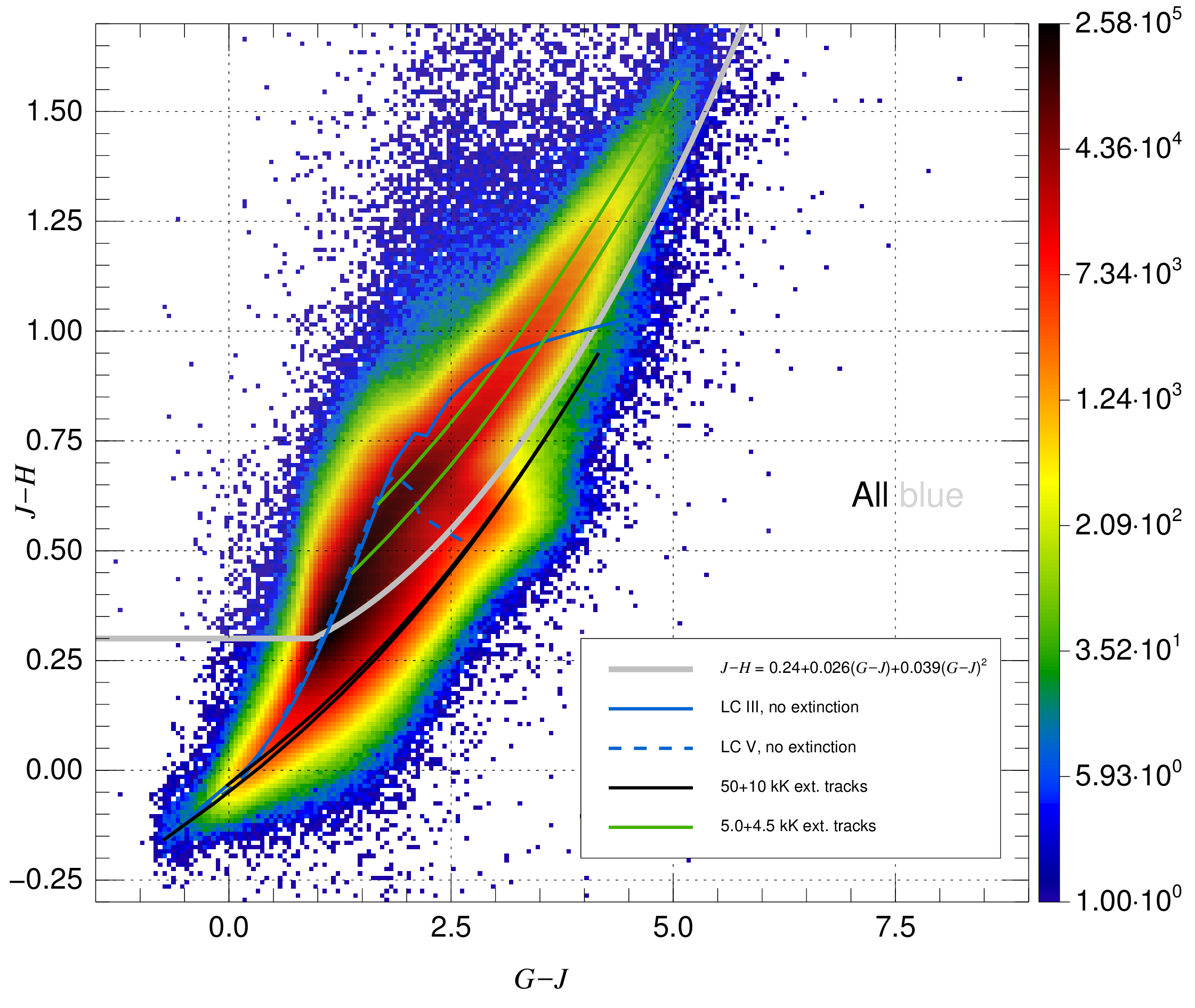}
  \\           
  \includegraphics*[width=0.49\linewidth]{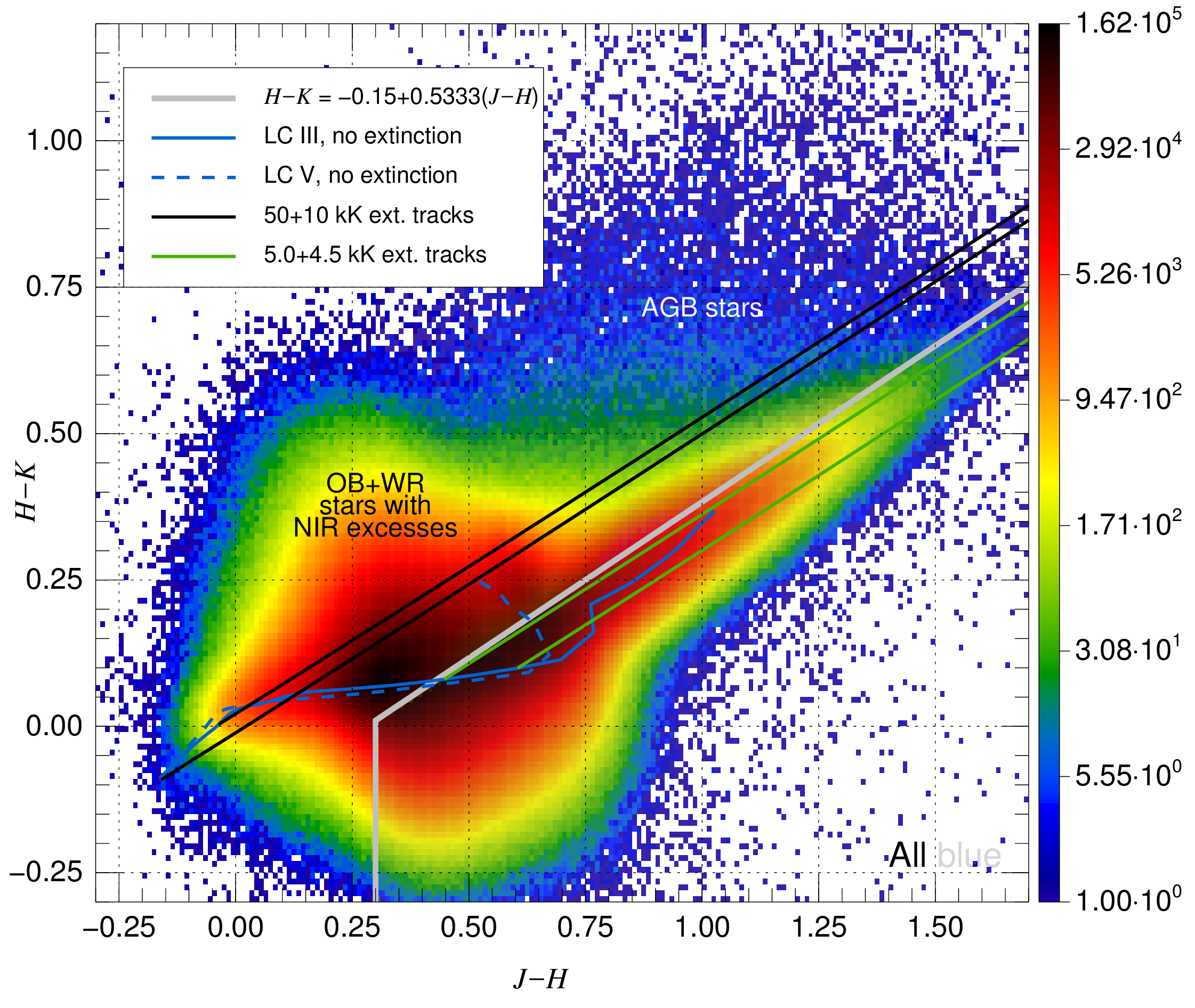} \
  \includegraphics*[width=0.49\linewidth]{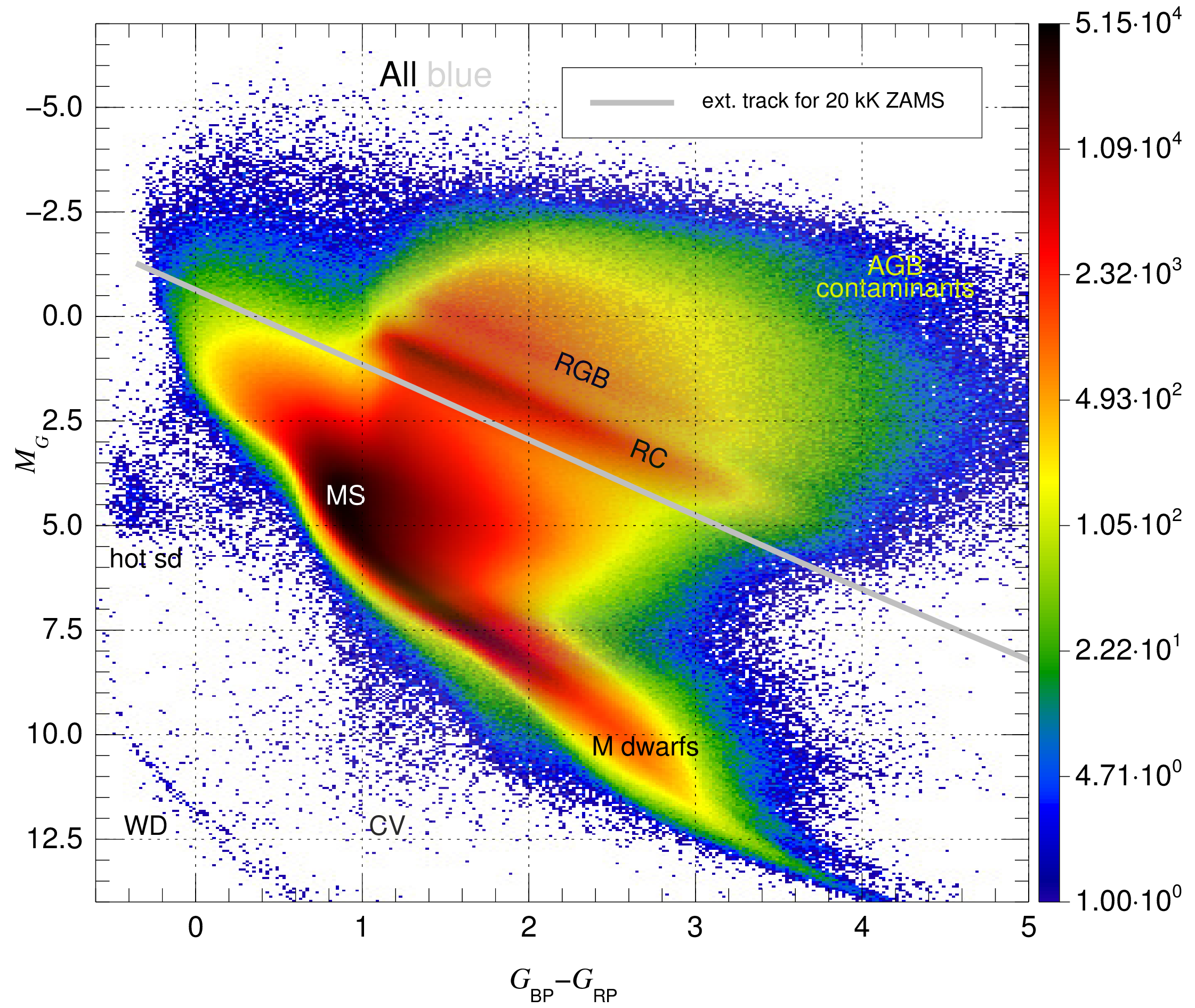}  
  \end{minipage}
  \end{animateinline}
             }
 \caption{Colour-colour selection diagrams and CAMD output for the full sample. On an {\cre Adobe-compatible viewer} the buttons at the
          bottom activate the animation and allow the reader to cycle between the full, cool (red), and hot (blue) samples.}
 \label{fig1}   
\end{figure}

The criteria were developed in the three colour-colour planes of Fig.~\ref{fig1} using as reference the ALS survey
\citep{Pantetal21,Pantetal25b} of hot massive stars and more specifically the second of those papers
(ALS~III). To satisfy the first condition (low false positive rate) it became clear that the criteria from all three planes had to
be satisfied simultaneously, as otherwise the false positive rate sample increased. The second condition required that the criteria
followed trajectories quasi-parallel to the extinction tracks, for which we used the combined SED grid of \citet{Maiz13a} and the
extinction laws of \citet{Maizetal14a}. The second condition was relaxed at the blue
end to accommodate the absence of cool stars with blue colours. To satisfy the third condition, we used the ALS~III sample to
determine the exact location of the colour-colour criteria, minimising the false negative rate from the ALS~III sample without 
significantly increasing the false positive rate from the first condition. The resulting three criteria required to include a star
in the hot sample, plotted in Fig.~\ref{fig1} as grey lines, are:

$\,\!$

\vfill

\eject

\begin{itemize}
 \item $J-H \le 0.30$ or $J-H <  0.08 + 0.27(\BPRP) + 0.003(\BPRP)^2$
 \item $J-H \le 0.30$ or $J-H <  0.24 + 0.026(G-J)  + 0.039(G-J)^2$
 \item $J-H \le 0.30$ or $H-K > -0.15 + 8(J-H)/15$
\end{itemize}


\begin{figure}
 \centerline{
  \begin{animateinline}[loop,controls,buttonsize=1em]{0.4}
  \begin{minipage}{\linewidth}
  \includegraphics*[width=0.49\linewidth]{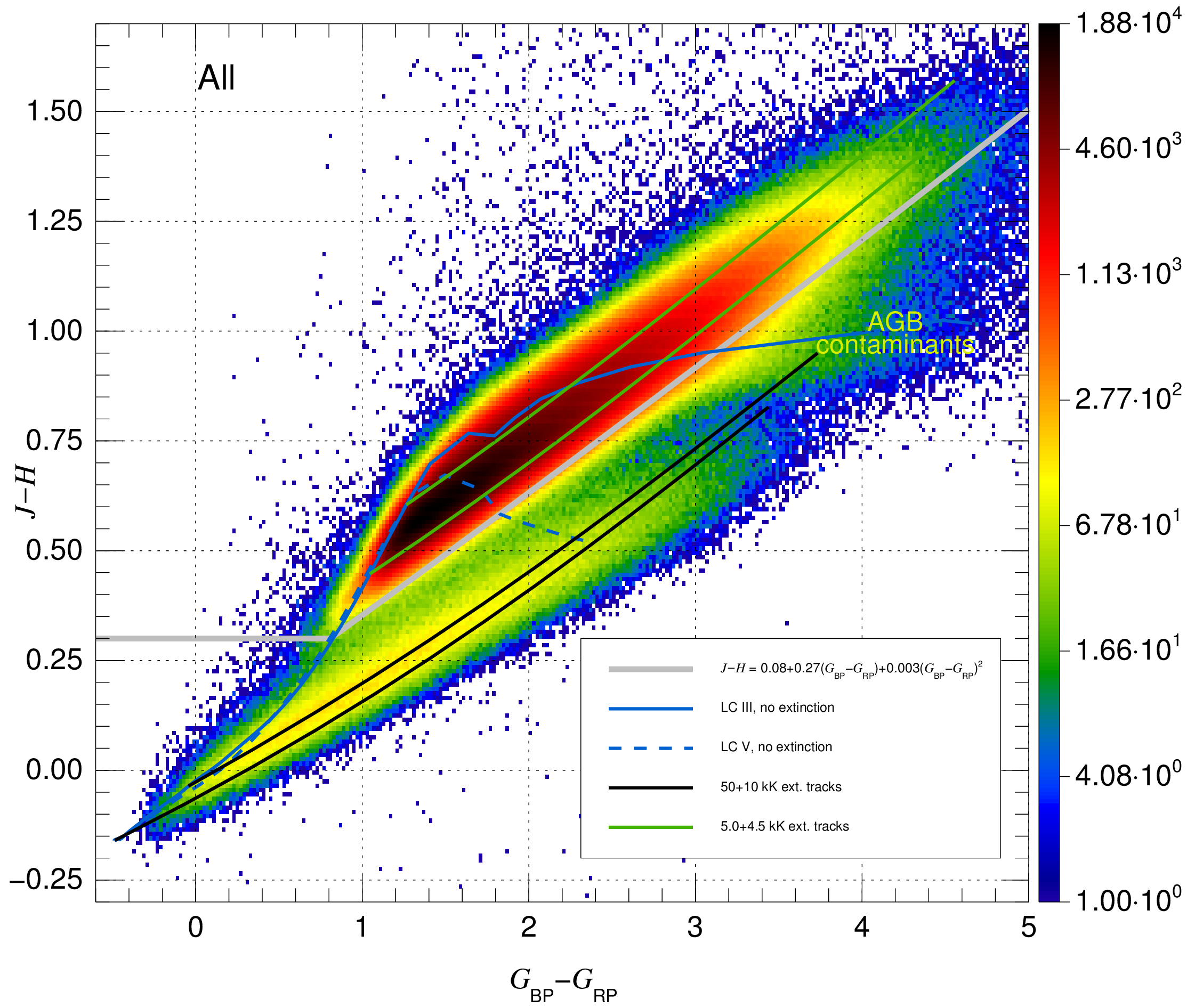} \
  \includegraphics*[width=0.49\linewidth]{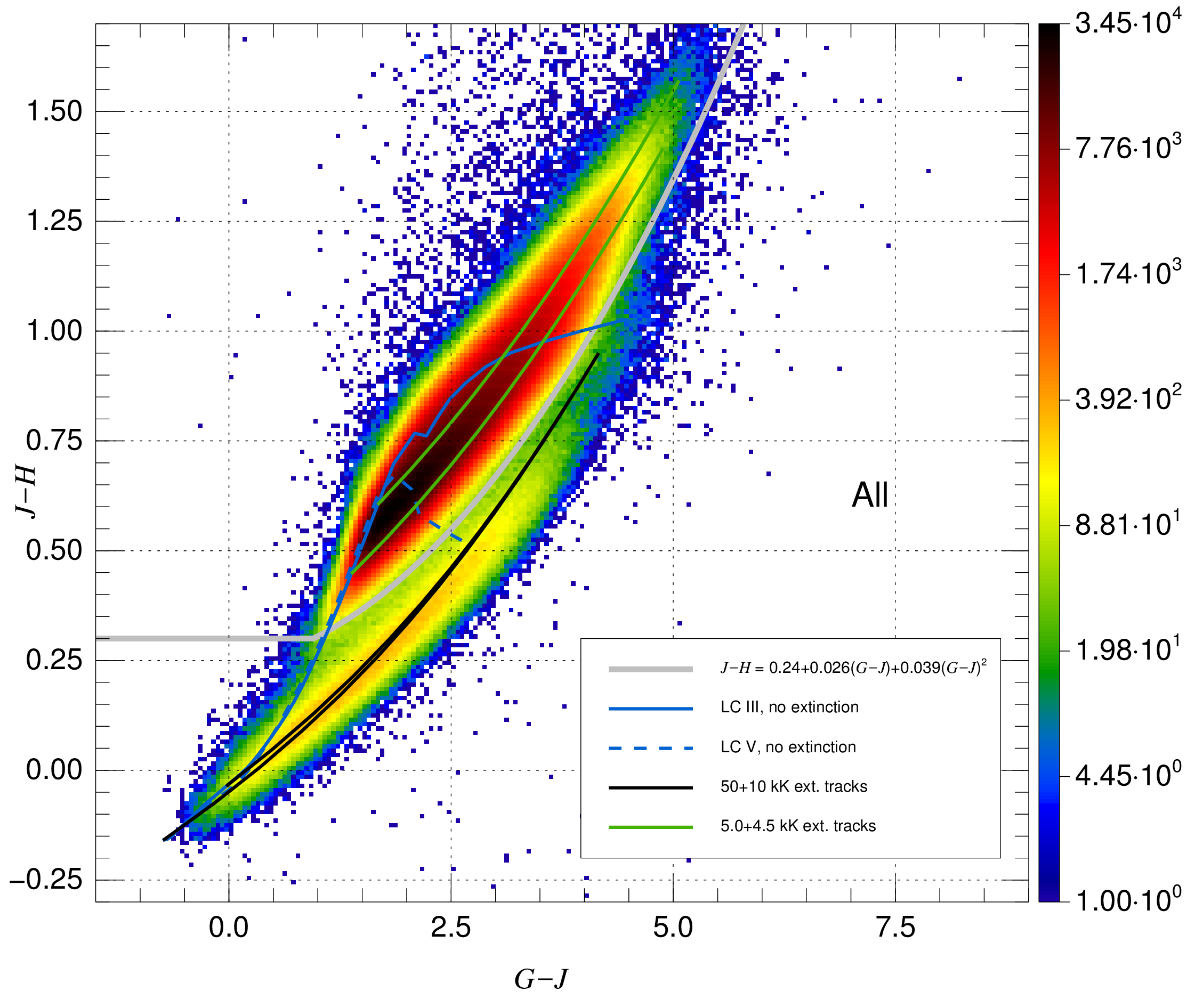}
  \\           
  \includegraphics*[width=0.49\linewidth]{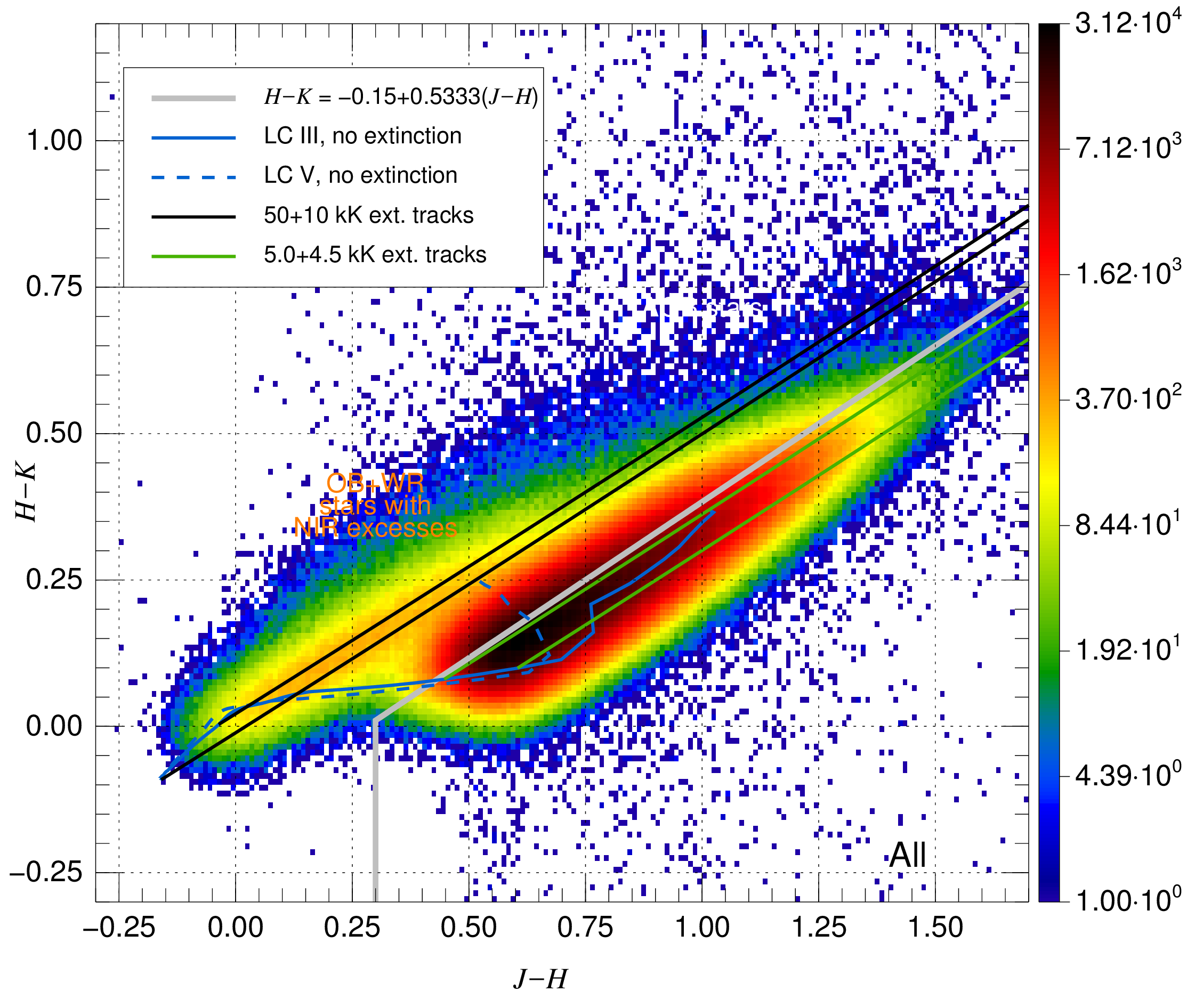} \
  \includegraphics*[width=0.49\linewidth]{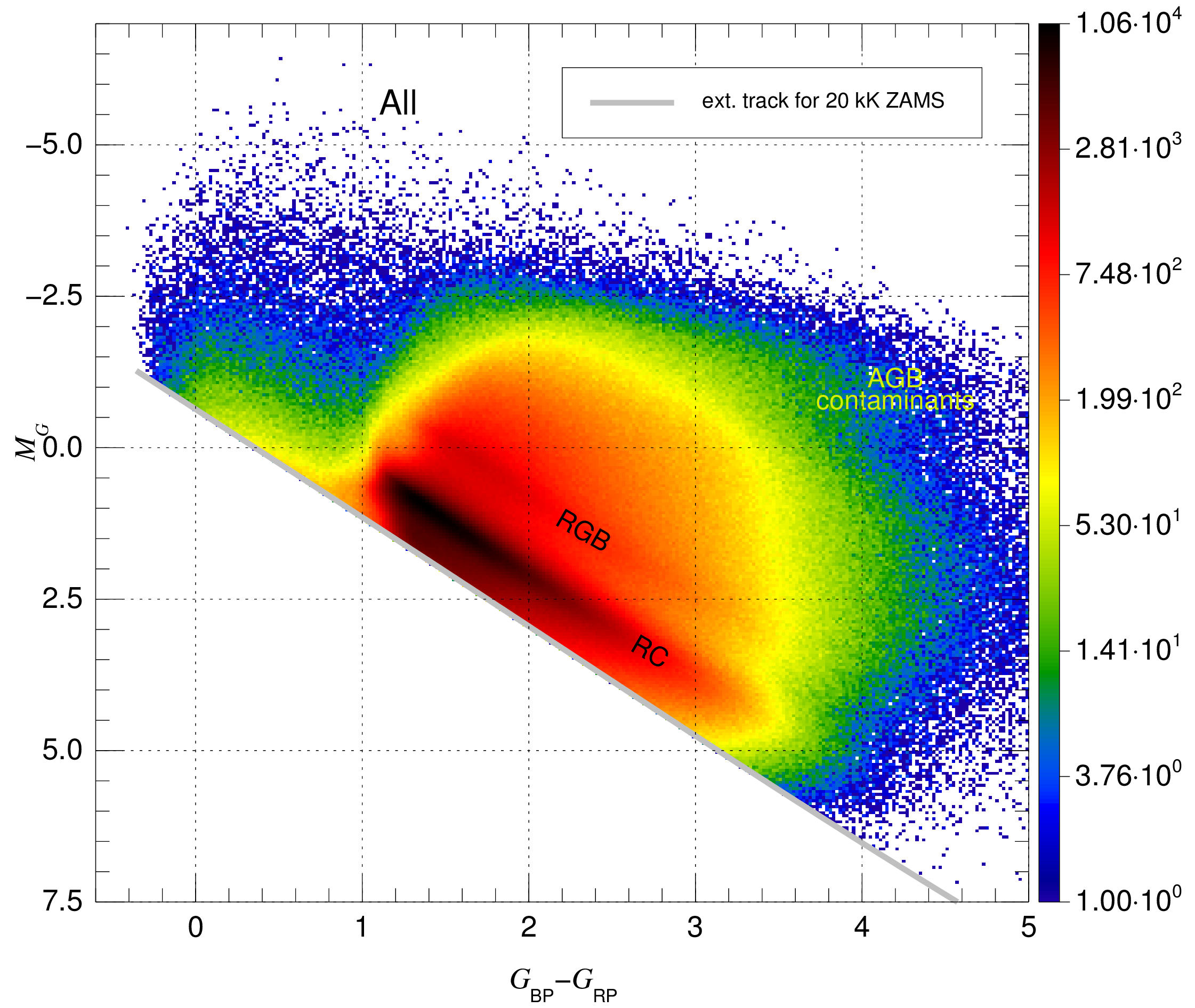}  
  \end{minipage}
  \newframe[5]
  \begin{minipage}{\linewidth}
  \includegraphics*[width=0.49\linewidth]{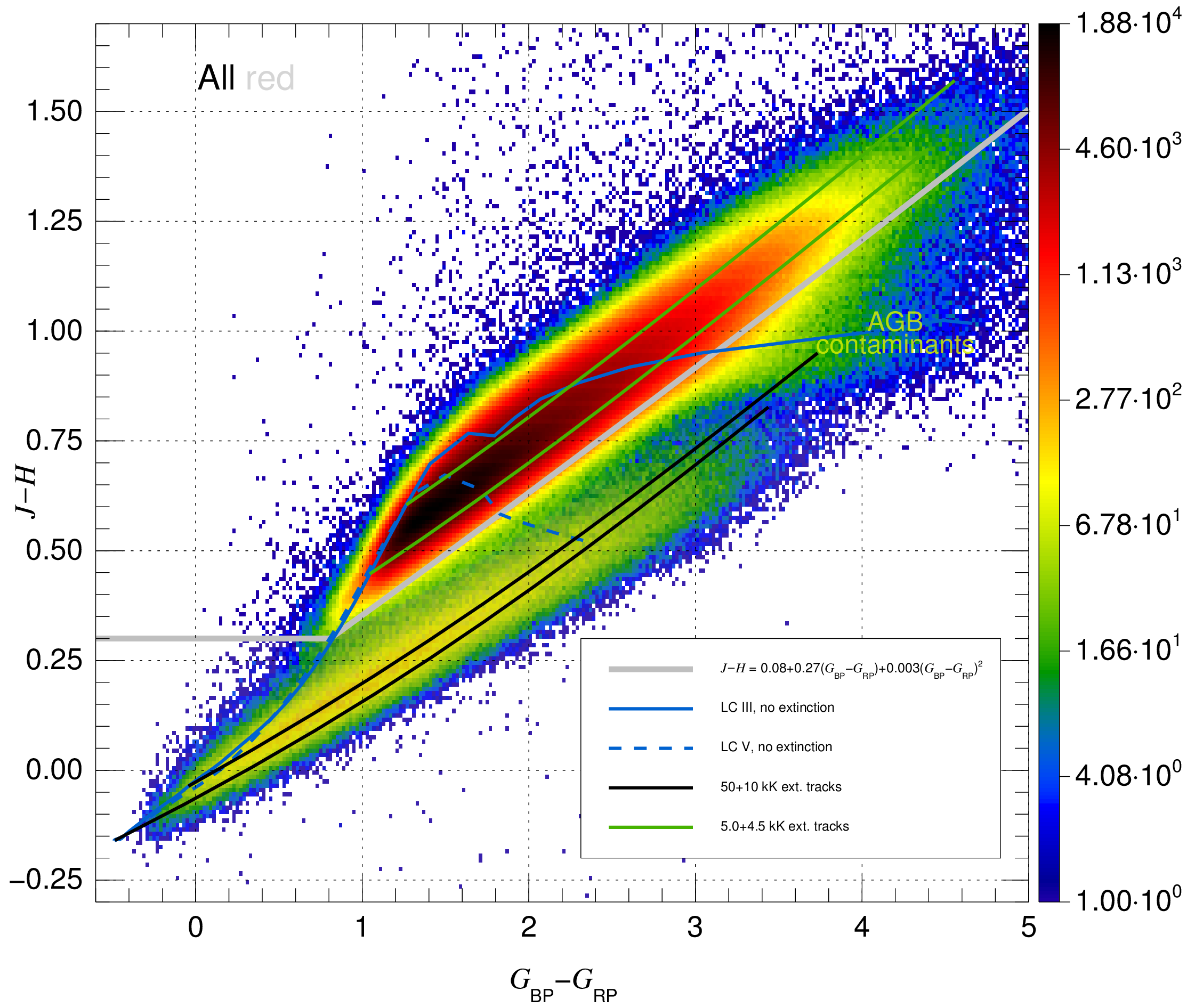} \
  \includegraphics*[width=0.49\linewidth]{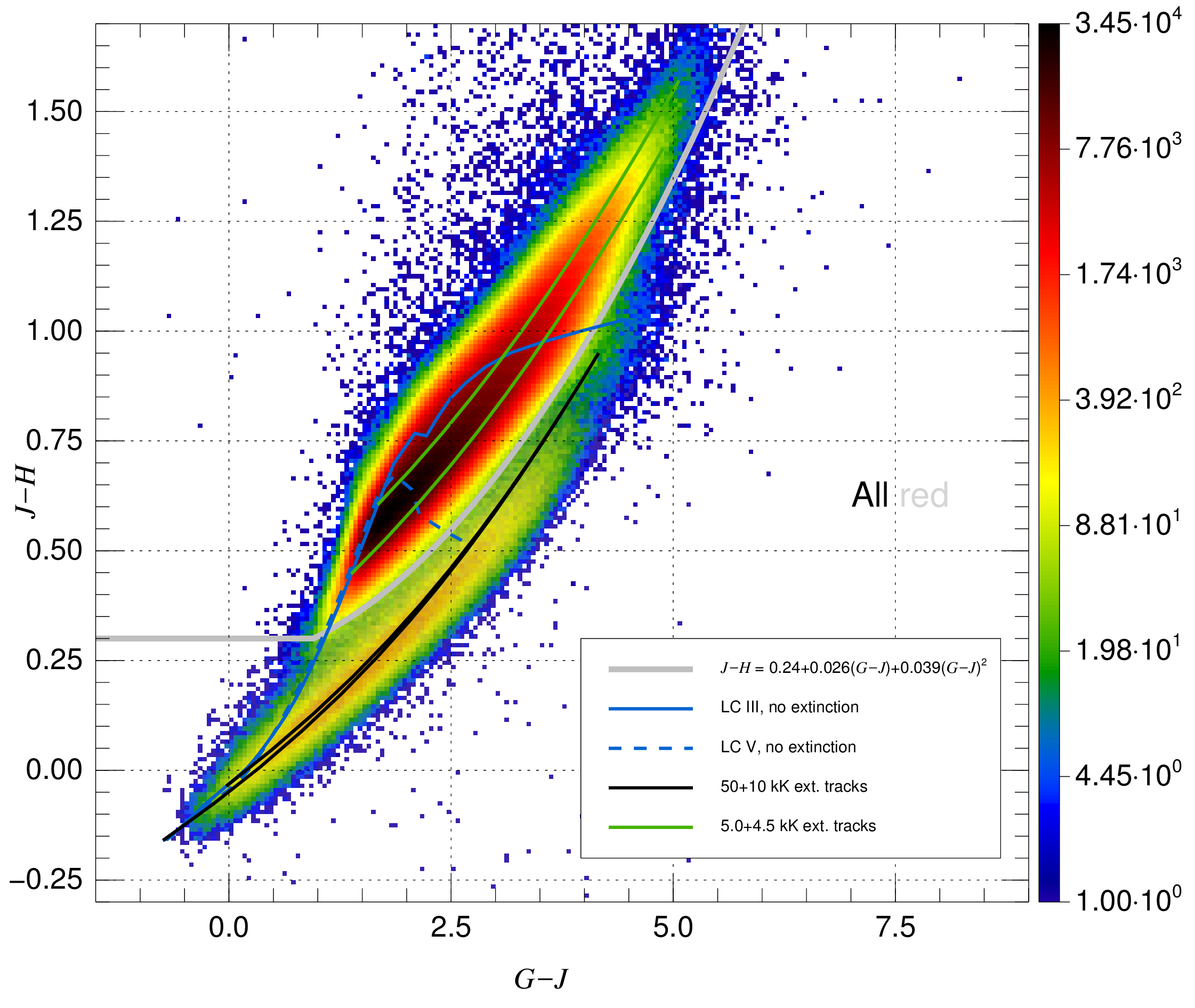}
  \\           
  \includegraphics*[width=0.49\linewidth]{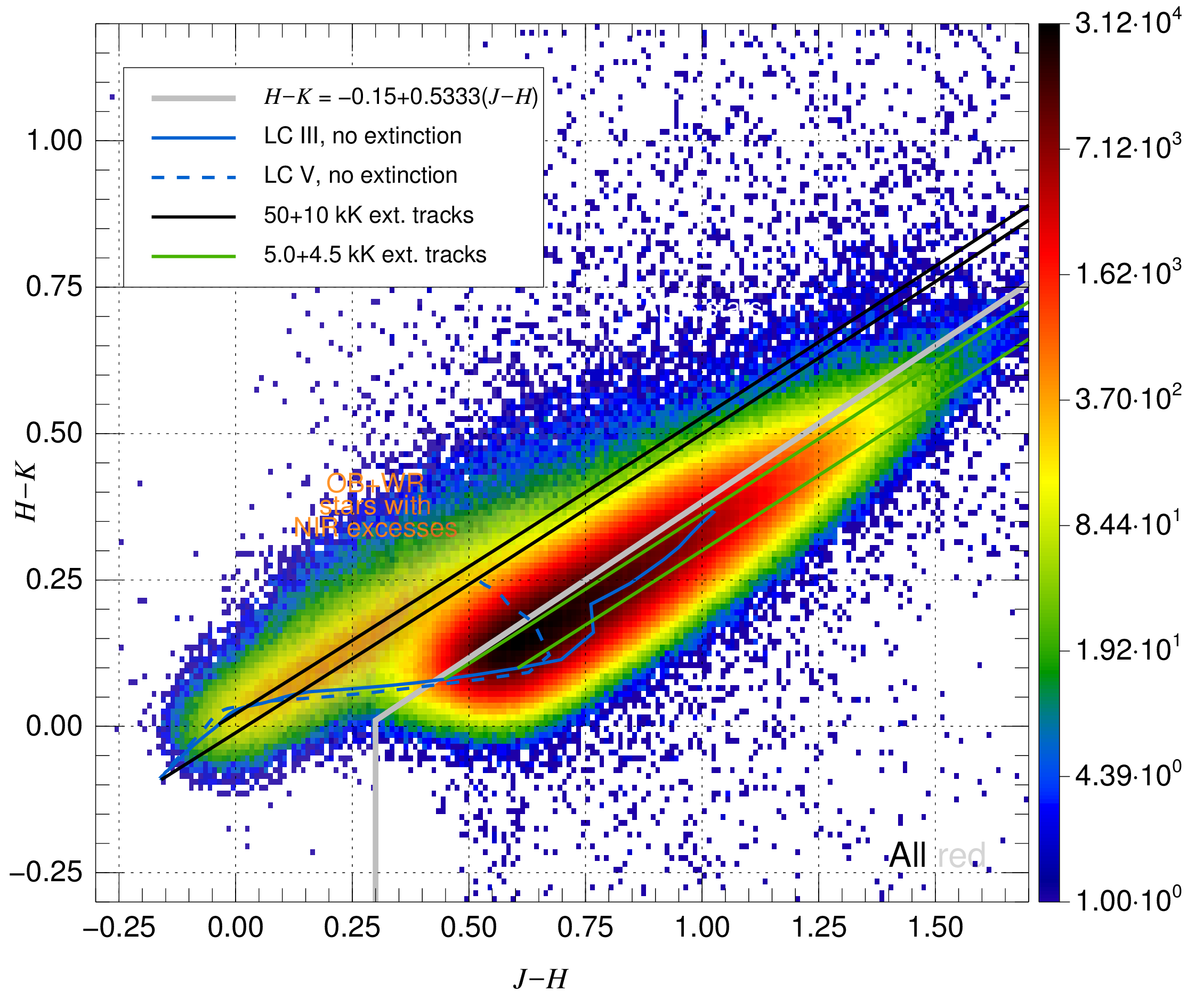} \
  \includegraphics*[width=0.49\linewidth]{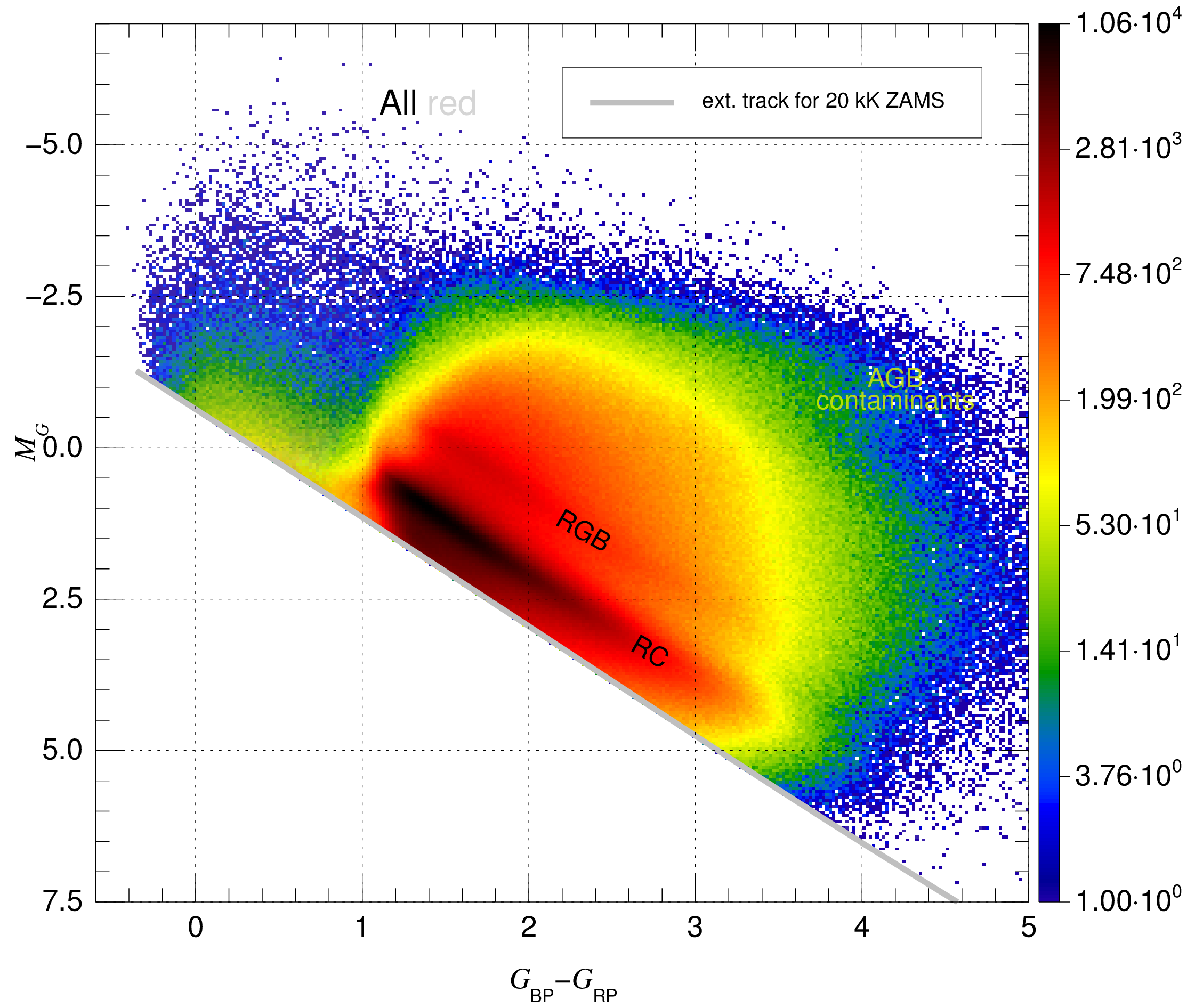}  
  \end{minipage}
  \newframe
  \begin{minipage}{\linewidth}
  \includegraphics*[width=0.49\linewidth]{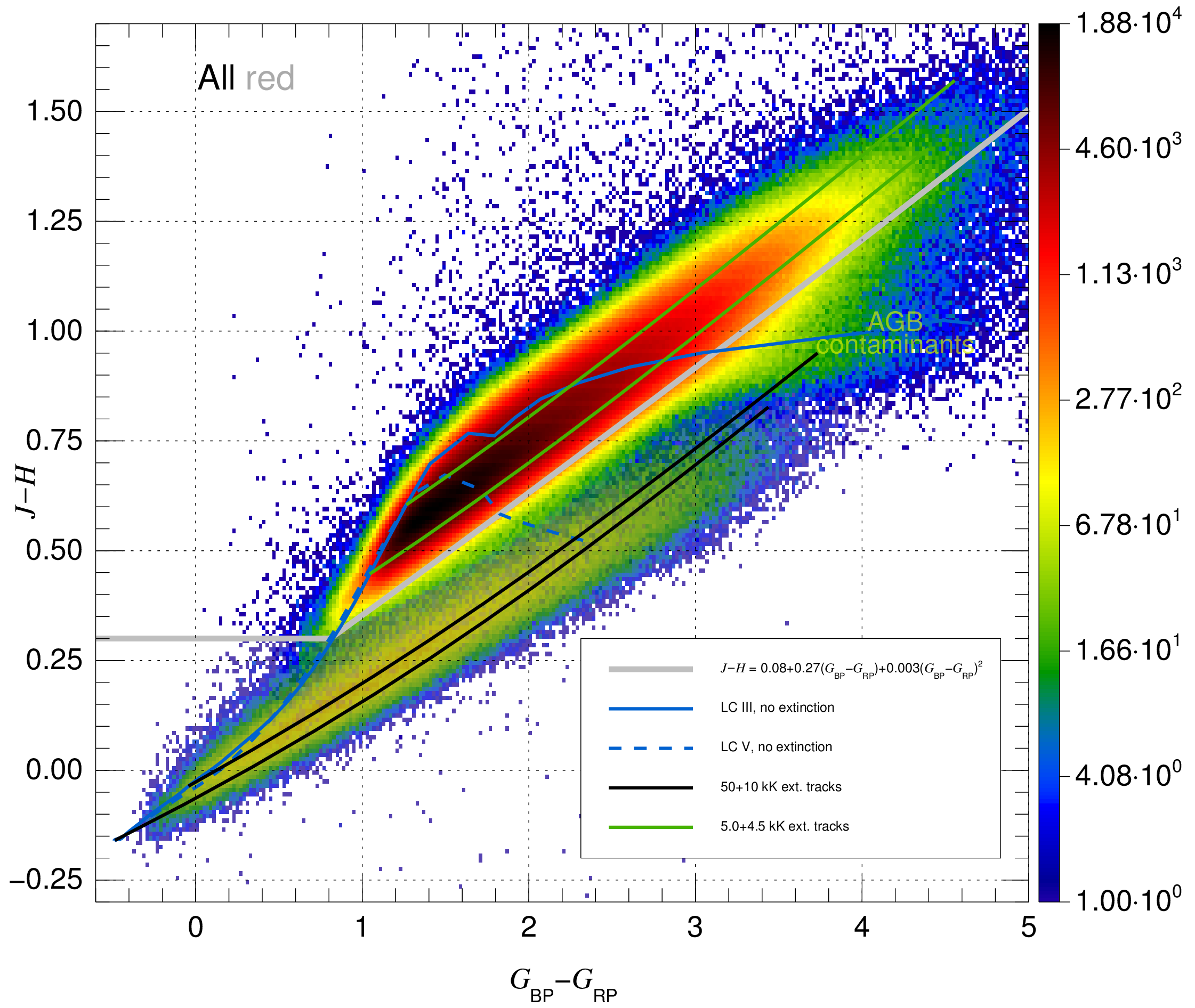} \
  \includegraphics*[width=0.49\linewidth]{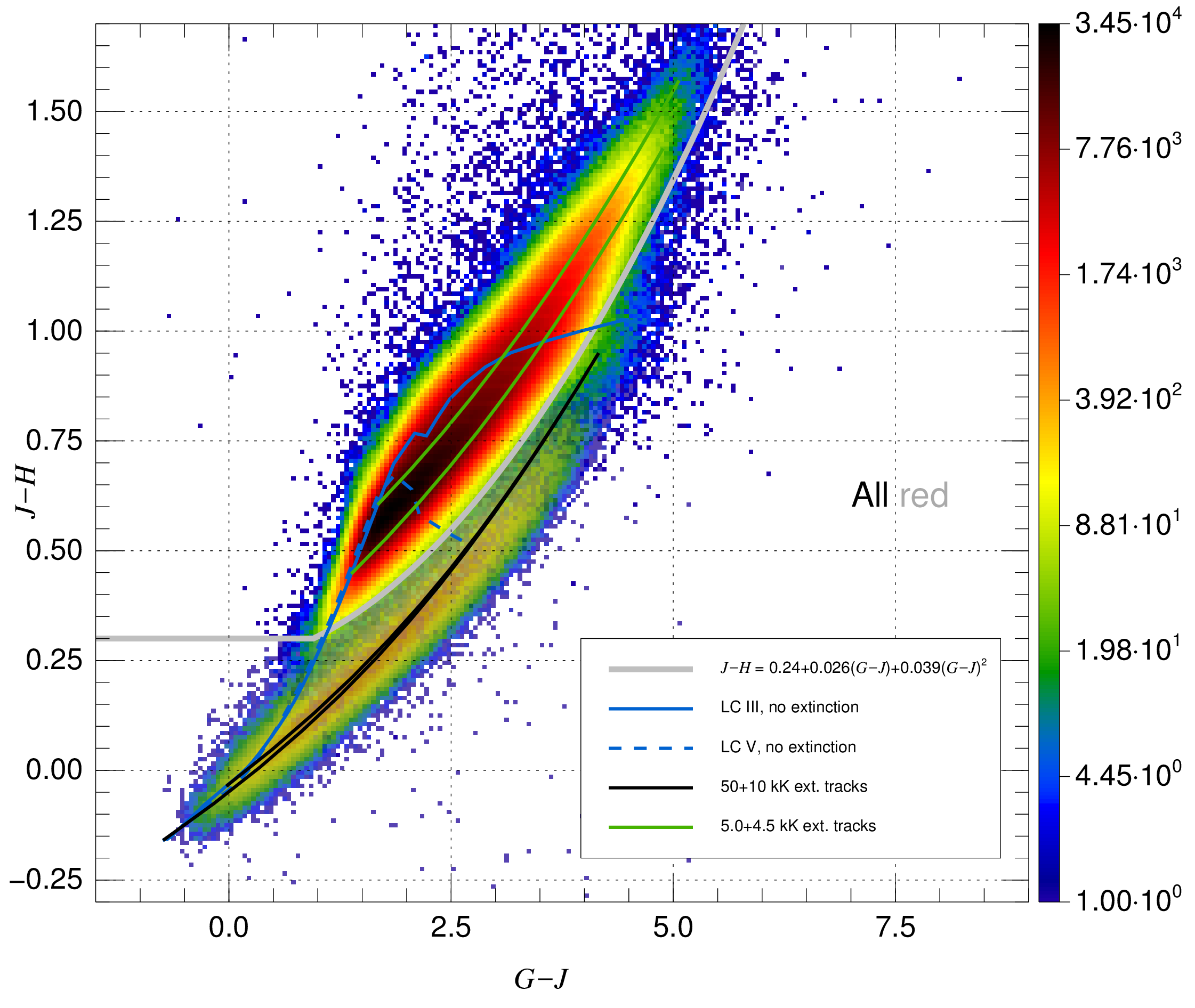}
  \\           
  \includegraphics*[width=0.49\linewidth]{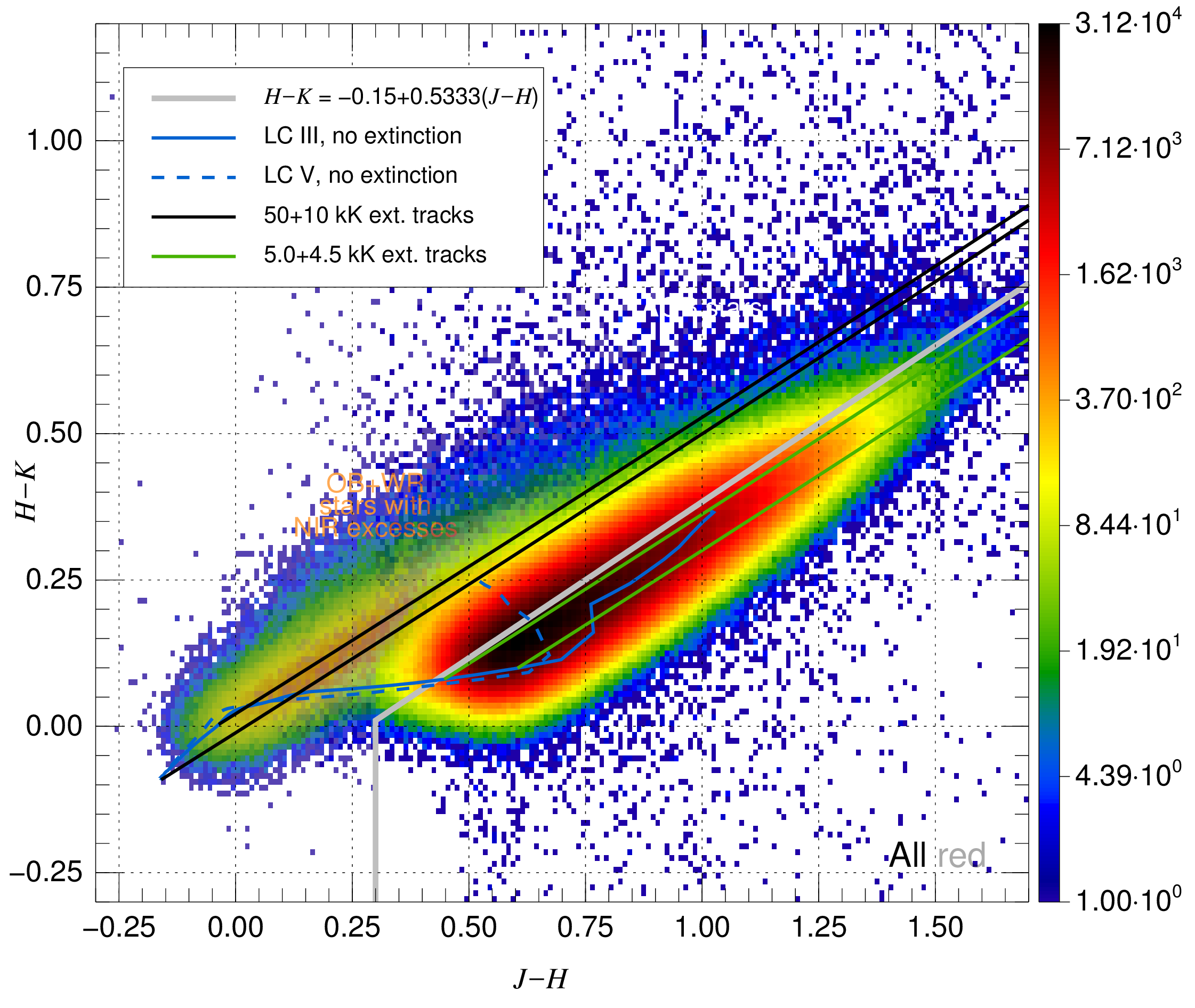} \
  \includegraphics*[width=0.49\linewidth]{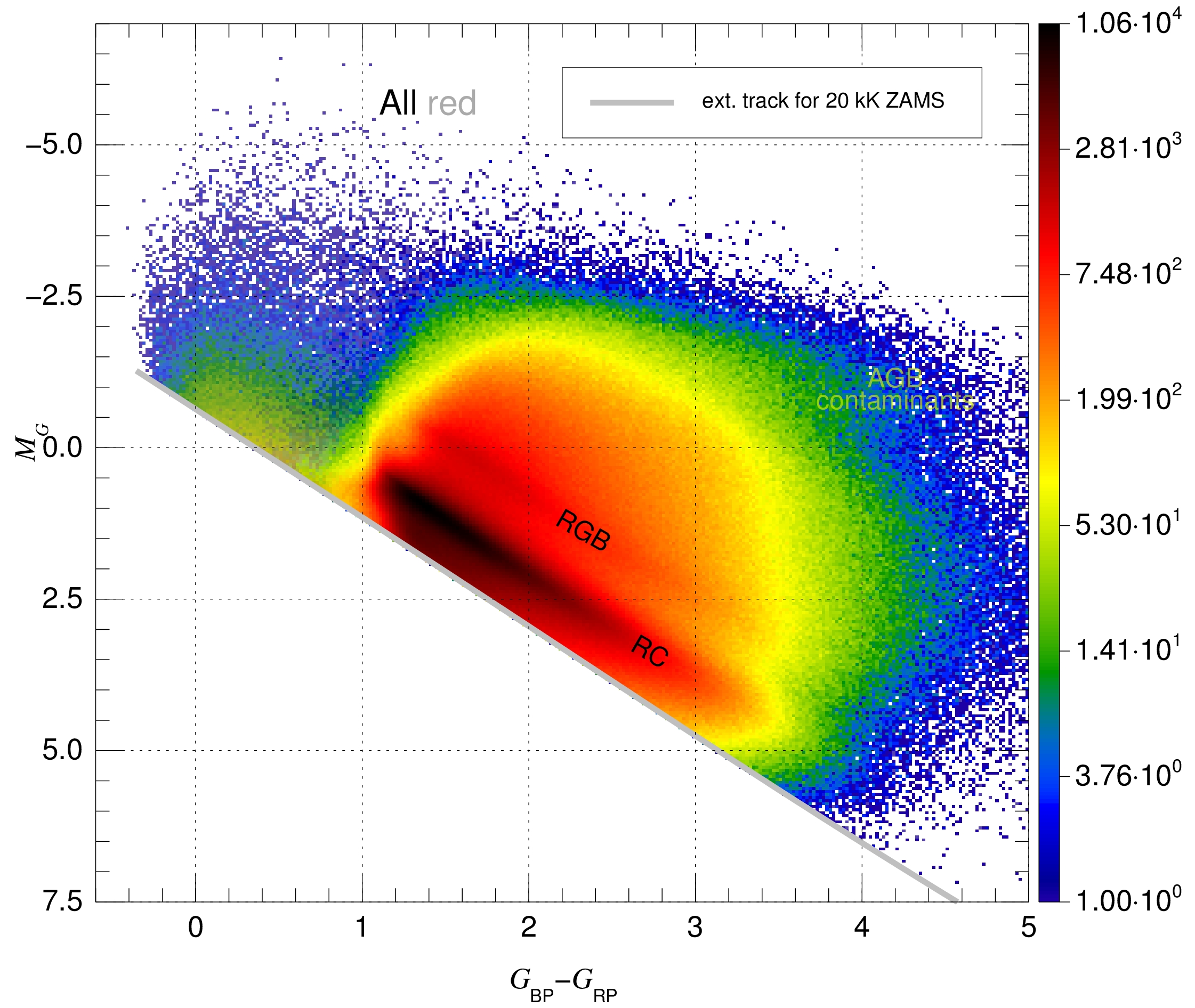}  
  \end{minipage}
  \newframe
  \begin{minipage}{\linewidth}
  \includegraphics*[width=0.49\linewidth]{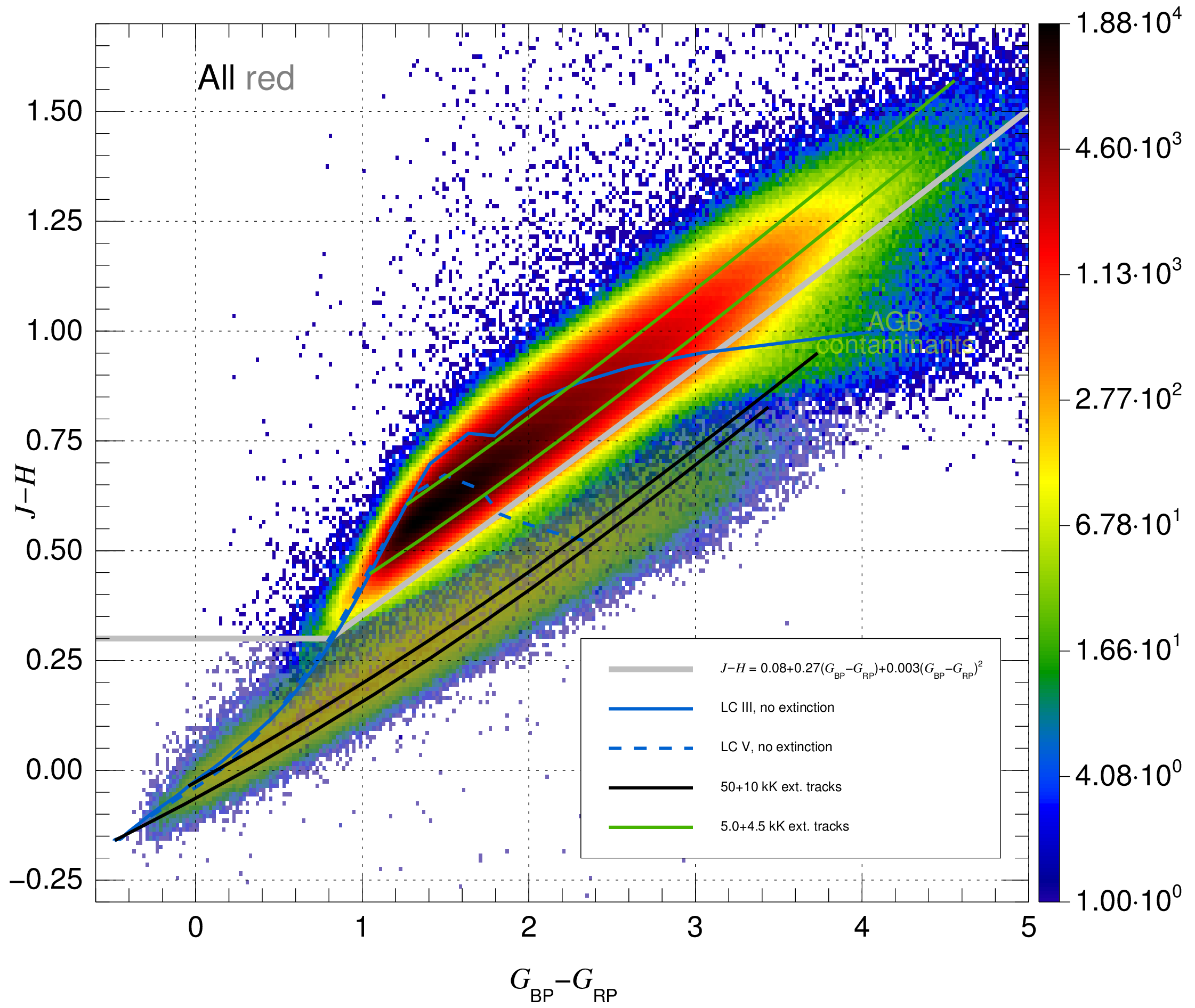} \
  \includegraphics*[width=0.49\linewidth]{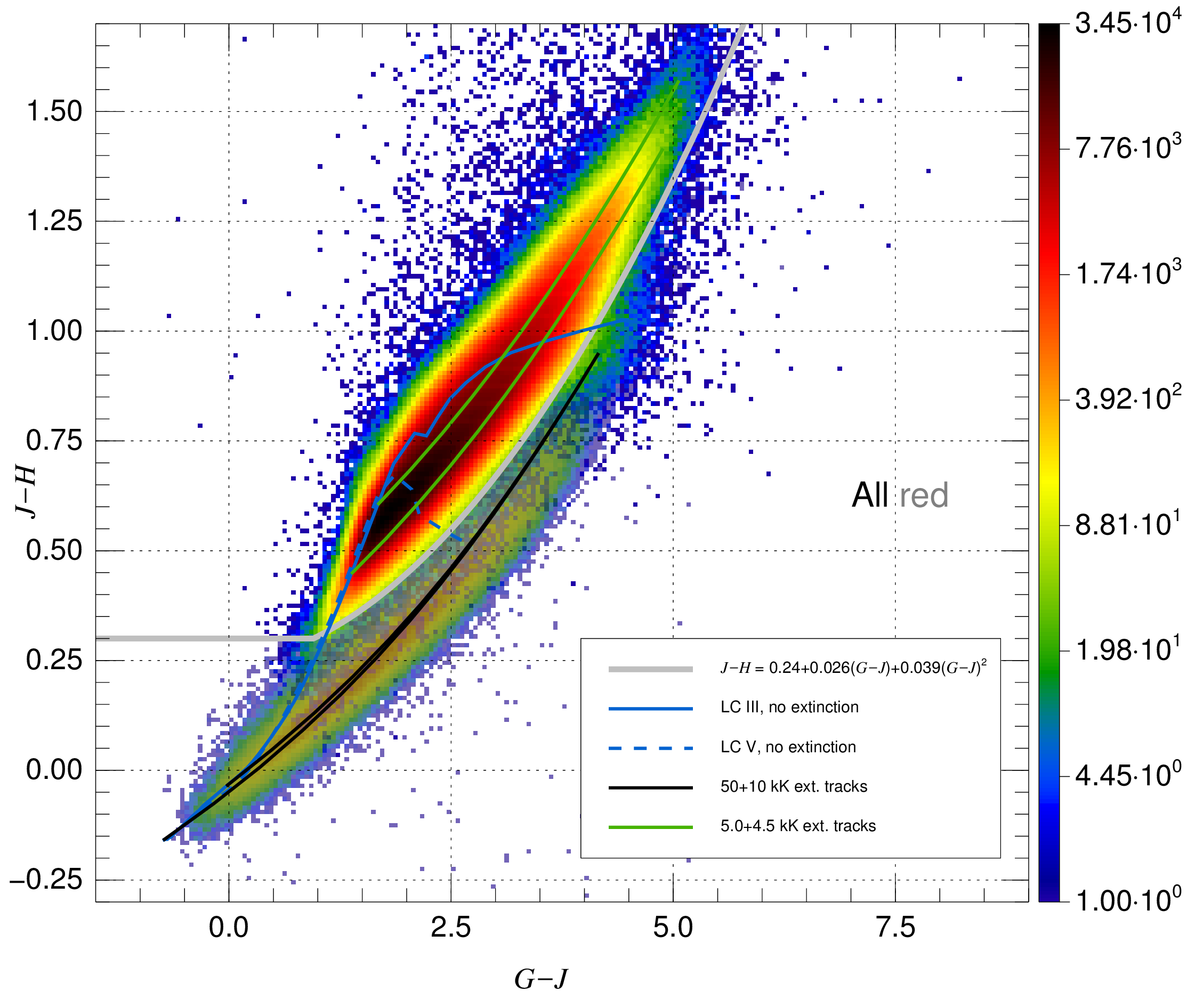}
  \\           
  \includegraphics*[width=0.49\linewidth]{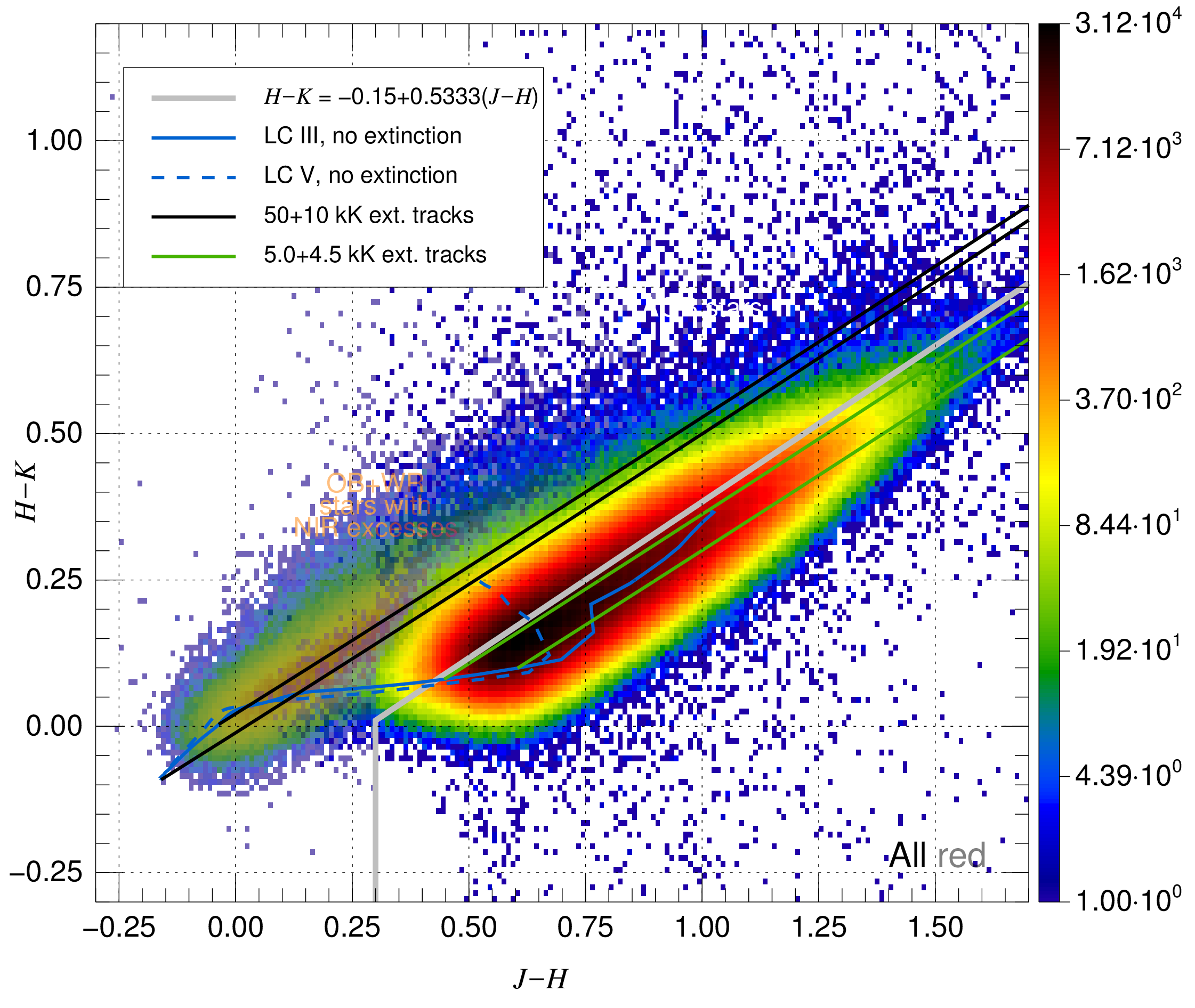} \
  \includegraphics*[width=0.49\linewidth]{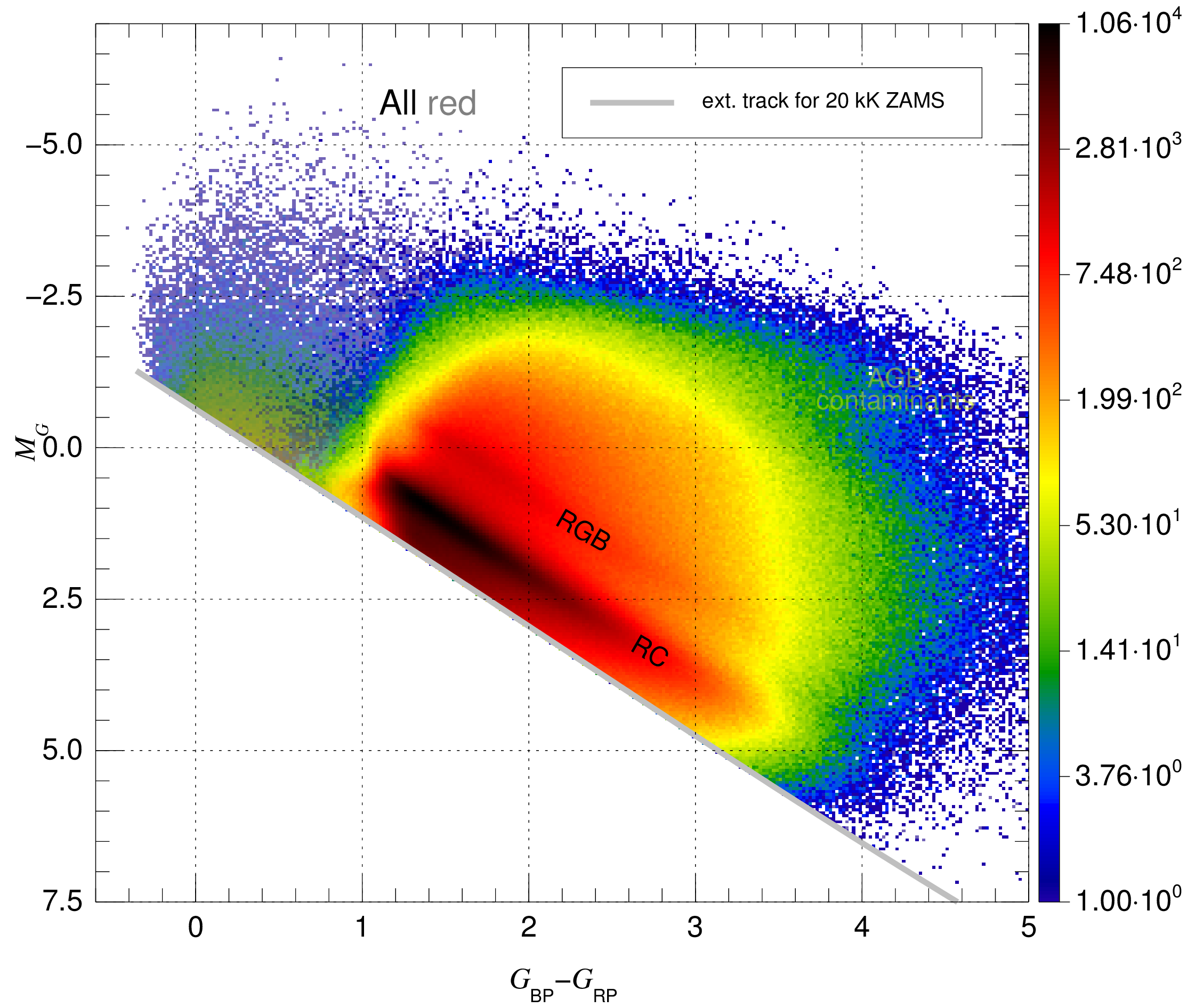}  
  \end{minipage}
  \newframe
  \begin{minipage}{\linewidth}
  \includegraphics*[width=0.49\linewidth]{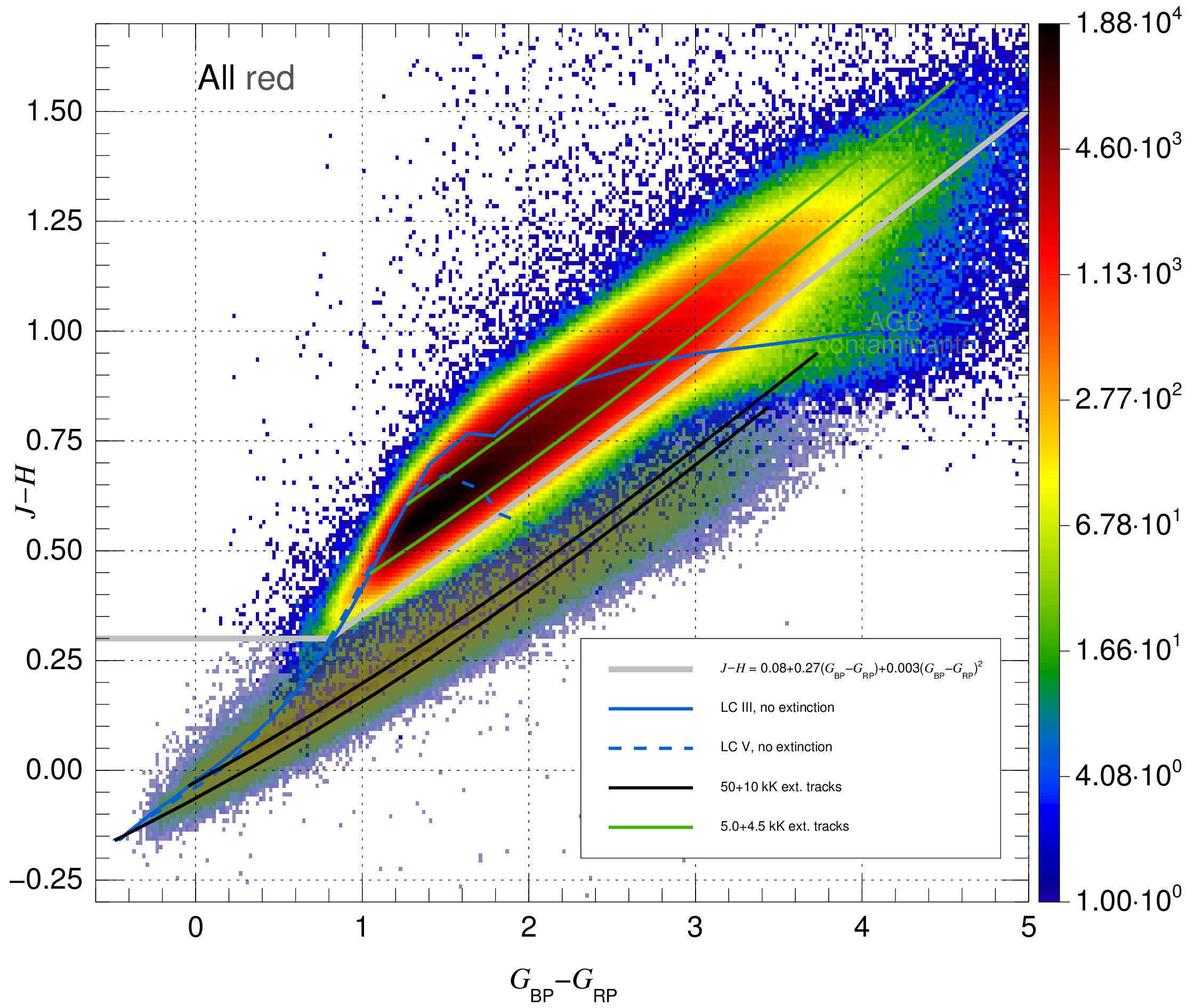} \
  \includegraphics*[width=0.49\linewidth]{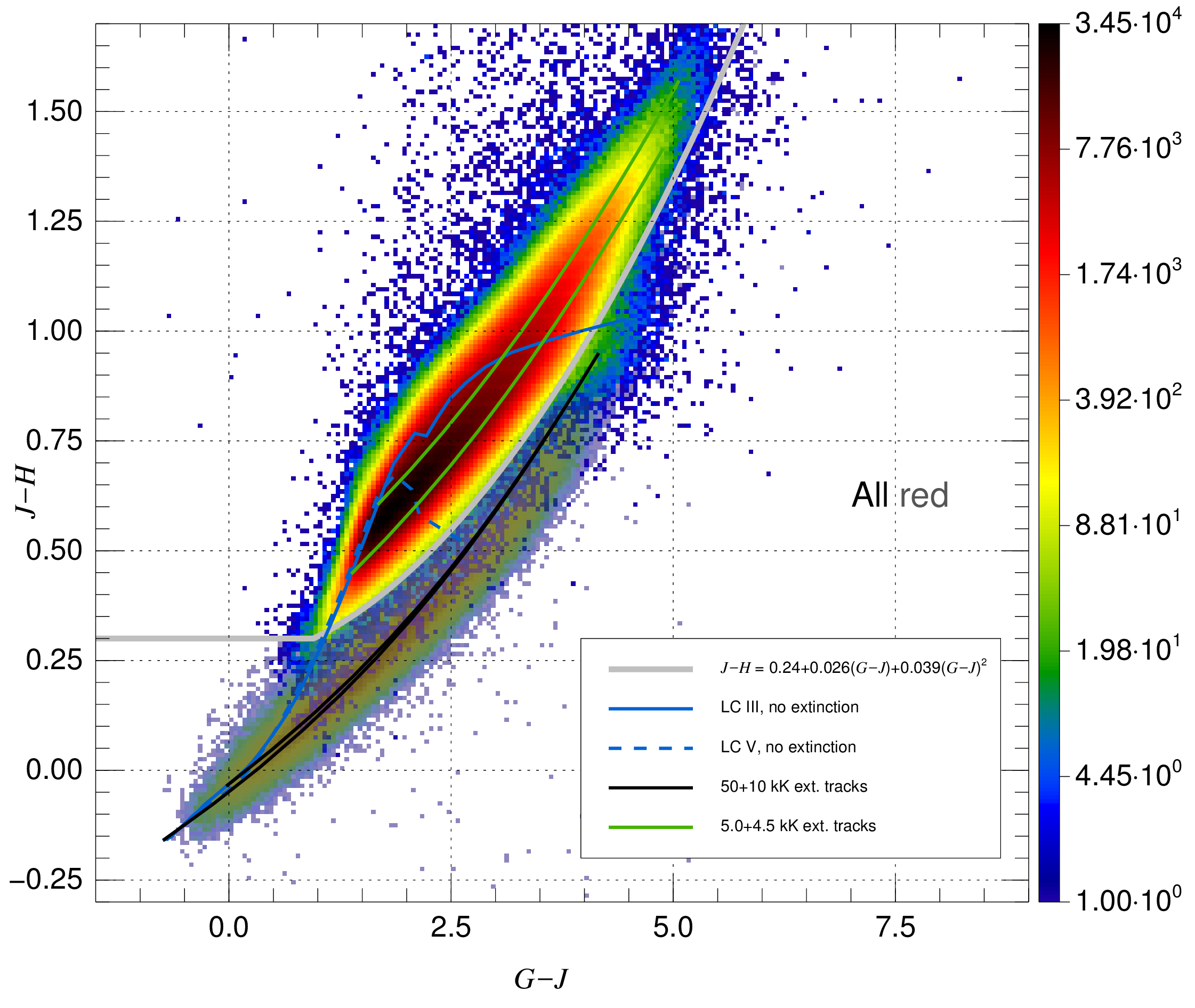}
  \\           
  \includegraphics*[width=0.49\linewidth]{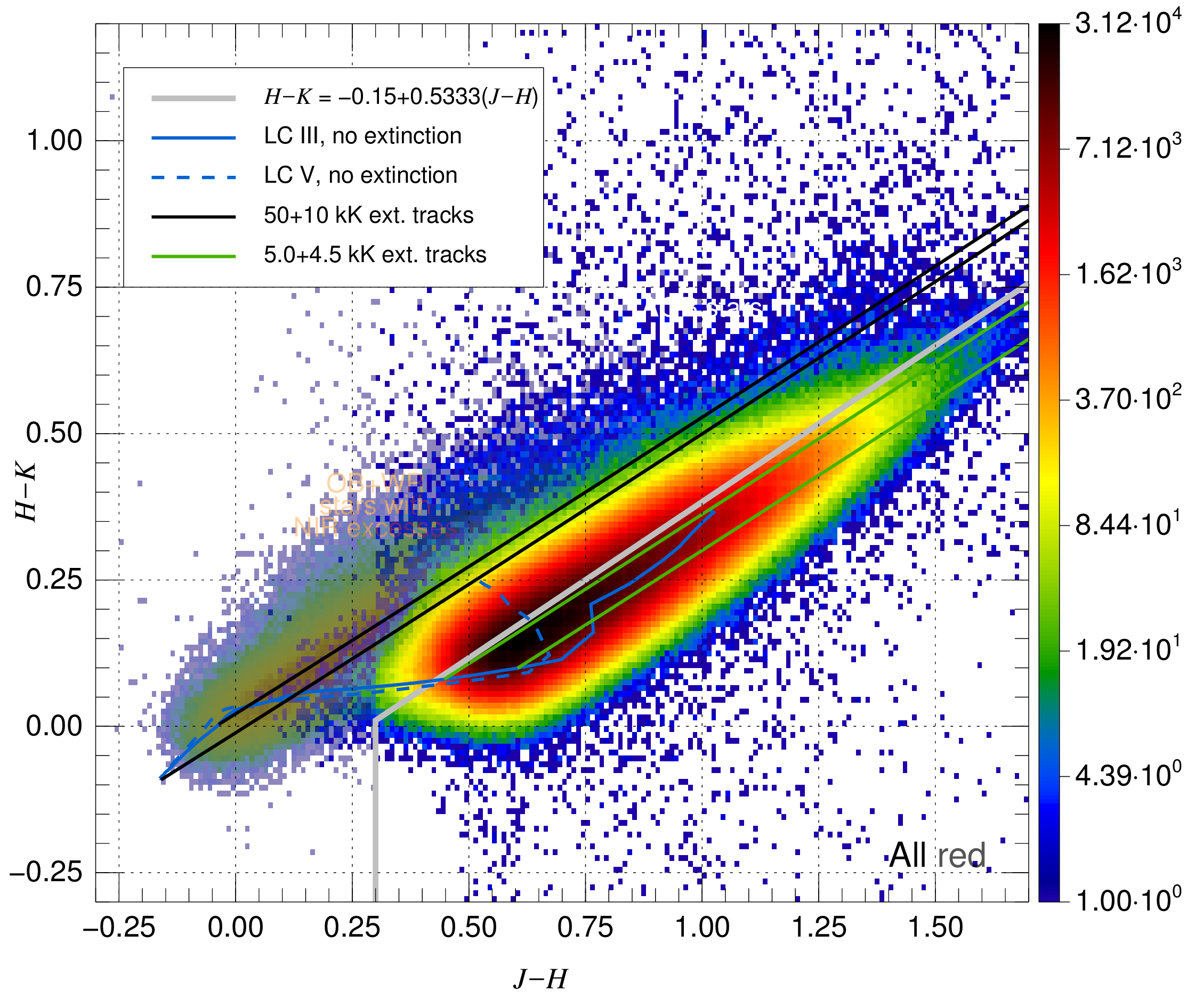} \
  \includegraphics*[width=0.49\linewidth]{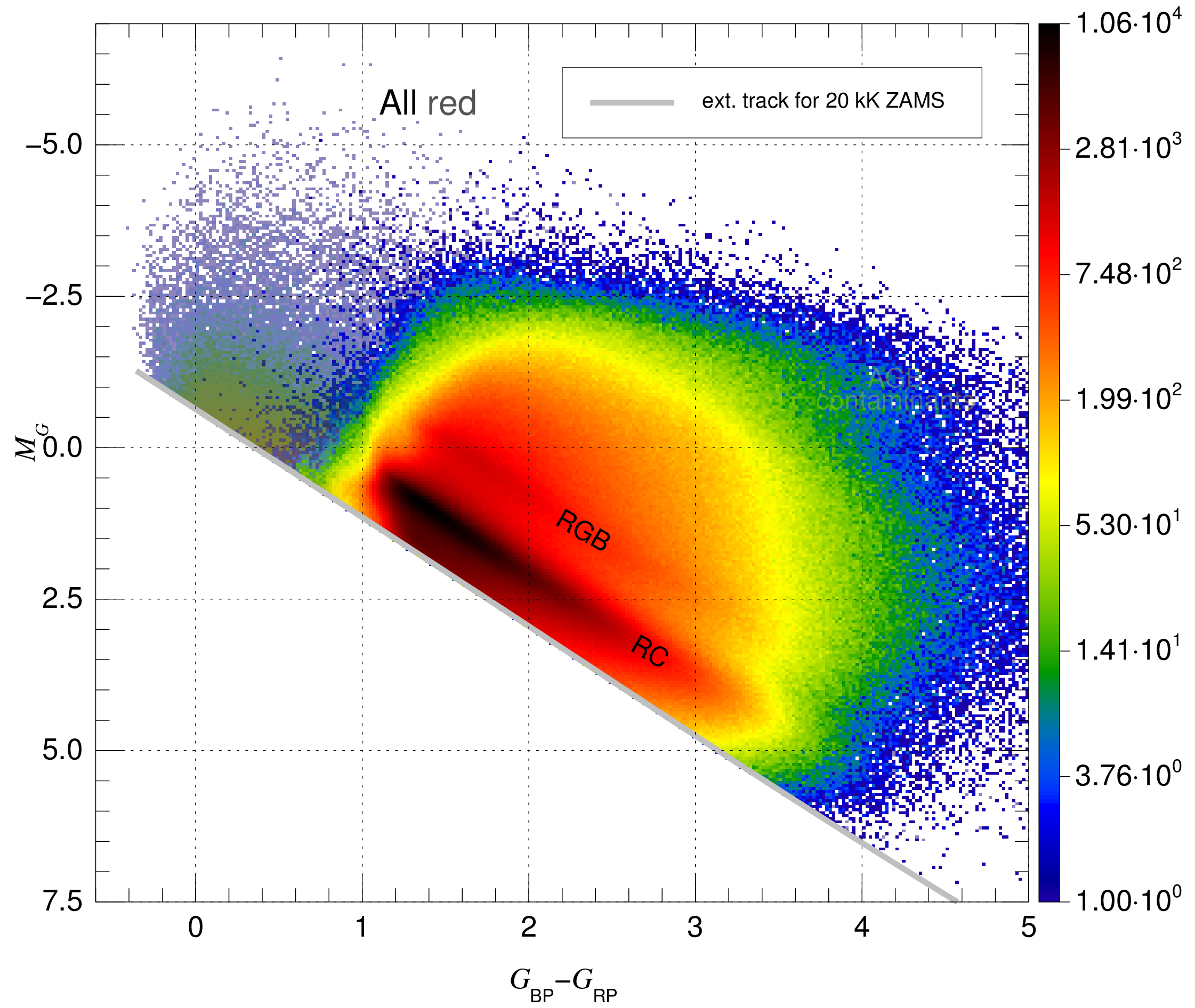}  
  \end{minipage}
  \newframe
  \begin{minipage}{\linewidth}
  \includegraphics*[width=0.49\linewidth]{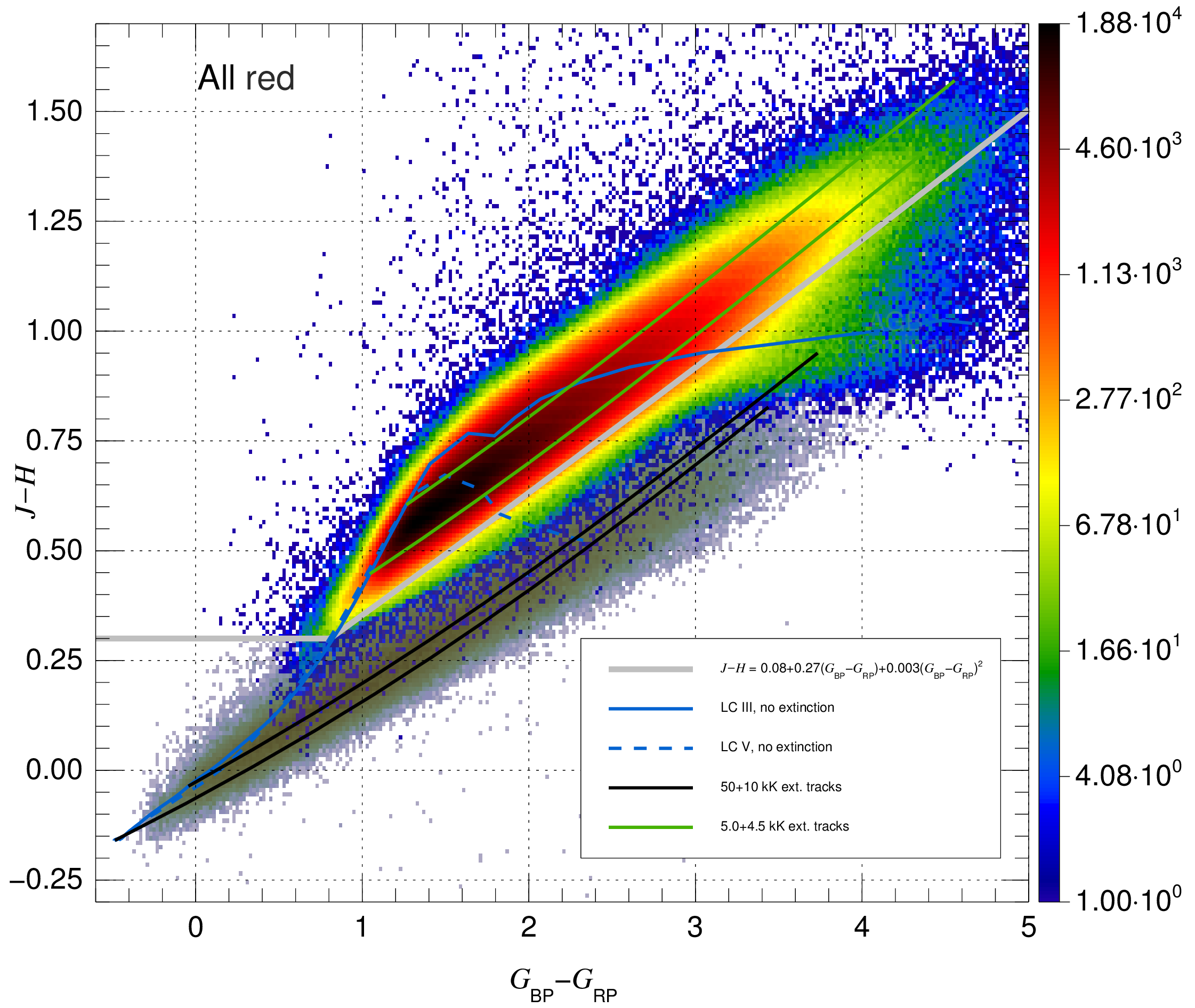} \
  \includegraphics*[width=0.49\linewidth]{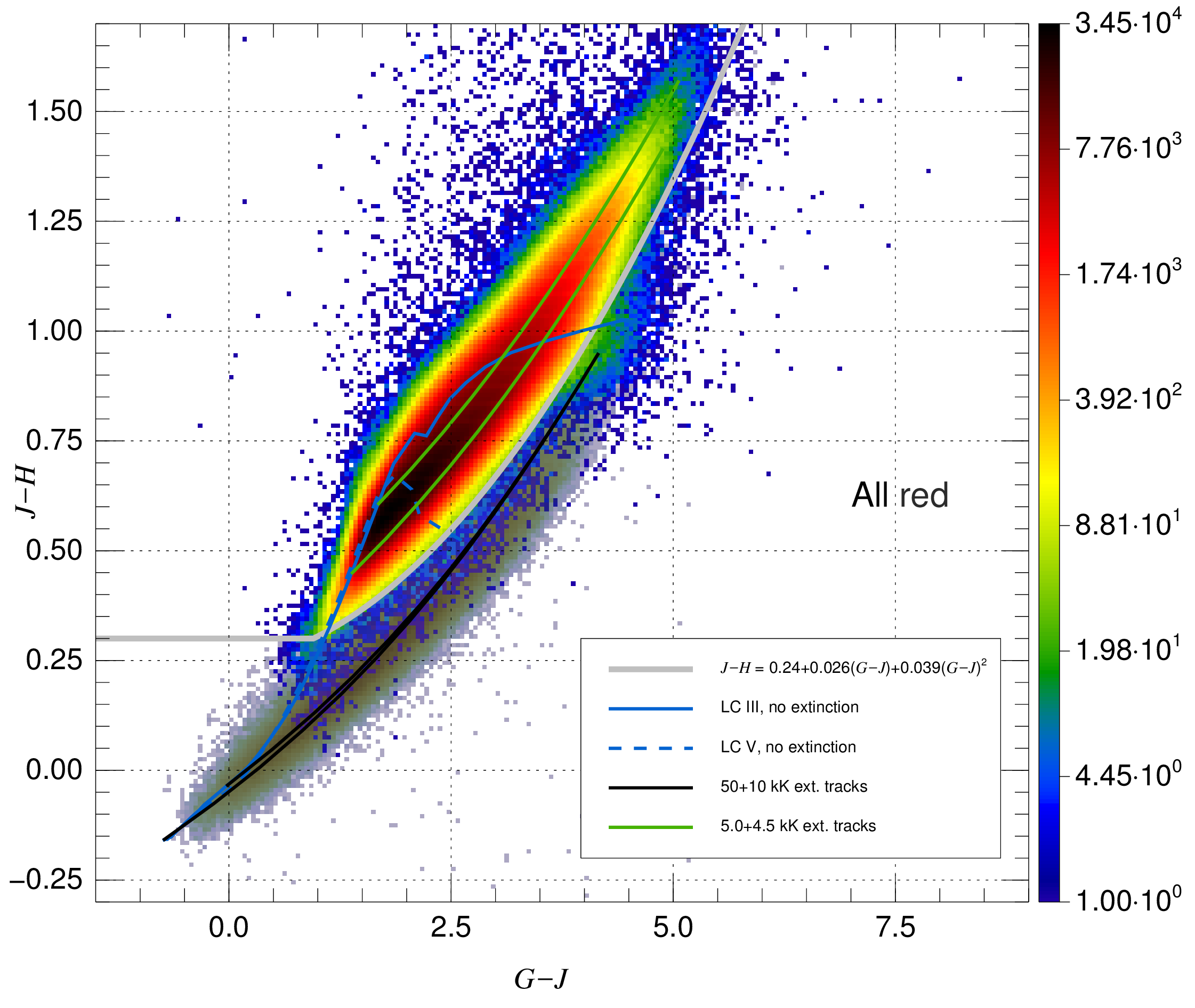}
  \\           
  \includegraphics*[width=0.49\linewidth]{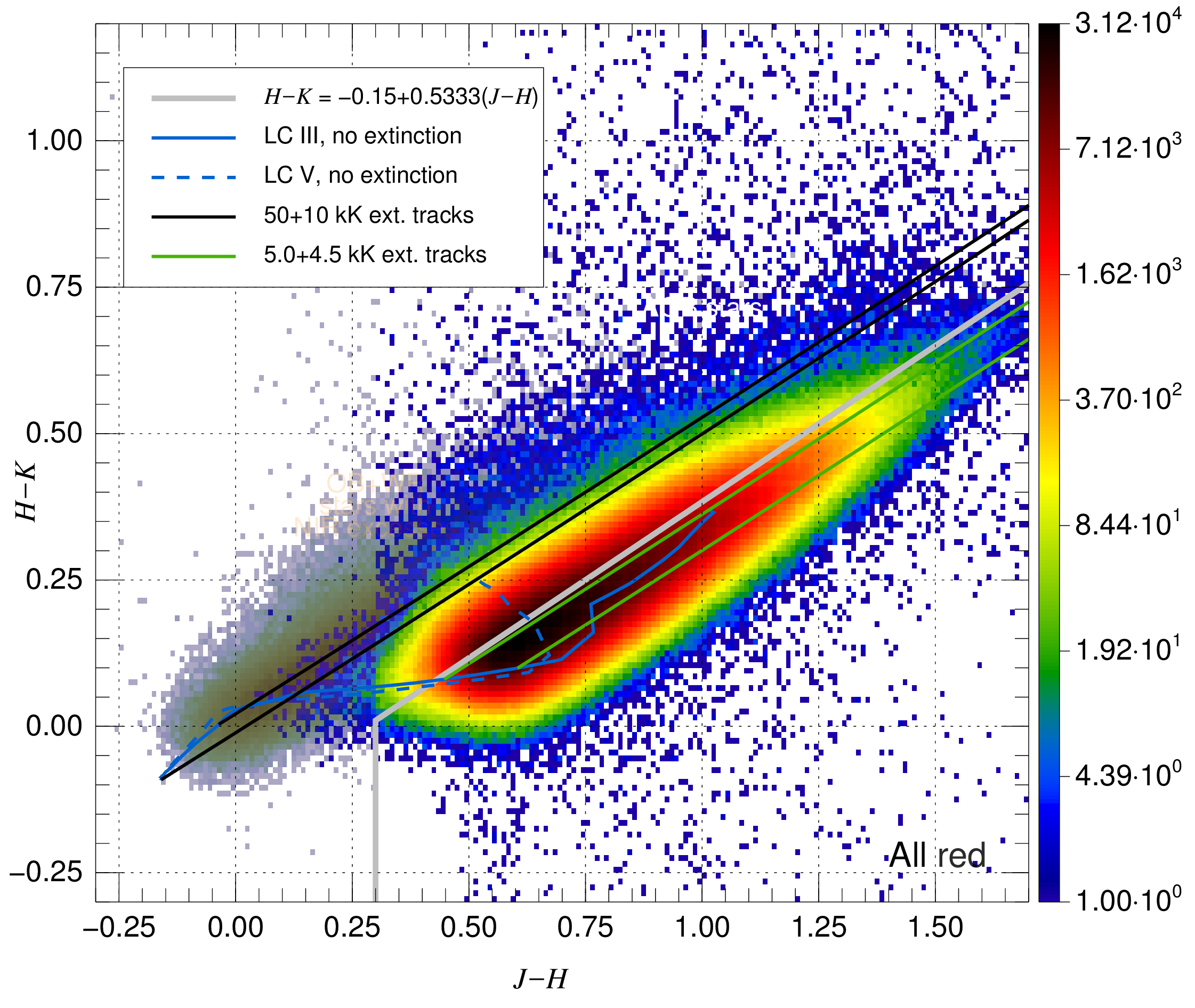} \
  \includegraphics*[width=0.49\linewidth]{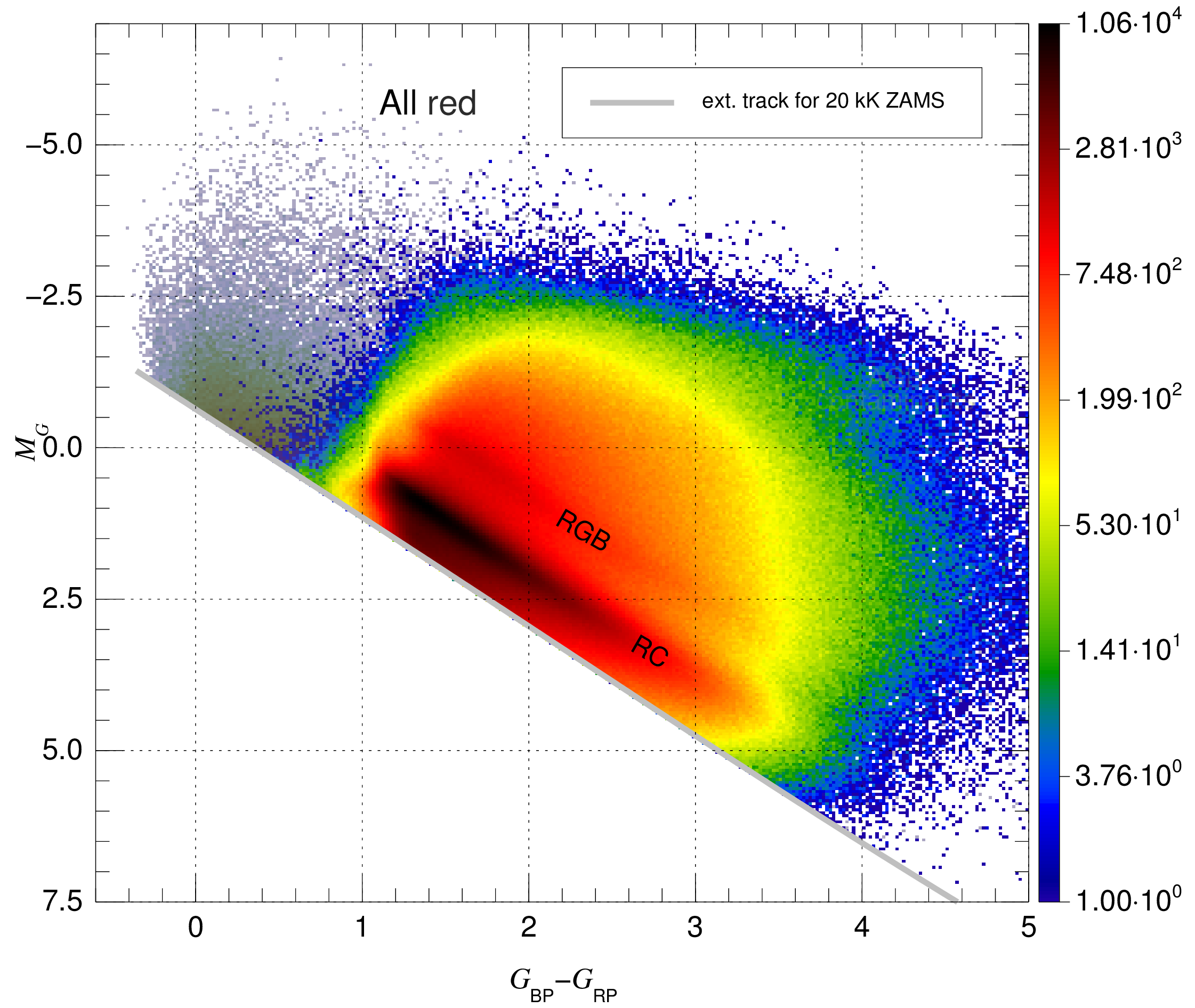}  
  \end{minipage}
  \newframe[0.4]
  \begin{minipage}{\linewidth}
  \includegraphics*[width=0.49\linewidth]{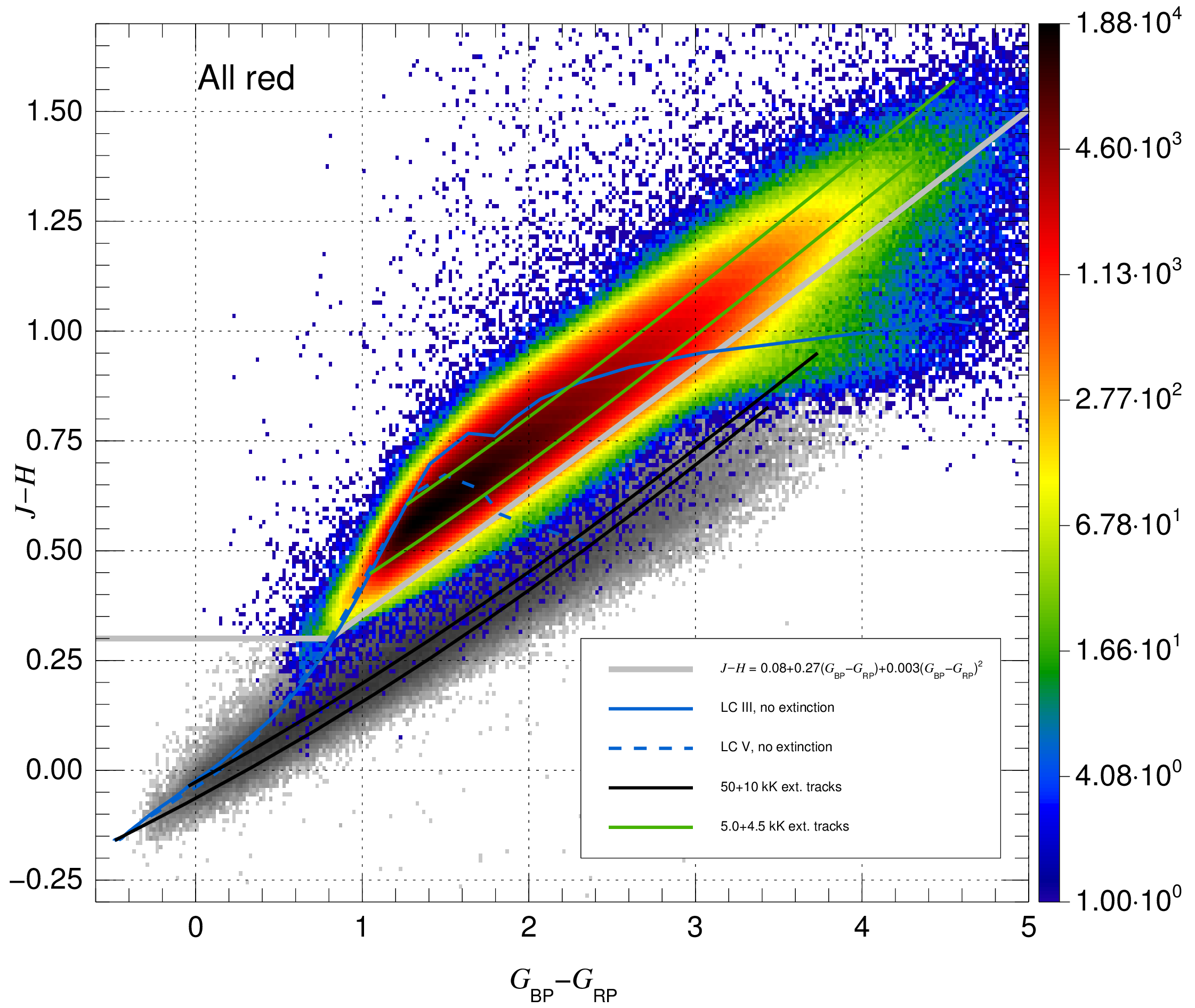} \
  \includegraphics*[width=0.49\linewidth]{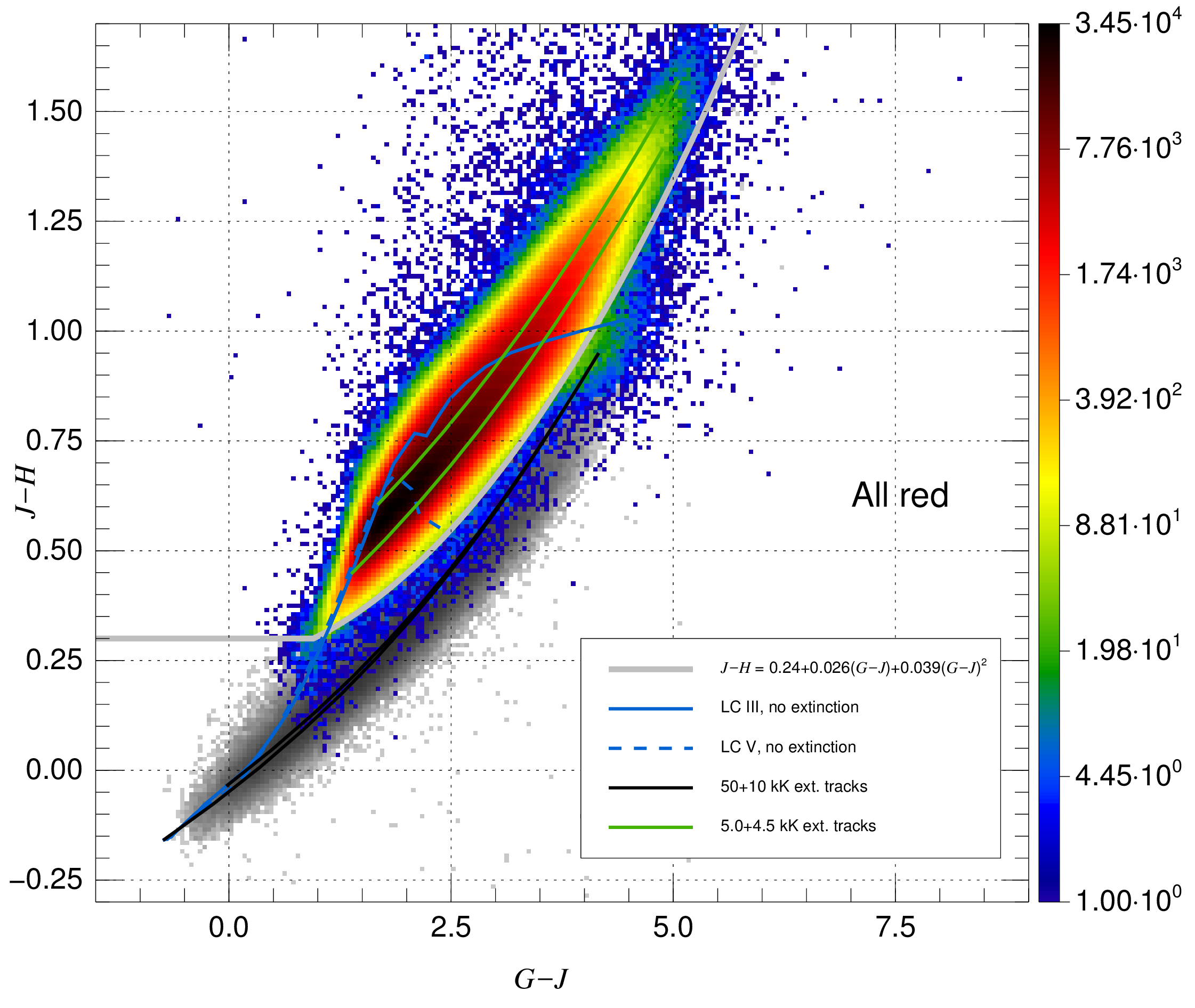}
  \\           
  \includegraphics*[width=0.49\linewidth]{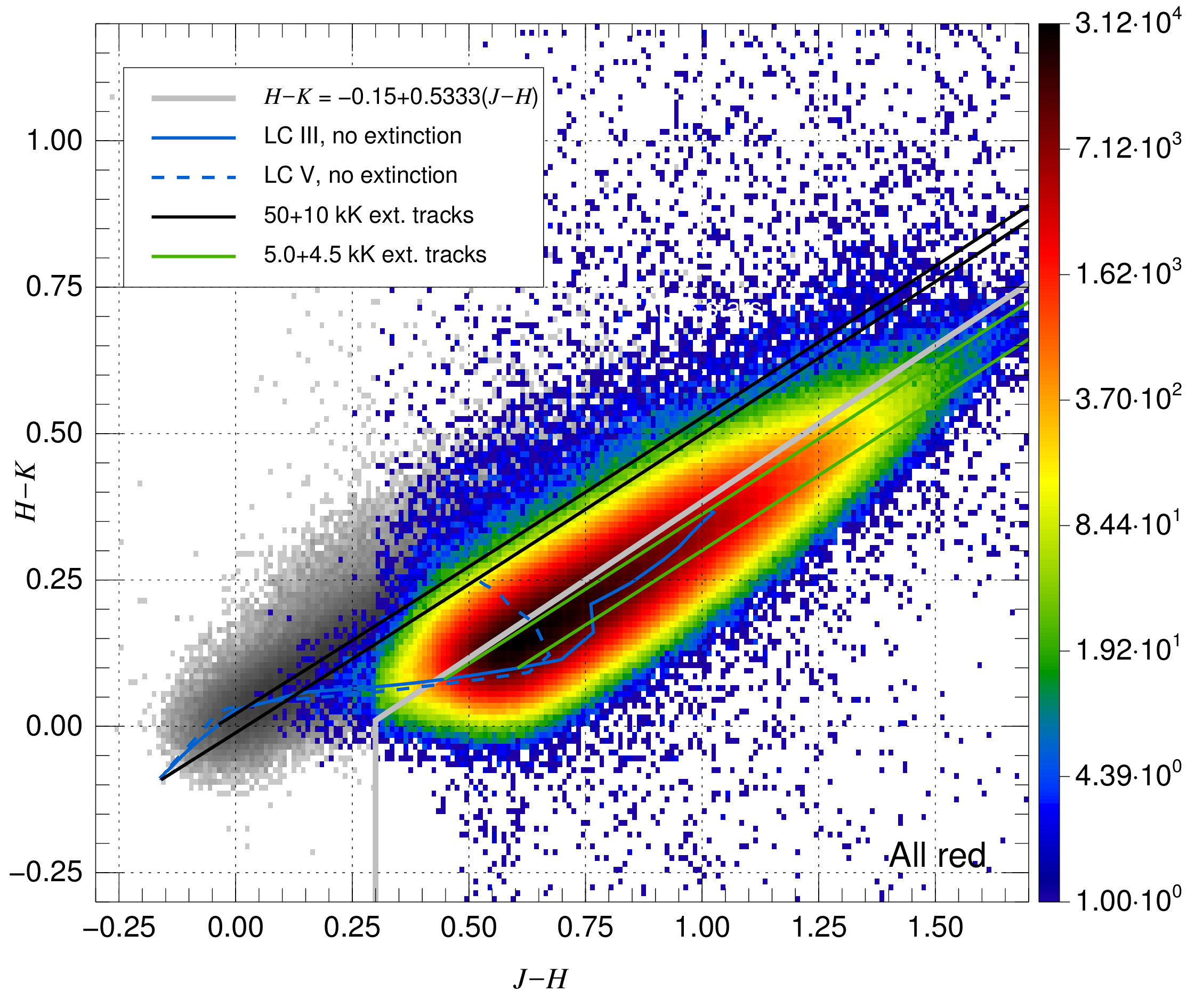} \
  \includegraphics*[width=0.49\linewidth]{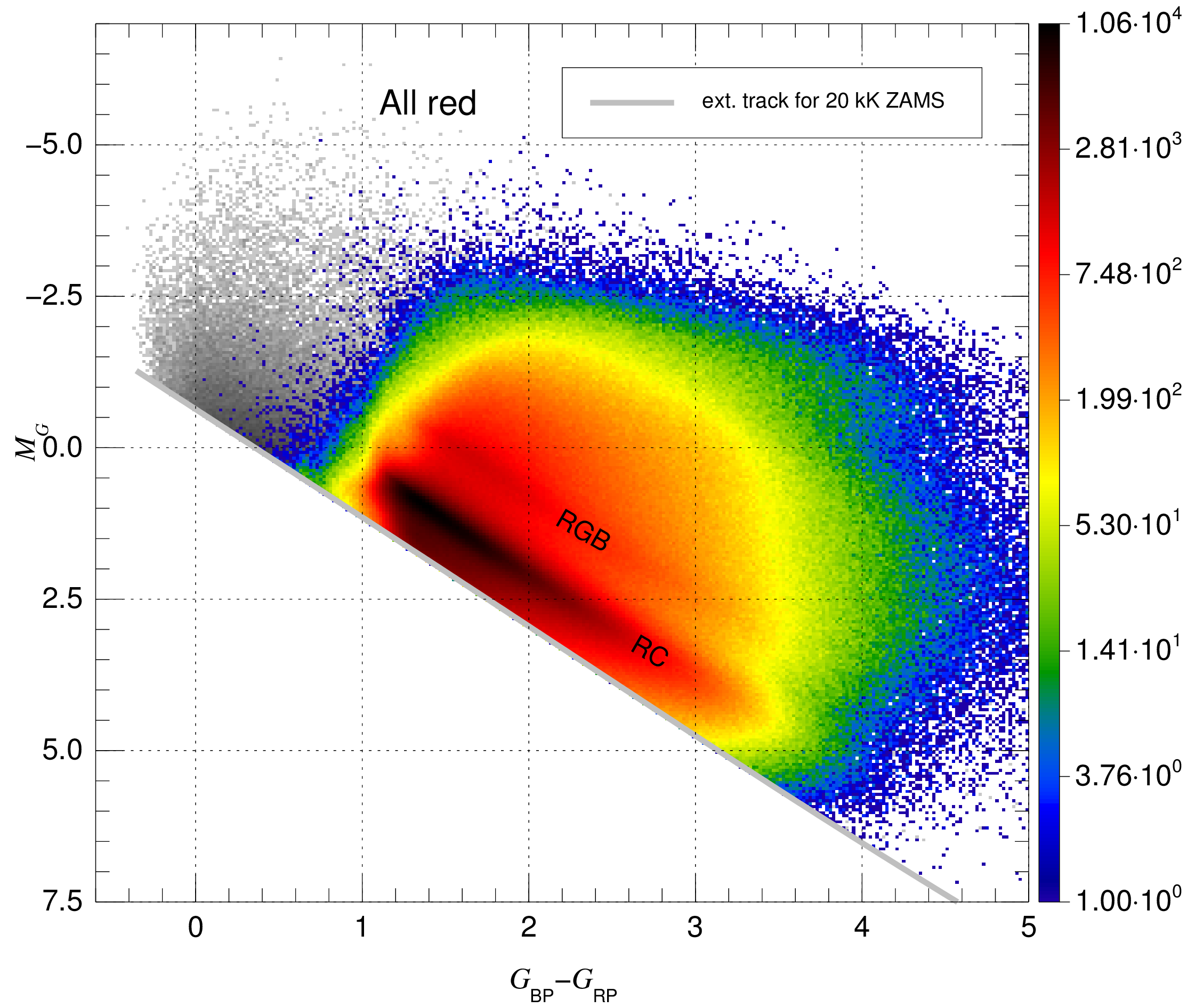}  
  \end{minipage}
  \newframe[5]
  \begin{minipage}{\linewidth}
  \includegraphics*[width=0.49\linewidth]{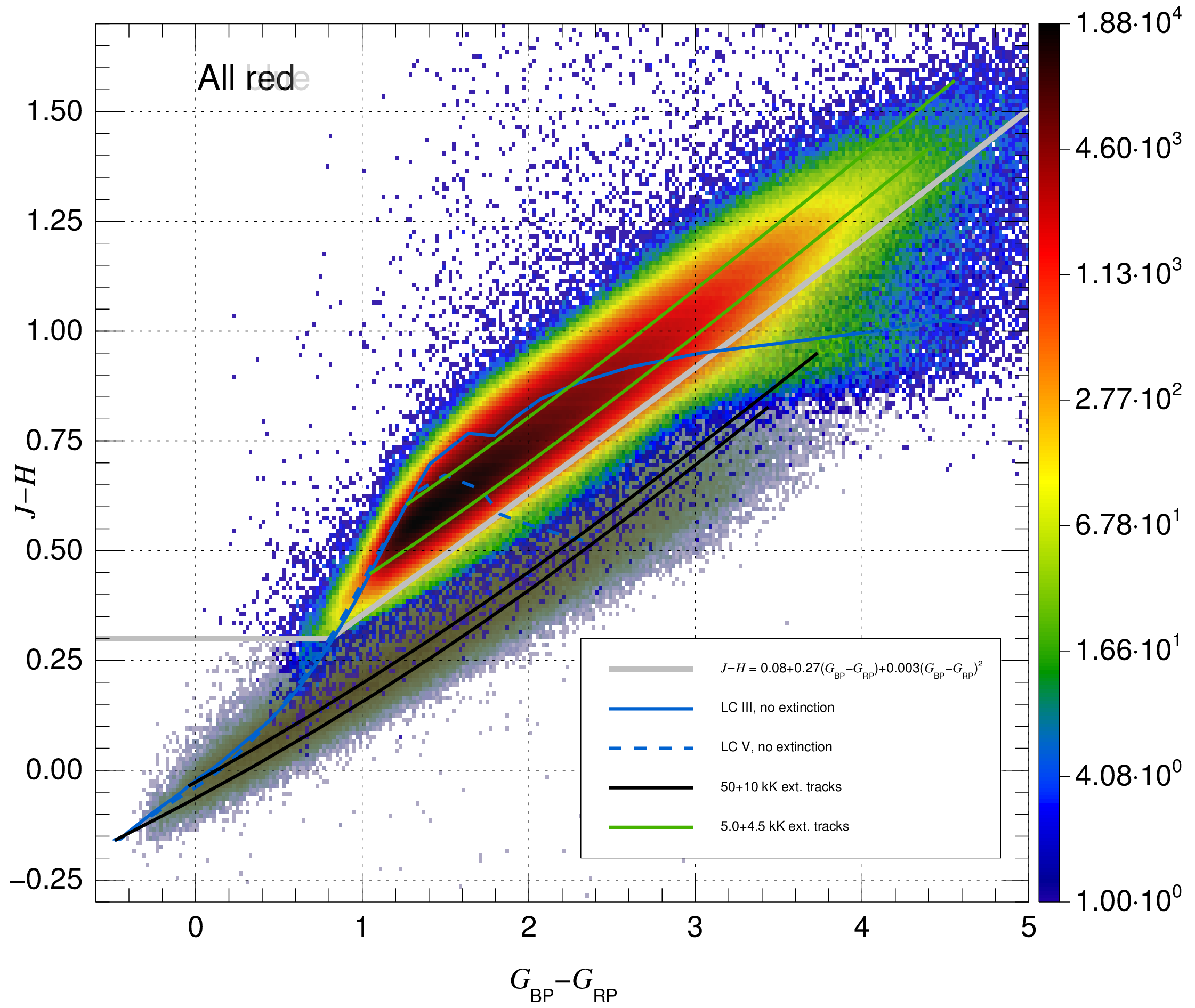} \
  \includegraphics*[width=0.49\linewidth]{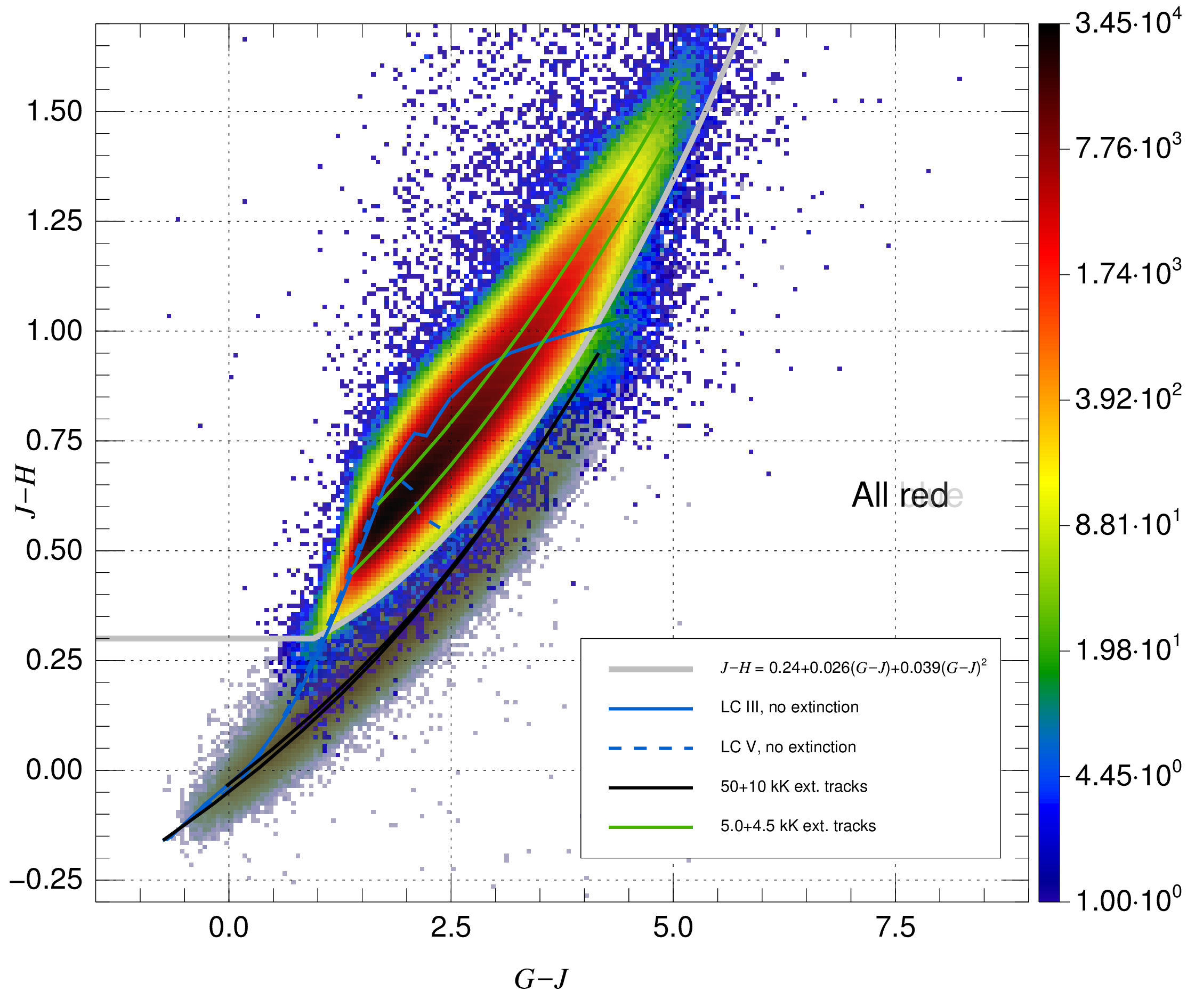}
  \\           
  \includegraphics*[width=0.49\linewidth]{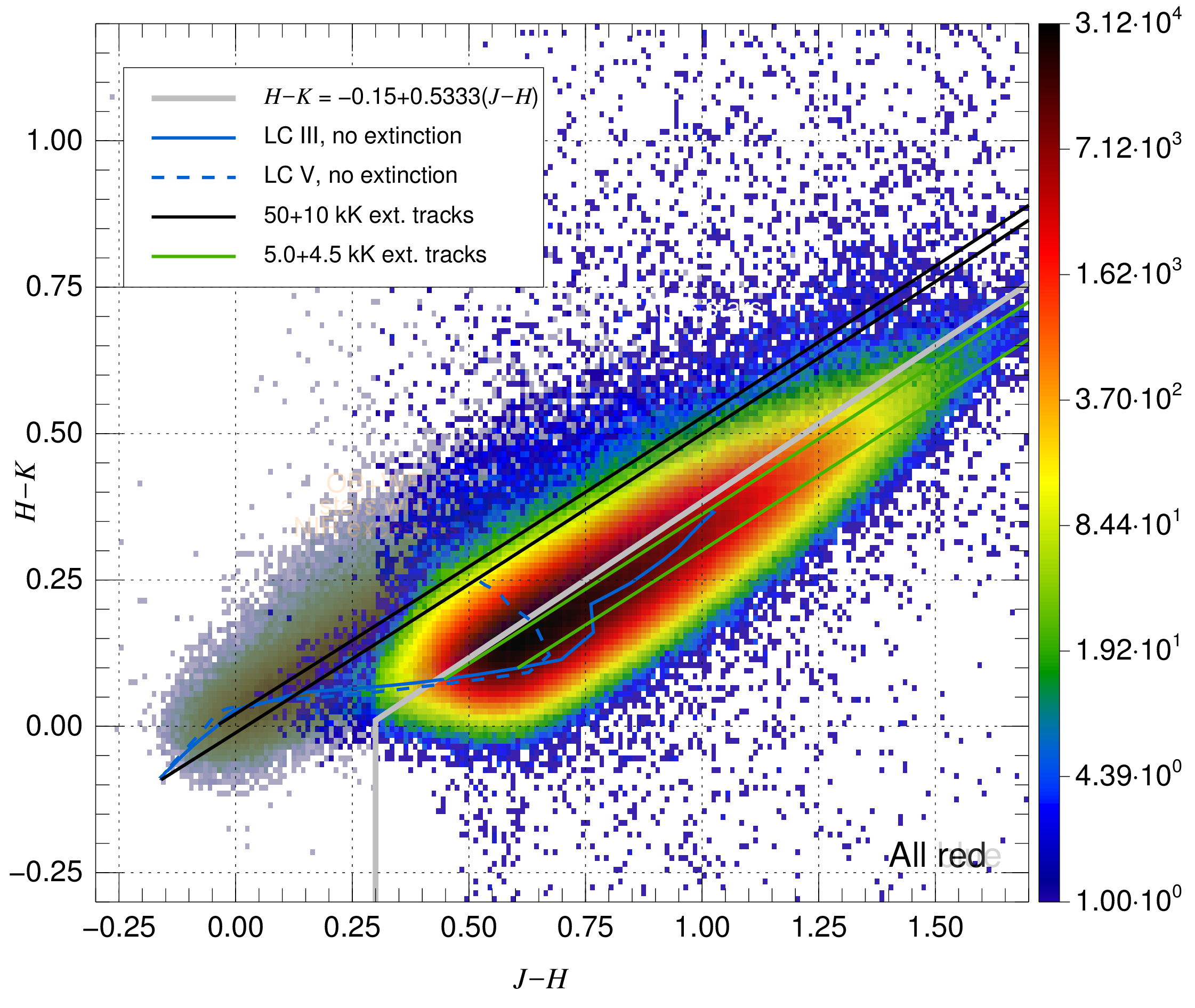} \
  \includegraphics*[width=0.49\linewidth]{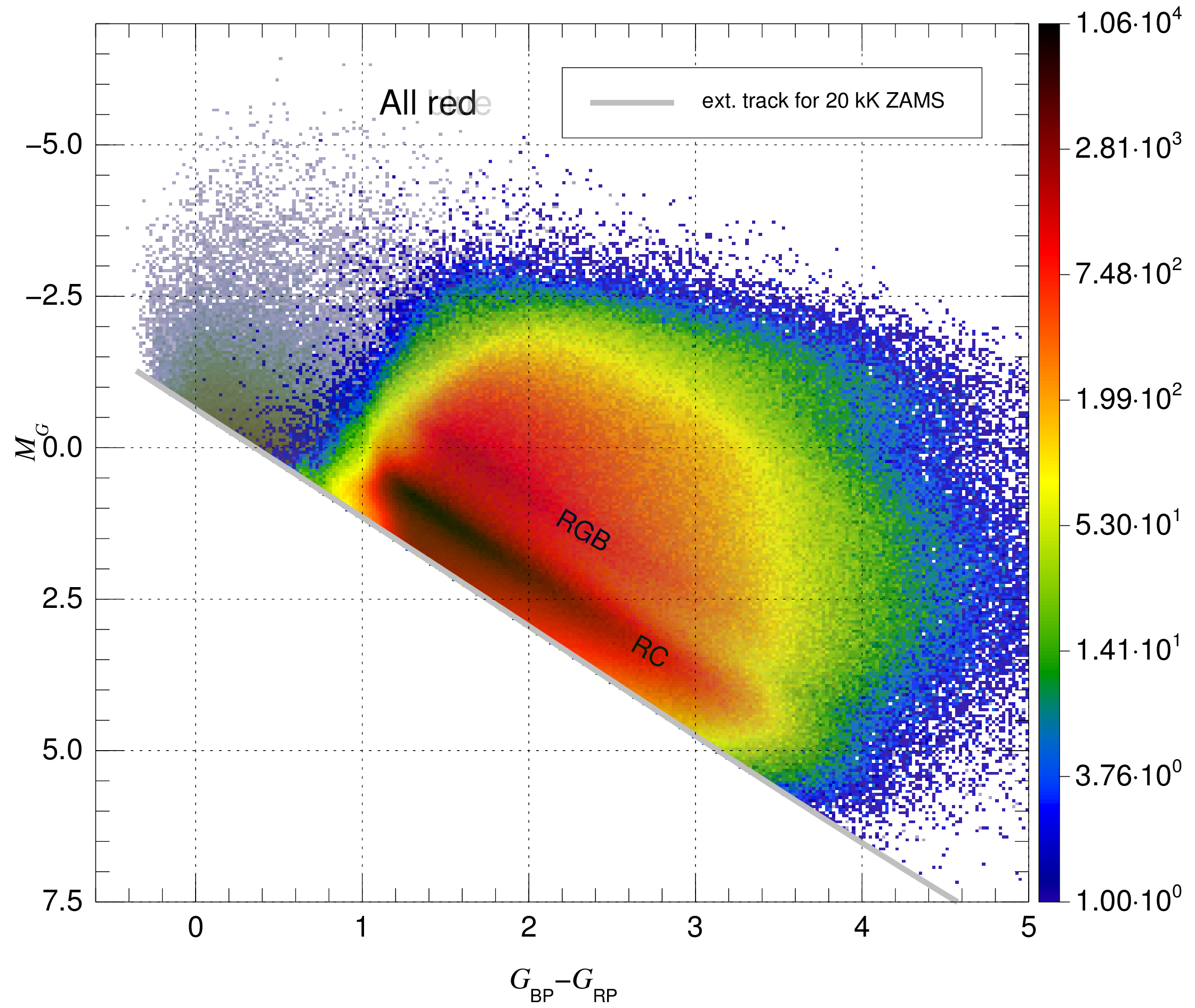}  
  \end{minipage}
  \newframe
  \begin{minipage}{\linewidth}
  \includegraphics*[width=0.49\linewidth]{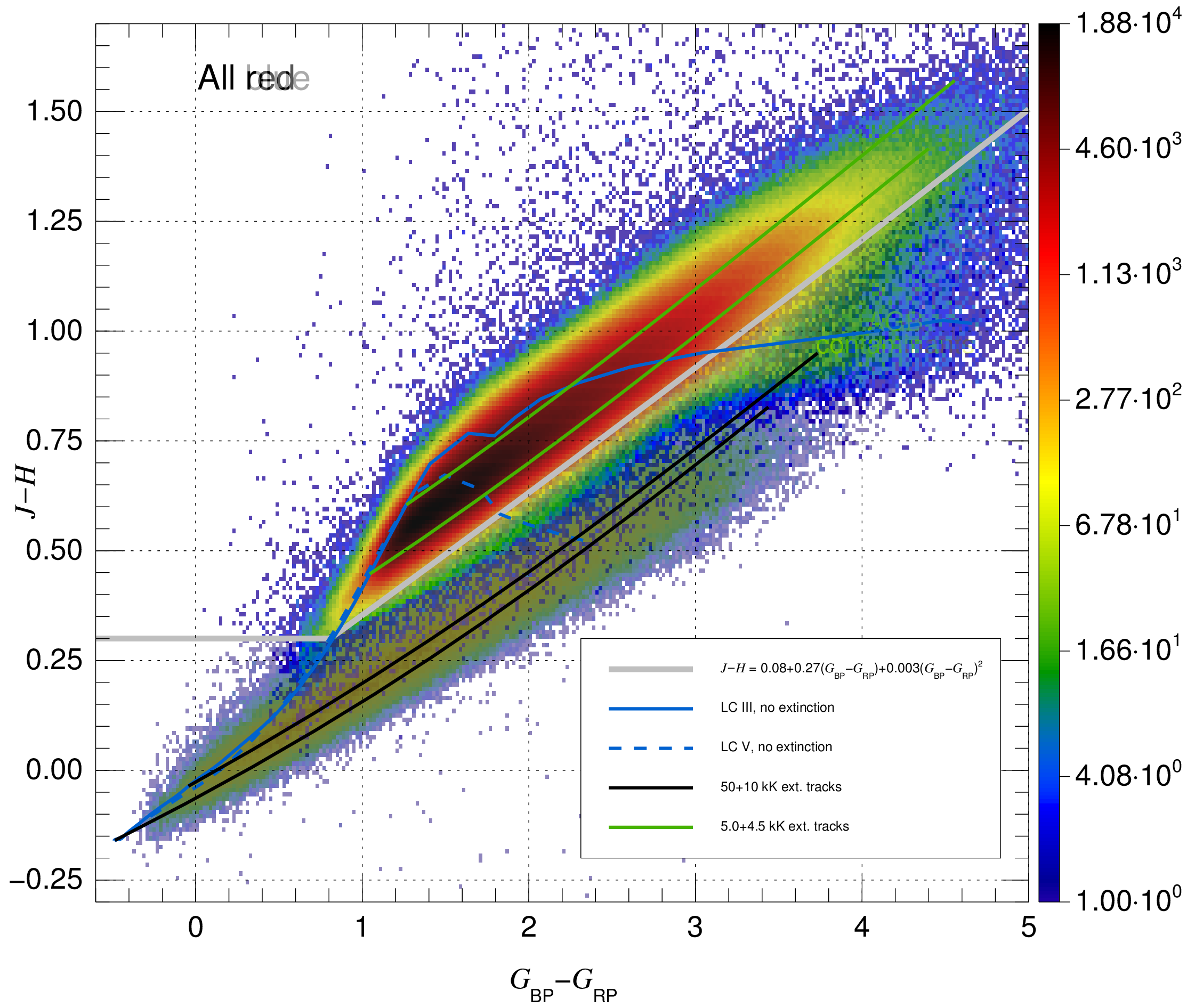} \
  \includegraphics*[width=0.49\linewidth]{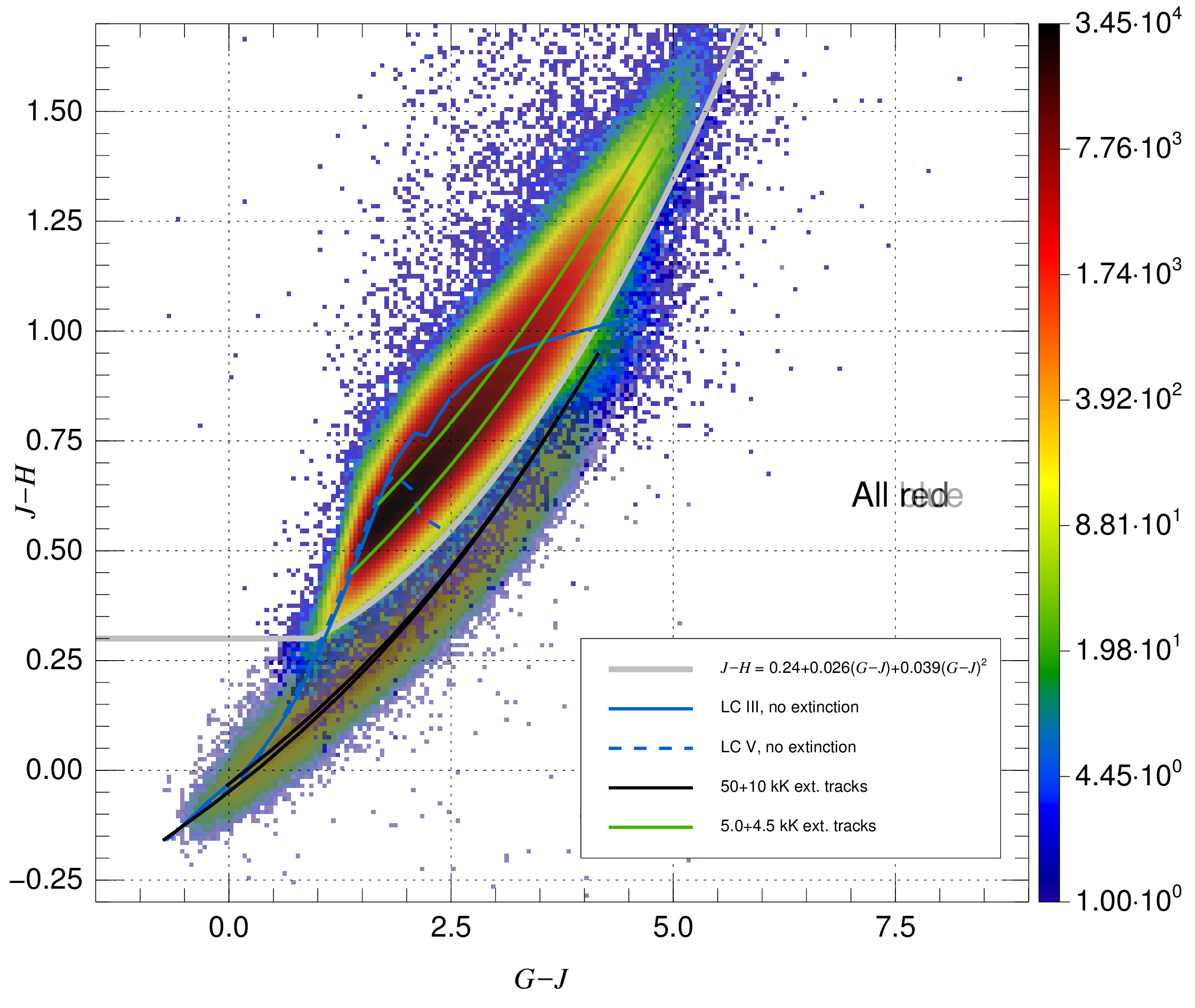}
  \\           
  \includegraphics*[width=0.49\linewidth]{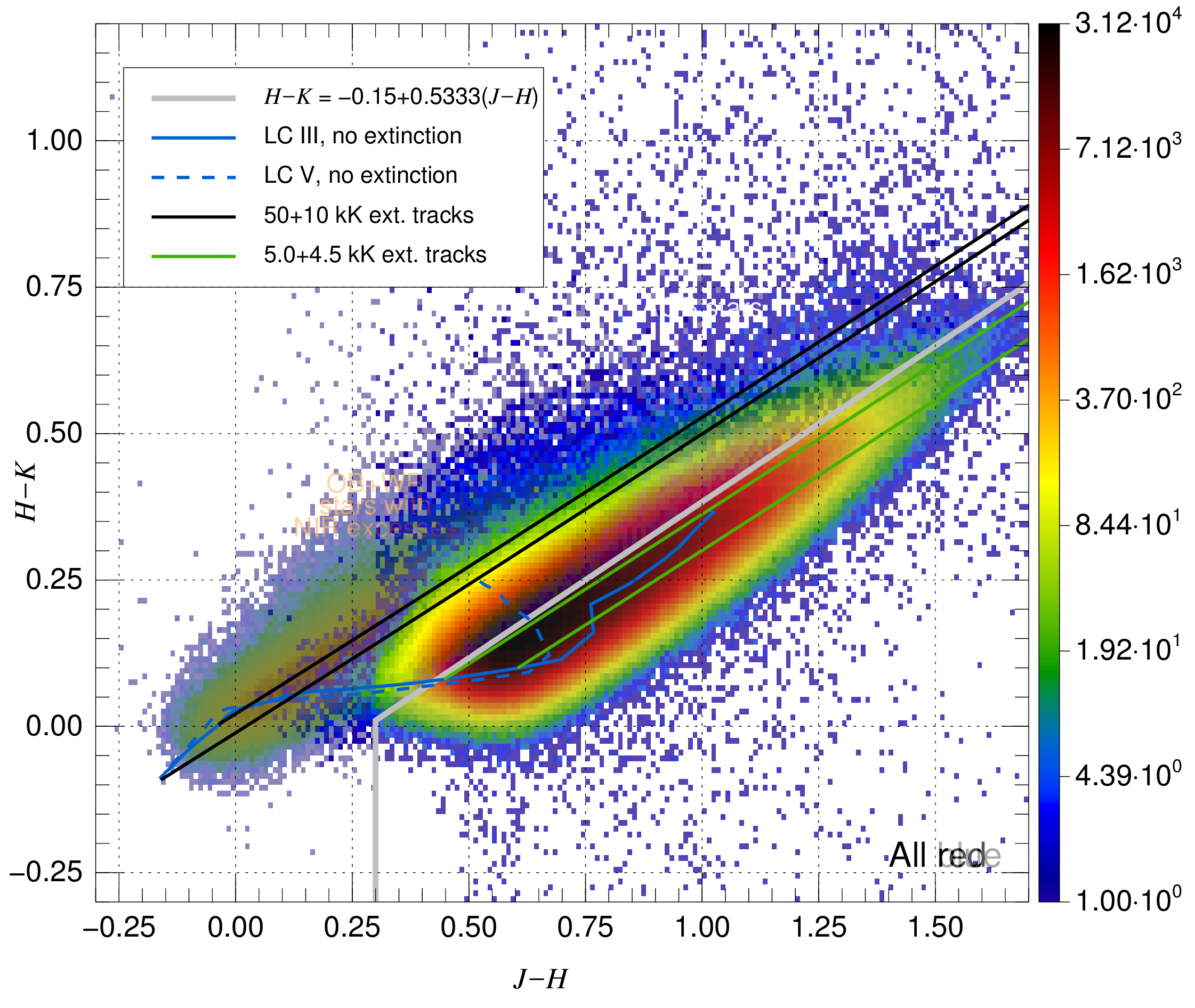} \
  \includegraphics*[width=0.49\linewidth]{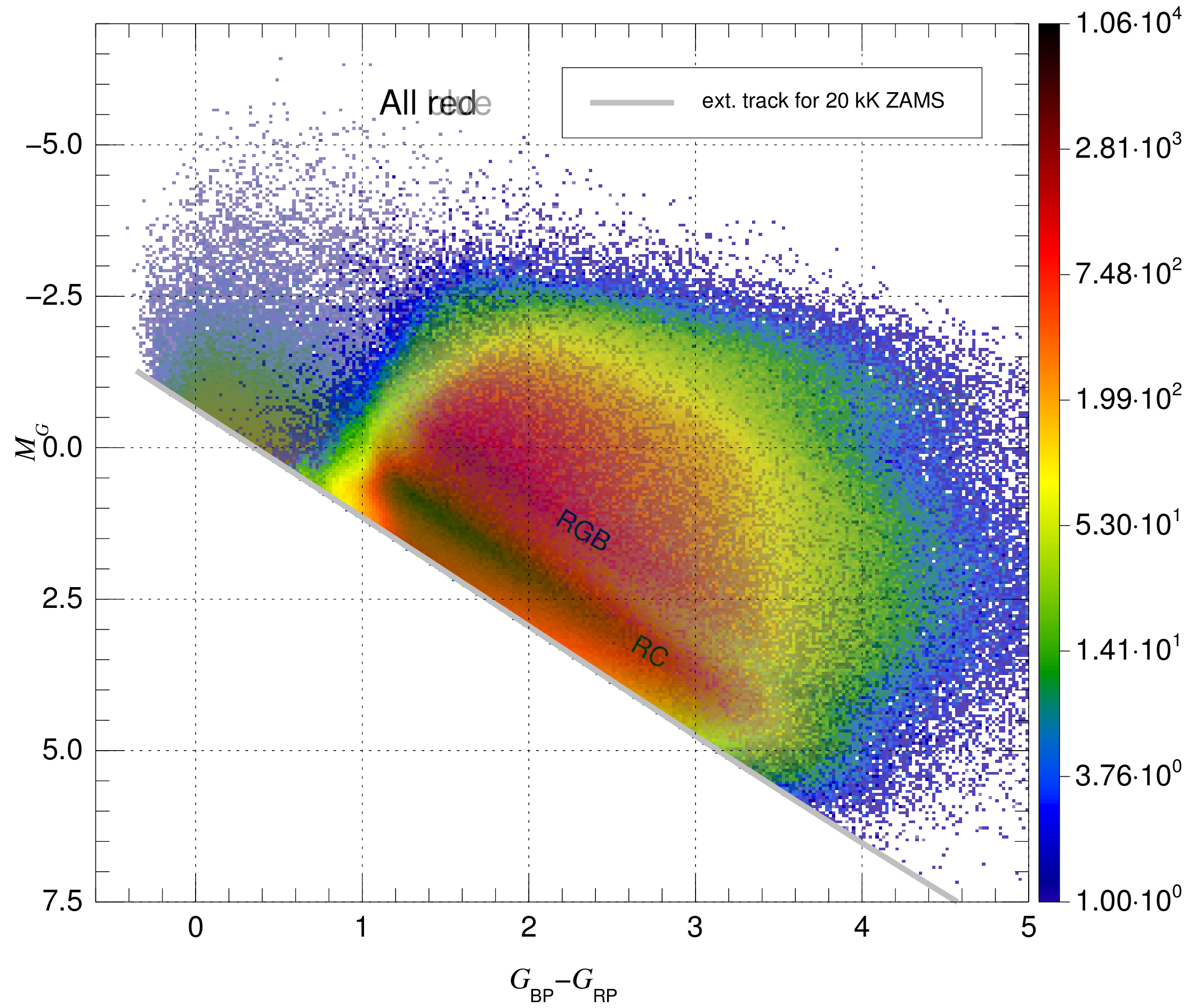}  
  \end{minipage}
  \newframe
  \begin{minipage}{\linewidth}
  \includegraphics*[width=0.49\linewidth]{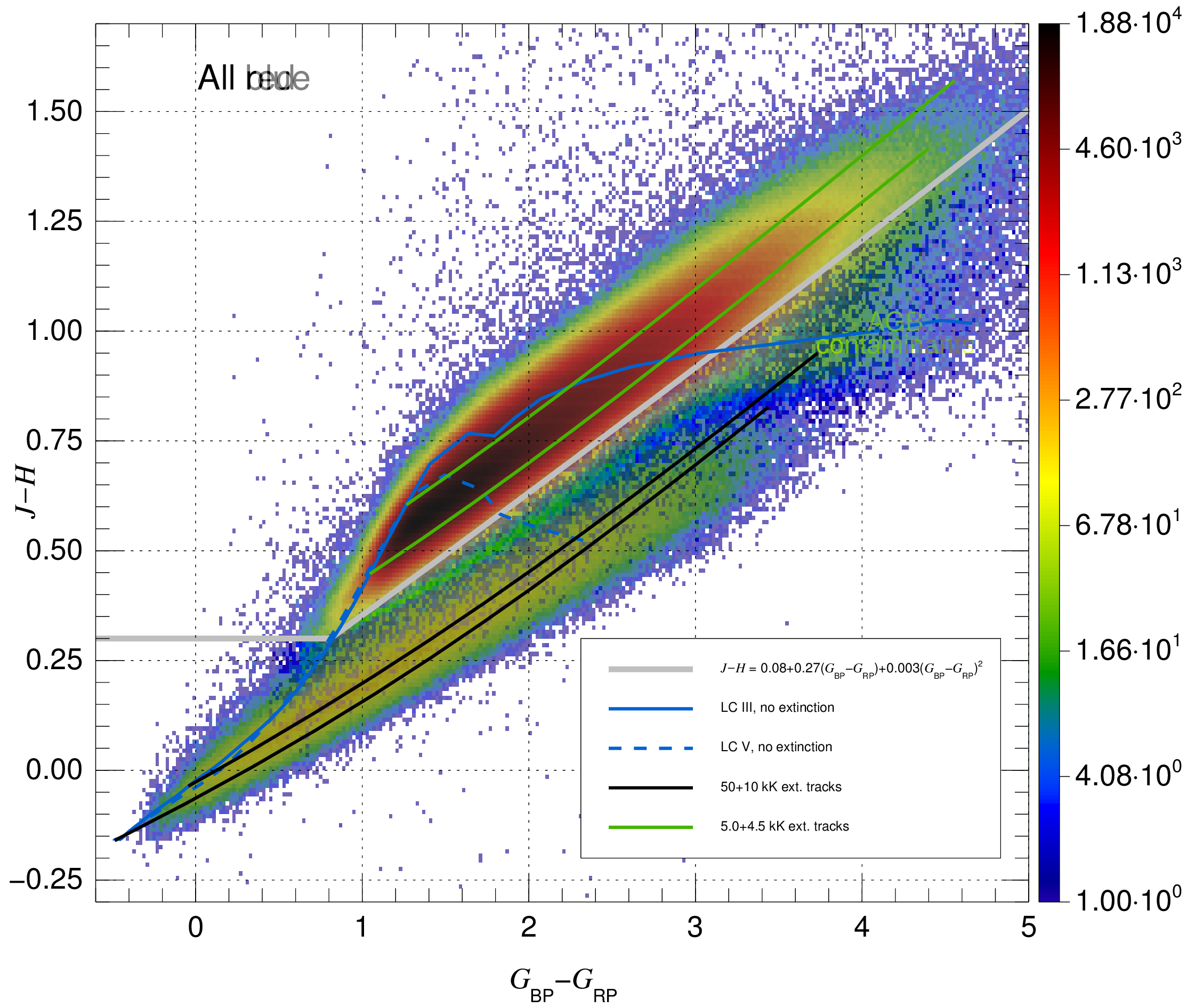} \
  \includegraphics*[width=0.49\linewidth]{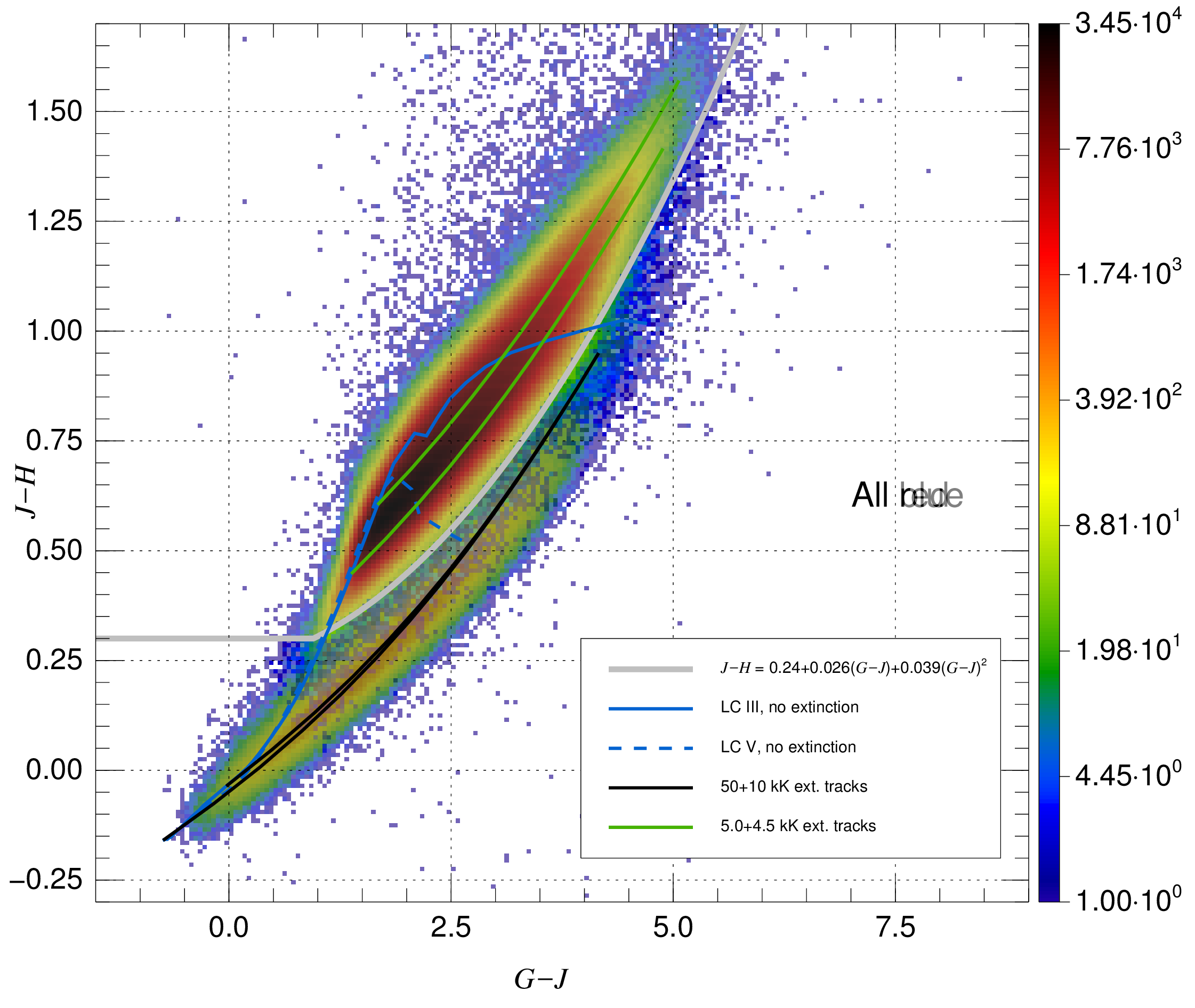}
  \\           
  \includegraphics*[width=0.49\linewidth]{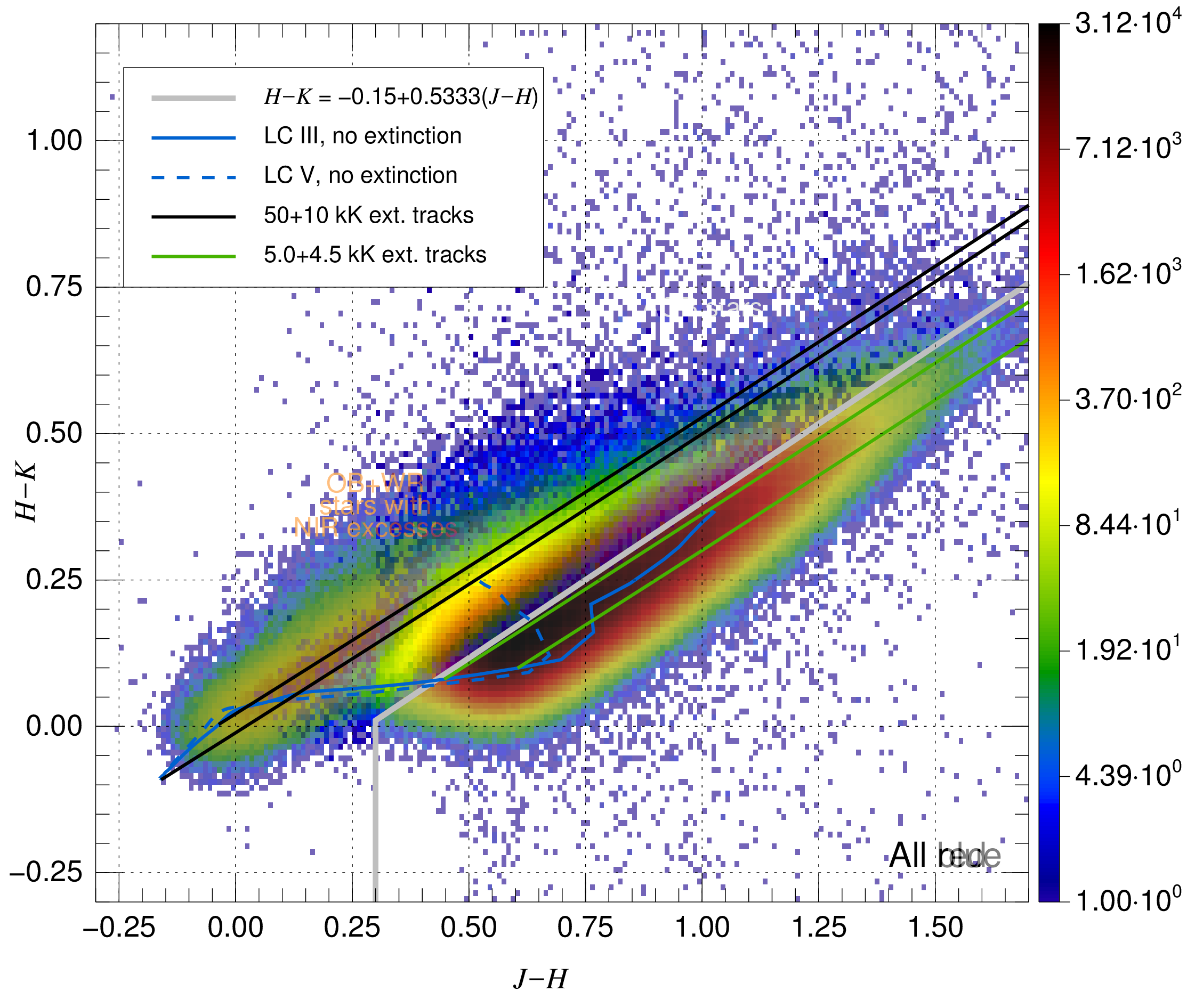} \
  \includegraphics*[width=0.49\linewidth]{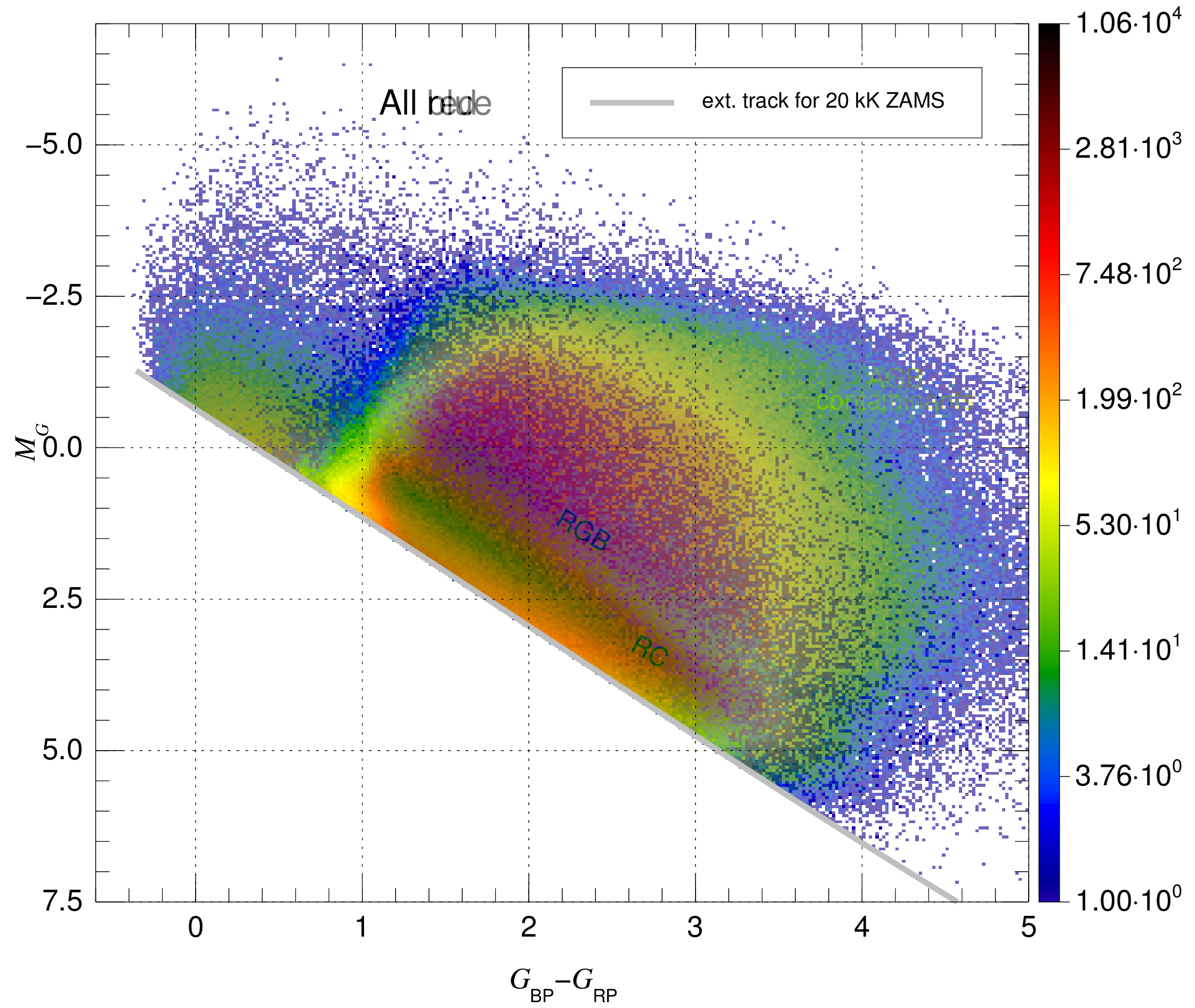}  
  \end{minipage}
  \newframe
  \begin{minipage}{\linewidth}
  \includegraphics*[width=0.49\linewidth]{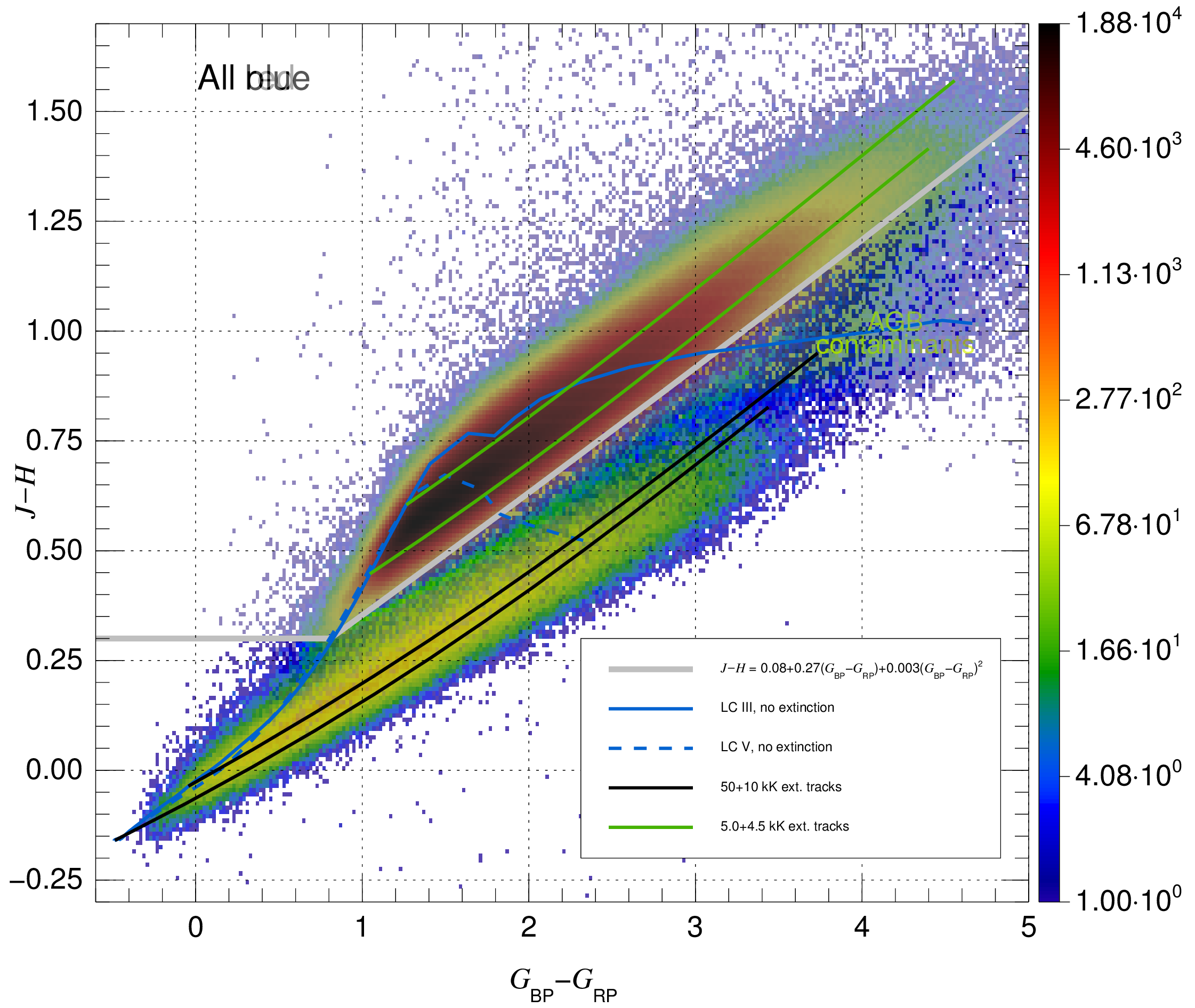} \
  \includegraphics*[width=0.49\linewidth]{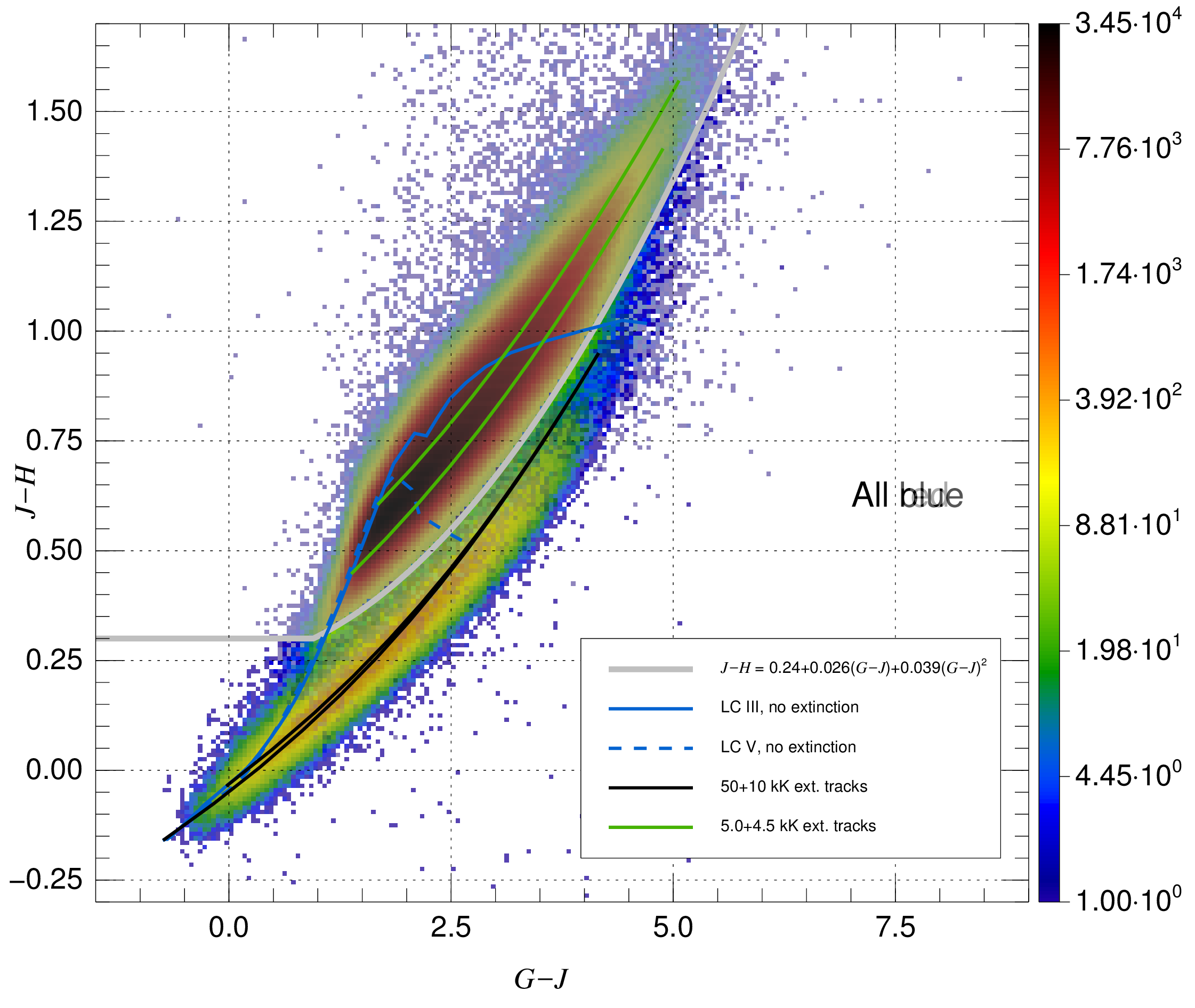}
  \\           
  \includegraphics*[width=0.49\linewidth]{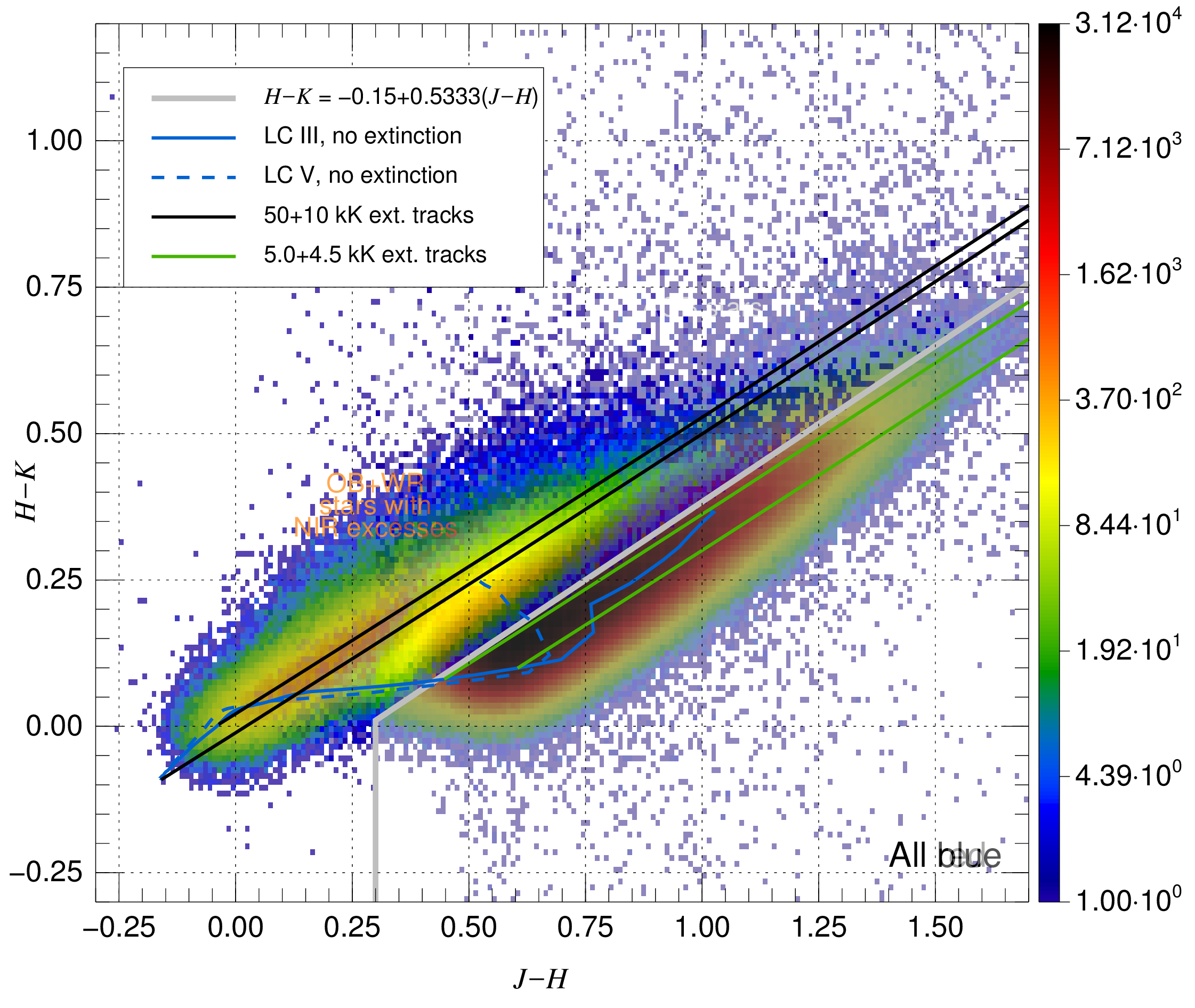} \
  \includegraphics*[width=0.49\linewidth]{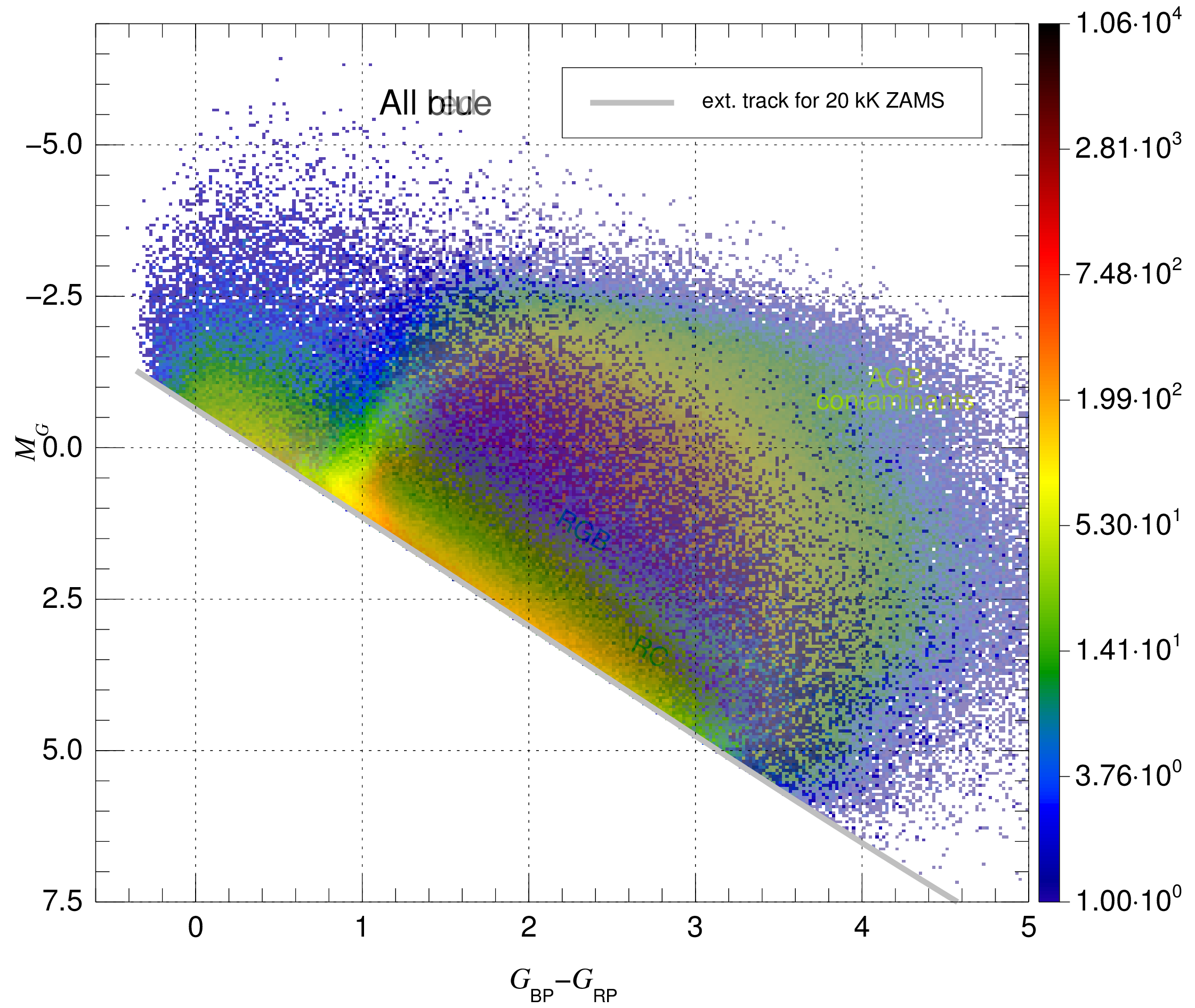}  
  \end{minipage}
  \newframe
  \begin{minipage}{\linewidth}
  \includegraphics*[width=0.49\linewidth]{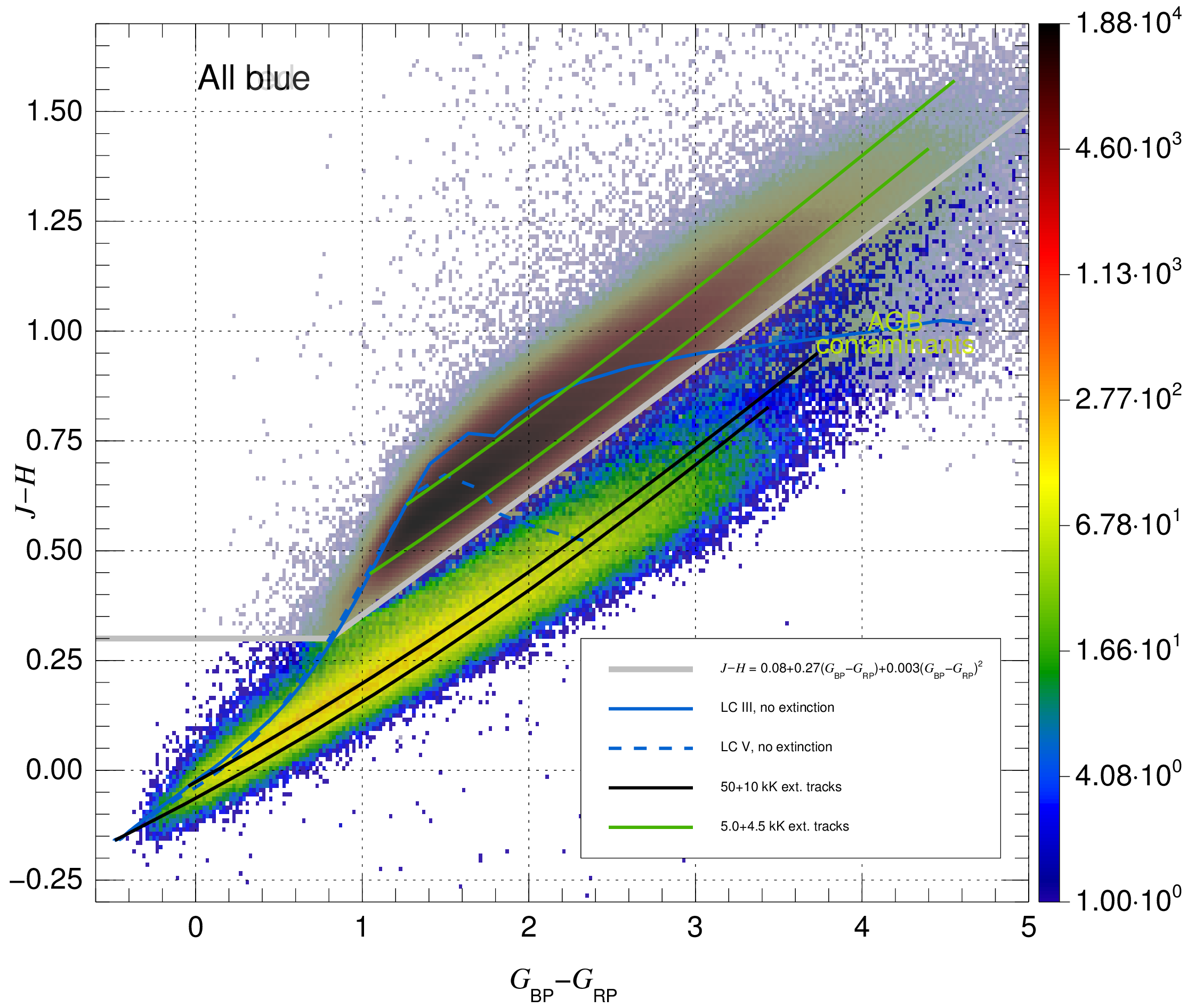} \
  \includegraphics*[width=0.49\linewidth]{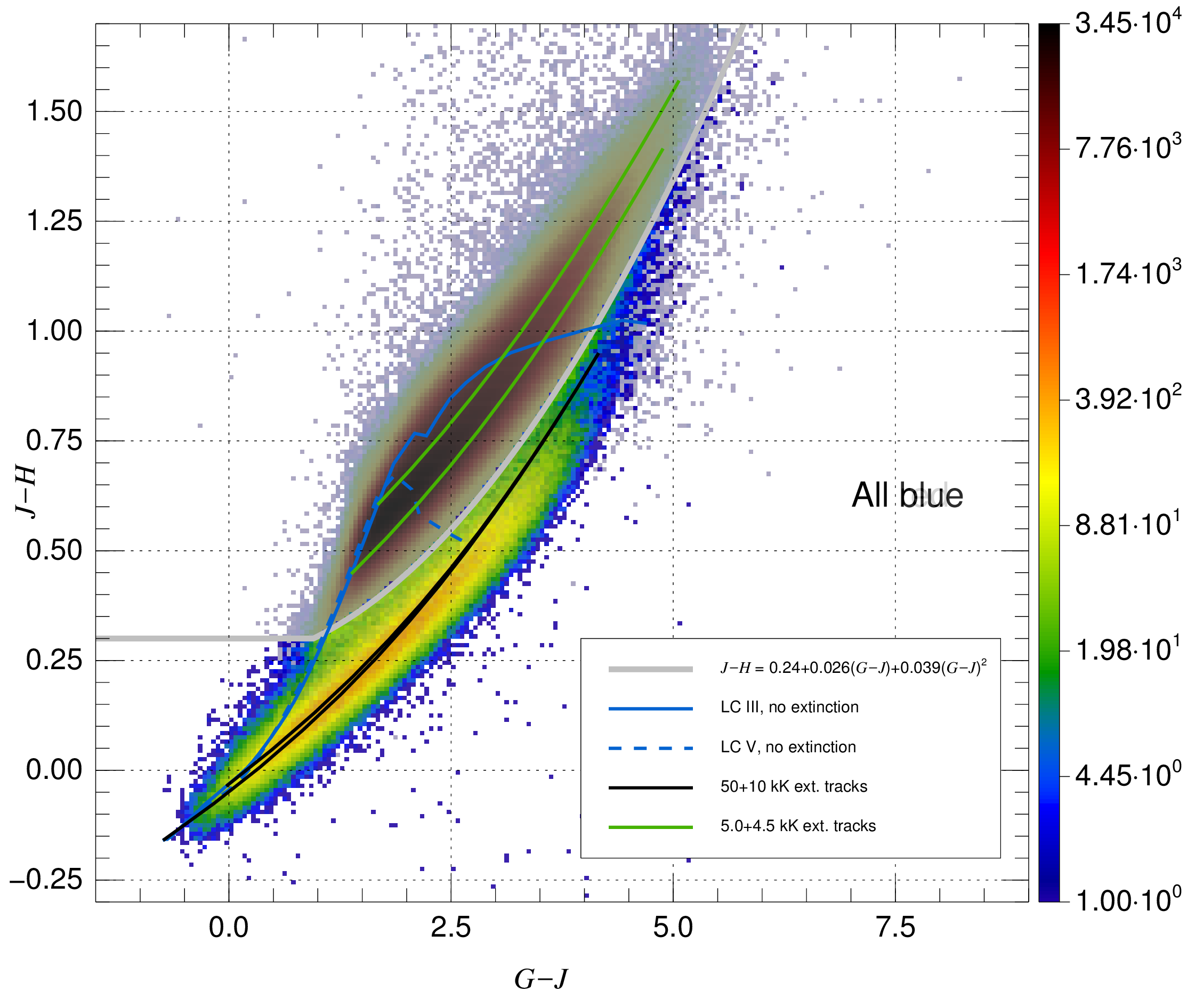}
  \\           
  \includegraphics*[width=0.49\linewidth]{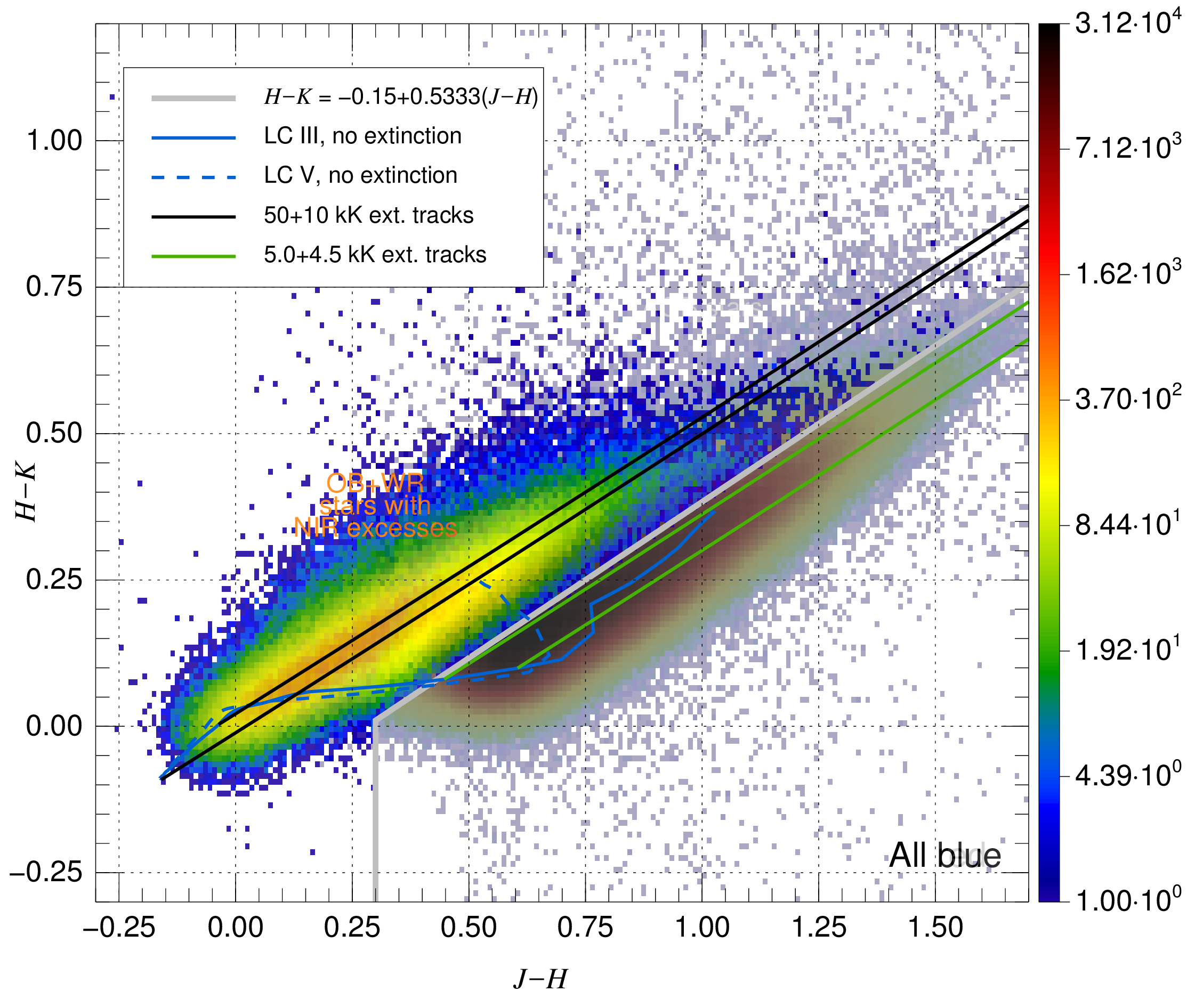} \
  \includegraphics*[width=0.49\linewidth]{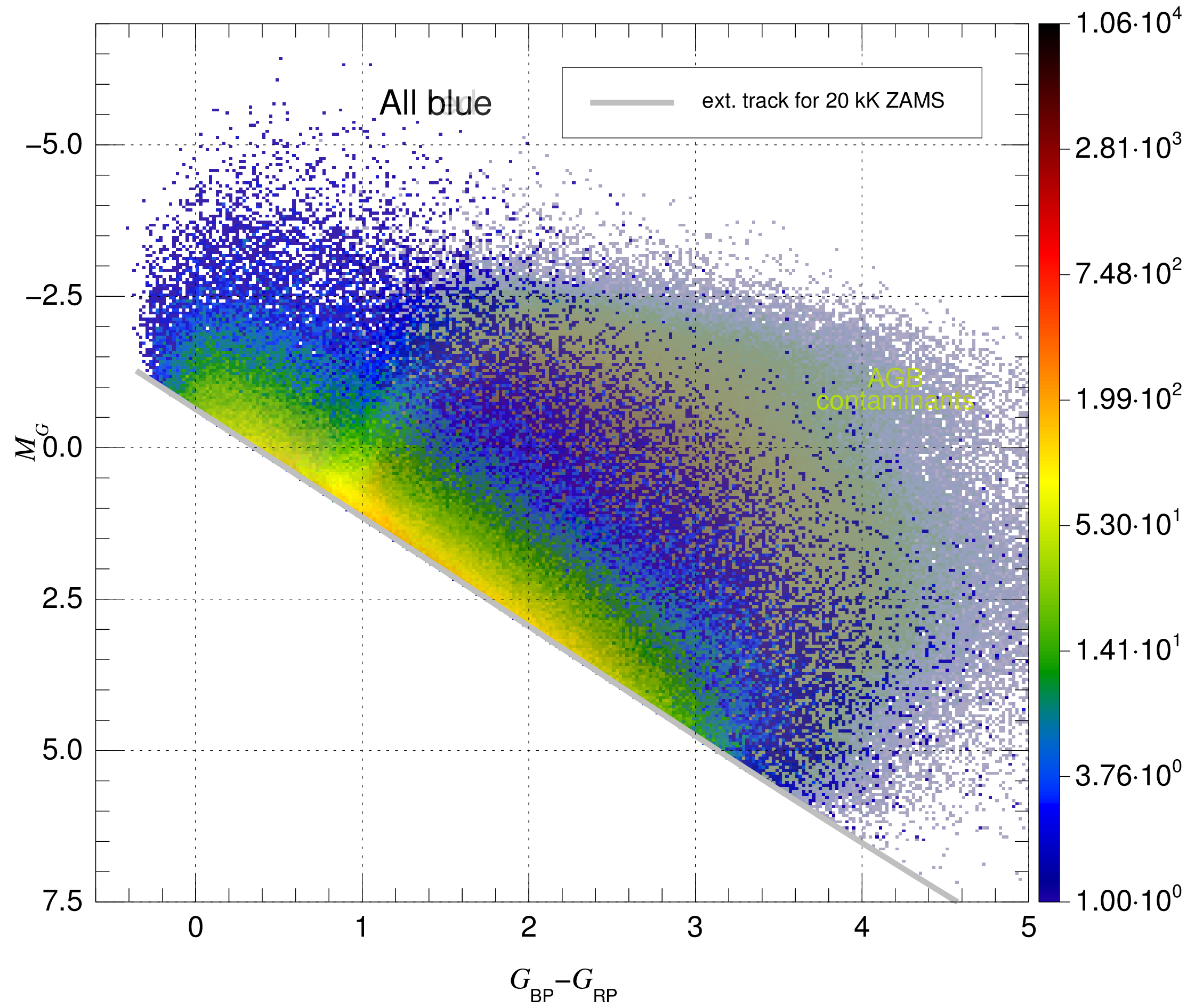}  
  \end{minipage}
  \newframe[0.4]
  \begin{minipage}{\linewidth}
  \includegraphics*[width=0.49\linewidth]{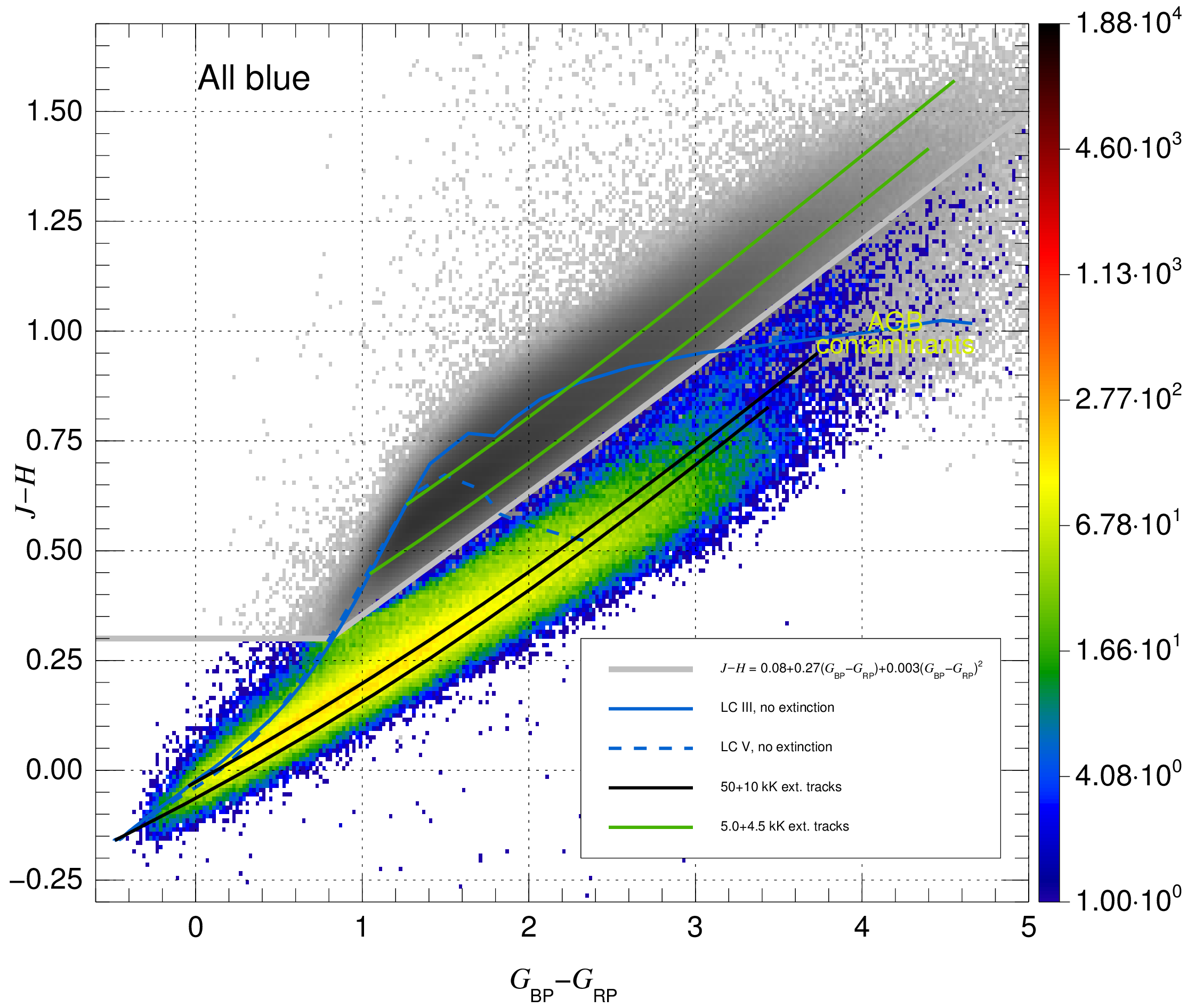} \
  \includegraphics*[width=0.49\linewidth]{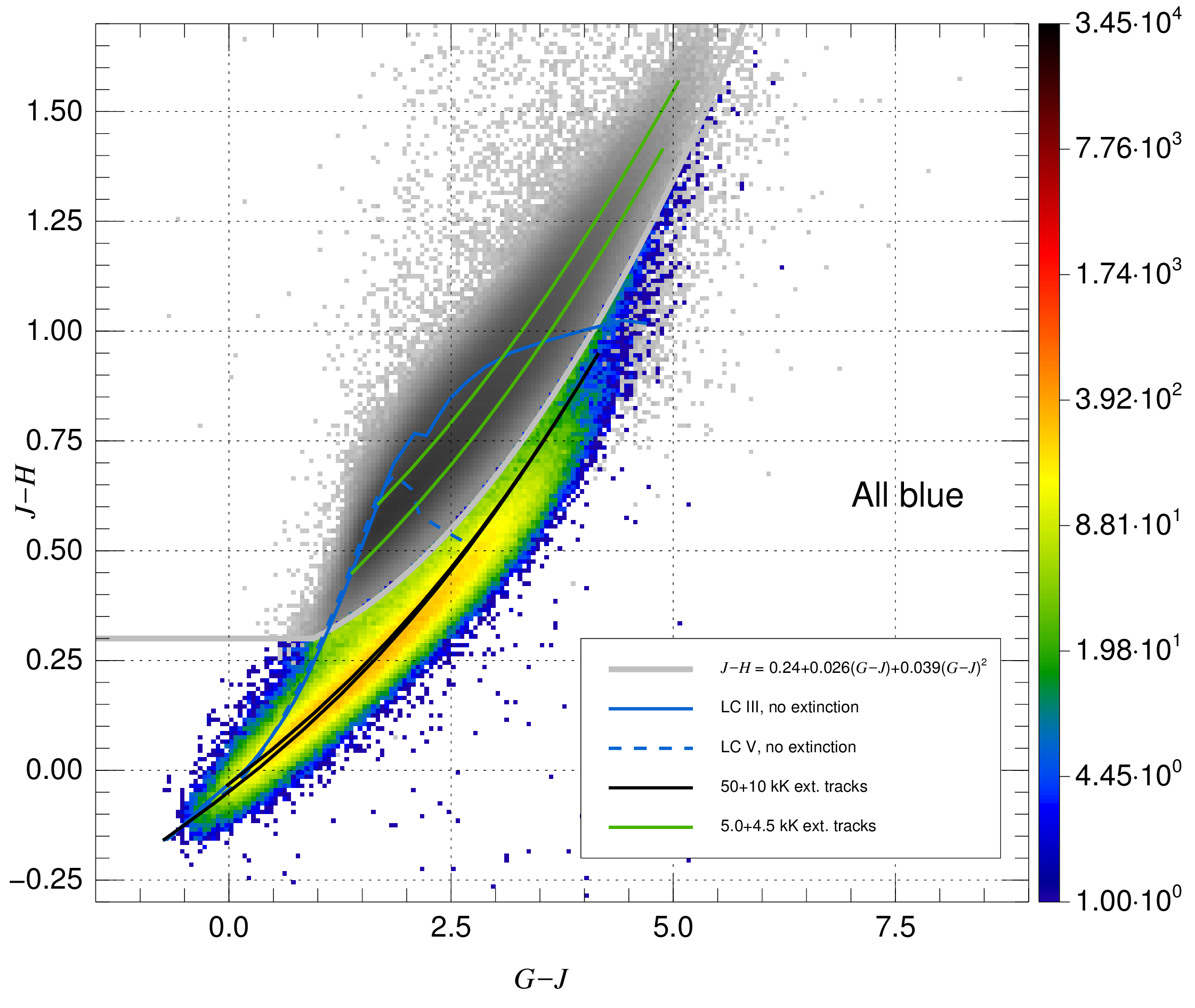}
  \\           
  \includegraphics*[width=0.49\linewidth]{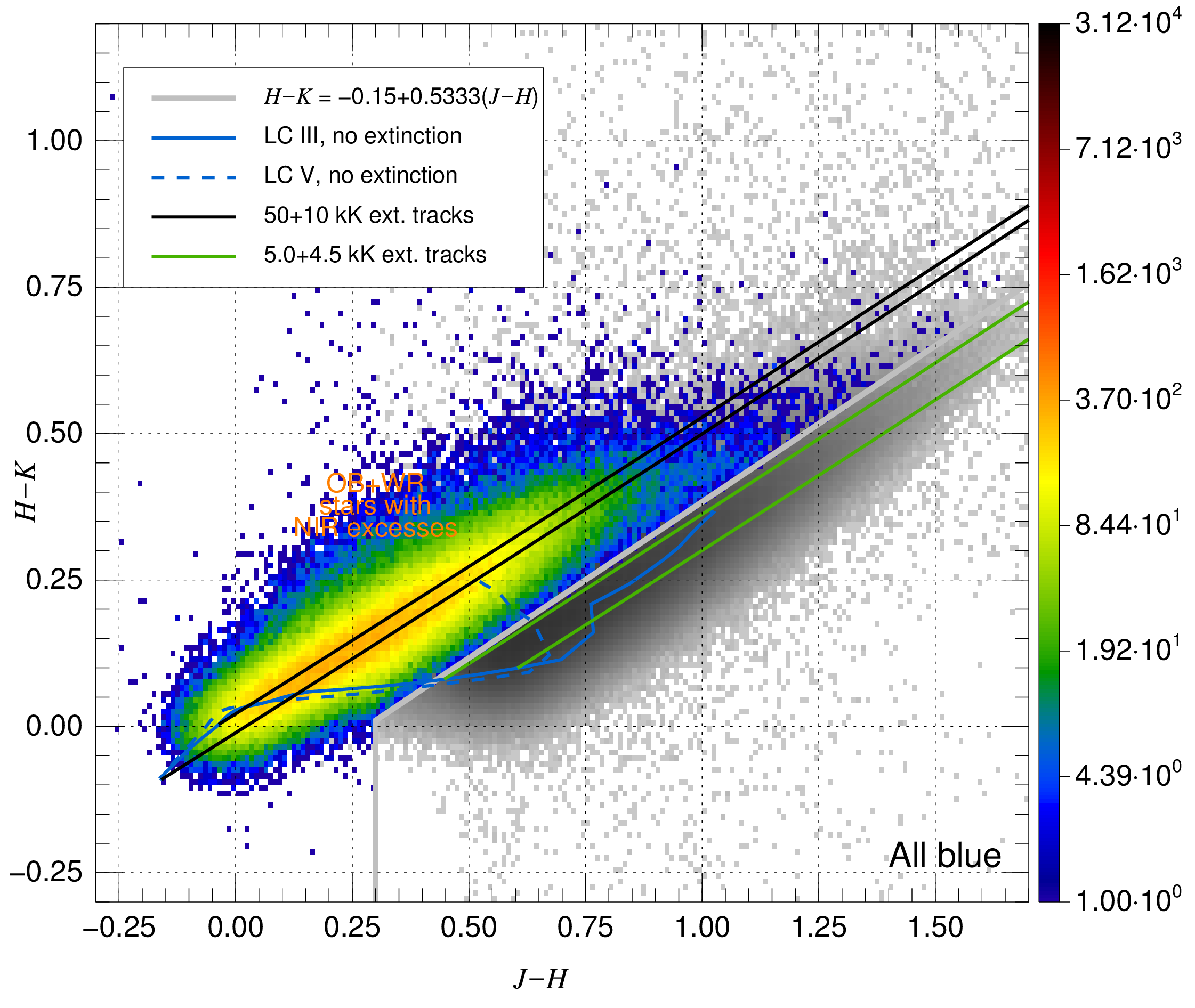} \
  \includegraphics*[width=0.49\linewidth]{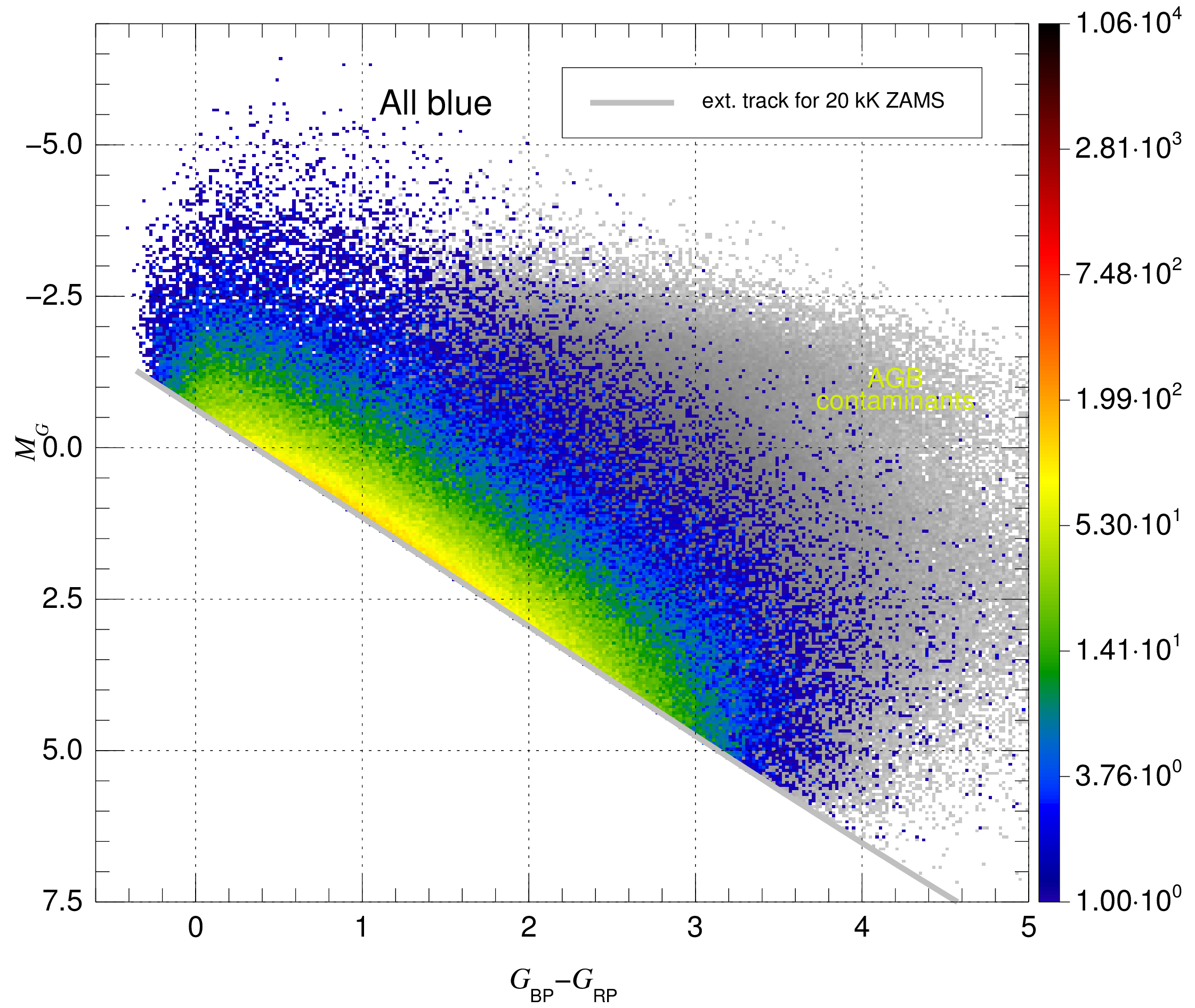}  
  \end{minipage}
  \newframe[5]
  \begin{minipage}{\linewidth}
  \includegraphics*[width=0.49\linewidth]{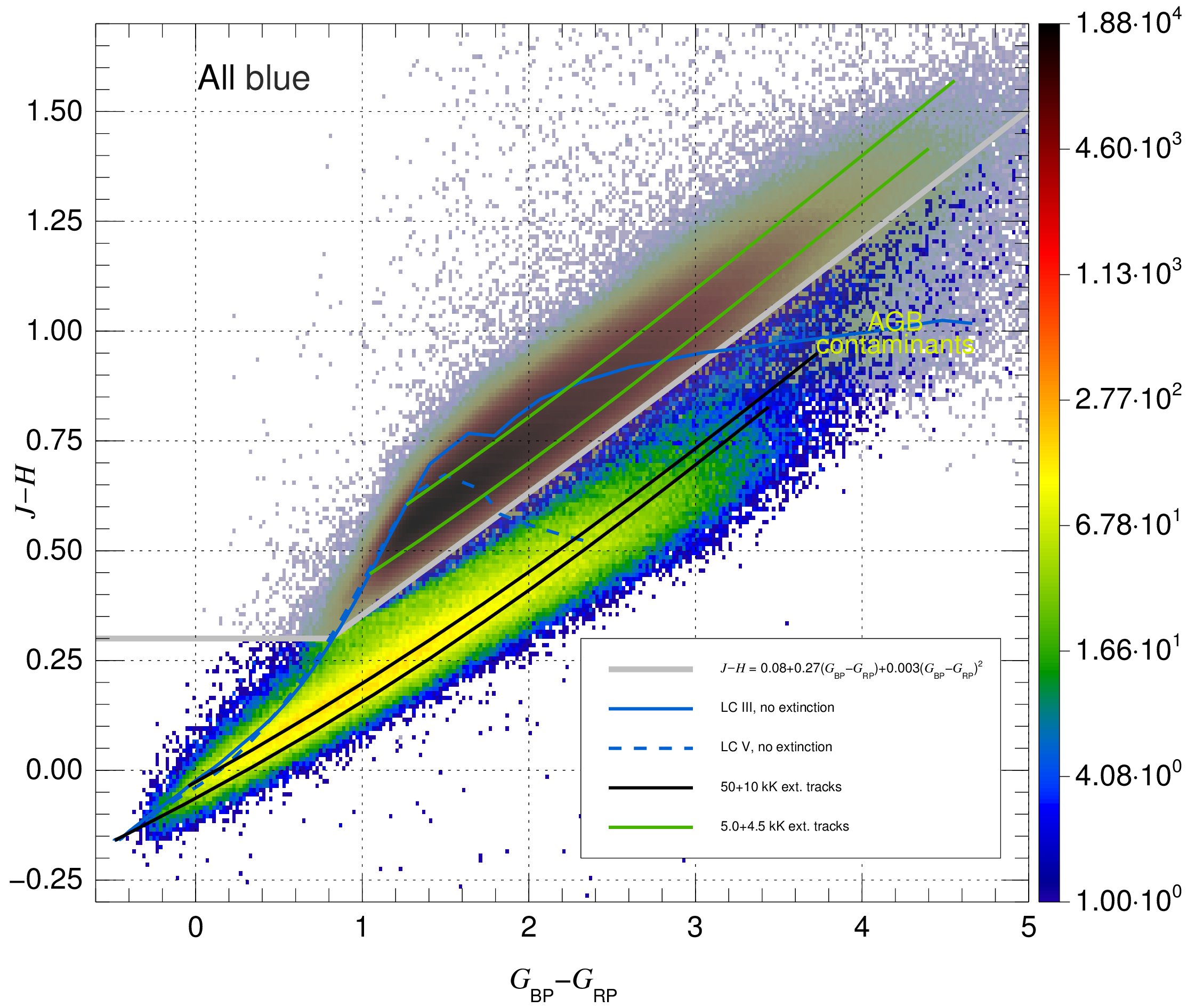} \
  \includegraphics*[width=0.49\linewidth]{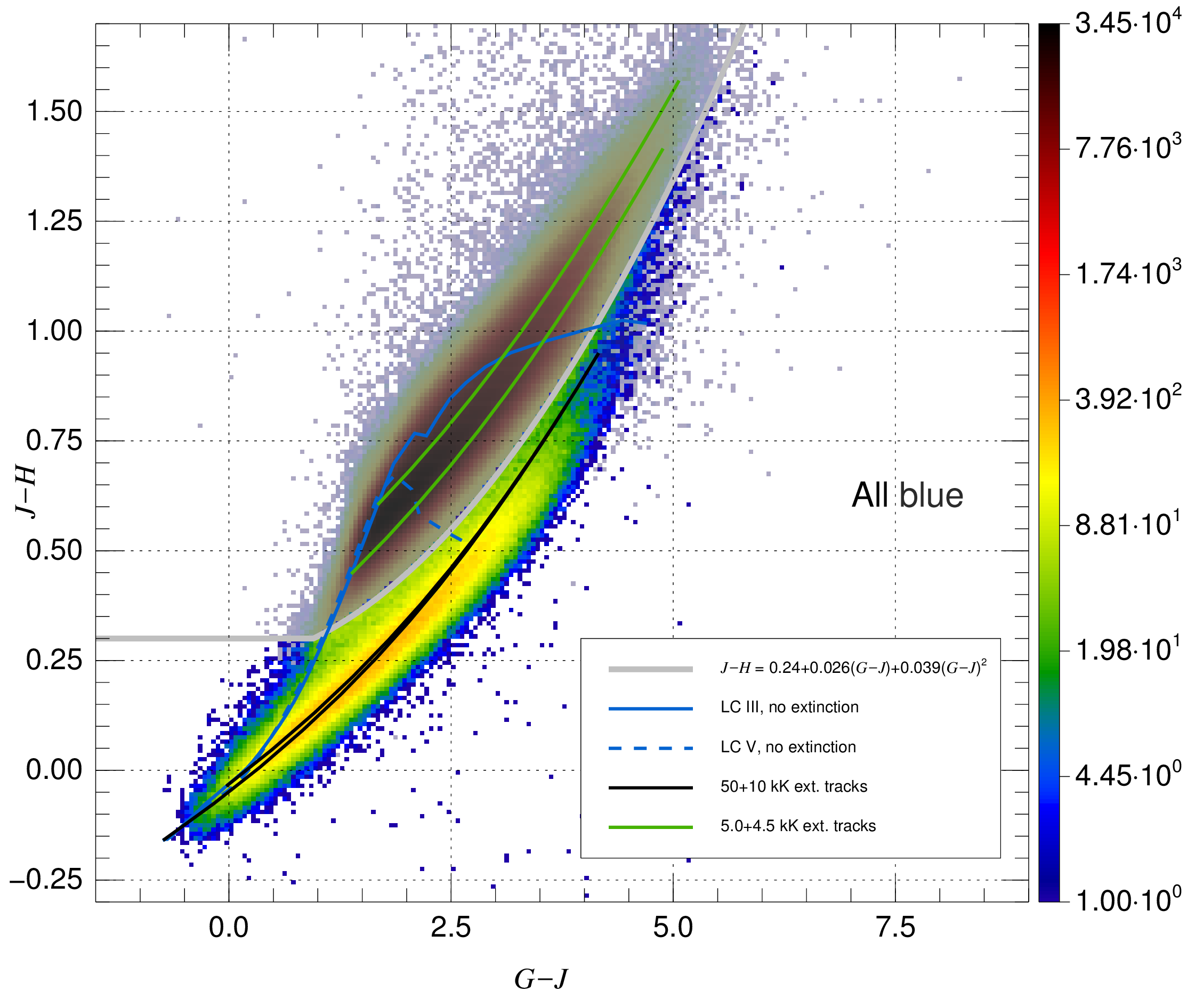}
  \\           
  \includegraphics*[width=0.49\linewidth]{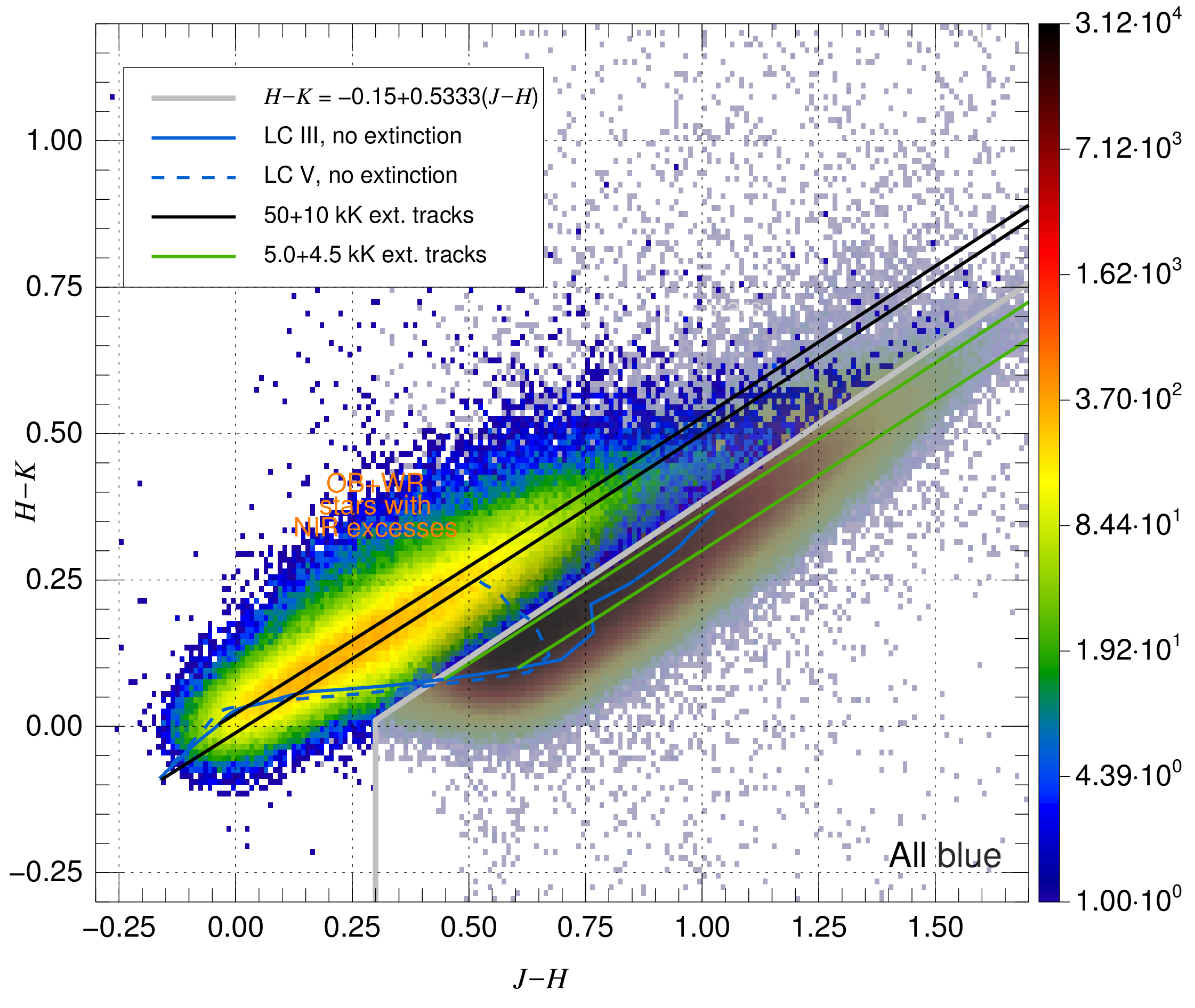} \
  \includegraphics*[width=0.49\linewidth]{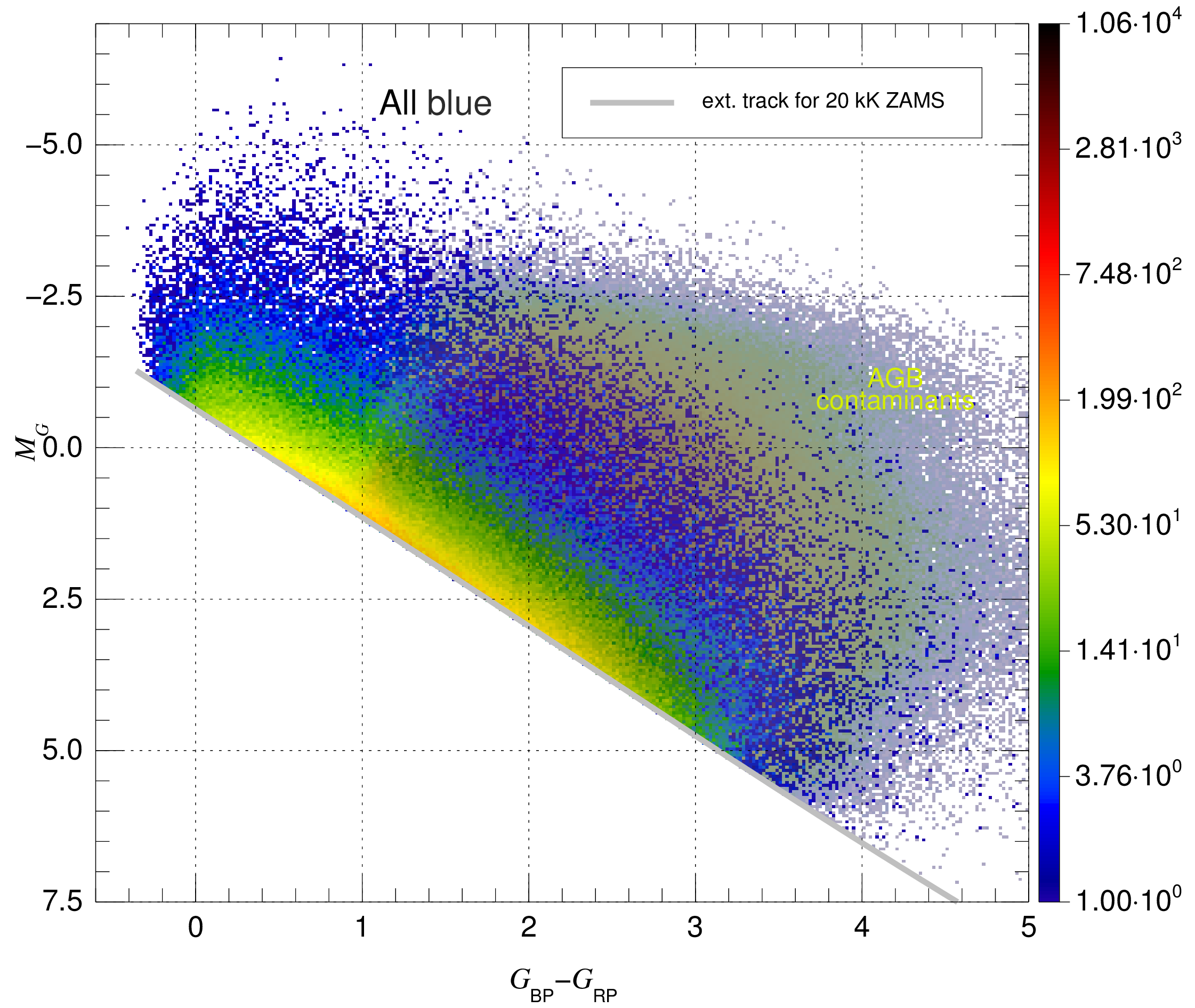}  
  \end{minipage}
  \newframe
  \begin{minipage}{\linewidth}
  \includegraphics*[width=0.49\linewidth]{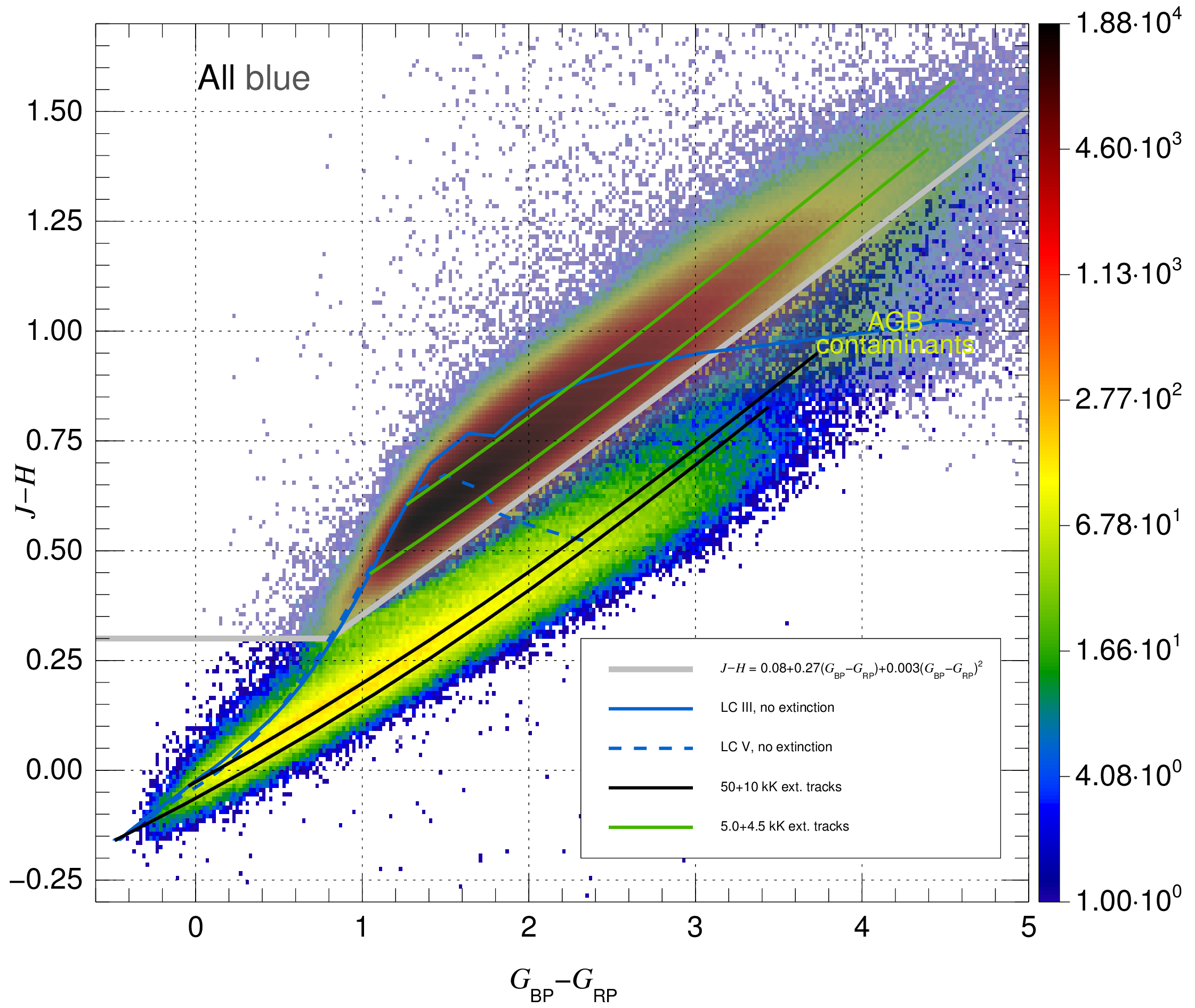} \
  \includegraphics*[width=0.49\linewidth]{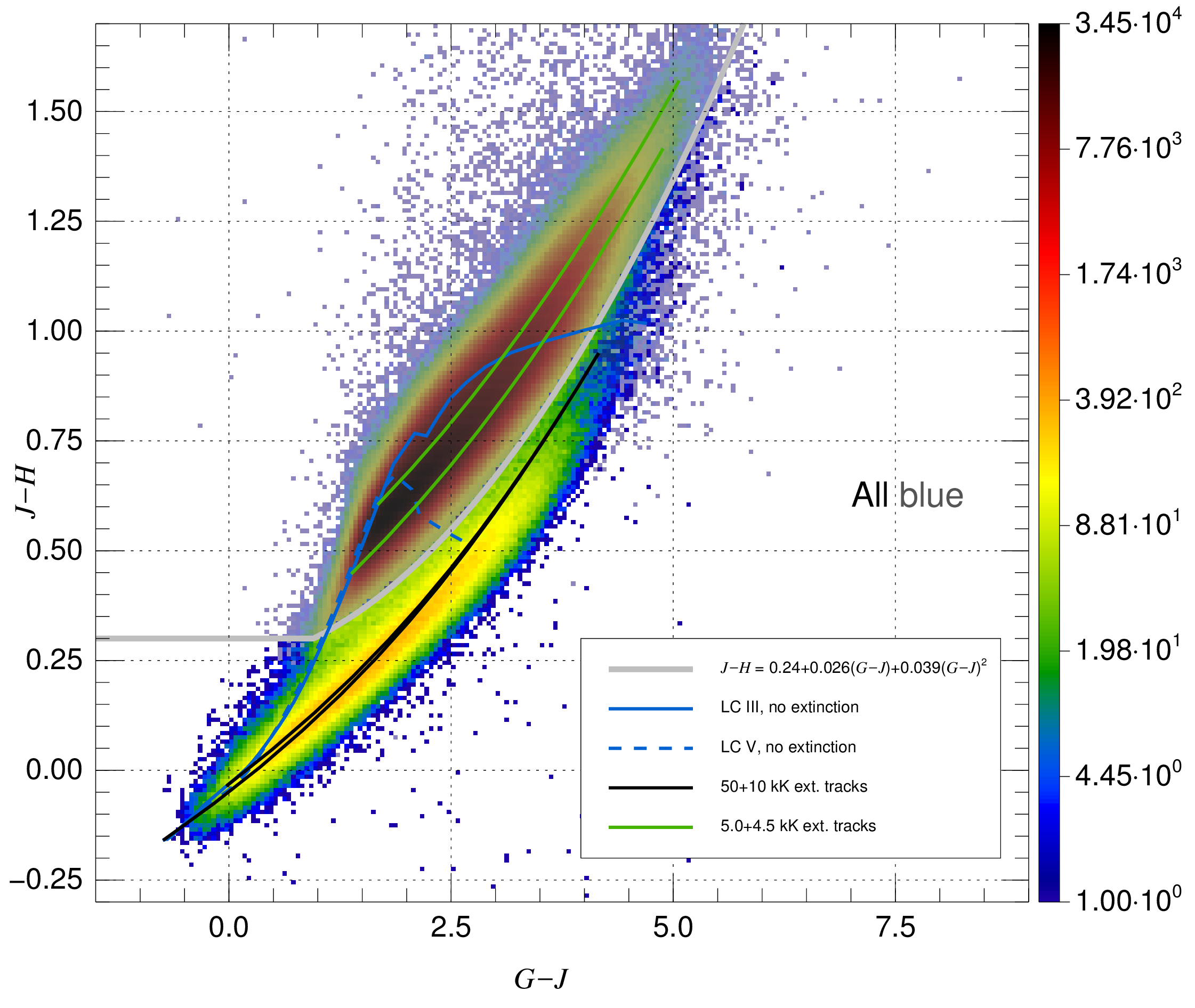}
  \\           
  \includegraphics*[width=0.49\linewidth]{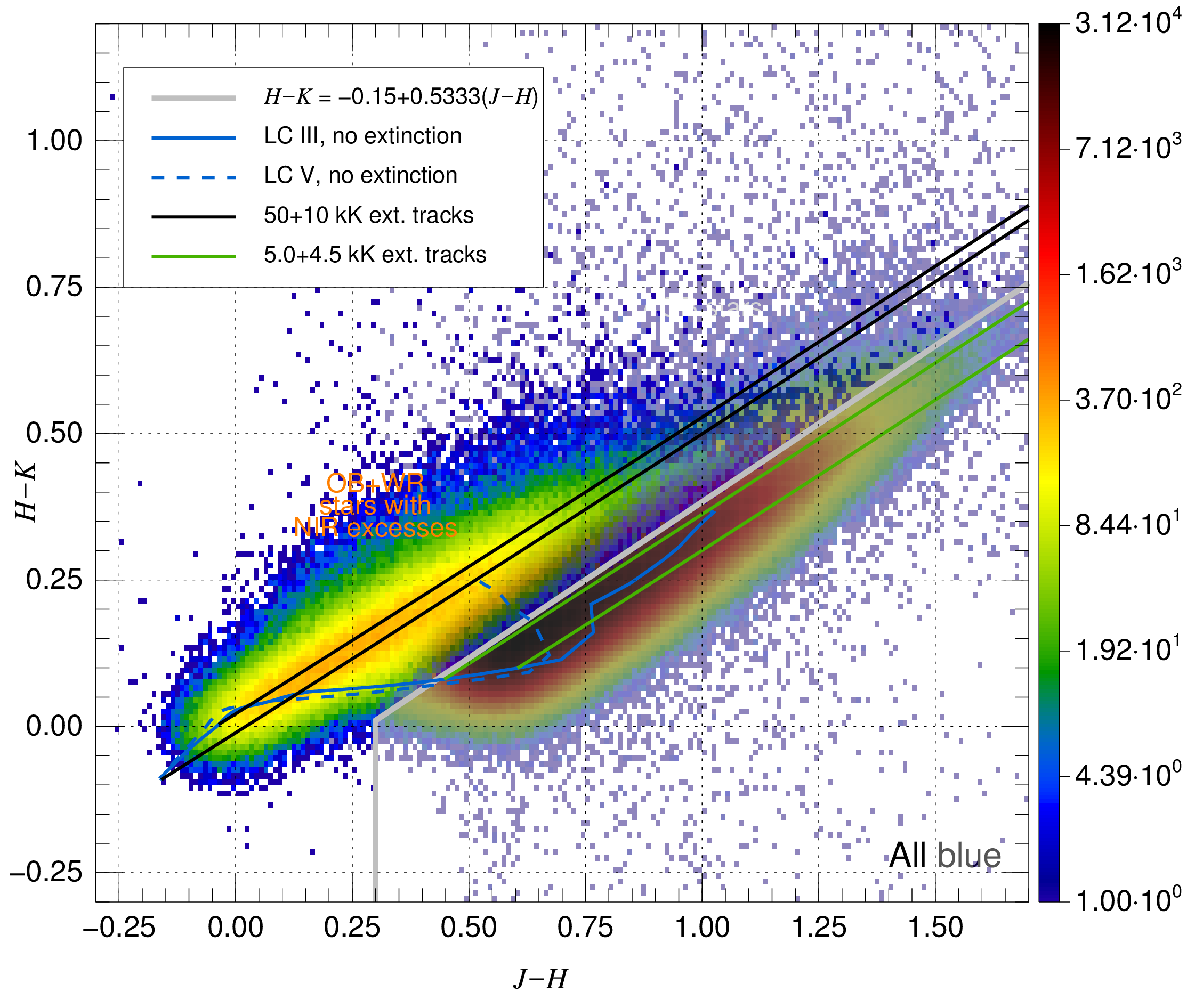} \
  \includegraphics*[width=0.49\linewidth]{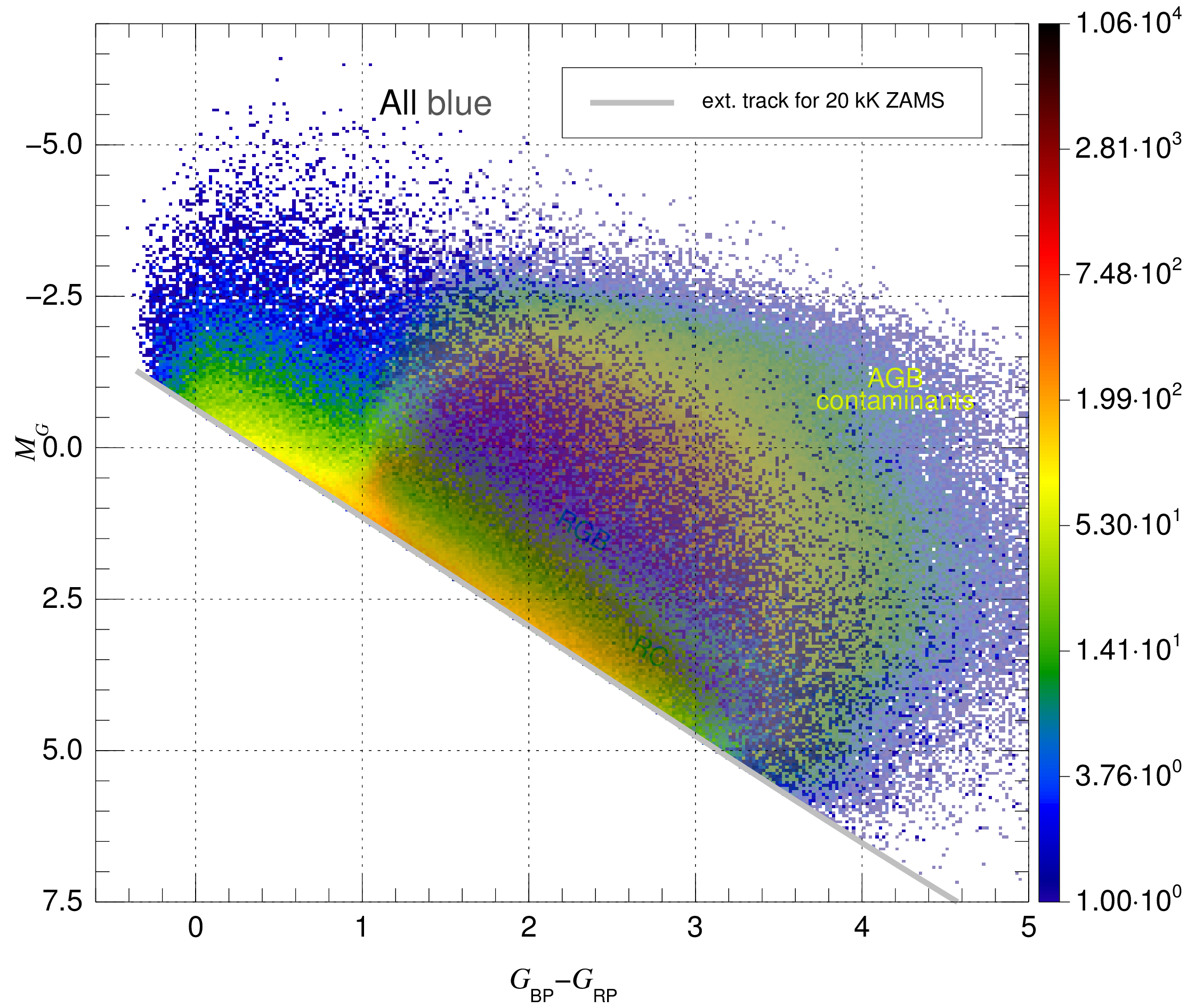}  
  \end{minipage}
  \newframe
  \begin{minipage}{\linewidth}
  \includegraphics*[width=0.49\linewidth]{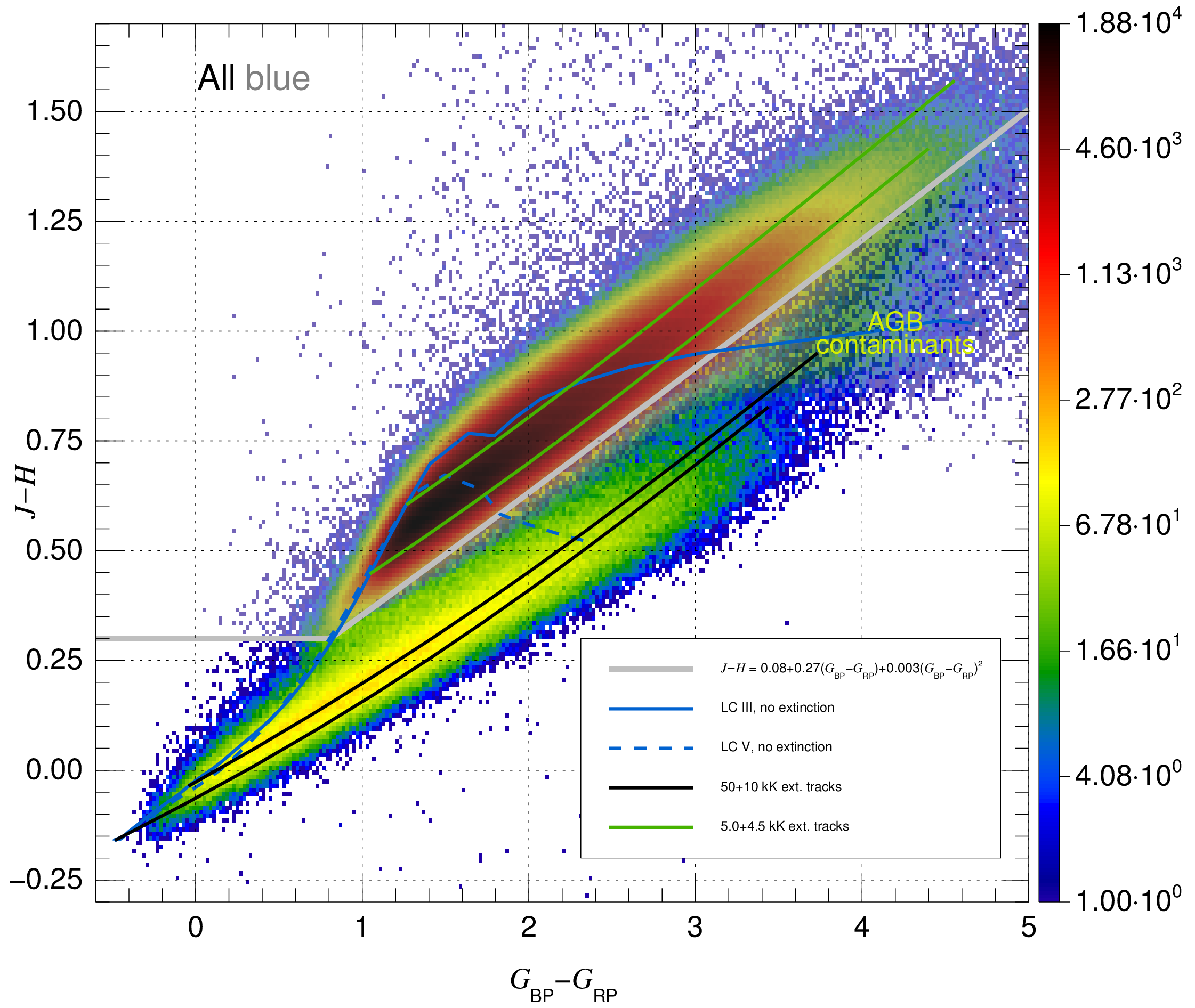} \
  \includegraphics*[width=0.49\linewidth]{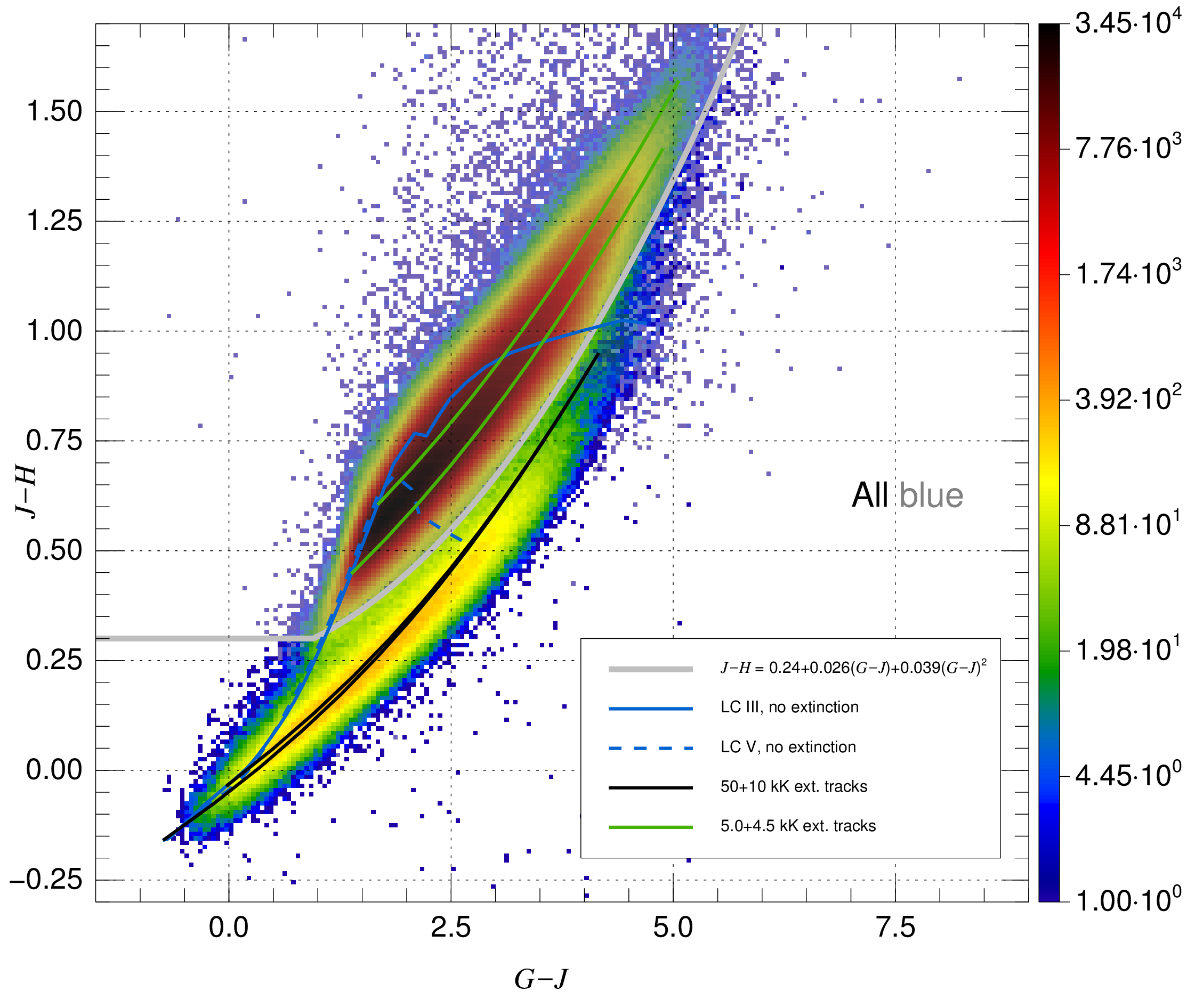}
  \\           
  \includegraphics*[width=0.49\linewidth]{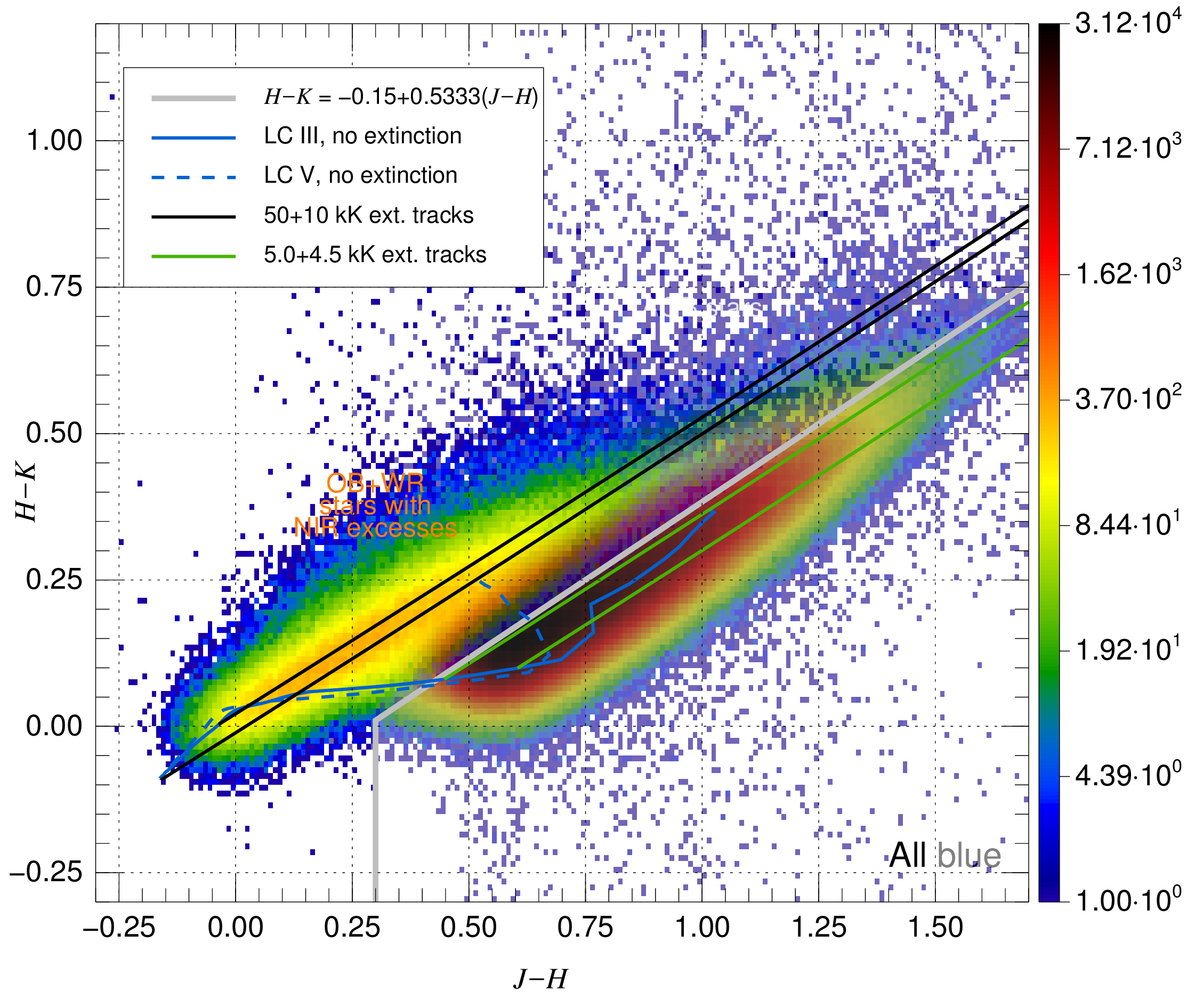} \
  \includegraphics*[width=0.49\linewidth]{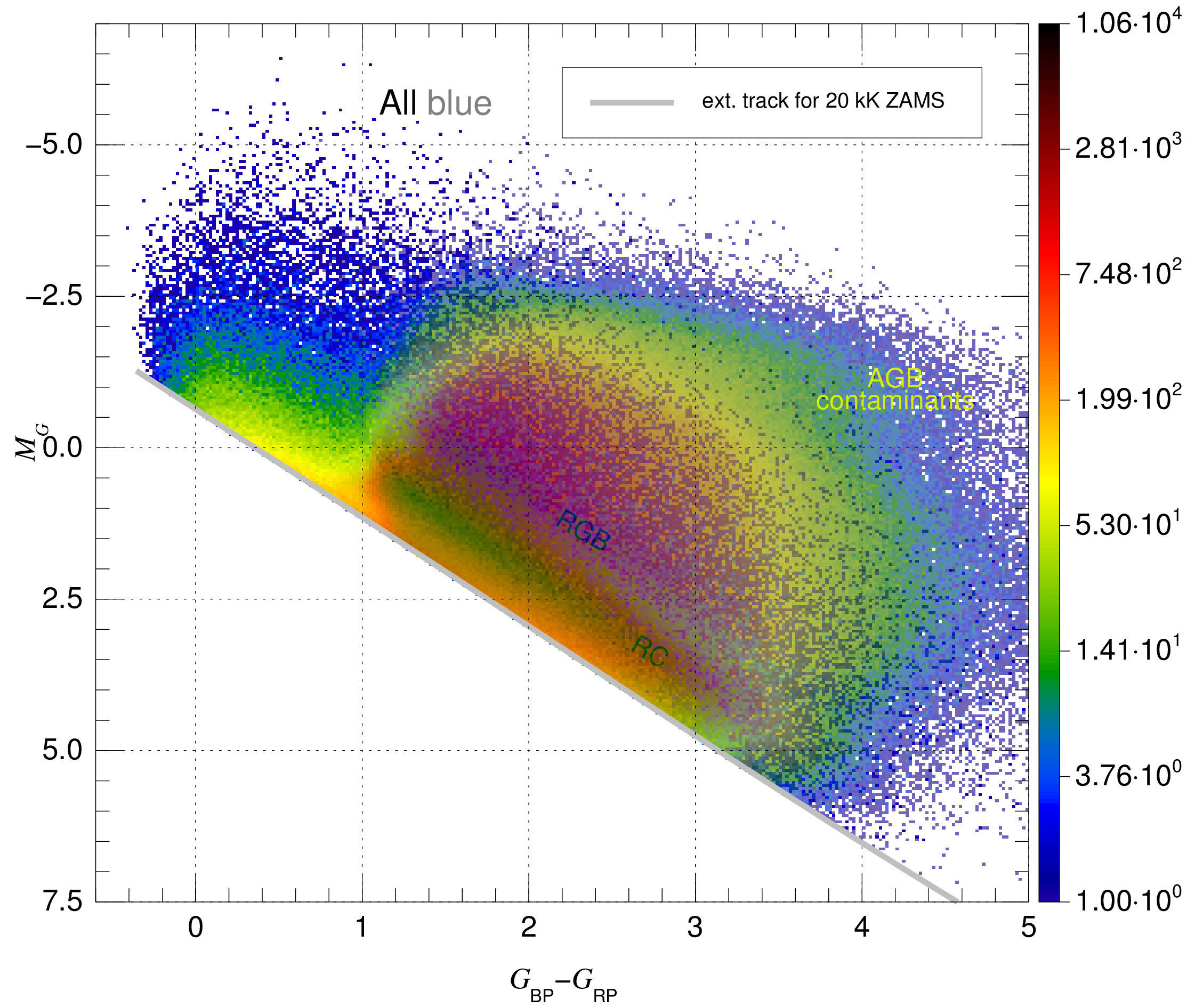}  
  \end{minipage}
  \newframe
  \begin{minipage}{\linewidth}
  \includegraphics*[width=0.49\linewidth]{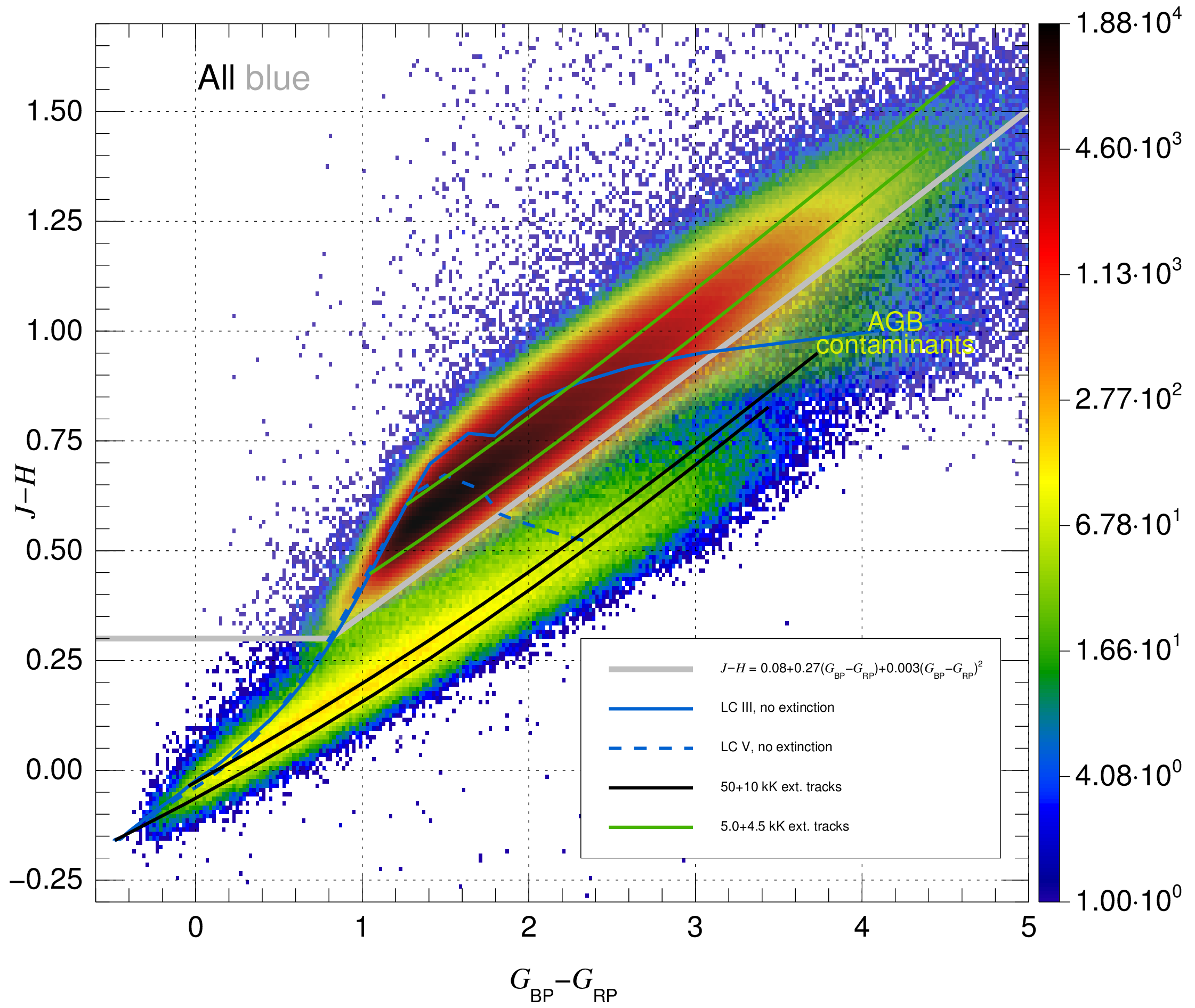} \
  \includegraphics*[width=0.49\linewidth]{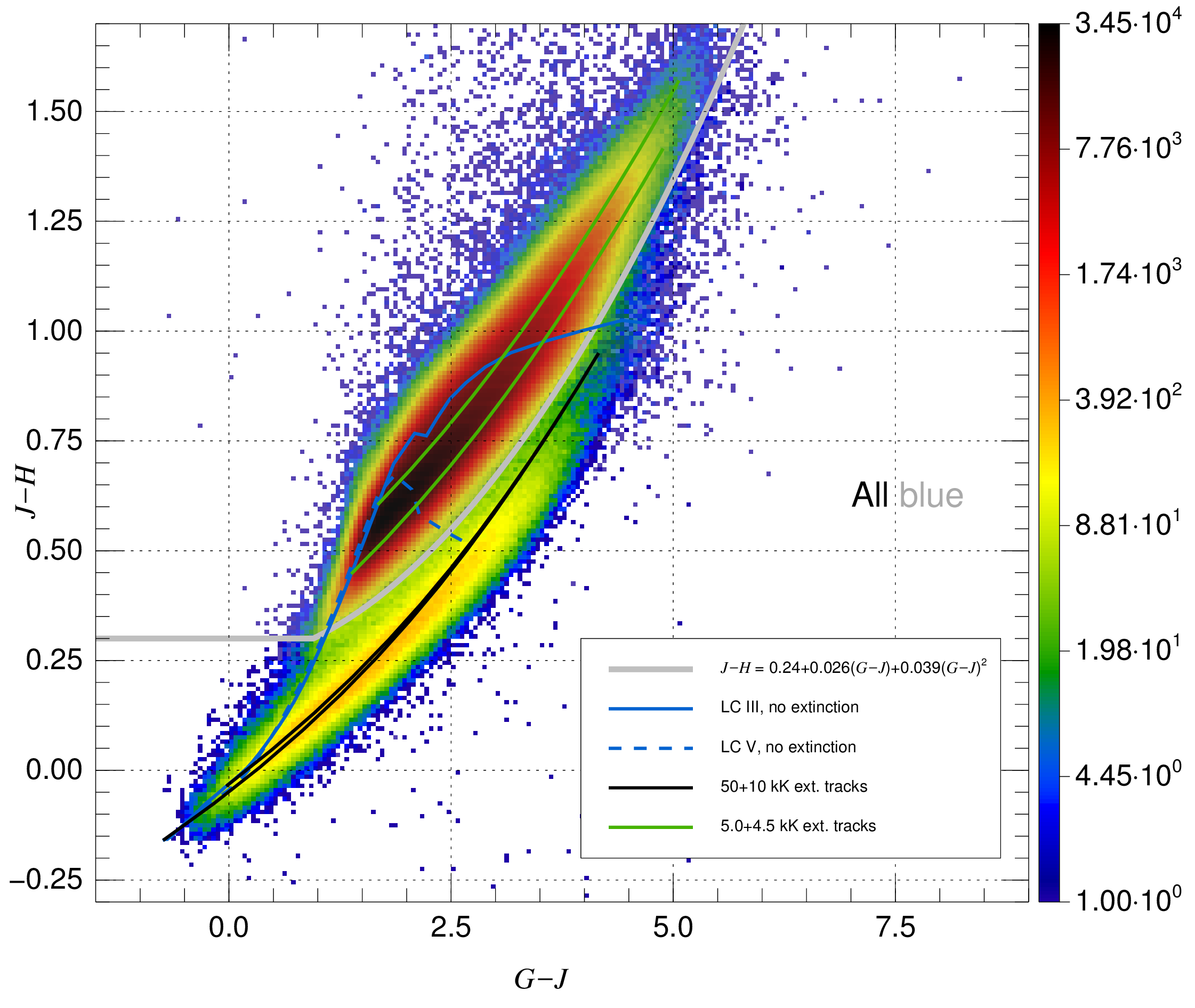}
  \\           
  \includegraphics*[width=0.49\linewidth]{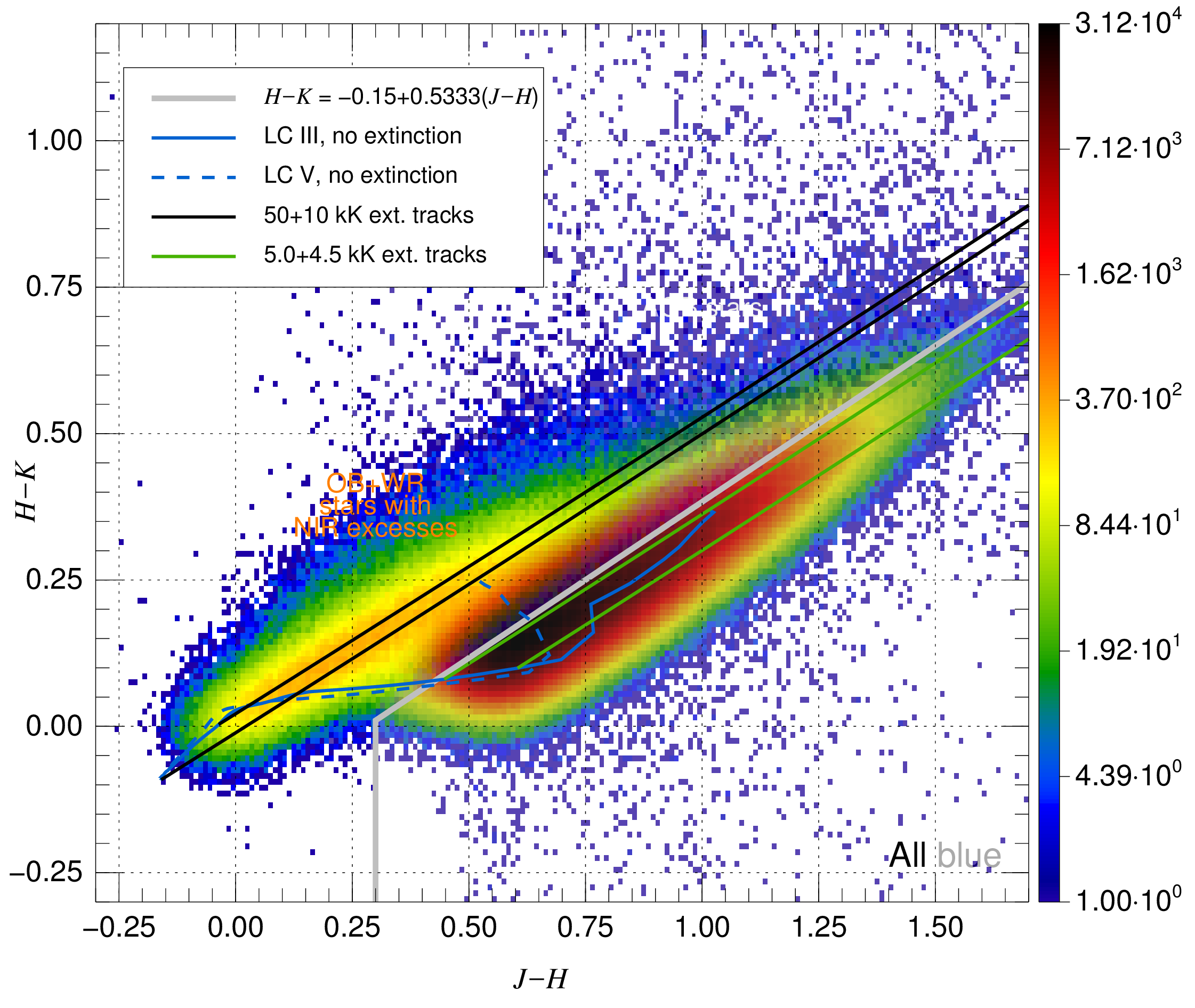} \
  \includegraphics*[width=0.49\linewidth]{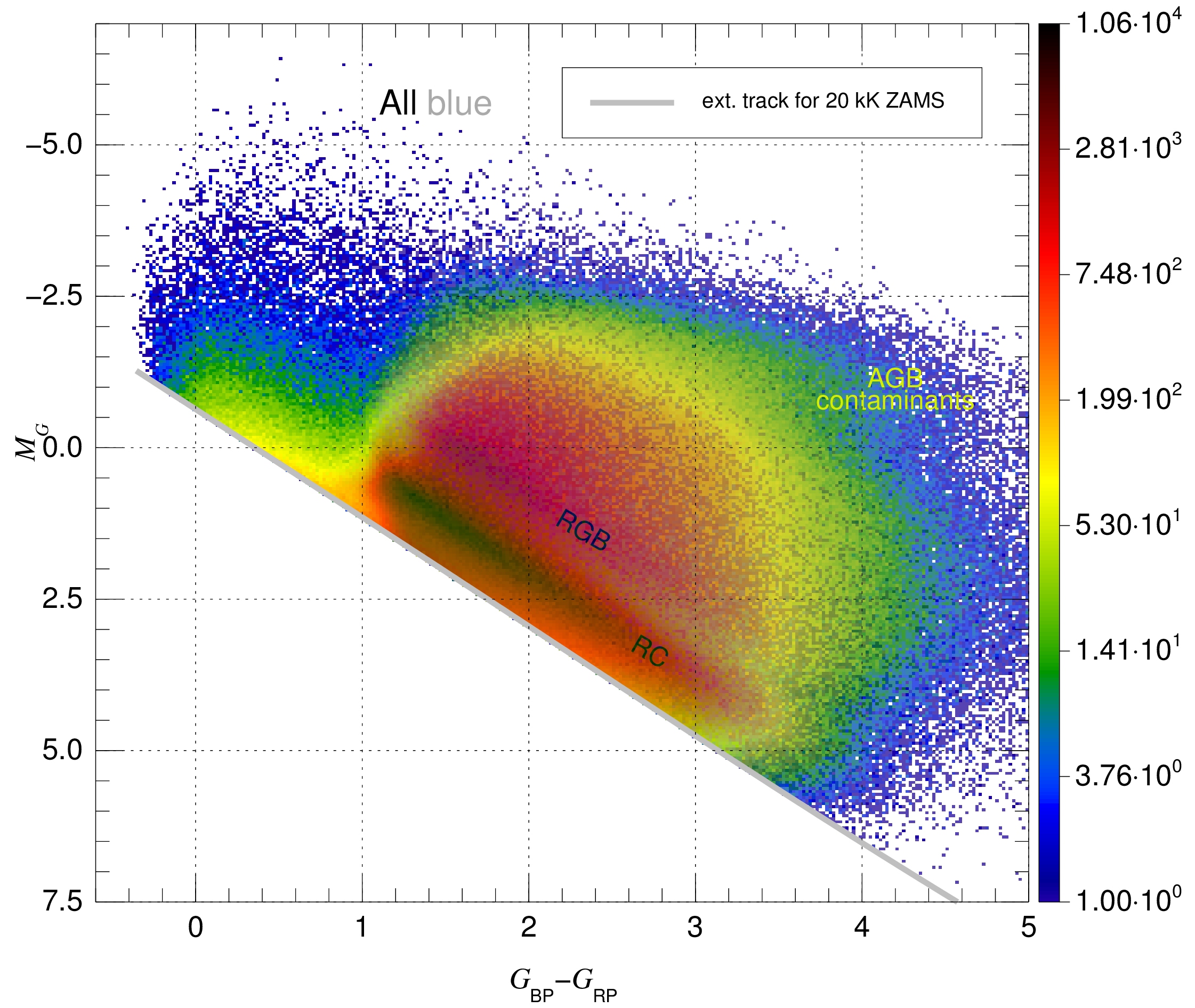}  
  \end{minipage}
  \newframe
  \begin{minipage}{\linewidth}
  \includegraphics*[width=0.49\linewidth]{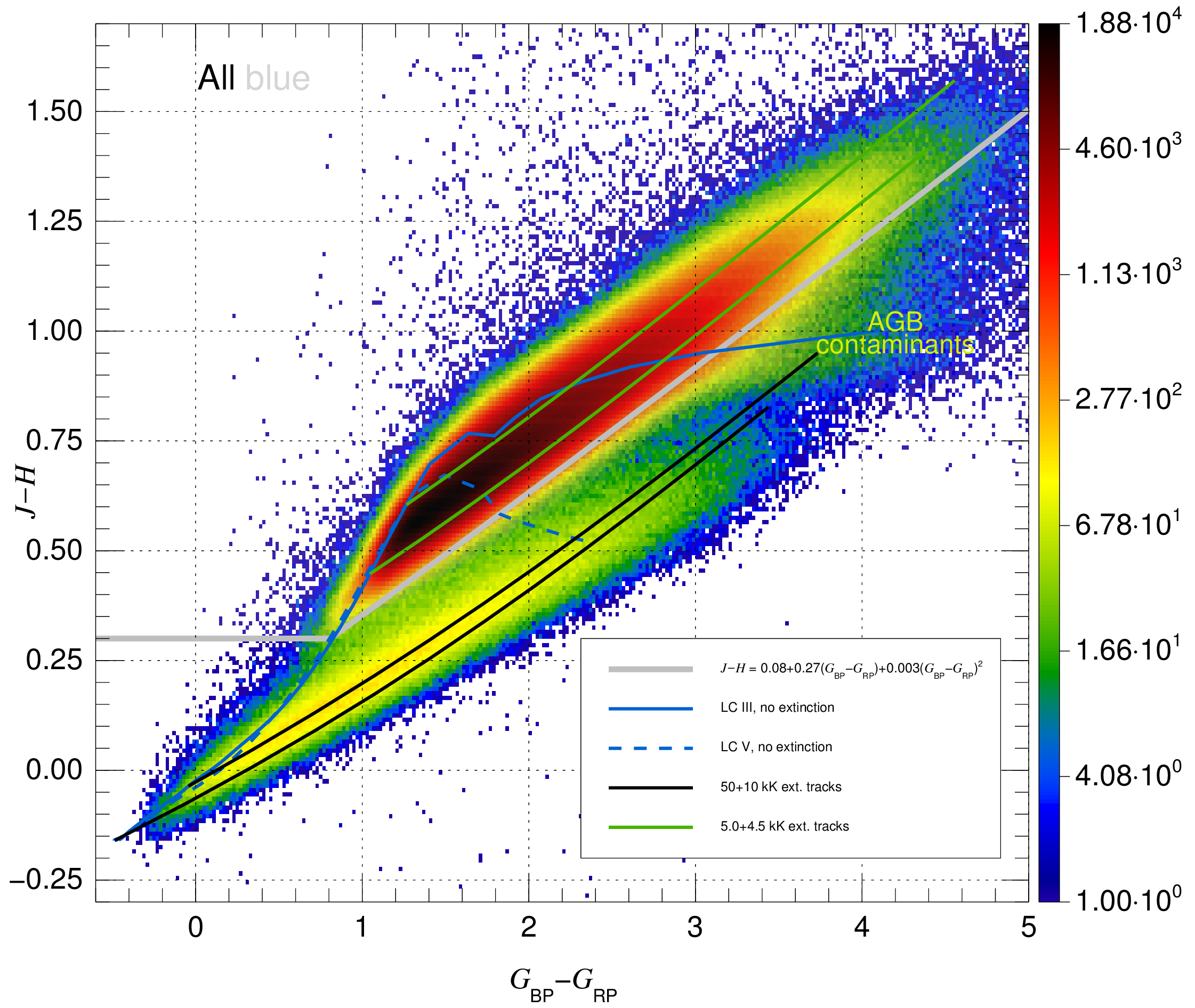} \
  \includegraphics*[width=0.49\linewidth]{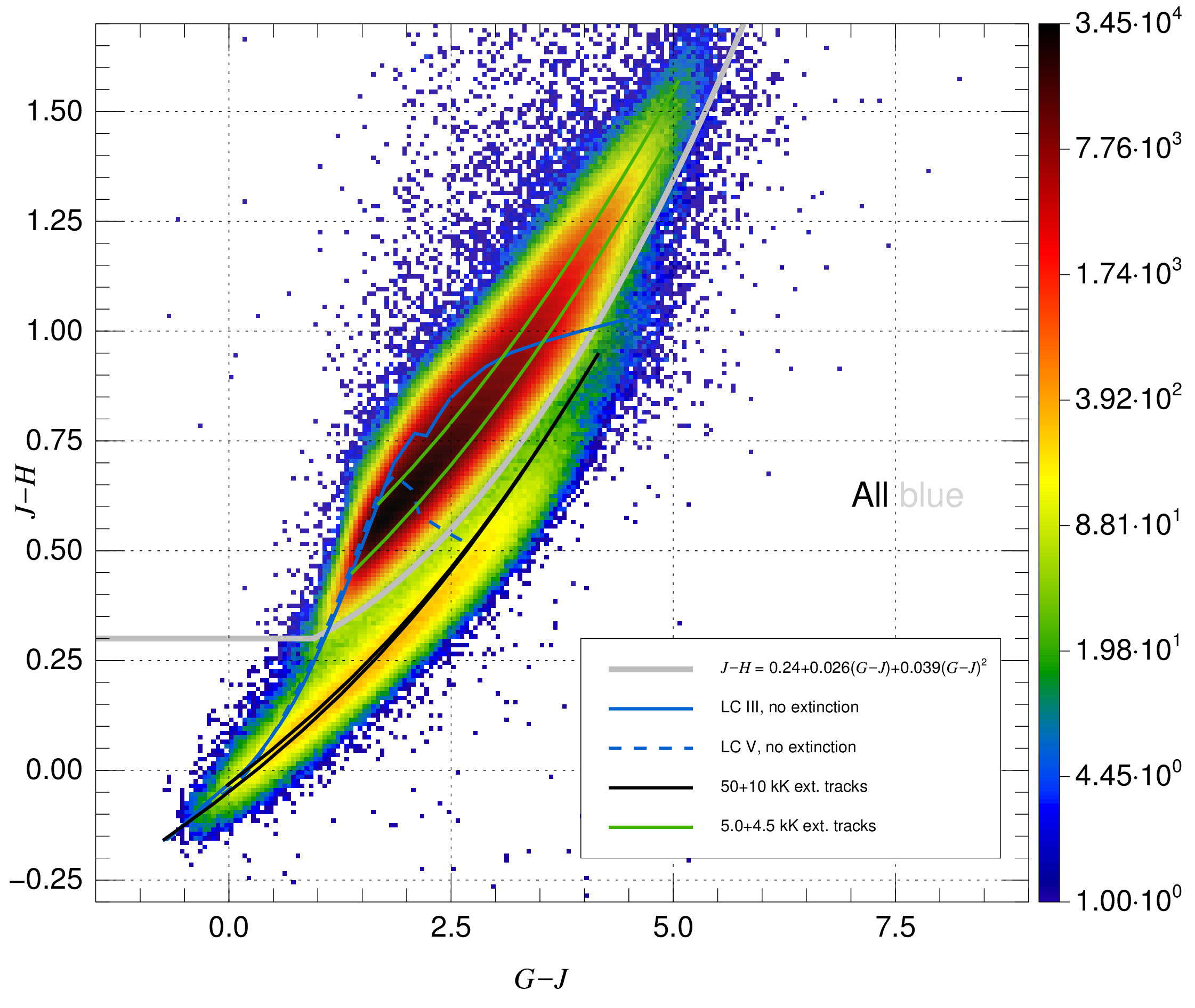}
  \\           
  \includegraphics*[width=0.49\linewidth]{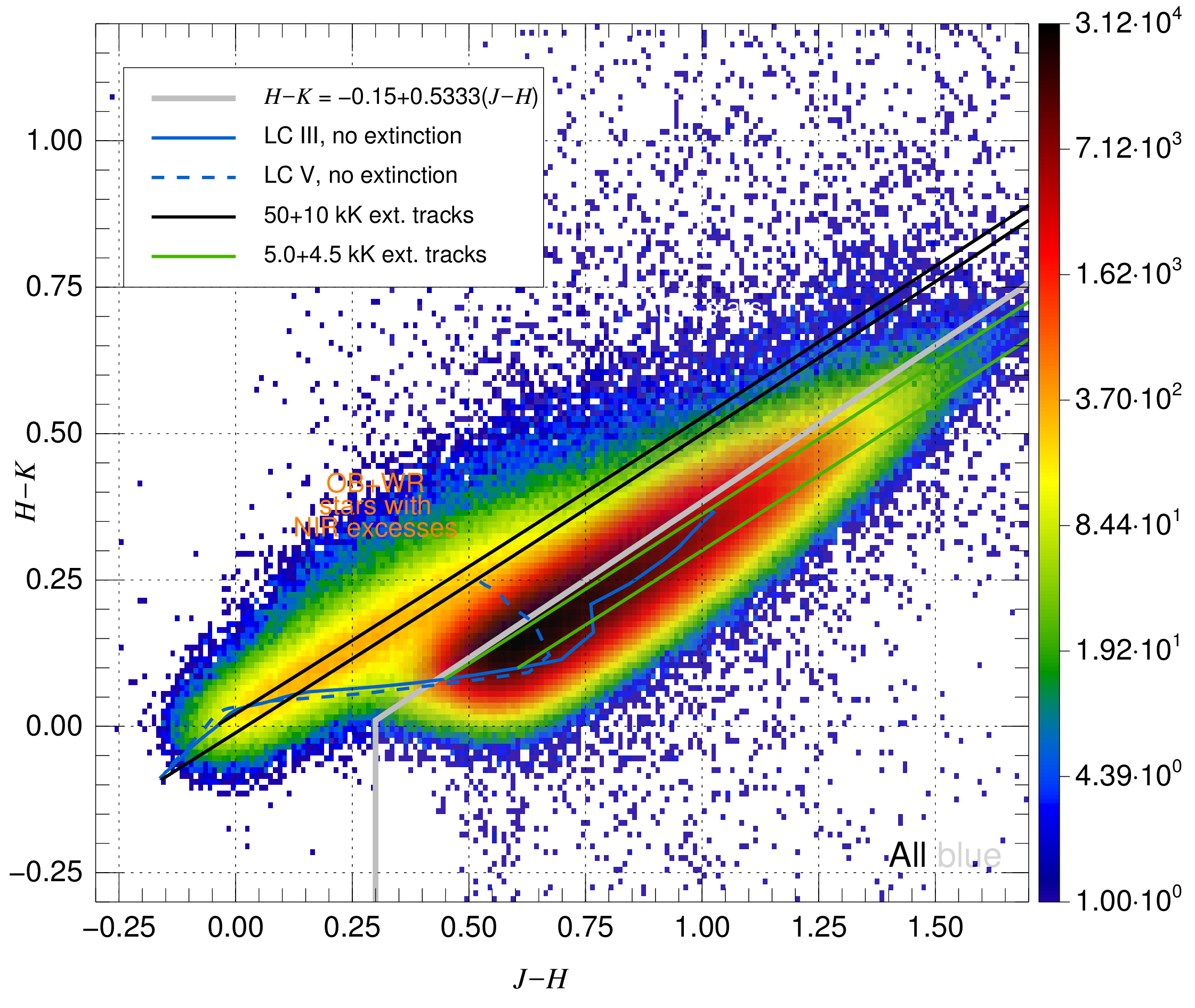} \
  \includegraphics*[width=0.49\linewidth]{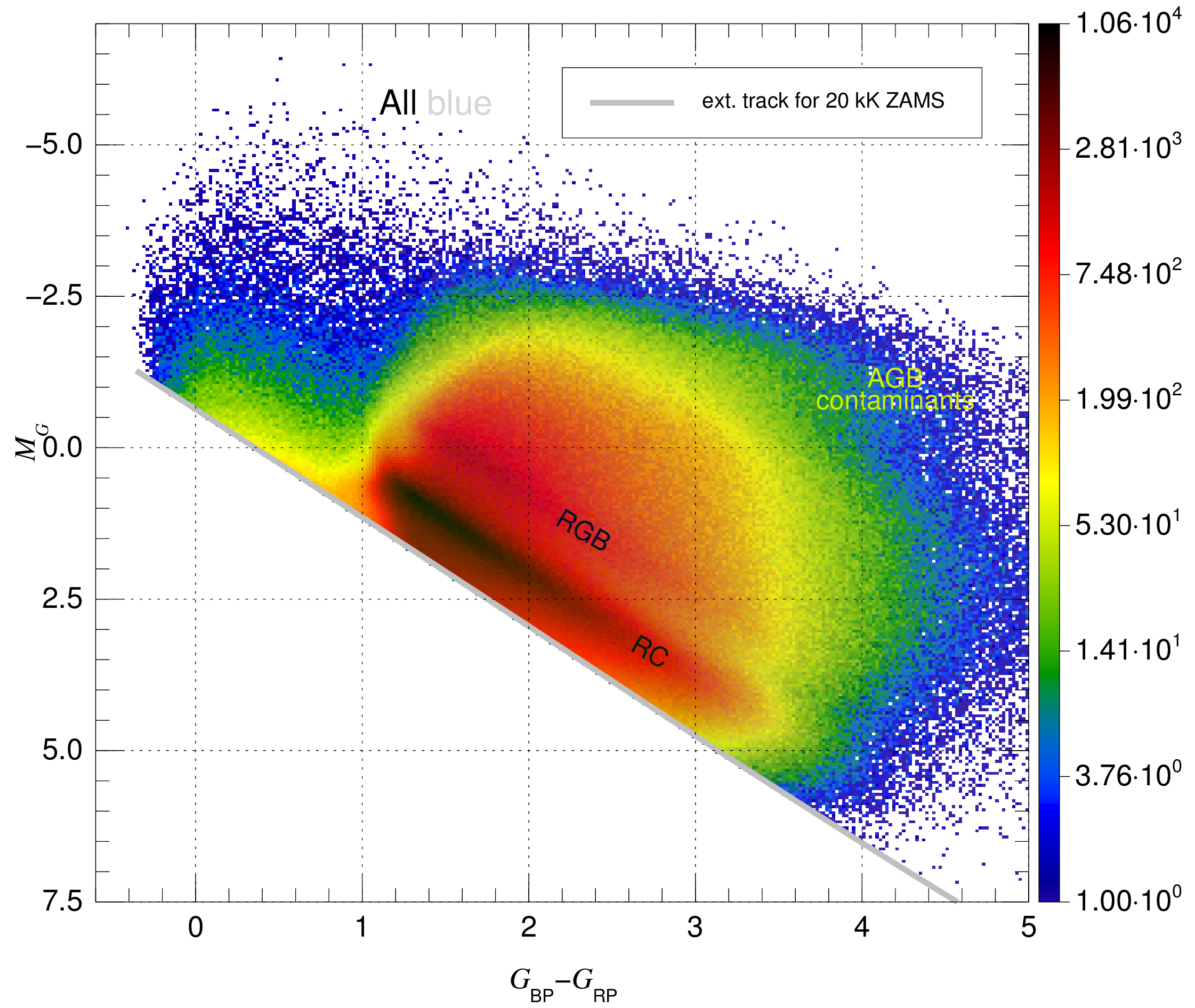}  
  \end{minipage}
  \end{animateinline}
             }
 \caption{Colour-colour selection diagrams and CAMD output for the luminous sample,
          to be compared with the equivalent Fig.~\ref{fig1} for the full sample. On an {\cre Adobe-compatible viewer} the buttons at the
          bottom activate the animation and allow the reader to cycle between the full luminous, cool (LR/red), and hot (OB/blue) samples.}
 \label{fig2}   
\end{figure}


\section{Analysing the resulting hot and cool samples}      

$\,\!$\indent The application of the three criteria results in a hot sample of \num{23254785} stars and a complementary cool
one (that does not satisfy one or more criteria) of \num{62150112} stars. The animation in Fig.~1 cycles through the full, cool/red
 and hot/blue samples (with red and blue meant as intrinsic, not extinguished, colours).

The cool-sample CAMD reveals: 

\begin{enumerate}
 \item A low-extinction MS that starts around $\BPRP\sim 0.7$~mag.
 \item A reddened population of red luminous stars (RGB, RC, and AGB) elongated along the extinction tracks.
 \item A variety of other red populations such as PMS stars and CVs.
\end{enumerate}

On the other hand, the hot-sample CAMD can be divided in four groups:

\begin{enumerate}
 \item The largest group corresponds to the hot MS (extended to evolved OB stars) reddened along the extinction track, with more
       luminous stars reaching higher extinctions due to the magnitude limitation.
 \item Two small groups of low-extinction WDs and hot subdwarfs.
 \item A significantly large group of M dwarfs.
 \item A small group of AGB contaminants in the upper right quadrant.
\end{enumerate}

The two first groups are the expected result of the process while the last two are false positives or contaminants. 
The explanation for the presence of the latter lies in the three colour-colour diagrams in Fig.~1, more conspicuously in
the upper right one ($\BPRP$ vs. $J-H$). The extreme of the no-extinction tracks for luminosity class V (M dwarfs) and luminosity
class III (coolest or extreme AGB stars) move away from the locus where most cool stars are and towards higher values of \BPRP\ 
($G-J$ or $J-K$ in the other two diagrams) while maintaining a near-constant value of $J-H \sim 0.55$~mag for M dwarfs and 
$J-H \sim 0.95$~mag for extreme AGB stars. This makes them intercept the tracks of extinguished OB stars, making their selected
colours indistinguishable. In other words, it is not possible to use \textit{Gaia}+2MASS photometry alone to separate those 
populations and one needs to resort to other methods.

For the case of M dwarfs, if one has accurate parallaxes the separation is relatively easy, at least in a sample
limited in magnitude to $\GG = 17$. Just using the CAMD in Fig.~\ref{fig1} it is clear that there is a division between M dwarfs
and extinguished hotter ones (likely of F spectral type) around $M_G = 8$. In addition, extinguished F dwarfs should lie close to
the $J-H =  0.08 + 0.27(\BPRP) + 0.003(\BPRP)^2$ line in the top left panel while M dwarfs can extend to significantly larger
values of \BPRP\ (which could be used as an additional criterion for fainter samples).


\section{The luminous stars and the OB sample}

$\,\!$\indent Most of the \num{85404897} stars of the full sample are of low luminosity, as in the battle for selection between 
the IMF+MS lifetime advantage of non-evolved low/intermediate-mass stars and the longer distance reach of the high-mass and evolved
low/intermediate-mass stars, the first advantage dominates. As a result, if we use as a luminosity-separation criterion that includes
the extinction effect the 20 kK ZAMS extinction track with $\RV = 3.0$ \citep{Maizetal14a} (plotted in the CAMD of Fig.~\ref{fig1}),
we find that \num{10618668} (12.4\%) lie above it, defining our luminous sample. Of those, \num{10422928} are cool stars and 
\num{195740} are hot stars according to our previous criteria.

As we mentioned before, the hot star sample is contaminated by M dwarfs and AGN stars. The former are not an issue for luminous stars
but the latter are. To eliminate the AGB (and other possible late-type) 
contaminants we use two criteria.  First, we eliminate the \num{5668} stars marked as being of variable type LPV, CEP, RR, RCB, or
SYST in \citet{Rimoetal23}. Second for the stars in the upper right corner of the CAMD [defined as $\BPRP \ge 3$~mag and 
$M_G \le -0.5+1.6(\BPRP-3)$~mag] we eliminate the 187 stars with $s_{G_{\rm BP}} > 70$~mmag in \citet{Maizetal23}. This leaves us
with a sample of \num{189885} massive early-type (mostly OB) candidates and \num{10428783} luminous red (LR) stars. Therefore, the
OB stars are just 1.8\% of the luminous stars in our sample, explaining the often used comparison of needles in a haystack used for
the search of massive stars in the Milky Way. The three luminous samples are shown in Fig.~\ref{fig2}. In the animation
(the pdf requires an Adobe-compatible viewer) frame that corresponds to the OB sample, no jump is seen around $\BPRP\sim 1.0$, where
RC stars are found, a sign of the effectiveness of the cleaning process.

The OB-candidate sample here is the largest Galactic sample of that type ever built by an order of magnitude if one uses the 
classical definition that excludes mid/late B dwarfs and late B giants because they are of intermediate mass (see the comparison 
between surveys in ALS~III). Given the extinction track criterion used, it should not be a completely clean sample, as it should be 
contaminated (mostly by non-ZAMS mid/late B dwarfs and late B giants). However, even if the contamination is of the order of 50\%, the
cleaned sample would still be much larger than any other equivalent sample ever assembled.

\begin{figure}
 \centerline{
 \includegraphics*[width=\linewidth]{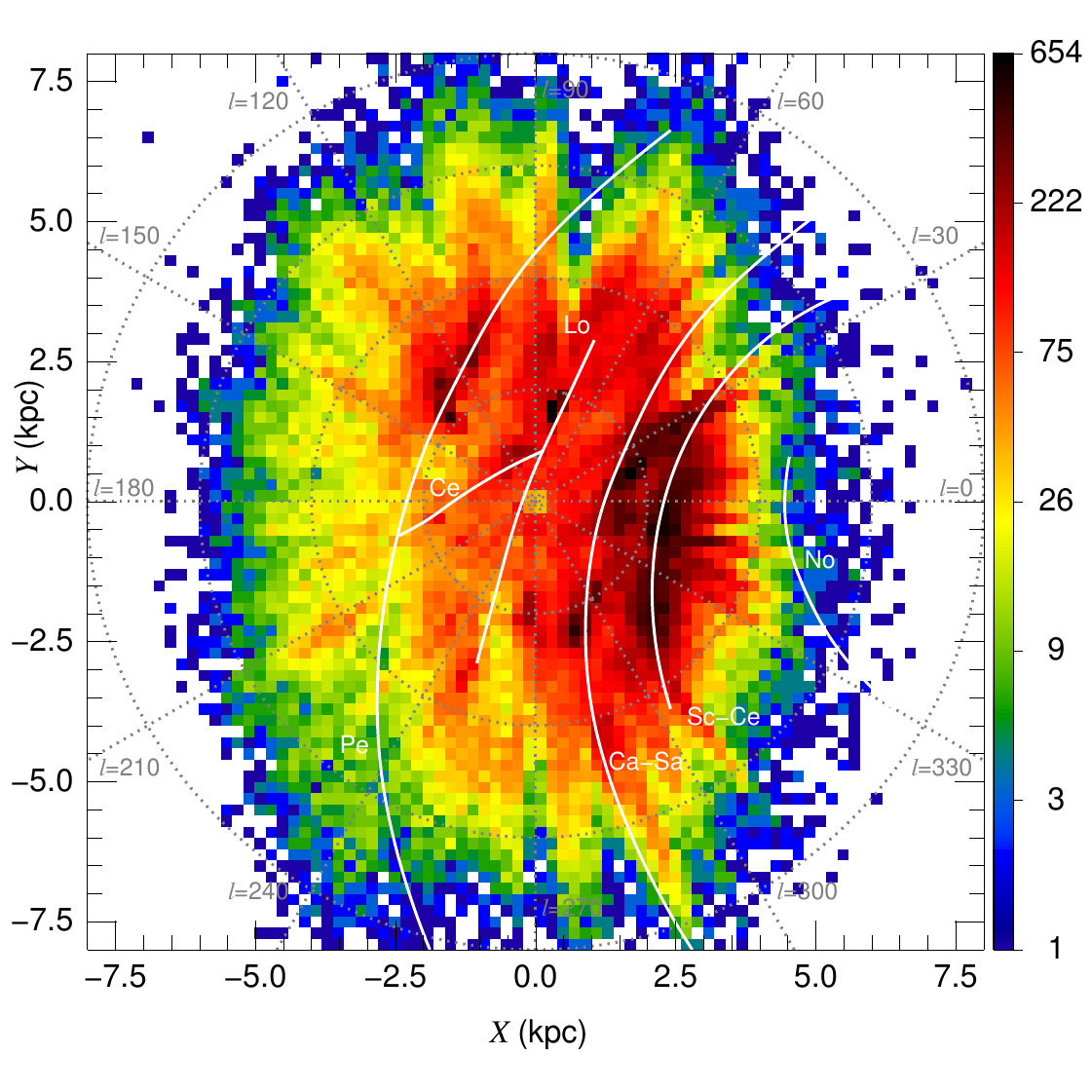}
 \vspace{-5mm}
            }
 \caption{Spatial distribution of the final massive early-type sample projected onto the Galactic Plane. The marked structures 
          correspond to the \textbf{Pe}rseus, \textbf{Lo}cal, \textbf{Ca}rina\textbf{-Sa}gittarius, 
          \textbf{Sc}utum\textbf{-Ce}ntaurus, and \textbf{No}rma spiral arms and the \textbf{Ce}pheus spur.}
 \label{fig3}   
\end{figure}

\begin{figure}
 \centerline{
 \includegraphics*[width=\linewidth]{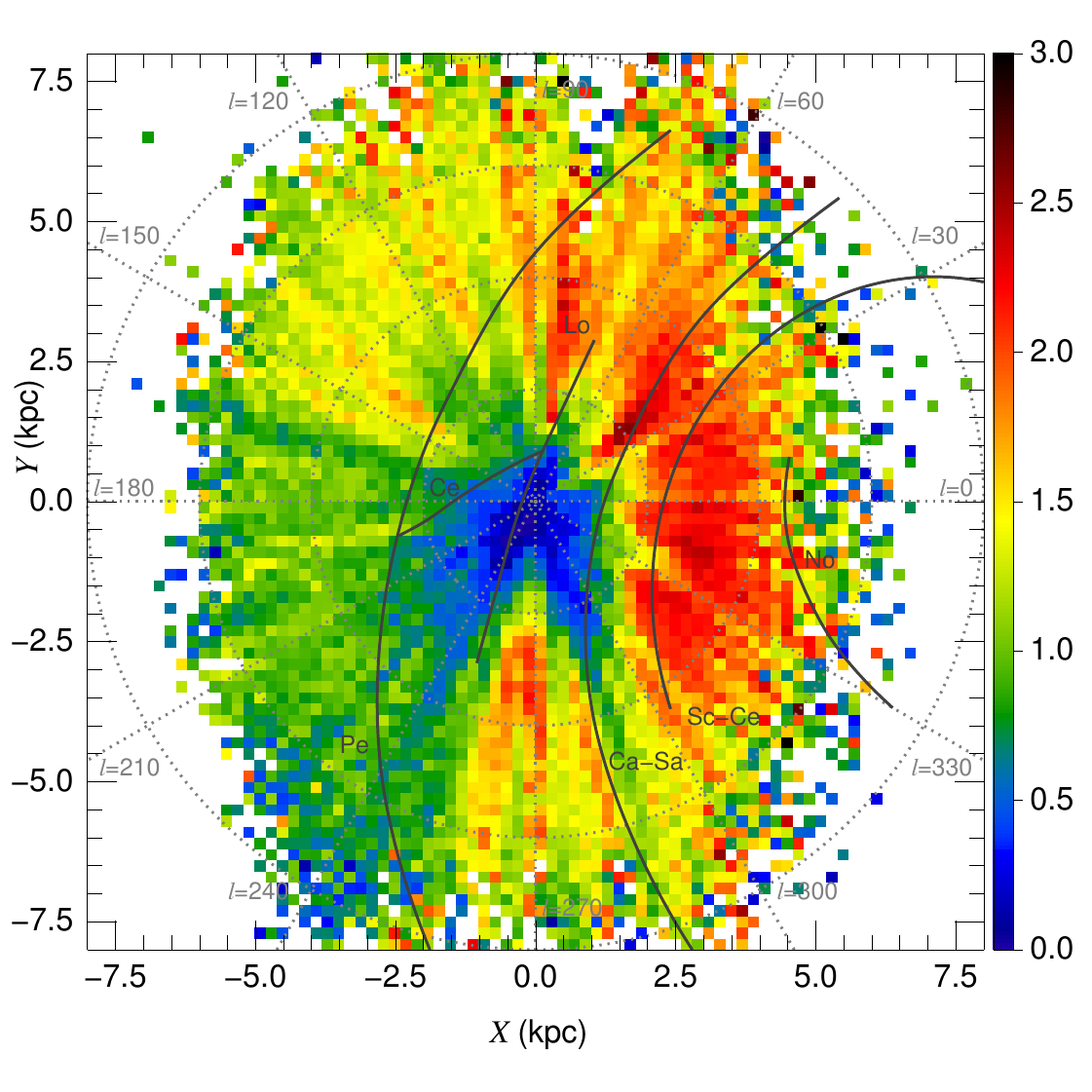}
 \vspace{-5mm}
            }
 \caption{Mean \BPRP\ (a proxy for reddening for OB stars) of the final massive early-type sample projected onto the Galactic Plane. 
          The marked structures are the same as for Fig.~\ref{fig3}.}
 \label{fig4}   
\end{figure}

Figures~\ref{fig3}~and~\ref{fig4} show the spatial distribution and average \BPRP\ of the OB sample. In the spatial distribution, 
four spiral arms (Perseus, Local, Carina-Sagittarius, and Scutum-Centaurus) and the Cepheus spur can be seen. Some stars are located 
at the position of the Norma arm but not enough to make it visible. The average \BPRP, as a proxy for reddening, shows the local 
minimum and windows, the Galactic radial extinction gradient, and some prominent extinction shadows, such as the ones associated 
with the Cygnus-X and the Carina Nebula massive-star-forming complexes.

We plan to use our technique with \textit{Gaia}~DR4 to create a significantly larger OB sample. This would be possible because the
improvement in parallax errors (random and systematic) should allow us to reach fainter magnitudes with $\pic/\spic \ge 5.0$. In that
way, it may be possible to approach a sample of $\sim 10^6$ OB candidates, of which half may be real massive stars. Stay tuned!


\section{Conclusions}

\begin{itemize}
 \item We have been able to obtain a sample of \num{23254785} Galactic hot stars wth $G > 17$~mag based on \textit{Gaia}+2MASS
       photometry alone.
 \item With the addition of variabilty information we have selected a subsample of \num{189885} massive early-type candidates, 
       the largest Galactic sample of that type ever built.
\end{itemize}


\small{\bibliographystyle{aa} 
\bibliography{general}\vspace{0.75in}} 

\end{document}

%% file: mlaeff.tex
\def\hoy{\number\day \space de \space\ifcase\month\or
 Enero\or Febrero\or Marzo\or Abril\or Mayo\or Junio\or
 Julio\or Agosto\or Septiembre\or Octubre\or Noviembre\or Diciembre\fi
 \space de \number\year}
\def\ii/{\'{\i}}
\def\cion/{ci\'on}
\def\cao/{\c c\~ao}
\def\utw{\smash{\rlap{\lower5pt\hbox{$\sim$}}}}
\def\udtw{\smash{\rlap{\lower6pt\hbox{$\approx$}}}}

\def\tens#1{\ifmmode\mathchoice{\mbox{$\sf\displaystyle#1$}}
{\mbox{$\sf\textstyle#1$}}
{\mbox{$\sf\scriptstyle#1$}}
{\mbox{$\sf\scriptscriptstyle#1$}}\else
\hbox{$\sf\textstyle#1$}\fi}
\def\vec#1{\ifmmode\mathchoice{\mbox{\boldmath$\displaystyle#1$}}
{\mbox{\boldmath$\textstyle#1$}}
{\mbox{\boldmath$\scriptstyle#1$}}
{\mbox{\boldmath$\scriptscriptstyle#1$}}\else
\hbox{\boldmath$\textstyle#1$}\fi}
\def\bbbc{{\mathchoice {\setbox0=\hbox{$\displaystyle\rm C$}\hbox{\hbox
to0pt{\kern0.4\wd0\vrule height0.9\ht0\hss}\box0}}
{\setbox0=\hbox{$\textstyle\rm C$}\hbox{\hbox
to0pt{\kern0.4\wd0\vrule height0.9\ht0\hss}\box0}}
{\setbox0=\hbox{$\scriptstyle\rm C$}\hbox{\hbox
to0pt{\kern0.4\wd0\vrule height0.9\ht0\hss}\box0}}
{\setbox0=\hbox{$\scriptscriptstyle\rm C$}\hbox{\hbox
to0pt{\kern0.4\wd0\vrule height0.9\ht0\hss}\box0}}}}
\def\bbbq{{\mathchoice {\setbox0=\hbox{$\displaystyle\rm
Q$}\hbox{\raise
0.15\ht0\hbox to0pt{\kern0.4\wd0\vrule height0.8\ht0\hss}\box0}}
{\setbox0=\hbox{$\textstyle\rm Q$}\hbox{\raise
0.15\ht0\hbox to0pt{\kern0.4\wd0\vrule height0.8\ht0\hss}\box0}}
{\setbox0=\hbox{$\scriptstyle\rm Q$}\hbox{\raise
0.15\ht0\hbox to0pt{\kern0.4\wd0\vrule height0.7\ht0\hss}\box0}}
{\setbox0=\hbox{$\scriptscriptstyle\rm Q$}\hbox{\raise
0.15\ht0\hbox to0pt{\kern0.4\wd0\vrule height0.7\ht0\hss}\box0}}}}
\def\bbbt{{\mathchoice {\setbox0=\hbox{$\displaystyle\rm
T$}\hbox{\hbox to0pt{\kern0.3\wd0\vrule height0.9\ht0\hss}\box0}}
{\setbox0=\hbox{$\textstyle\rm T$}\hbox{\hbox
to0pt{\kern0.3\wd0\vrule height0.9\ht0\hss}\box0}}
{\setbox0=\hbox{$\scriptstyle\rm T$}\hbox{\hbox
to0pt{\kern0.3\wd0\vrule height0.9\ht0\hss}\box0}}
{\setbox0=\hbox{$\scriptscriptstyle\rm T$}\hbox{\hbox
to0pt{\kern0.3\wd0\vrule height0.9\ht0\hss}\box0}}}}
\def\bbbs{{\mathchoice
{\setbox0=\hbox{$\displaystyle     \rm S$}\hbox{\raise0.5\ht0\hbox
to0pt{\kern0.35\wd0\vrule height0.45\ht0\hss}\hbox
to0pt{\kern0.55\wd0\vrule height0.5\ht0\hss}\box0}}
{\setbox0=\hbox{$\textstyle        \rm S$}\hbox{\raise0.5\ht0\hbox
to0pt{\kern0.35\wd0\vrule height0.45\ht0\hss}\hbox
to0pt{\kern0.55\wd0\vrule height0.5\ht0\hss}\box0}}
{\setbox0=\hbox{$\scriptstyle      \rm S$}\hbox{\raise0.5\ht0\hbox
to0pt{\kern0.35\wd0\vrule height0.45\ht0\hss}\raise0.05\ht0\hbox
to0pt{\kern0.5\wd0\vrule height0.45\ht0\hss}\box0}}
{\setbox0=\hbox{$\scriptscriptstyle\rm S$}\hbox{\raise0.5\ht0\hbox
to0pt{\kern0.4\wd0\vrule height0.45\ht0\hss}\raise0.05\ht0\hbox
to0pt{\kern0.55\wd0\vrule height0.45\ht0\hss}\box0}}}}
\def\bbbz{{\mathchoice {\hbox{$\sf\textstyle Z\kern-0.4em Z$}}
{\hbox{$\sf\textstyle Z\kern-0.4em Z$}}
{\hbox{$\sf\scriptstyle Z\kern-0.3em Z$}}
{\hbox{$\sf\scriptscriptstyle Z\kern-0.2em Z$}}}}
\def\diameter{{\ifmmode\mathchoice
{\ooalign{\hfil\hbox{$\displaystyle/$}\hfil\crcr
{\hbox{$\displaystyle\mathchar"20D$}}}}
{\ooalign{\hfil\hbox{$\textstyle/$}\hfil\crcr
{\hbox{$\textstyle\mathchar"20D$}}}}
{\ooalign{\hfil\hbox{$\scriptstyle/$}\hfil\crcr
{\hbox{$\scriptstyle\mathchar"20D$}}}}
{\ooalign{\hfil\hbox{$\scriptscriptstyle/$}\hfil\crcr
{\hbox{$\scriptscriptstyle\mathchar"20D$}}}}
\else{\ooalign{\hfil/\hfil\crcr\mathhexbox20D}}%
\fi}}
\def\sq{\ifmmode\squareforqed\else{\unskip\nobreak\hfil
\penalty50\hskip1em\null\nobreak\hfil\squareforqed
\parfillskip=0pt\finalhyphendemerits=0\endgraf}\fi}
\def\squareforqed{\hbox{\rlap{$\sqcap$}$\sqcup$}}